\documentclass[3p]{elsarticle}

\pdfoutput=1
\usepackage{natbib} 
\usepackage[utf8]{inputenc} 
\usepackage[T1]{fontenc}
\usepackage{hyperref}       
\usepackage{url}
\usepackage{amsmath}
\usepackage[english]{babel}
\usepackage{dcolumn}
\usepackage{bm}
\usepackage{booktabs}
\usepackage{verbatim}

\newcommand{\be}{\begin{equation}}
\newcommand{\ee}{\end{equation}}
\newcommand{\bea}{\begin{eqnarray}}
\newcommand{\eea}{\end{eqnarray}}
\newcommand{\non}{\nonumber}
\newcommand{\bie}{\begin{itemize}}
\newcommand{\ben}{\begin{enumerate}}
\newcommand{\ie}{\item}
\newcommand{\eie}{\end{itemize} }
\newcommand{\een}{\end{enumerate}}

\journal{Physics Reports}

\begin{document}

\begin{frontmatter}



%
\selectlanguage{english}
\title{%
Tetraquarks and pentaquarks in lattice QCD with light and heavy quarks}


\author[1]{Pedro Bicudo}
\ead{bicudo@tecnico.ulisboa.pt}
\address[1]{organization={CeFEMA, Departamento de Física, Instituto Superior Técnico, Universidade de Lisboa},
            addressline={Av. Rovisco Pais}, 
            city={Lisboa},
            postcode={1049-001}, 
            country={Portugal}}

\begin{abstract}
We review how lattice QCD can contribute to the prediction and the comprehension of tetraquarks, pentaquarks and related exotic hadrons such as hybrids, with at least one heavy quark. We include all families of exotic hadrons, except for the quarkless glueballs, and the hexaquarks which are related to nuclear physics.

Since the discovery of quarks and the development of the QCD theory,  there has been a large interest in exotic hadrons, initiated by the tetraquark models developed by Jaffe in 1977. Lattice QCD, being a first principle approach to solve non-perturbative QCD, has been crucial not only to compute precise results, but also to motivate and inspire research in hadronic physics, with particular interest in exotic hadrons.

In the new millennium, this interest exploded with several experimental discoveries of tetraquark and pentaquark resonances with heavy quarks, starting with the $Z_c$ and $Z_b$. So far, lattice QCD has not yet been able to comprehend this $Z$ class of tetraquarks, and is developing new methods to determine their masses, decay widths and decay processes. 

The interest in tetraquarks was also fuelled by the lattice QCD prediction of a second class of tetraquarks such as the $T_{bb}$, boundstates in the sense of having no strong decays. Very recently,  the $T_{cc}$ tetraquark first predicted with quark models in 1982 by Richard et al, was observed experimentally. We expect the lattice QCD community will be able to explore this $T$ class of tetraquarks in more detail and with very precise results.

We report on all the different direct and indirect approaches that lattice QCD, so far with most focus on tetraquarks, has been employing to study exotic hadrons with at least one heavy quark. We also briefly review the experimental progress in observing tetraquarks and pentaquarks, and the basic theoretical paradigms of tetraquarks, including three different types of mechanisms (diquark, molecular and s pole), comparing them with the results of lattice QCD.  We aim to show the journey of Lattice QCD  in the exploration of these fascinating and subtle hadrons.
\end{abstract}



\end{frontmatter}

\tableofcontents

\section{Introduction}\label{sec:intro}

\subsection{
The first proposals for exotic tetraquarks and pentaquarks}

After the quarks were proposed in the sixties by Gell-Mann \cite{Gell-Mann:1964ewy} and Zweig 
\cite{Zweig:1964ruk,Zweig:1964jf},
the Pandora box opened and the physics of quarks and gluons exploded in the seventies.  The charm quark was predicted in 1970 
\cite{Glashow:1970gm}
and observed in 1974
\cite{SLAC-SP-017:1974ind,E598:1974sol}. Quantum Chromodynamics  (QCD) was first shown to be renormalizable with Feynman diagrams 
\cite{tHooft:1972tcz}, 
and then assymtotic free in 1973
\cite{Gross:1973id,Politzer:1973fx}. 
Soon after, in 1974, a discrete and numerical version of QCD, lattice QCD was presented by Wilson
\cite{Wilson:1974sk}. 
Exotic four-quark hadrons, the tetraquarks, beyond the common two-quark mesons and three-quark baryons were proposed and worked out by Jaffe within the bag model in 1977 
\cite{Jaffe:1976ig,Jaffe:1976ih}.

However, after the lightning fast seventies, the progress in QCD slowed down. An initial surge of interest in exotics was motivated by the proposal of the dibaryon $H$
\cite{Jaffe:1976yi}. On the experimental side, it took many decades for exotic hadrons to be observed experimentally. On the theoretical side, QCD is fundamentally non-perturbative and definite theoretical results are difficult to achieve. While models are important to develop ideas and test them, expensive fist principle techniques such as lattice QCD are necessary to validate the theory of exotic hadrons.
Meanwhile some claims of light exotics turned out to either be complicated such as the light crypto-exotics - which are difficult to separate from ordinary mesons - or remained unconfirmed experimentally after many experimental searches. 

\begin{figure}[t]
\begin{centering}
\includegraphics[width=0.45\columnwidth]{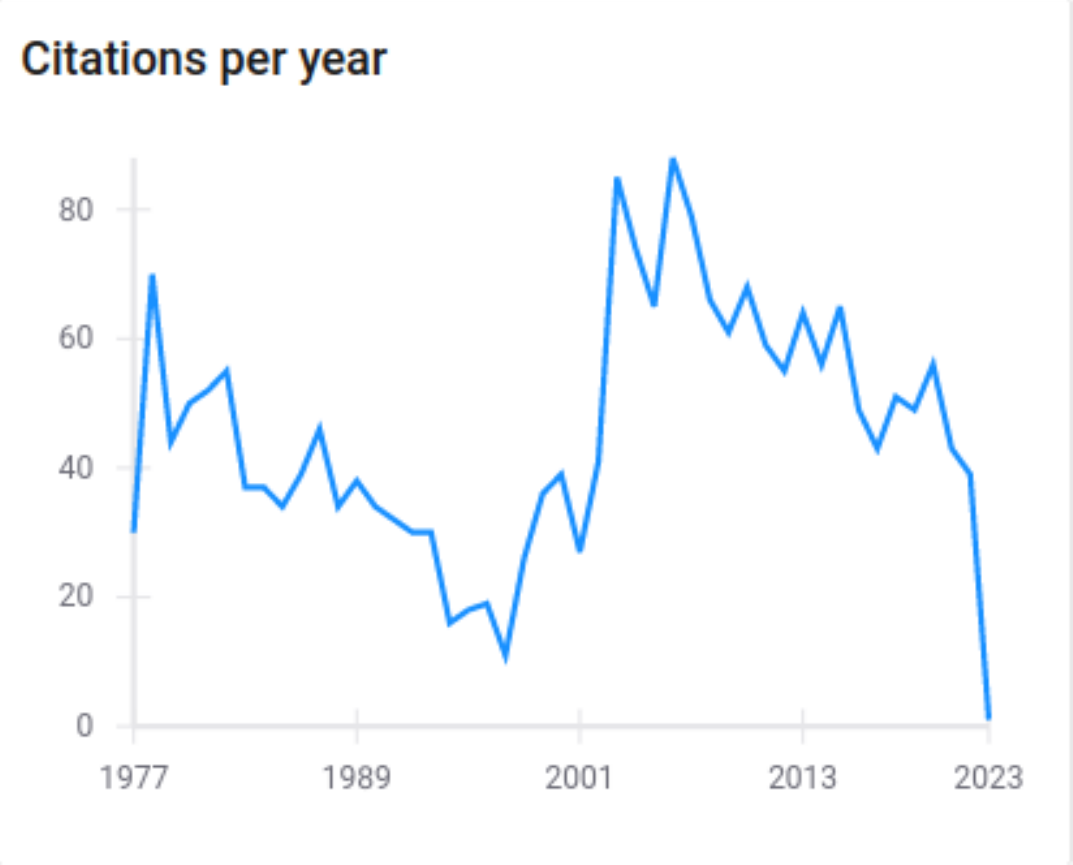}
\includegraphics[width=0.45\columnwidth]{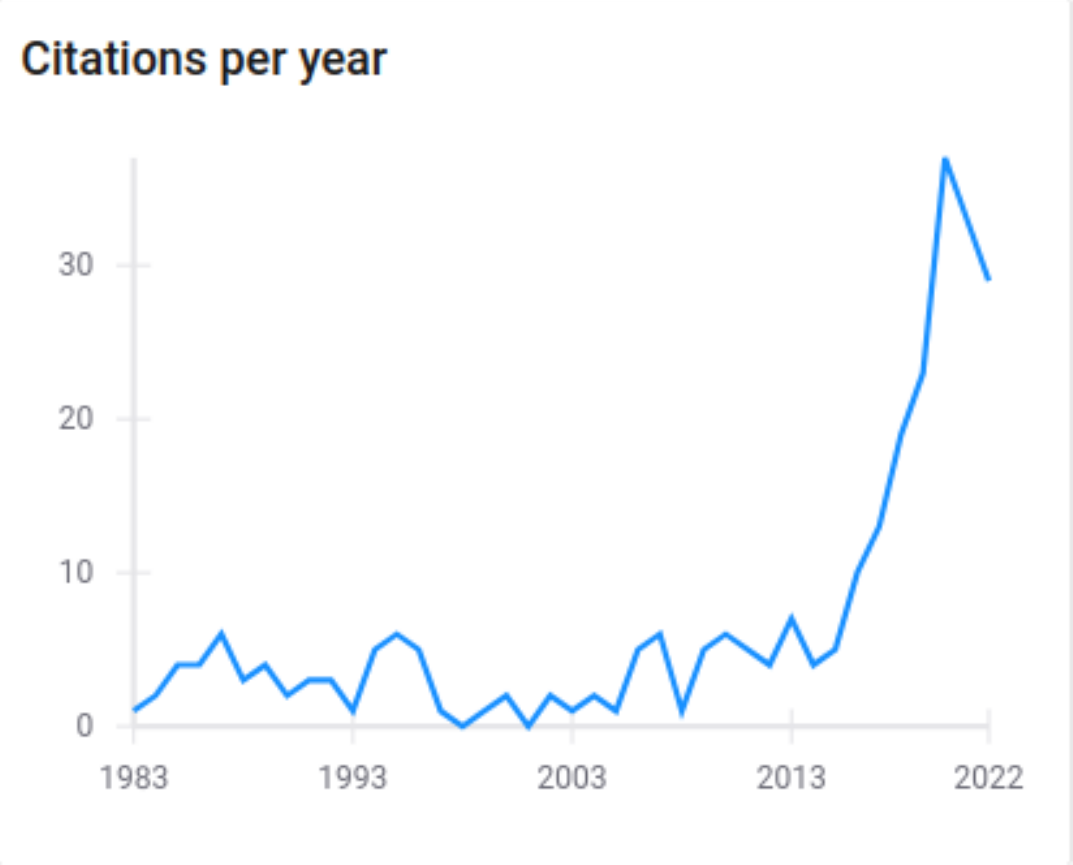}
\caption{Time evolution of the citations, extracted from the database inspirehep.net, of Ref. \cite{Jaffe:1976ig} (left) who proposed multiquarks in general of and Ref.  \cite{Ader:1981db} (right) who proposed the $T_{cc}$ tetraquark: the interest was decreasing since the onset of QCD, but the citations of Ref. \cite{Jaffe:1976ig} were boosted by the experimental evidences of tetraquarks and pentaquarks while the citations of Ref.  \cite{Ader:1981db} was boosted by lattice QCD new results, confirming the prediction of a $T_{bb}$.
\label{fig:citing}}
\end{centering}
\end{figure}

In the new millennium, the interest in exotics peaked again with searches of the pentaquark $\theta^+$ 
\cite{Diakonov:1997mm,LEPS:2003wug}, 
leading to many theoretical and experimental works. This also coincided with the planning of the experimental facility FAIR in Darmstadt, who will soon be contributing to this field. While the $\theta^+$ remained unconfirmed, the interest in exotic was maintained by the consistent discovery of a new class of exotic multiquarks. This is clear in the bumps in the citation evolution of two seminal multiquark papers, shown in Fig. \ref{fig:citing}.

%

Finally, there is one direction in the exotic world where exotic hadrons are found: multiquarks with some heavy quarks. 
In the early eighties, Richard 
\cite{Ader:1981db,Ballot:1983iv,Zouzou:1986qh,Gignoux:1987cn}
and colleagues were already proposing a mechanism for tetraquark stability. Using the unbalance in the quark masses, they realized realized $T_{cc}$, and other tetraquarks with two heavy quarks, were the most promising ones for future discoveries.

Indeed it took many decades for our knowledge of QCD to mature and prove Richard was right. At last, in the new millennium, exotic tetraquarks with heavy quarks have finally been confirmed experimentally, starting by the $Z_c$ and the $Z_b$ who were produced in different types of $B$ meson factories. Meanwhile, lattice QCD also has been progressing, and was able to show from first principles that the tetraquark $T_{bb}$ should form a boundstate. This already increased the literature interest in tetraquarks, see Fig. \ref{fig:citing}. Presently there are more than 40 different experimental evidences of different tetraquarks and pentaquarks, most of them with two heavy quarks, including the very recently observed 
\cite{LHCb:2021vvq,LHCb:2021auc}
$T_{cc}$. This led to a more consistent theoretical interest in tetraquarks, pentaquarks and other exotics.

\subsection{
Present status of the experimental evidences of tetraquarks and pentaquarks}

\begin{table}[t!]
\begin{small}
\begin{center}
\begin{tabular}{cc|ccccc} 
\toprule[1pt]
State             & qnumber & mass (MeV)   & width (MeV)  & decay mode & significance & Experiment \\
\midrule[1pt]
  $\chi_{c1}(3872)$ & $c \bar c (u \bar u + d \bar d) ~ ?$ 
  & $3872 \pm 1.1$ & $< 2.3$ & $\pi^+ \pi^- J/\psi$ & $10 \sigma$
  & Belle \cite{Belle:2003nnu}  
\\
 & 
 & $3871.3 \pm 0.7$ & $4.9 \pm 0.7$ & $\pi^+ \pi^- J/\psi$ & $11.6 \sigma$
 & CDF \cite{CDF:2003cab}
 \\
 & 
 & $3872$ & - & $\pi^+ \pi^- J/\psi$ & $4.2 \sigma$
 & CMS \cite{CMS:2021znk}
 \\
  & 
 & $3872$ & - & $\pi^+ \pi^- J/\psi$ & $7.1 \sigma$
 & LHCb
 \cite{LHCb:2022bly}
  \\[3pt]
\midrule[1pt]
  $\chi_{c1} (4140)$
  &  $c \bar c s \bar s ~ ?$
  & $4143.0 \pm 4.1$ & $11.7 \pm 12$ & $J / \psi \phi$ & $3.8 \sigma$ & CDF \cite{CDF:2009jgo}
\\
  &
  & $4159.0 \pm 10.9$ & $19.9 \pm 20.6$ & $J / \psi \phi$ & $3.1 \sigma$ & D0 \cite{D0:2013jvp}
\\
  &
  & $4148.00 \pm 8.7$ & $28 \pm 34$ & $J / \psi \phi$  & $5 \sigma$ & CMS \cite{CMS:2013jru}
\\
  & $c \bar c s \bar s ~ 1^{++}$
  & $4146.5 \pm 9.1$ & $83 \pm 42$ & $J / \psi \phi$ & $8.4 \sigma$ & LHCb \cite{LHCb:2016axx}
\\
  $\chi_{c1} (4274)$  &  $c \bar c s \bar s ~ 1^{++}$
  & $4273.3 \pm 25.5$ & $56 \pm 22$ & $J / \psi \phi$  & $6.0 \sigma$ &LHCb \cite{LHCb:2016axx}
\\
  $\chi_{c0} (4500)$ &  $c \bar c s \bar s ~ 0^{++}$
  & $4506 \pm 26$  & $92 \pm 42$ & $J / \psi \phi$  & $6.1 \sigma$ & LHCb \cite{LHCb:2016axx}
\\
  $\chi_{c0} (4700)$ &  $c \bar c s \bar s ~ 0^{++}$
  &  $4704 \pm 34$ & $ 120 \pm 73 $ & $J / \psi \phi$  & $5.6 \sigma$ & LHCb \cite{LHCb:2016axx}
\\
  $X (4630)$  &  $c \bar c s \bar s ~1^-$ 
  &  $4626 \pm 126$  & $174  \pm 161$ & $J / \psi \phi$ & $5.5 \sigma$
  & LHCb \cite{LHCb:2021uow}
\\
  $X (4685)$  &  $c \bar c s \bar s ~1^+$ 
  &  $4684 \pm 23$ &  $126 \pm 56$ & $J / \psi \phi$ & $15 \sigma$
  & LHCb \cite{LHCb:2021uow}
\\
  $X (4740)$ & $c \bar c s \bar s ~ ? $ 
  &  $4741 \pm 12$ & $53 \pm 26$ & $J / \psi \phi$ & $ 5.3 \sigma$
  & LHCb \cite{LHCb:2020coc}
  \\[3pt]
\midrule[1pt]
  $X(6900)$ &  $c \bar c  c \bar c ~ ?$  
  & $6905 \pm 18$      & $80 \pm 32$   & $J/ \psi J/ \psi$  &  $5.25 \pm 0.15 \sigma $ &  LHCb \cite{LHCb:2020bwg} 
\\
                           &  
  & $6886 \pm 22$      & $168 \pm 102$   & $J/ \psi J/ \psi$  NRSPS  &   $5.25 \pm 0.15 \sigma $ &  LHCb \cite{LHCb:2020bwg}  
\\
                           &  
  & $6927 \pm 10$      & $117 \pm 24$   & $J/ \psi J/ \psi$  &   $9.4 \sigma $ &  CMS \cite{CMS:2022yhl} 
\\
                          &  
  & $6918 \pm 10$      & $187 \pm 40$   & $J/ \psi J/ \psi$  NRSPS  &   $9.4 \sigma  $ &  CMS \cite{CMS:2022yhl} 
\\
  $X(6600)$ &  $c \bar c  c \bar c ~ ?$  
  & $6552 \pm 22$      & $124 \pm 63$   & $J/ \psi J/ \psi$  &  $6.5 \sigma$ &  CMS \cite{CMS:2022yhl}
  \\
  \midrule[1pt]
  $\Upsilon(10753)$  &  $b \bar b (u \bar u + d \bar d) 1^- ?$ & $10752.7 \pm 7.0$ & $35.5 \pm 21.5$ &  $\Upsilon (nS) {\pi^+} {\pi^-}$ & $5.2 \sigma$ & BELLE \cite{Belle:2019cbt}
\\  
\bottomrule[1pt]
\end{tabular}
\end{center}
\caption{We review the flavour singlet tetraquarks candidates, with hidden heavy quarks, observed experimentally. The quantum numbers are not exotic, but they may be considered tetraquarks because they don't fit the spectrum of quark-antiquark mesons in the quark model. In the q-number we list flavour and $J^P$. They have been detected from 2003 to 2022.
\label{tab:cryptoTetra}}
\end{small}
\end{table}

\begin{table}[t!]
\begin{small}
\begin{center}
\begin{tabular}{cc|ccccc} 
\toprule[1pt]
State             & qnumber & mass (MeV)   & width (MeV)  & decay mode & significance & Experiment \\[3pt]
\midrule[1pt]
$\pi_1$ &   $ u \bar d g$ or $u \bar d q \bar q$ $1^{-+}$ & $1564 \pm 110$ &  $492 \pm 156 MeV$    &  $\pi \eta^{(')}$  & ?  & \cite{Ketzer:2012vn} JPAC \cite{JPAC:2018zyd}
  \\[3pt]
\midrule[1pt]
  $X_0(2900)$ & $\bar c d u \bar s ~ 0$
 & $2866 \pm 9$ & $57 \pm 16$ & $D^- K^+$ & $3.9 \sigma$
 & LHCb \cite{LHCb:2020bls,LHCb:2020pxc} 
\\
  $X_1(2900)$   & $\bar c d u \bar s ~ 1$ 
 & $2904 \pm 6$ & $110 \pm 15$ & $D^- K^+$ & $3.9 \sigma$
 & LHCb \cite{LHCb:2020bls,LHCb:2020pxc}   \\[3pt]
  \midrule[1pt]
  $Z_{cs} (3985)$  & $c \bar c u \bar s ~ 1^+$ 
  & $3982.5 \pm 4.2 $ & $12.8 \pm 8.3$ & $D_s^{-} D^{*0} , D_s^{*-} D^{0}$ & $5.3 \sigma$
  & BESIII ~\cite{BESIII:2020qkh} \\
  & 
  & $3985.2 \pm 3.8 $ & $13.8 \pm 13.0 $ & $D_s^{-} D^{*0} , D_s^{*-} D^{0}$ & $ 4.6 \sigma$
  & BESIII ~\cite{BESIII:2022qzr} \\
  $Z_{cs} (3985)^0$  & $c \bar c d \bar s ~ 1^+$ 
  & $3992.2 \pm 3.3 $ & $7.7 \pm 8.4$ & $D_s D^{*} , D_s^{*} D$ & $ 4.6 \sigma$
  & BESIII ~\cite{BESIII:2022qzr} \\
  $Z_{cs} (4000)$  & $c \bar c u \bar s ~1^+$ 
  & $4003 \pm 20$ & $131 \pm 31$ & $J / \psi K^+$ & $15 \sigma$
  & LHCb \cite{LHCb:2021uow}
\\
  $Z_{cs} (4220)$  & $c \bar c u \bar s~1^+$ 
  &  $4216 \pm 54$ &  $233 \pm 149$  & $J/\psi K^+$ & $5.9 \sigma$
  & LHCb \cite{LHCb:2021uow}
  \\[3pt]
\midrule[1pt]
   ${Z_c(3900)}$ &   $c \bar c  u \bar d ~ ? $  
   & $3899.0\pm8.5$ &  $46\pm 30$    &  $J/\psi\, \pi^+$  & $8 \sigma$  & BESIII \cite{BESIII:2013ris} \\
                  &  & $3894.5\pm 11.1$ &  $63 \pm 50$    &  $J/\psi\, \pi^+$   & $5.2 \sigma$ & Belle  \cite{Belle:2013yex} \\
  &
  & $3886 \pm 6$       &  $37 \pm 12$      &  $J/\psi\, \pi^+$ & $5 \sigma $ & CLEO-c \cite{Xiao:2013iha} \\
   ${Z_c^0(3900)}$ &   $c \bar c  (u \bar u - d \bar d) ~ ? $ 
  & $3901 \pm  4$ & $58 \pm 27$  &  $J/\psi\, \pi^0$ & $3.5 \sigma$ & CLEO-c \cite{Xiao:2013iha} \\
  ${Z_c(3885)}$     & $c \bar c  u \bar d, \, c \bar c d \bar u ~1^+$ 
  & $3881.7\pm3.2$ &  $26.6\pm 4.1$&  $(D\bar{D}^{*})^{\mp}$  & $10 \sigma$ & BESIII \cite{BESIII:2013qmu,BESIII:2015pqw} \\
  ${Z_c(4020)}$     & $c \bar c  u \bar d, c \bar c d \bar u ~ 1^-$ 
  & $4022.9 \pm 3.5$ &  $7.9 \pm 5.3$ &  $h_c\, \pi^{+-}$ & $8.9 \sigma$  & BESIII \cite{BESIII:2013ouc} \\
  ${Z_c(4025)}$     & $c \bar c  u \bar d ~ ?1^+$ 
  & $4026.3\pm 6.3$ &  $24.8\pm 13.3$&  $(D^*\bar{D}^{*})^+$ & $10 \sigma$ & BESIII \cite{BESIII:2013mhi} \\
  $Z_c (4050)$ & $c \bar c  u \bar d ~ ?$
  & $4051 \pm 55$ & $82 \pm 68$ & $\pi^+ \chi _{c1}$ & $5 \sigma$ & Belle \cite{Belle:2008qeq} \\
  &
  & $4248 \pm 224$ & $177 \pm 370$ & $\pi^+ \chi _{c1}$ & $5 \sigma$ & Belle \cite{Belle:2008qeq} \\
  $X (4100)$  & $c \bar c  u \bar d ~ ?$
  & $4096 \pm 42 MeV$ & $152 \pm 118$ & $\eta_c(1S) \pi^-$ & $3 \sigma$ & LHCb \cite{LHCb:2018oeg} \\
  $Z_c (4200)$  & $c \bar c  u \bar d ~ 1^+$
  & $4196 \pm 58$ & $370 \pm 202$ & $J / \psi \pi^+$ & $6.2 \sigma$ & Belle \cite{Belle:2014nuw} \\
  $Z_c^+(4430)$ & $c \bar c  u \bar d ~ ?$ 
  & $4433 \pm 6 2$               &  $45 \pm 48$   &   $\pi^+ \psi(3686)$ &   $6.5 \sigma$ & Belle \cite{Belle:2007hrb}        \\
  & 
  & $4443 \pm 34$             & $107 \pm 160$  &   $\pi^+ \psi(3686)$ &  $6.4 \sigma$ & Belle \cite{Belle:2009lvn}        \\
  &    $c \bar c  u \bar d ~ 1^+$
  & $4485 \pm 50 $  & $200 \pm 81$  & $\pi^- \psi(3686)$ & $6.4 \sigma$ &   Belle \cite{Belle:2013shl,Belle:2014nuw}    \\
  & $c \bar c  u \bar d ~ 1^+$
  & $4475 \pm 32$  & $172 \pm50$  &   $\pi^- \psi(3686)$ &  $13.9 \sigma$ & LHCb \cite{LHCb:2014zfx,LHCb:2015sqg} 
    \\[3pt]
\midrule[1pt]
  $R_{c0}(4240)$  & $c \bar c d \bar u ~ 0^-$ 
  & $ 4239 \pm 63$ & $ 220 \pm 182$ &  $\pi^- \psi(3686)$ & $8 \sigma$ & LHCb  \cite{LHCb:2014zfx}
\\
\midrule[1pt]
  $Z_b(10610)$ &  $b \bar b  u \bar d ~1^+$  
  & $10611 \pm 7$      &$22.3 \pm 11.7$   & $\pi^\pm\Upsilon(1S)$   & $16.0 \sigma$ & Belle \cite{Belle:2011aa} \\
                           &                                          
  & $10609 \pm 5$      &$24.2 \pm 6.1$   & $\pi^\pm\Upsilon(2S)$   & $16.0 \sigma$ & Belle \cite{Belle:2011aa} \\
                          &                                          
  & $10608 \pm 53$      &$17.6 \pm 6.0$         & $\pi^\pm\Upsilon(3S)$   & $16.0 \sigma$ & Belle \cite{Belle:2011aa} \\
                          &                                          
  & $10605 \pm 5$  &$11.4 \pm 6.6$   & $\pi^\pm h_b(1P)$    & $16.0 \sigma$ & Belle \cite{Belle:2011aa} \\
                          &                                          
  & $10599 \pm 11$    &$13 \pm 19$            & $\pi^\pm h_b(2P)$    & $16.0 \sigma$ & Belle \cite{Belle:2011aa}  \\
                          &                                          
  & $10607.2 \pm 2.0$        &$18.4 \pm 2.4$                 & Averaged                & $16.0 \sigma$ & Belle \cite{Belle:2011aa}  \\[1pt]
$Z_b(10650)$ &  $b \bar b  u \bar d ~1^+$    
  & $10657 \pm 9$     & $16.3 \pm 15.8$ & $\pi^\pm\Upsilon(1S)$    & $16.0 \sigma$ & Belle \cite{Belle:2011aa} \\
                          &                                          
  & $10651 \pm 5$     & $13.3 \pm 7.3$ & $\pi^\pm\Upsilon(2S)$   & $16.0 \sigma$ & Belle \cite{Belle:2011aa} \\
                          &                                          
  & $10652 \pm 3$     & $8.4  \pm 4.0$      & $\pi^\pm\Upsilon(3S)$   & $16.0 \sigma$ & Belle \cite{Belle:2011aa} \\
                          &                                          
  & $10654 \pm 5$ & $20.9 \pm 10.4$ & $\pi^\pm h_b(1P)$   & $16.0 \sigma$ & Belle \cite{Belle:2011aa}  \\
                          &                                         
   & $10651 \pm 5$   & $19 \pm 18$ & $\pi^\pm h_b(2P)$    & $16.0 \sigma$ & Belle \cite{Belle:2011aa}  \\
                          &                                          
  & $10652.2 \pm 1.5$       & $11.5 \pm 2.2$ & Averaged             & $16.0 \sigma$ & Belle \cite{Belle:2011aa}    
 \\
\bottomrule[1pt]
\end{tabular}
\end{center}
\caption{We review the exotic tetraquark (or hybrid) resonances observed experimentally. In the qnumber we list flavour and $J^P$. They have been detected from 2007 to 2022.
\label{tab:Tetra}}
\end{small}
\end{table}

\begin{table}[t!]
\begin{small}
\begin{center}
\resizebox{1.0\textwidth}{!}{
\begin{tabular}{cc|ccccc} \toprule[1pt]
  state  & qnumber            & $\delta$mass            & width                    &   decay mode   & significance & experiment / lattice  \\
\midrule[1pt]
  $Tcc(3874)$ &  $u d  \bar c \bar c ~? $  
  & $-360 \pm 44$ KeV    & $48 \pm16 $ KeV  & virtual $D^0D^{*+}$ &  $15.5 \pm 6.5 \sigma $ &  experimental LHCb \cite{LHCb:2021vvq,LHCb:2021auc} \\
  & $u d  \bar c \bar c ~1^+$
  &   $-23 \pm 11$ MeV & 0 & - &  - & dynamical lattice QCD \cite{Junnarkar:2018twb} \\
  &
  &   vBS $\sim -9 $ MeV & 0 & - &  - & scattering lattice QCD \cite{Padmanath:2022cvl} \\
\midrule[1pt]
  $Tccs$ & $us  \bar c \bar c~1^+$
  &  $-8 \pm 8$ MeV & 0 & - & - & dynamical lattice QCD \cite{Junnarkar:2018twb} \\
\midrule[1pt]
 $Tbc$ & $us  \bar b \bar c~1^+$, $0^+$
  &  $\sim -40 \pm 50$ MeV & 0 & - & - & heavy quark lattice QCD \cite{Wagner:2022qwg,Pflaumer:2022lgp} \\
\midrule[1pt]
  $Tbb$ & $u d  \bar b \bar b ~1^+$
  &  $-90 \pm 43$ MeV & 0 & - & - & static lattice QCD \cite{Bicudo:2012qt,Bicudo:2015vta,Bicudo:2015kna,Bicudo:2017szl,Bicudo:2021qxj} \\
  &
  &  $-59 \pm 38$ MeV & 0 & - & - & $2 \times 2$ static lattice QCD \cite{Bicudo:2016ooe} \\
  & 
  &   $ -189 \pm 13 $ MeV & 0 & - & - & heavy quark lattice QCD \cite{Francis:2016hui} \\
  & 
  &   $\sim -113 $ MeV & 0 & - & - &  heavy quark lattice QCD \cite{Francis:2018jyb,Hudspith:2020tdf}  \\
  & 
  &   $ -143 \pm 34 $ MeV & 0 & - & - & heavy quark lattice QCD \cite{Junnarkar:2018twb} \\
  & 
  &   $ -128 \pm 34 $ MeV & 0 & - & - & heavy quark lattice QCD \cite{Leskovec:2019ioa} \\
  & 
  &   $\sim  -120 $ MeV & 0 & - & - & heavy quark lattice QCD  \cite{Colquhoun:2022dte} \\
  & 
  &   $ -154.8 \pm 37.2 $ MeV & 0 & - & - & scattering lattice QCD  \cite{Aoki:2022lat} \\
   & 
  &   $ -83.0 \pm 30.2 $ MeV & 0 & - & - & scattering lattice QCD  \cite{Aoki:2022lat} \\
    & 
  &  $ -103 \pm 8$ MeV & 0 & - & -   &  scattering lattice QCD \cite{Wagner:2022bff,Pflaumer:2022lgp} \\
  & $u d  \bar b \bar b ~ 0^+$
  &  $-50.0 \pm 5.1$ MeV & 0 & - & - & static lattice QCD \cite{Brown:2012tm} \\
  &
  & $-5 \pm 18$ MeV & 0 & - & - &  heavy quark lattice QCD \cite{Junnarkar:2018twb} \\
\midrule[1pt]
  $Tbbs$ & $us  \bar b \bar b, ds  \bar b \bar b~1^+$
  &  $-98 \pm 10$ MeV & 0 & - & - &  heavy quark lattice QCD \cite{Francis:2016hui} \\
  &
  &  $ \sim - 36$ MeV & 0 & - & -  &  heavy quark lattice QCD \cite{Francis:2018jyb,Hudspith:2020tdf} \\
  &
  & $-87 \pm 32$ MeV & 0 & - & - & heavy quark lattice QCD \cite{Junnarkar:2018twb} \\
  & 
  &  $ \sim - 80$ MeV & 0 & - & -   &  heavy quark lattice QCD \cite{Pflaumer:2021ong} \\
  & 
  &  $ - 86 \pm 32$ MeV & 0 & - & -   &  scattering lattice QCD \cite{Wagner:2022qwg,Pflaumer:2022lgp} \\
  \midrule[1pt]
  $Tbbc$ & $uc  \bar b \bar b~1^+$
  &  $-6 \pm 11$ MeV & 0 & - & - &  heavy quark lattice QCD \cite{Junnarkar:2018twb} \\
\midrule[1pt]
  $Tbbcs$ & $sc  \bar b \bar b~1^+$
  &  $-8 \pm 3$ MeV & 0 & - & - &  heavy quark lattice QCD \cite{Junnarkar:2018twb} \\
\bottomrule[1pt]
\end{tabular}
}
\end{center}
\caption{The tetraquark boundstates, or very narrow resonances. Experimentally, only the $T_{cc}$ has been observed. In this table, we include the lattice QCD predictions as well, where we detail the approach to address the heavy quarks (static, heavy quark effective theory, dynamical or scattering).  These results are from 2012 to 2022.  
\label{tab:boundTetra}}
\end{small}
\end{table}

\begin{figure}[!t]
  \centering   
  \includegraphics[width=\textwidth]{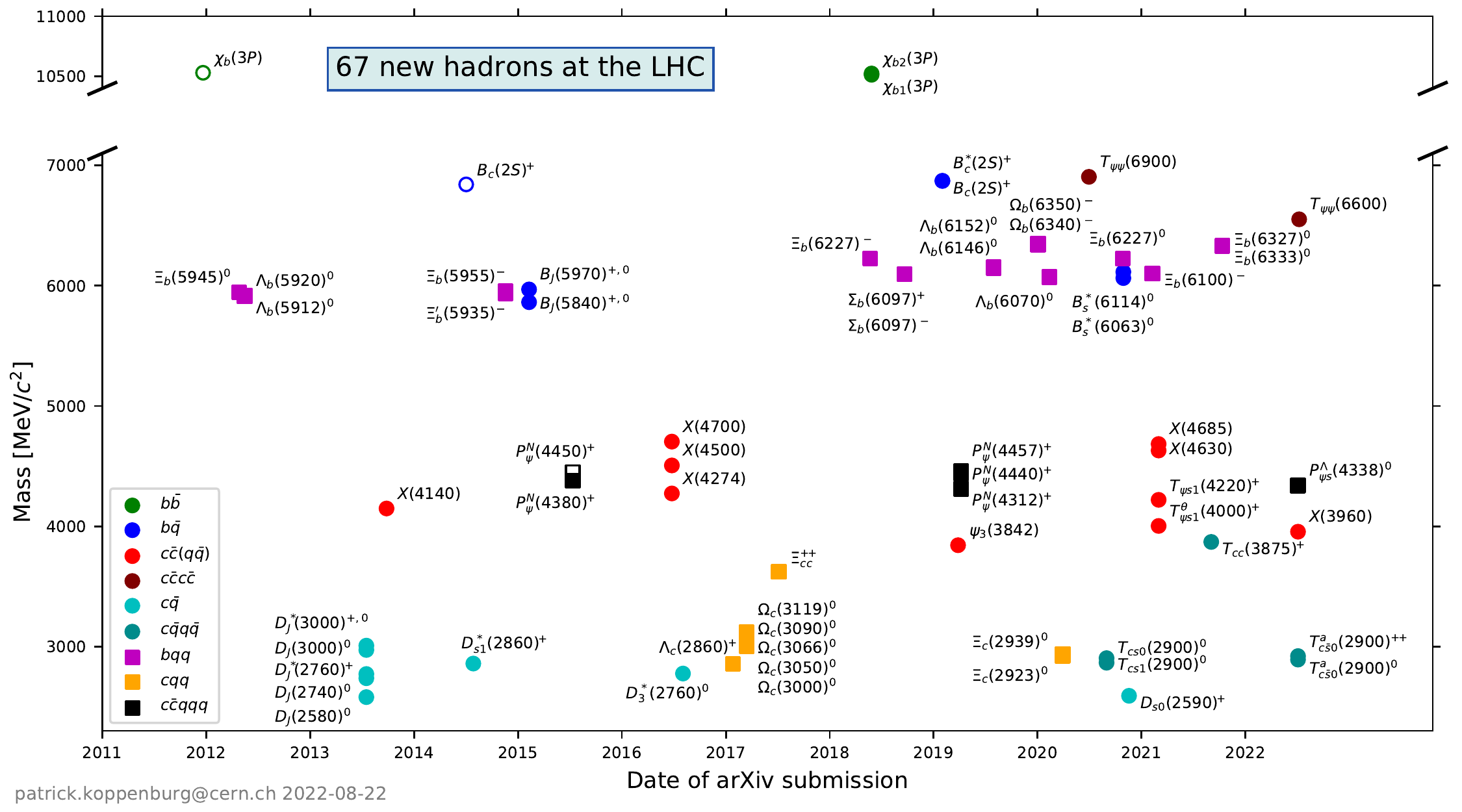}   
  \caption{
    New hadrons discovered at the LHC, including baryons, tetraquarks and pentaquarks, plotted as mass versus preprint submission date~\cite{LHCparticlesurl}. 
    Only states observed with significance exceeding $5\sigma$ are included.
  }
  \label{fig:masses} 
\end{figure}

We briefly review the landscape of multiquark resonances so far observed, in more than forty different experiments and lattice QCD computations, for instance Fig. \ref{fig:masses} shows the large number of resonances discovered at just one of the particle accelerators, the LHC \cite{Gershon:2022xnn,LHCparticlesurl}. Notice LHC is proposing a new naming convention. However we still follow the Particle Data Group (PDG) naming convention 
\cite{ParticleDataGroup:2022pth}.

Since there are many tetraquark resonances  we divide them in different tables, classifying them with their flavour content and their stability.

In Table \ref{tab:cryptoTetra},
we review the flavour singlet, tetraquark candidates observed experimentally. These tetraquarks are not explicitly exotic, they have been coined crypto-exotic \cite{Jaffe:1976ig}, but they may be considered tetraquarks because they don't fit the spectrum of quark-antiquark mesons in the quark model. The first state of this type is the $\chi_{c1}(3872)$, observed at Belle \cite{Belle:2003nnu} in 2003, has been extensively discussed in the literature. Presently there are twelve different such states, 
with flavour  $c \bar c (u \bar u + d \bar d)$,  $c \bar c s \bar s$, $b \bar b (u \bar u + d \bar d)$ and the very recently observed  $c \bar c c \bar c$.
The latter, $X(6900)$ and $X(6600)$ resonances with flavour $c \bar c c \bar c$, were observed at LHCb in 2020 \cite{LHCb:2020bwg} 
and CMS in 2022 
\cite{CMS:2022yhl}, 
and could belong to the next group of tetraquarks as well.

Then we list in Table  \ref{tab:Tetra} the unambiguous tetraquark resonances. The first one discovered was the $Z_c^+(4330)$, at Belle \cite{Belle:2007hrb}.  Some of these resonances have a heavy quark antiquark pair and a light quark-antiquark pair, we then denote them $l\bar l Q \bar Q$. These cases are not strictly flavour exotic, since the $Q \bar Q$ pair has no flavour or charge, the heavy flavour is hidden 
\cite{Chen:2016qju}, but nevertheless its masses clearly require a $Q \bar Q$ pair and then the $l\bar l$ has isospin and charge only compatible with an exotic. Several of these states have been observed at $B$ meson factories, others in colliders.
The B meson weak decay produces frequently a $c \bar c$ pair plus quarks and leptons, and the $Z_c$ tetraquarks are for instance observed in two-meson correlations in hadronic three-body decays.  Nineteen states of this class have been observed by several collaborations, with flavours $\bar c d u \bar s$, $c \bar c u \bar d$, $c \bar c u \bar s$, $b \bar b u \bar d$ and relates ones. 
The $Z_b$ tetraquarks has only been observed by the Belle collaboration, in 2011, among the decay products of the $\Upsilon(5S)$. Belle is a $B$ meson factory at the KEK electron positron collider which is tuned to produce bottomonium and $B \bar B$ meson pairs. In this case, the other experimental collaborations have not been able to address these resonances. Nevertheless, the Belle observation has a high significance, see Table \ref{tab:Tetra}.
In this table we also include is the light exotic $\pi_1$, which is most likely a hybrid, observed in 2012 at COMPASS and analysed at JPAC \cite{Ketzer:2012vn,JPAC:2018zyd} with exotic parity and charge conjugation, $J^{PC} (I^{G})= 1^{-+}(1^{-})$, but which quantum numbers are also compatible with a tetraquark.

We also include in a the next Table \ref{tab:boundTetra} the only tetraquarks which are boundstates or very narrow resonances. So far, they turn out to be absolutely flavour exotic.  
These tetraquark, have two heavy quarks (or antiquarks), and we denote them $l l \bar Q \bar Q$ or $T_{QQ}$. They have been predicted theoretically since 1981 \cite{Ader:1981db}.  From 2012 \cite{Bicudo:2012qt}, they started to be been computed from first principles in lattice QCD, with the $b$ quark as the heavy quark. This class of tetraquarks was just discovered in 2021, at LHCb, with flavour $\bar c \bar c u d$ \cite{LHCb:2021vvq,LHCb:2021auc}.

Notice that, in terms of flavour, not all tetraquarks in Table  \ref{tab:Tetra} are of the $Z$ type with a heavy $Q \bar Q$ pair. The $X_0(2900)$ and $X_1(2900)$ only have one heavy quark, but they have a strange quark, the less light quark pair they have is a $\bar c \bar s$, in this sense they can be related to the $T_{QQ}$ tetraquarks of Table \ref{tab:boundTetra}. Table \ref{tab:Tetra} also includes the tetraquarks $X(6600)$ and $X(6900)$ denominated by LHC as $T_{\psi\psi}$, with four heavy quarks, which is both of the $T$ and $Z$ tetraquark type.

Finally we list in table \ref{tab:Penta} the $P_c$ and $P_{cs}$ pentaquarks, first observed at LHCb in 2015 \cite{LHCb:2015yax}.
These pentaquarks, as most of the so far observed tetraquarks, are hidden heavy flavour states, in that sense they are an extension of the $Z_c$ tetraquarks.

\subsection{
Theoretical paradigms of hadrons and three types of tetraquarks  \label{sec:theory}}

We now review some of the main ideas in hadronic models who may impact multiquark systems. Because the literature is quite vast, we are not able to be any comprehensive at all in this brief review. We opt to follow a historical panorama, relating some of the different types of hadronic models with simple concepts.

We also describe three types of mechanisms driving tetraquarks, since the tetraquarks are the simplest multiquarks and what binds a tetraquark may also bind a pentaquark or a hexaquark. Tetraquarks may be molecular meson-meson systems with Yukawa-like meson exchange interactions similar to the ones of nuclear physics. Besides, tetraquarks may have a novel diquark-antidiquark core producing the necessary attraction to bind them. Moreover tetraquarks may appear as non-perturbative poles in the scattering matrix with meson-meson and quarkonium coupled channels, well beyond the Born approximation, where quark-antiquark annihilation and creation produces the main driving effect.  We respectively denominate these systems {\em molecular} tetraquarks, {\em diquark} tetraquarks and {\em s pole} tetraquarks.

\subsubsection{
Nuclear physics  and meson-meson {\em molecular} tetraquarks \label{sec:nuclearTh}}

As soon as mesons were proposed in the thirties by Yukawa \cite{Yukawa:1935xg} to understand the $NN$ interaction, it would be plausible that meson-meson molecules could exist as well. The long range part of the $NN$ strong interaction \cite{Reid:1968sq}
according to the Nijmegen, Paris, Bonn, Argonne, potentials
\cite{Nagels:1975fb,Lacombe:1980dr,Machleidt:1987hj,Wiringa:1994wb},
is due to meson exchange, in particular at large distances the one pion exchange potential (OPEP) dominates. The pion, which is an isovector and a pseudoscalar, couples to the nucleon with a flavour Pauli matrix $\bm \tau$ and with a spin-derivative coupling $\bm \sigma \cdot \bm p$, and the long distance part of the OPEP potential is,
\be
V_{\pi N N}(r)= {f_{\pi N N}}^2 \ {e^{- m_\pi r} \over r} \ (\bm \tau_1 \cdot \bm \tau_2 ) (\bm \sigma_1 \cdot \hat r ) ( \bm \sigma_2 \cdot \hat r)
\label{eq:OPEP}
\ee
In this perspective, with two mesons with flavour $I=1/2$ , say $u \bar c$ and $d \bar c$ as in the recently observed 
\cite{LHCb:2021vvq,LHCb:2021auc}
$T_{cc}$ a boundstate similar to a deuteron composed  by a  proton $p=uud$ and a neutron $n=ddu$, e.g. which have the same flavour $I=1/2$, could exist as well. Notice the flavour term in OPEP, $\bm \tau_1 \cdot \bm \tau_2$, could produce an attraction between a $DD$ or a $BB$ similar to the one in the $NN$ system. Thus several authors are proposing the observed experimental resonances to be {\em molecular} tetraquarks \cite{Guo:2017jvc}.

\begin{table}[t!]
\centering
\begin{small}
\begin{tabular}{lc|ccccc}
\toprule[1pt]
	state  
	& qnumber  &mass (MeV) & width (MeV) & decay & significance & experiment
  \\[1mm]
  \hline 
  \\[-2mm]
  $P_c (4312)$  & $c \bar c u u d ~ ?$
  & $ 4311.9 \pm 7.5$  & $9.8 \pm 7.2$ &  $J/\psi p$ & $7.3 \sigma$ & LHCb \cite{LHCb:2019kea}
\\
  $P_c (4380)$   & $c \bar c u u d ~ 3/2^-$
  & $4380 \pm 37$ & $205 \pm 104$ & $J/\psi p$ & $9 \sigma$ & LHCb \cite{LHCb:2015yax}
\\  
  $P_c (4440)$   & $c \bar c u u d  ~ ?$
  & $4440.3 \pm 6.0$ & $20.6 \pm 15.0$ & $J/\psi p$ & $5.4 \sigma$ & LHCb \cite{LHCb:2019kea}
\\
  $P_c (4450)$
 & $c \bar c u u d  ~5 /2^+$
  & $4449.8 \pm 4.2 MeV$ & $39 \pm 24$ & $J/\psi p$ & $12 \sigma$ & LHCb \cite{LHCb:2015yax}
\\
  $P_c (4457)$  & $c \bar c u u d  ~ ?$   
  & $4457.3 \pm 4.7$ & $6.4 \pm 7.7$ & $J/\psi p$ & $5.4 \sigma$ & LHCb \cite{LHCb:2019kea}
\\
  $P_c (4357)$   & $c \bar c u u d ~ ?$
  & $4337 \pm 9 $ & $29 \pm 40$ & $3.7 \sigma$ & $J/\psi p$ & LHCb \cite{LHCb:2021chn}
  \\[3pt]
\midrule[1pt]
  $P_{cs}(4459)$   & $c \bar c u d s  ~ ?$
  & $4458.8 \pm 7.6$ & $17.3 \pm 6.2$ & $J / \psi \Lambda$ & $3.1 \sigma$ & LHCb \cite{LHCb:2020jpq} 
\\
\bottomrule[1pt]
\end{tabular}
\caption{We review the exotic pentaquarks observed experimentally. In the qnumber we list flavour and $J^P$. They have been detected from 2015 to 2022.
\label{tab:Penta}}
\end{small}
\end{table}

\subsubsection{
Spontaneous chiral symmetry breaking \label{sec:chisb}}

Spontaneous symmetry breaking is crucial to all areas of theoretical physics. In early sixties, Nambu and Jona-Lasinio 
\cite{Nambu:1961tp}
addressed the problem of having a light boson, the pion, composed of a pair of much heavier fermions, nucleons. They succeeded by devising a model for the spontaneous chiral symmetry breaking (S$\chi$SB). In the initial Lagrangian, the fermions are massless and the theory is chiral invariant. When S$\chi$SB occurs, the fermions acquire a mass and the lightest boson is massless, it is a Nambu-Goldstone boson
\cite{Goldstone:1961eq}.
The sigma model was also developed by Gell-Mann and Levy, with the same  S$\chi$SB
\cite{Gell-Mann:1960mvl}
but using only bosons, and in this case it is the scalar boson sigma who acquires a mass.
This scientific revolution was completed by the non-linear topological model of Skyrme,
who was able to produce fermions from bosons
\cite{Skyrme:1961vq}.
S$\chi$SB leads to  several theorems of current algebra and of Partially Conserved Axial Currents (PCAC).
S$\chi$SB is also one of the building blocks of effective chiral Lagrangians
\cite{Weinberg:1991um}.
Such an important phenomenon is of course relevant for the hadron masses and their interactions, and affects any multiquark with light quarks.

After quarks were discovered, the Nambu and Jona-Lasinio model was successfully applied to quark systems.  This mechanism stabilizes the vacuum, and in doing so provides mass to the quarks. In the chiral limit, pseudoscalar mesons like the pion would be massless.  Moreover the sigma model can be supplemented with exotic hadrons. These models for S$\chi$SB have a caveat: they are non-renormalizable. A more fundamental renormalizable theory is necessary. 

In the case of hadronic physics, the solution to this problem is QCD. It is a fully renormalizable theory and in principle it includes S$\chi$SB, which is verified in lattice QCD.
Notice S$\chi$SB may be one of the reasons why lattice QCD is computable: the preconditioning with heavier quark masses, mentioned in Subsection \ref{sec:latticeQCD}, may work because the constituent quark mass is much heavier than the bare quark mass.

\subsubsection{
Quark confinement, hadronic strings, and {\em hybrid} mesons}

Then,  in the mid sixties, the hadronic physics focus shifted to confining strings. Notice flux tubes and strings have been studied for a long time in type II superconductivity, where the magnetic field is confined. 
In 1911 Kamerlingh-Onnes 
\cite{Kamerlingh-Onnes:1911}
discovered superconductivity, the Meissner-Ochsenfeld 
\cite{Meissner:1933}
effect was discovered in 1933.
In 1935, Rjabinin and Shubnikov 
\cite{Rjabinin:1935}
experimentally discovered the Type-II superconductors. In 1950, Landau and Ginzburg
\cite{Ginzburg:1950sr}, 
continued by Abrikosov
\cite{Abrikosov:1956sx}, 
arrived at superconductor vortices, or flux tubes. Interestingly, superconductors (and ferromagnets) also include the spontaneous  symmetry breaking of Subsection \ref{sec:chisb}. 

 When confinement was proposed for quarks inside hadrons in 1964 by Gell-Mann 
 \cite{Gell-Mann:1964ewy} 
 and Zweig 
\cite{Zweig:1964ruk,Zweig:1964jf},
the analogy with flux tubes also led to an explosion of the number of publications in the literature on the quantum excitations of strings.
Lattice QCD  by  Wilson in 1974 was partly inspired in strings 
\cite{Wilson:1974sk}. 
It was also realized that experimentally, the meson spectra lie in linear $j(M^2)$ Regge trajectories, and this is characteristic of strings. It is interesting that the similar slope of radial and angular excitations, so far can only be understood with strings. Moreover a string-like confinement implies for heavy quarks a linear confinement plus a Coulomb term due to the string quantum vibrations, the Lüscher term, and this is the backbone of the potentials used in most quark models,
\be
V_{Q \bar Q} (r) = \sigma r - {\pi \over 12} { 1 \over r} \ .
\label{eq:string}
\ee

The interest in strings returned in 1997 when Maldacena 
\cite{Maldacena:1997re}
and others explored the  AdS/CFT correspondence.
The AdS/CFT and Holography has also been used as a model to compute spectra in hadronic physics.
Besides, strings continue to be relevant to the study of exotics. Strings already are expected to form exotic systems, denominated hybrids, when the quantum numbers of the strings are excited.
While the simplest possible strings do not form tetraquarks, QCD strings with junctions may bind tetraquarks.

\subsubsection{
Quark models and {\em diquark} tetraquarks \label{sec:quark}}

Since the seventies, as soon as quarks were discovered  it was expected that tetraquarks and other exotics should exist. The first models were the bag models  
\cite{Chodos:1974je,Bardeen:1974wr,Friedberg:1978sc,Nielsen:1978tr}
where the quarks are assumed to dig a bubble (similar to a broad string) in the vacuum, and are confined by the vacuum pressure. Since the eighties, the dominant quark models in the literature have been the potential quark models, with many different variants of the original Ref. 
\cite{DeRujula:1975qlm} and its spin dependent terms,
\be
V_{ij}(\bm r_{ij}) \propto {1 \over r_{ij}} - {\pi \over 2} \delta^3(\bm r_{ij})\left( {1 \over {m_i}^2}+ {1 \over {m_j}^2}+{16 \, \bm S_i \cdot \bm S_ j \over 3 \, m_i m_j} \right) + \cdots
\label{eq:qmOGEP}
\ee
including the standard relativized quark model of Ref. 
\cite{Godfrey:1985xj}. 
At short distances the quark models rely on an extension of the one gluon exchange potential (OGEP) of Eq. (\ref{eq:qmOGEP}) and at large distances confinement is provided by a thin string, as in Eq. (\ref{eq:string}), in agreement with the results of lattice QCD. 

This is an adiabatic approach in the sense the gluons are supposed to be integrated out, and their role is to provide the potential for the quarks. It is also generally supposed the short range part of the potential is vector-like, from OGEP while the long part range is scalar-like, from the string
\cite{Henriques:1976jd}. 

\begin{figure}[t!]
\begin{centering}
\includegraphics[width=0.5\columnwidth,trim=50pt 50pt 100pt 70pt, clip]{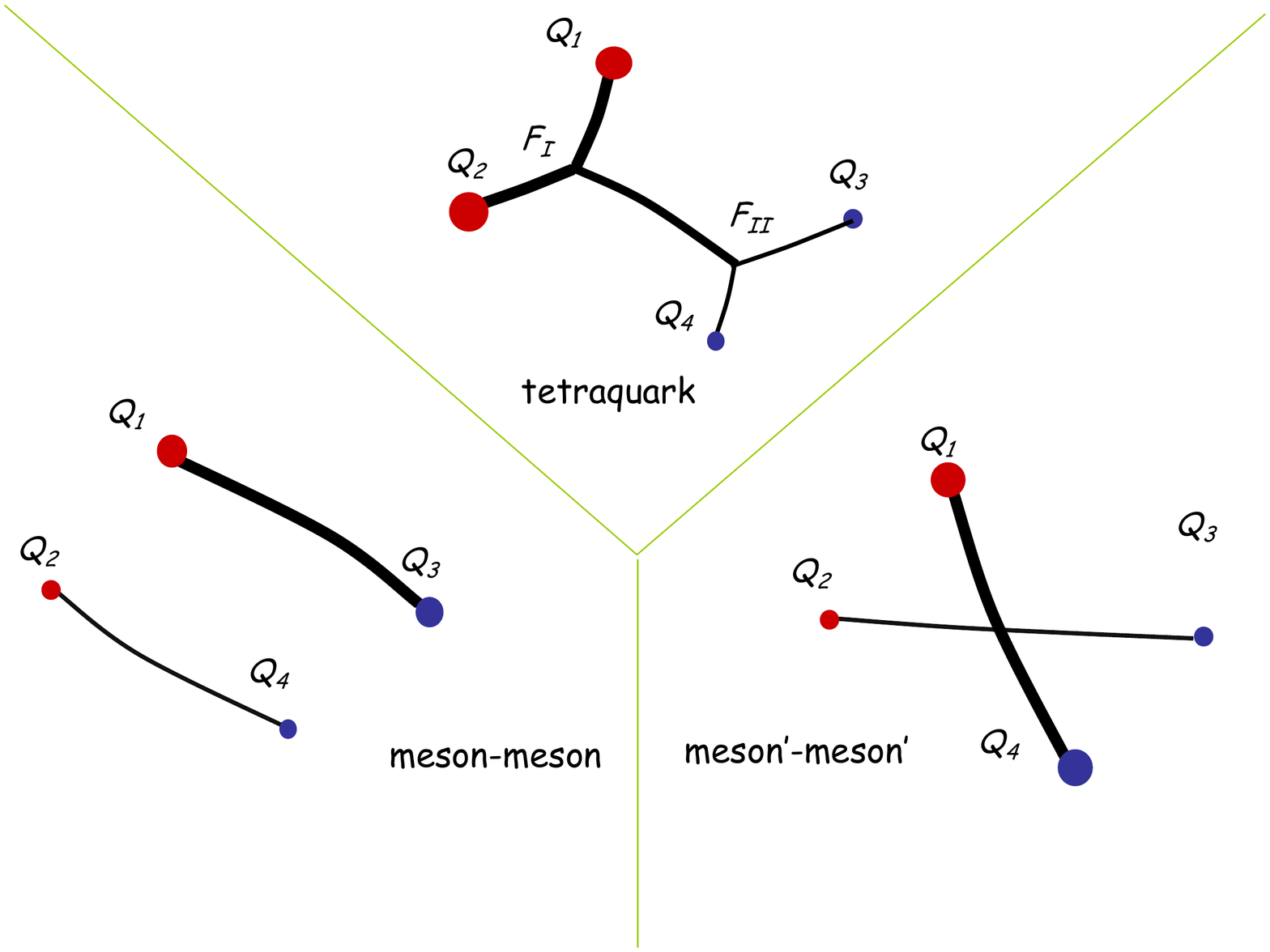}
  \caption{Triple string flip-flop Potential potential illustrated by Ref. \cite{Bicudo:2015bra}. 
\label{fig:poteflop}
}
\end{centering}
\end{figure}

Quarks are coloured particles, in a colour triplet 3 representation of $SU(3)$.
Colour-wise, with four quarks, two independent colour singlets may exist. A possible doublet is the orthogonal basis $1 1$ and $8 8$, another possible doublet is the orthogonal basis  $\bar 3  3$ and $6 \bar 6$. Notice these two possible bases are different but not orthogonal, $1 1 $ overlaps with $\bar 3 3$. Moreover with flavour, spin and space many more independent combinations of four quarks may exist. Nevertheless, depending on the geometry of the system, either one of the two meson-meson colour combinations $11$ or the {\em diquark} tetraquark  $\bar 3  3$ may be the lowest energy state, see Fig. \ref{fig:poteflop}. In this case it is plausible  {\em diquark} tetraquarks  $\bar 3  3$ may exist. There is a wide interest in the literature on whether the observed tetraquark systems are {\em diquark} tetraquarks with colour $\bar 3  3$ described by quark models
\cite{Jaffe:2003sg,Maiani:2004vq}
, or {\em molecular} meson-meson tetraquarks with colour $11$ described by models of the nuclear physics type. The models with string-flop potentials 
\cite{Lenz:1985jk}, defined in Subsection \ref{sec:flipflop} explore the combination of both colours. However these models still need improvement, the binding of $T_{bb}$ using flip-flop is still too large in some quantum numbers 
\cite{Bicudo:2015bra}, 
possibly because it does not yet have spin effects.

\subsubsection{
Quark impact on nuclear physics, chiral symmetry, strings and the scattering matrix  \label{sec:quarkfeedback}}

Notice that the discovery of the quarks, fermions with SU(3) colour, had an important feedback on our understanding of nuclear physics, chiral symmetry breaking and string theory. 

\begin{figure}[t!]
\begin{centering}
\includegraphics[width=0.3\columnwidth]{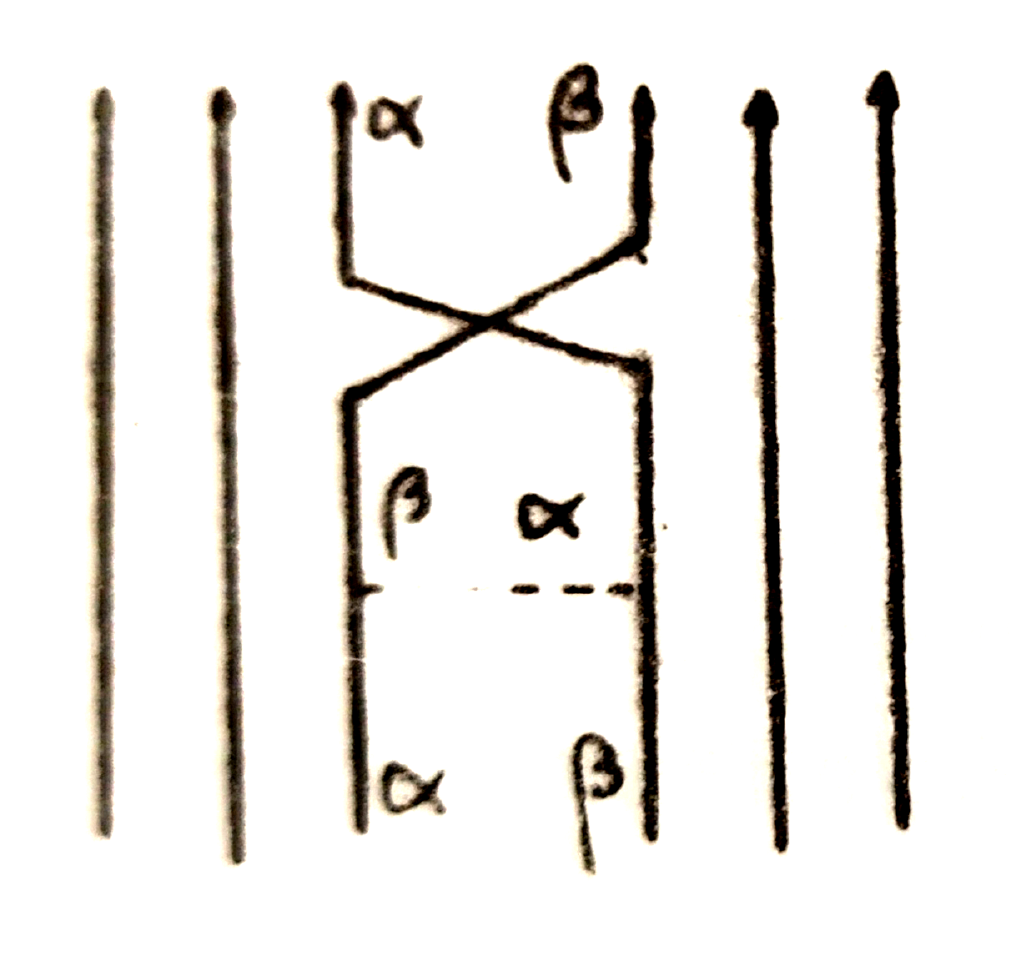}
\includegraphics[width=0.5\columnwidth]{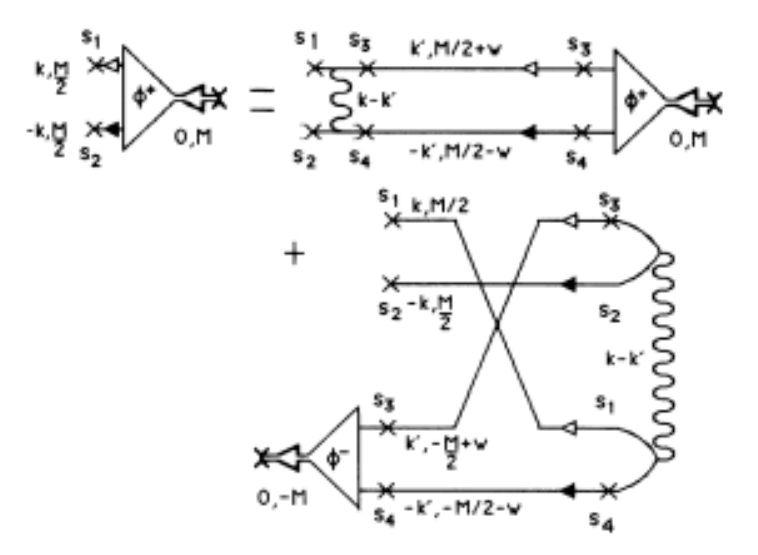}
\caption{(left) Example of a diagram for the microscopic calculation of the repulsive core in the elastic nucleon-nucleon scattering, described in Ref. \cite{Ribeiro:1978gx}.
(right) One of the Salpeter equations for the pion, to include spontaneous $\chi SB$ in the microscopic studies of hadrons, from Ref. \cite{Bicudo:1989si}.}
\label{fig:withXiSB}
\end{centering}
\end{figure}

In nuclear physics, Ribeiro 
\cite{Ribeiro:1978gx} 
used the quark content of the nucleon to derive the $NN$ repulsion at short distances. The $NN$ repulsive core can be computed from quark models, applying the Resonating Group Method (RGM, devised by Wheeler for Nuclear Physics) 
\cite{Wheeler:1937zz}
to the quarks and their interactions. The main mechanism is the interplay of the Pauli principle 
for the  fermionic quarks and the hyperfine quark-quark interaction, present in the potential of Eq. (\ref{eq:qmOGEP}) as illustrated in Fig. \ref{fig:withXiSB}.
The same principle can be applied to any other hadronic interactions, for instance to $KN$ exotic scattering \cite{Bicudo:1987tz}. This work is very technical since the short range interaction of hadrons involves microscopically 4 particles for a tetraquark, 5 particles for a pentaquark or 6 particles for a hexaquark, with complicated potentials and colour, flavour, spin and Jacobi coordinates.

The spontaneous breaking of chiral symmetry has also been included in confining quark models in the eighties, 
see Fig. \ref{fig:withXiSB},
by Adler and Davis, le Yaouanc et al and Bicudo and Ribeiro.
The first interest of this approach was to compute with a Bogoliubov-Valatin transformation 
\cite{Adler:1984ri,Bicudo:1989sh} 
the quark condensate that breaks the chiral symmetry in the QCD vacuum, composed of $^3P_0$ scalar quark-antiquark pairs, computed with the non-linear mass gap equation. 
Then the light states have a more complicated boundstate equation, say a least a Salpeter or a Bethe-Salpeter equation, with more coupled channels due to negative energy components
\cite{LeYaouanc:1984ntu,Bicudo:1989si}, 
leading to a vanishing pseudoscalar meson mass in the chiral invariant limit. This explains the  large splitting between the $\pi$ and $\rho$ mesons. Equivalent to this is the Dyson-Schwinger  approach
\cite{Roberts:1987xc,Williams:1989tv,vonSmekal:1997ohs,Feuchter:2004mk}, 
a truncated version of diagrammatic QCD. These approaches can be also extended to the gluon sector of QCD
\cite{Szczepaniak:1995cw}. 
Moreover the authors of Ref. 
\cite{Bicudo:1989sh}
 were interested to study how S$\chi$SB impacts on the microscopic derivation of hadron-hadron interactions. 
The strength of the hadron-hadron interactions decreases, even more so for the interactions of the light mesons who are suppressed by the Adler zero. 
This led to very consistent works, proving that quark models can comply with the PCAC theorems \cite{Bicudo:2001jq,Bicudo:2003fp} 
including the Weinberg theorem for $\pi-\pi$ scattering 
\cite{Weinberg:1966kf}. 
However S$\chi$SB makes the quark models very technical, and harder to calibrate since few parameters are used but their impact on the results is non-linear.

With the S$\chi$SB, it is then possible to also include quark-antiquark annihilation and creation in the $RGM$, compute the hadron couplings to derive coupled channel equations for hadrons and the respective decay channels. The T matrix can be computed with the Lippmann-Schwinger series. Then the s matrix, defined by Wheeler
\cite{Wheeler:1937zz},
 can be analytically continued to the Argand space and its poles can be determined
\cite{Bicudo:1989sj}, providing the mass and decay width of hadronic resonances.

In what concerns string physics, quarks are attached to the ends of open strings (without quarks we may only have closed strings). Moreover the QCD strings may bifurcate and we have the $y$ string in baryons, and more complicated strings with two junctions in tetraquarks and with three junctions in pentaquarks. Thus the flux tubes of $SU(3)$ QCD correspond to more complex world sheets than the simplest string theories. They will be detailed in Subsection \ref{sec:static}.

\subsubsection{
Non-perturbative dynamical {\em s pole} tetraquark resonances   \label{sec:nonpert}}

A third type of mechanism produces resonant poles in the scattering matrix, due to meson exchange in the channel $s$ Mandelstam variable, forming crypto-exotic  systems, with a dominant tetraquark component. 

\begin{figure}[t]
\begin{centering}
\includegraphics[width=0.7\columnwidth]{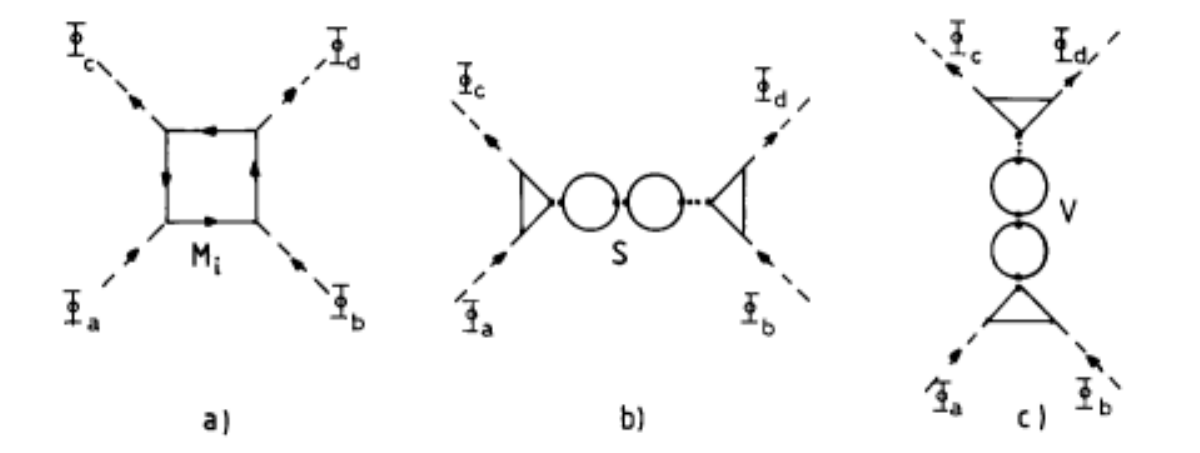}
\caption{Meson-meson scattering in the NJL model. $\phi_{a.b.c.d}$ denotes any pseudoscalar meson ($\pi, K, \eta$). (a) gives the box diagram, with $M_i$
a constituent quark of flavor i( i = u, d, s ). (b) and (c) display the scalar (S) and vector (V) exchange in the $s$ and $t$ channel, respectively. from Ref. \cite{Bernard:1990ye}.
\label{fig:sigmabubble}}
\end{centering}
\end{figure}

The threes types of strong interaction mechanisms able to drive the formation of tetraquarks are then,
\bie
\ie
quark-quark and antiquark-antiquark forces of Subsection \ref{sec:quark} at the QCD scale for {\em diquark} tetraquarks, in the case advocated by Richard 
\cite{Ader:1981db,Ballot:1983iv,Zouzou:1986qh,Gignoux:1987cn}, 
requiring an imbalance for the masses,
with two heavier quarks (or antiquarks) and two light antiquarks (or quarks), is certainly exotic;
however the diquark-antidiquark approach may also lead to crypto-exotic mesons
\cite{Maiani:2004vq};
\ie 
nuclear-physics like forces of Subsection \ref{sec:nuclearTh} with attractive meson exchange and repulsive short range for {\em molecular} tetraquarks, this is expected to produce resonances just below the threshold of the molecular system which may be exotic or crypto-exotic, for instance the OPEP attraction favours I=0; in this case the attraction is due to a $t$ channel exchange, illustrated in Fig. \ref{fig:sigmabubble} c),
\ie
non-perturbative coupled channel equations, driven by the coupling of a pair of mesons to quarkonium; when it is strong enough, it 
originates non-perturbatively poles in the s matrix, we denominate this third type  {\em s pole} tetraquark since is corresponds to an $s$ channel exchange, illustrated in Fig. \ref{fig:sigmabubble} b),
\eie 
The intrinsic QCD part of the forces, may partly be illustrated in Fig. \ref{fig:sigmabubble} a). It is clear in this figure that these three mechanisms are not exclusive, they may act conjointly to produce a tetraquark. It is important to understand, for each tetraquark, what is the dominant mechanism.

An example of the third type {\em s pole} tetraquark is the $\sigma$ meson. It was already advocated by Jaffe in 1974 as a crypto-exotic. The $\sigma$ meson took many decades to be understood because it is a wide resonance with a much lower mass that the scalar quarkonium groundstate, but presently it is confirmed 
\cite{Garcia-Martin:2011iqs} 
by detailed analyses, with a large $\pi \pi$ component.

The $\sigma$ meson is related to most of the points discussed in this subsection. It was anticipated as a meson, t channel exchanged, in the $N-N$, necessary to bind the deuteron. Then it appeared in the first models of S$\chi$SB as the massive partner of the pion \cite{Goldstone:1961eq} which is massless in the chiral limit. Effective meson models, with a coupled channel of two pions and a scalar meson, where the pion pair can annihilate into a scalar meson  \cite{Tornqvist:1982yv,vanBeveren:1986ea}, find an extra, dynamical pole in the s matrix, corresponding to the $\sigma$ meson. The inverse mechanism to the meson pair annihilation is string breaking. The coupling can also be computed microscopically in a quark model \cite{Ribeiro:1981fk,vanBeveren:1982sj}, in an approach complying with the analyticity of the s matrix, and this case an extra full nonet of scalar mesons, with poles in the second Riemann sheet of the s matrix analytical continuation, is obtained \cite{vanBeveren:1986ea}.

Notice, if the coupling is small, then the Born approximation can be used to compute the decay width of the scalar meson, equivalent to the Fermi Golden rule. In this case there is a single pole in the s matrix, close to the energy of the scalar quarkonium meson. But when the coupling is strong, a new pole is dynamically produced in the s matrix, far from the scalar $q \bar q$ meson mass.  The content of this resonance has a small $q \bar q$ component and a large $q q \bar q \bar q$ component, it can be interpreted as a crypto-exotic tetraquark.

The S$\chi$SB can also be included  in a consistent way in this mechanism. Using the Nambu and Jona-Lasinio chiral invariant interaction, Bernard et al \cite{Bernard:1990ye} computed a series of bubble diagrams in Fig.  \ref{fig:sigmabubble} equivalent to the exchange of a scalar meson in the $s$ Mandelstam variable. They reproduced the Weinberg theorem \cite{Weinberg:1966kf} for $\pi-\pi$ scattering. This was generalized to any chiral invariant quark model and to the Dyson-Schwinger approach in Refs. \cite{Bicudo:2001jq,Bicudo:2003fp}.

\begin{figure}[t!]
\begin{center}
\includegraphics[width=0.6\columnwidth]{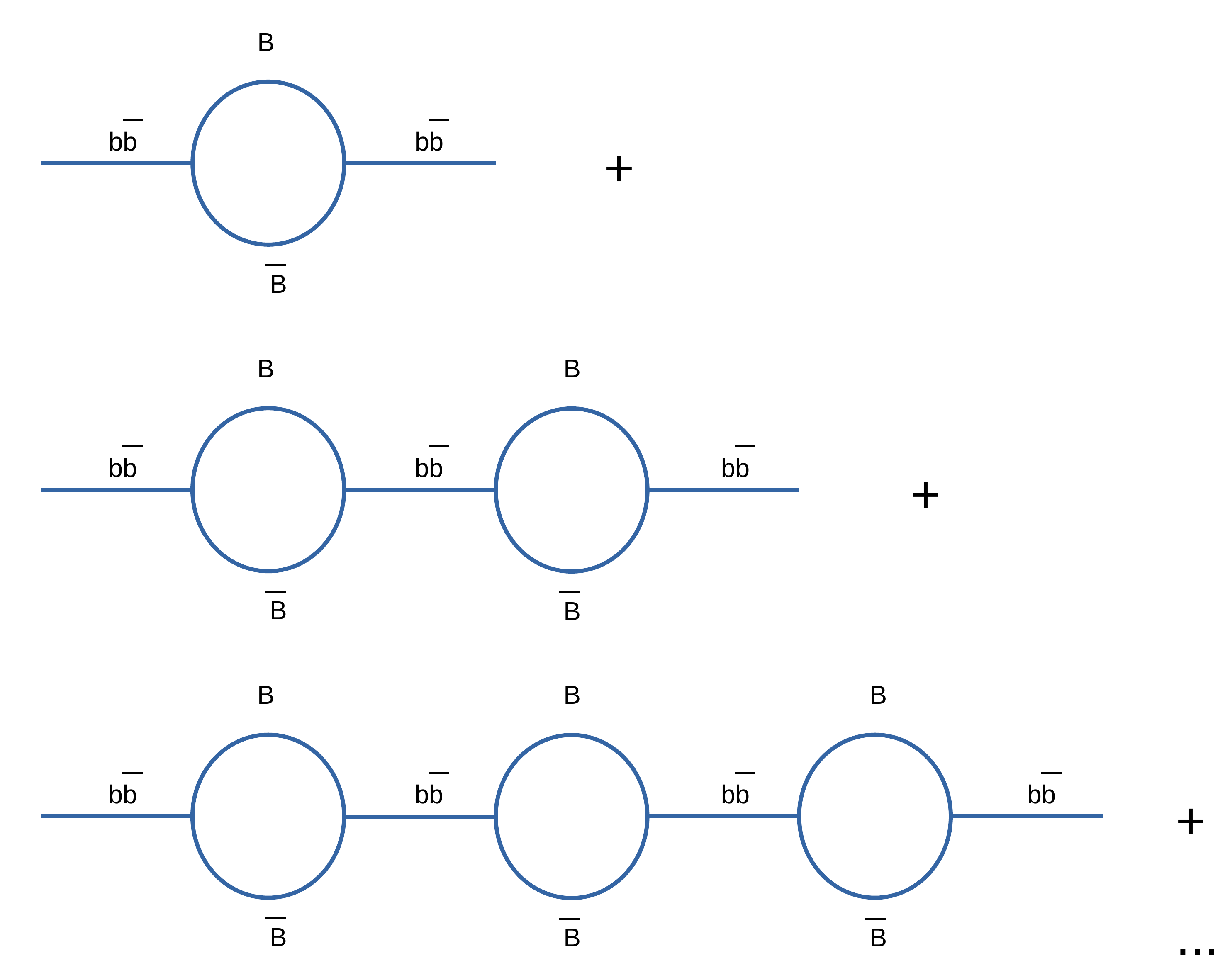}
\caption{Example of a Lippmann-Schwinger series for the $T$ matrix with bubble diagrams contributing to an {\em s pole} tetraquark. If the coupling between the quarkonium and the meson-meson channel is small, the first order term, equivalent to the Born approximation, suffices to compute the quarkonium decay width. However, if the coupling is large, a new pole may appear in the series, with a large meson-meson component: this is the {\em s pole} tetraquark.}
\label{fig:Spole}
\end{center}
\end{figure}

Then, the mechanism is the $s$ channel exchange of a scalar meson between the $\pi-\pi$ pair. This is allowed in meson-meson systems where a quark-antiquark pair may be annihilated, but not in nuclear physics where only $t$ channel meson exchange is expected. In the case of $\pi-\pi$ scattering, the pion pair partly annihilates into a scalar meson. This produces a series of bubble diagrams as in Fig. \ref{fig:Spole}, similar to the ones exchanged in the s channel in Fig.  \ref{fig:sigmabubble} b) but with effective mesons instead of quarks.  

This mechanism is expected to also hold when at most a quark and an antiquark are heavy, since we need a light quark-antiquark pair to be created from the quarkonium and to be annihilated again, repeatedly, in a series of bubble diagrams. For instance, most of the crypto-tetraquark candidates of Table \ref{tab:cryptoTetra} might be of the {\em s pole} tetraquark type.

%

\subsection{
Minimal overview of the lattice QCD techniques used here  \label{sec:latticeQCD}}

Historically, theoretical models have been crucial to understand the concepts of hadronic physics. Nature hides its fundamental hadronic building blocks, the quarks and gluons. They appear indirectly  in many different phenomena, who need particular perspectives to be understood. There is colour and its confinement, several different flavours with different quark mass scales, spontaneous partial chiral symmetry breaking, asymptotic freedom, the deconfinement crossover transition, nuclear physics, nuclear matter, etc , as partly reviewed in Subsection \ref{sec:theory}.

Hadronic models, say quark models or effective hadron models, are important to understand all these complex concepts and they can reproduce the experimental results up to a good approximation. However the larger the number of experiments they reproduce, the larger the number of needed parameters, and the most comprehensive models end up by being cumbersome to work out.

%

So far, the only first principle non-perturbative approach to hadronic physics is Lattice QCD
\cite{Wilson:1974sk}, 
using explicitly the QCD Lagrangian with no approximations except for its cutoffs
\cite{Creutz:1984mg,Gattringer:2010zz}.
This statement should be taken with a grain of salt, because there are other different approximations to QCD. But, nevertheless, lattice QCD is, so far, the only approach able to get arbitrarily close to QCD. The first computations were performed by Creutz in the late seventies
\cite{Creutz:1979zg,Creutz:1979dw,Creutz:1980zw,Creutz:1980wj,McLerran:1980pk}, 
and since then, with the exponential development of computers, the lattice QCD results have been converging to the experimental data.

Notice there are two different types of QCD lattices. Quenched lattice QCD only includes the gluon fields, in the links of the lattice.  Dynamical lattice QCD includes as well the quark fields, in the points of the lattice. For the gluonic observables, such as glueballs or static quarks cases both produce similar results, for instance for potentials, such as the string tension and the Coulomb potential.
In other phenomena, quenched and dynamical QCD are quite different. For instance in phase transitions, pure gauge QCD has a deconfinement transition of first order at temperature $T_d \sim 260$ K , whereas dynamical QCD has a deconfinement and chiral transition which is a crossover at a much smaller temperature of $T_d \sim 155$ K. 

In the Lagrangian of QCD 
\cite{Gross:1973id,Politzer:1973fx}
there are few physical parameters, just the current/bare quark masses because the coupling is dimensionless. 
While the analytic proof of confinement remains a major open problem in theoretical physics, in lattice QCD the confinement scale emerge from the numerical computations, since the quantisation on the lattice produces a dimensional transmutation.
When the computations are precise enough, with small lattice spacings $a_s , \, a_t$ and large space-time volume ${a_s}^3\, N_x\, N_y\, N_z \times a_t\, N_t$, lattice QCD gets results arbitrarily close to QCD, and so far the results also get arbitrarily close to the experimental ones. This is the main proof that QCD is indeed the fundamental theory of strong interactions.

However lattice QCD also has its technical problems. An obvious problem is that lattice QCD is technically very demanding. Monte Carlo techniques borrowed from statistical physics, derived from the path integral, require the full power of high performance computing both in computational power and in coding efficiency. Computations are so far limited to lattices of size $ \ll 100^4$. Thus there are lattice QCD groups only in rich enough countries, essentially in the same countries who are able to perform high energy physics experiments.

Moreover, contrary to models, in lattice
 QCD gluons are much easier to understand and cheaper to study than fermions which creators are Grassmann variables. Finite temperature is also easier to study than a vanishing one. In other words, the lightest the fermions and the smallest the temperature, the more expensive the computations. Nevertheless, there are different approaches to circumvent the difficulties of using fermions, and systems with up to six dynamical quarks have been studied with lattice QCD.

The third difficulty of lattice QCD resides in the necessity of an Euclidean space to have a positive definite probability density.  This space is reached with a Wick rotation. The probability is then the exponential of minus the action $e^{-S_E[\Phi]} $,
where we use the notation of Gattringer and lang \cite{Gattringer:2010zz}.
While some computations turn out to be difficult, with a sign problem in the action, the hadron spectra at zero fermion density can be computed in lattice QCD. This is the foundation stone of all the lattice QCD computations we are reporting here.

So far, the main techniques used to study tetraquarks in Lattice QCD are all based in computing the energy of observables, extracted from the evolution operator of correlation matrices in Euclidean space,
\bea
\langle O_2(t) O_1(0) \rangle_T&=& {1 \over Z_T} 
\int
{\cal D} [\Phi]
e^{-S_E[\Phi]} 
O_2[\Phi(\cdots n_t)]   O_1[\Phi(\cdots 0)] 
\non \\
Z_T&=&
\int
{\cal D} [\Phi]
  e^{-S_E[\Phi]} 
  \non \\
  {\cal D} [\Phi] &=& \prod_{n \in \Lambda} \text{d} \, \Phi(n)
\label{eq:correlation}
\eea
where $O_1$ and $O_2$ are two operators sharing similar quantum numbers, and $t$ is the Euclidean time. $T$ is the temporal size of the lattice, close to zero temperature it should be very large.
The probability density is utilized to generate an ensemble of configurations that capture the physics of the system we are studying. The results are obtained statistically, the larger the ensemble the smallest the error bar. In this sense the results of lattice QCD are similar to the experimental results in quantum physics, where small error bars result from many independent observations. But in lattice QCD we can ask different, more theoretical questions that in the real experiment. Thus lattice QCD complements both the experiments and the theoretical hadronic models.

%
\begin{figure}[t!]
\begin{center}
\includegraphics[width=0.35\columnwidth]{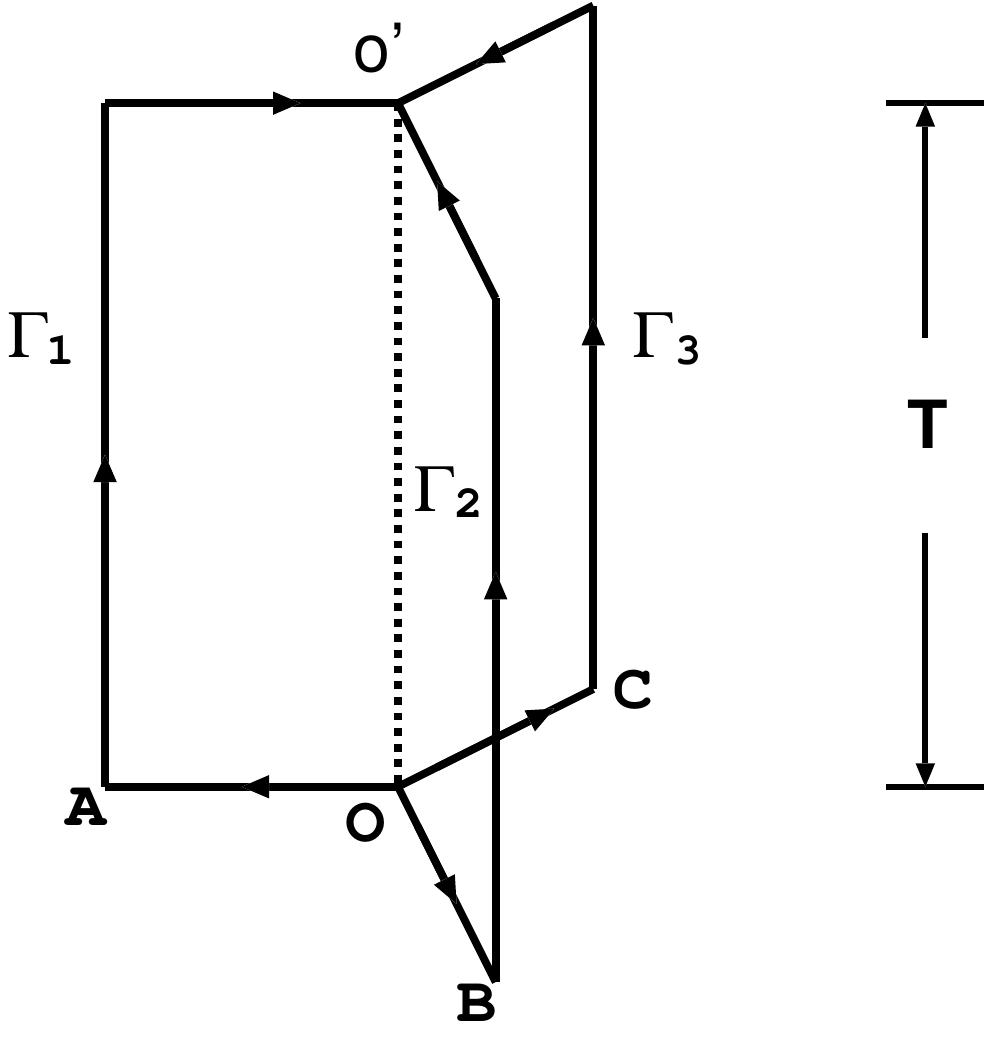}
\hspace{10pt}
\includegraphics[width=0.6\columnwidth]{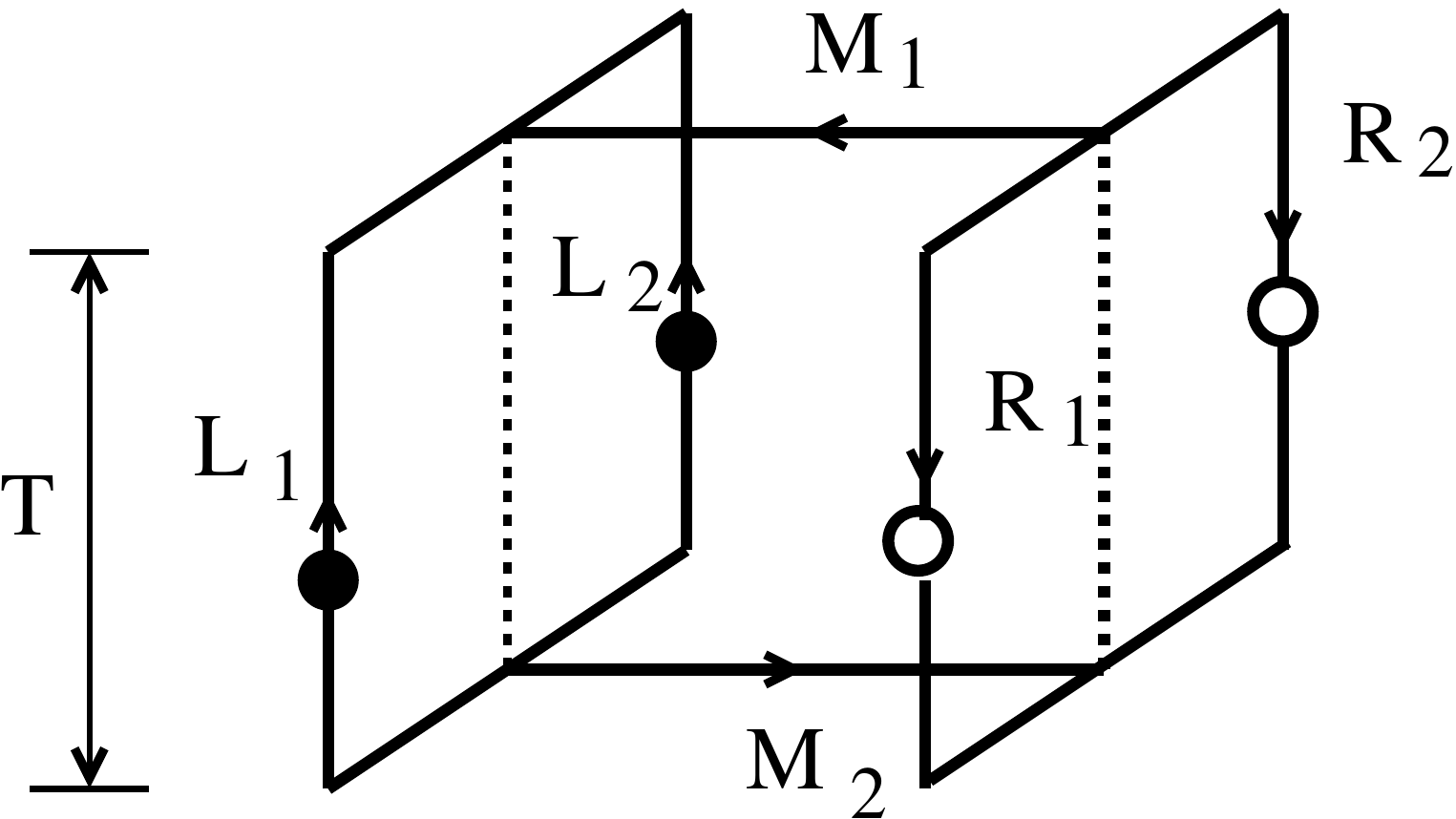}
\end{center}
\caption
{(left) Baryon (3Q) Wilson loop for the calculation of the static (3Q) potential from Refs. 
\cite{Takahashi:2000te,Takahashi:2002bw}.
(right) The tetraquark (4Q) Wilson loop 
for the calculation of the 4Q potential $V_{\rm 4Q}$, utilized in Ref. \cite{Okiharu:2004ve}, is obtained as an extension of the (3Q) Wilson loop, with two Levi-Civita junctions per operator (four in total in the Wilson loop).
The 3Q and 4Q gauge-invariant states are generated at $t=0$ 
and are annihilated at $t=T$.
\label{fig:Okiharu2004wy}}
\end{figure}

When the Wick rotation changes from the Minkowski space-time to the Euclidean space-time the time evolution operator also changes from an oscillating function to a decaying exponential,
\be
e^{ - i \widehat H t} \to e^{- \widehat H t}
\label{eq:time}
\ee
Thus, if we compute the evolution operator, it should decay exponentially. And fitting this exponential, we extract the energy matrix $\widehat H$ of our system.

Let us show this in the simplest case, where we just compute the groundstate energy. Denoting $| O_1(0) \rangle$  the physical quantum state corresponding to the operator $O_1(0)$, it can be decomposed in eigenvectors of the hamiltonian,
\be
|O_1 (0) \rangle = \sum_i c_i |v_i \rangle
\ee
and the matrix element between the same operator at different times is,
\be
\langle O_1 (t)| O_1(0) \rangle = \sum_i c_i^* c_i e^{- \lambda_i t}\ ,
\label{eq:project}
\ee
clearly for a time long enough the groundstate dominates this matrix element.
 Technically, we compute the effective mass,
\be
m_\text{eﬀ} (n_t + 1/2) = \ln {C(n_t ) \over C(n_t + 1)} \to \lambda_0
\label{eq:plateau}
\ee
and plot as a function of the time separation $n_t$, it to identify a plateau before the noise gets larger than the signal.

%
\begin{figure}[t!]
\begin{center}
    \includegraphics[width=0.7\columnwidth]{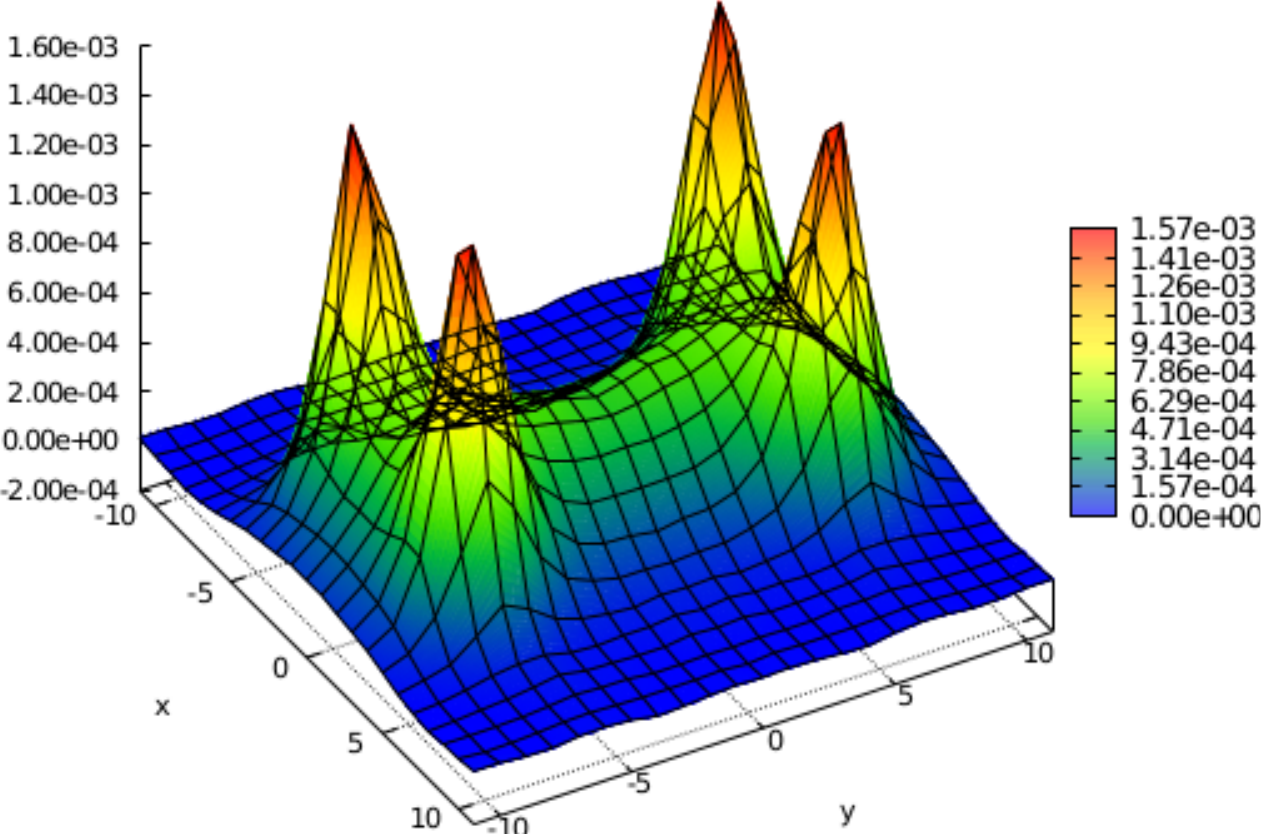}
\end{center}
    \caption{Lagrangian density 3D plot for a tetraquark with $r_1=8,  \, r_2=14$, from Ref. \cite{Cardoso:2011fq},
showing a clear a tetraquark double-Y flux tube. The results are presented in lattice spacing units (colour online).
    \label{fig:fields}}
\end{figure}

Thus, assuming the ensemble of configurations is large enough, and employing noise reduction techniques
\cite{Brower:1981vt,Parisi:1983hm,APE:1987ehd,Luscher:2001up,Cardoso:2013lla}, 
to have a good signal over noise ratio, what we have choose is a set of operators $O_1, O_2 \cdots$ to describe our system, and compute the correlation matrix between these operators at time $t$ and time $0$ to determine the time evolution of our operators.
If we are clever enough, we can study important theoretical structures such as potentials, a spectrum, the wavefunctions / Bethe-Salpeter amplitudes of the  groundstate and excited states, and scattering matrices.

We remark the correlation matrix elements must be gauge invariant (unless we want to use some gauge fixing) since the sum over the configurations also sums over the gauges, and gauge dependent correlations would have a vanishing mean value. To devise gauge invariant correlators, notice each lattice link (connecting two neighbour points) transforms under a local set of SU(3) matrices $\Omega(\bm x)$ depending on the position vector $\bm x$,
\be
U_{\bm \mu}(\bm x) \to \Omega( \bm x) U_{\bm \mu}(\bm x) \Omega^\dagger( \bm x + \bm \mu)
\ee
and fermion propagators transform in the same way, albeit they can connect any pair of points of the lattice,
\be
{S(\bm x, \bm y)^{\alpha \beta}}_{ab} \to \Omega(\bm x)_{aa'} {S(\bm x, \bm y)^{\alpha \beta}}_{a' b'} \Omega(\bm y)^\dagger_{b' b} .
\ee
We can get gauge invariant objects either using a closed loop of links, or with a path composed of gauge links and propagators. In this case, all the gauge transformation matrices contribute in products $\Omega^\dagger(\bm x) \cdot \Omega(\bm x)=1$.

Thus the simplest operator we can have is just a point $\bm x$ in space. To construct a gauge invariant correlator, it is connected with a path of links (or propagators) to other operators in the form,
\be
\cdots {U_{\bm \mu}}_{a\, c}^\dagger (\bm x) {U_{\bm \nu}}_{a\, b}(\bm x) \cdots
\label{eq:ginv2}
\ee
and if the path is closed in loops, we just need the group to be unitary for this to be gauge invariant. A simple example is the Wilson Loop, a closed rectangular loop composed by a chain of links. Another example is the Polyakov loop in a lattice with periodic boundary conditions, a purely temporal loop winding around the toroidal lattice.

Another interesting example of an operator, relevant for the study of tetraquarks, is the operator creating a baryon, used in Fig. \Ref{fig:Okiharu2004wy} (left). If three links (or $N_c$ links in the more general case of $SU(N_c)$ links converge or diverge at one point, then we can construct an operator with a Levi-Civita Symbol, connected with links in this form,
\be
\varepsilon_{a\, b \cdots} {U_{\bm \mu}}_{a\, a'}(\bm x) {U_{\bm \nu}}_{b\, b'}(\bm x) \cdots
\label{eq:ginv3}
\ee
which is gauge invariant if 
\be
\varepsilon_{a\, b \cdots} {\Omega}_{a\, a'}(\bm x) {\Omega}_{b\, b'}(\bm x) \cdots = \varepsilon_{a' \, b' \cdots}
\ee
where the indices are colour indices.
Let us start by the simplest particular case $a'=1, \, b'=2 \cdots = N_c$,
\be
\varepsilon_{a\, b \cdots} {\Omega}_{a\, 1}(\bm x) {\Omega}_{a\, 2}(\bm x) \cdots = \text{det} [ \Omega ]
\label{eq:det}
\ee
and since $\Omega$ is a $SU(N_c)$ matrix this determinant is $1=\varepsilon_{1\, 2 \cdots}$. Any other permutation of the indices $1\, 2 \cdots$ produces a $\pm 1$ phase consistent with the Levi-Civita symbol. And if any indices are identical, then Eq. (\ref{eq:ginv3}) vanishes. This demonstrates Eq. (\ref{eq:det}) and completes the proof of the gauge invariance of the baryon operator. Notice this operator needs the gauge group to be special (its matrices having determinant 1).

%
\begin{figure}[t]
\begin{centering}
\includegraphics[width=0.8\columnwidth]{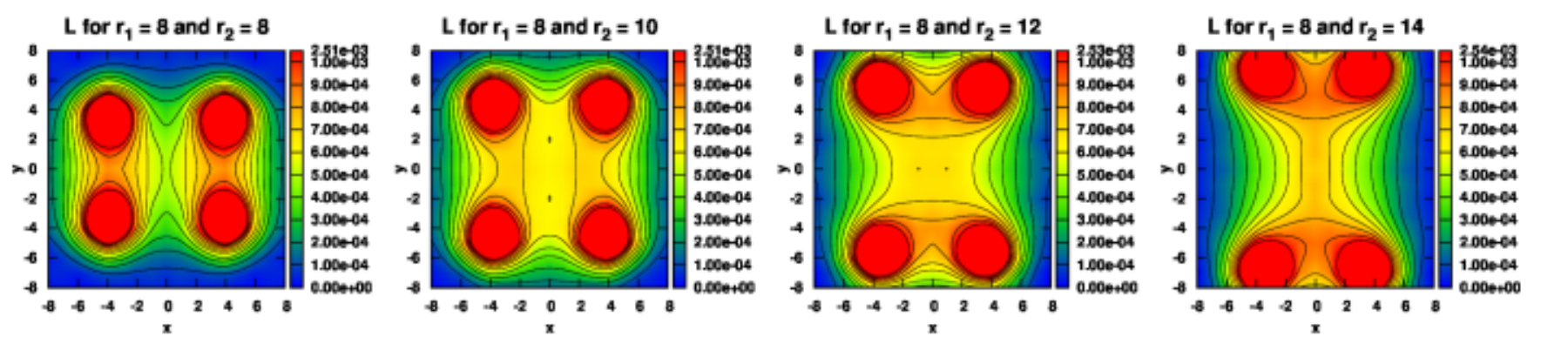}
\caption{Evidence of flip-flop of the flux tube in the colour field densities for the $Q Q \bar Q \bar Q$ system, as in Ref. \cite{Cardoso:2012uka}. For different positions of the quarks and antiquarks, we can either get flux tubes of two mesons or the flux tube of a diquark-antidiquark tetraquark.
\label{fig:flipflop}}
\end{centering}
\end{figure}

Starting from the operators in Eqs. (\ref{eq:ginv2}) and (\ref{eq:ginv3}), combining with more links to smear the source, or with Dirac matrices to connect with propagators in case we have dynamical quarks, we can build any operator and the respective correlators necessary to study hadrons. An example is shown in Fig. \Ref{fig:Okiharu2004wy} (right) where with four Levi-Civita Symbols and a triple loop, we can construct the generalized Wilson loop to study the tetraquark static potential, detailed in Subsection \ref{sec:static}.

There are three types of quarks used in the study of tetraquarks: the dynamical quarks, the non-relativistic quarks and the static quarks. The most expensive quarks are the fully dynamical ones. Quarks are fermions and their field operators are Grassman variables. The path integral summation implies the computation of the determinant of the Dirac fermion matrix operator (Euclidian),
\be 
\gamma_\mu ( \partial_\mu + i A_\mu ) + m 
\label{eq:dirac}
\ee
where the Dirac matrices are multiplied by a phase 
\cite{Gattringer:2010zz}
to make them Hermitean $\gamma_i^E=-i \gamma_iM$ and $\gamma_4^E=\gamma_0^M$. Here we don't detail the terms necessary to avoid chiral doublers \cite{Ginsparg:1981bj}. A solution to the problem of computing the determinant is in the use of pseudo-fermions, gauge fields with an action inverse to the fermion action. Moreover the generation of configurations with the hybrid Monte Carlo method is difficult with light quarks but this was solved using a  preconditioner with a heavier mass. Possibly this is enabled by spontaneous chiral symmetry breaking, who generated for the light quarks a  constituent mass much larger than its bare mass, as detailed in Subsection \ref{sec:quarkfeedback}. Moreover the deflation technique, factorizes the Dirac quark matrix in low energy modes and in high energy modes, where the low energy modes are the ones more sensitive to confinement and chiral symmetry breaking, and this also speeds up the necessary inversion of the fermion determinant. Then, for systems with light valence quarks we have to compute the quark propagators, inverse of the Dirac matrix in Eq. \ref{eq:dirac}.

For the bottom quark, it is not yet possible to apply the dynamical quark approach. The bottom quark is quite heavy and would need too small lattice spacings. But we can use non-relativistic QCD which is less expensive on the lattice.  In this case the Lagrangian is expanded in the velocity $v$, and only the leading orders are kept. Then the propagator is quadratic in the momentum and several of the dynamical quark problems are avoided. 
Davies et al 
\cite{Davies:1994mp}
 use an improved lattice NRQCD Hamiltonian \cite{Gray:2005ur}:
\begin{equation}
H_0=-\frac{\Delta^{(2)}}{2M^0}
\end{equation}
\begin{eqnarray}
\delta H &=& -c_1\frac{(\Delta ^{(2)})^2}{8(M^0)^3} + c_2\frac{ig}{8(M^0)^3}
({\Delta \cdot E - E \cdot \Delta})
-c_3\frac{ig}{8(M^0)^3}{\bf\sigma}\cdot({ {\Delta \times \tilde{E}} - {\tilde{E} \times \Delta}})\nonumber\\
&&-c_4\frac{g}{2M^0}{\bf\sigma}\cdot {\bf\tilde{B}} 
+ c_5\frac{a^2\Delta^{(4)}}{24M^0}
-c_6\frac{a(\delta^{(2)})^2}{16n(M^0)^2}.
\end{eqnarray}
We thus get a Schrödinger equation, with a quadratic kinetic energy, no doublers and we don't need to use a fine mesh lattice.

%
\begin{figure}[t]
\begin{centering}
\includegraphics[width=0.7\columnwidth]{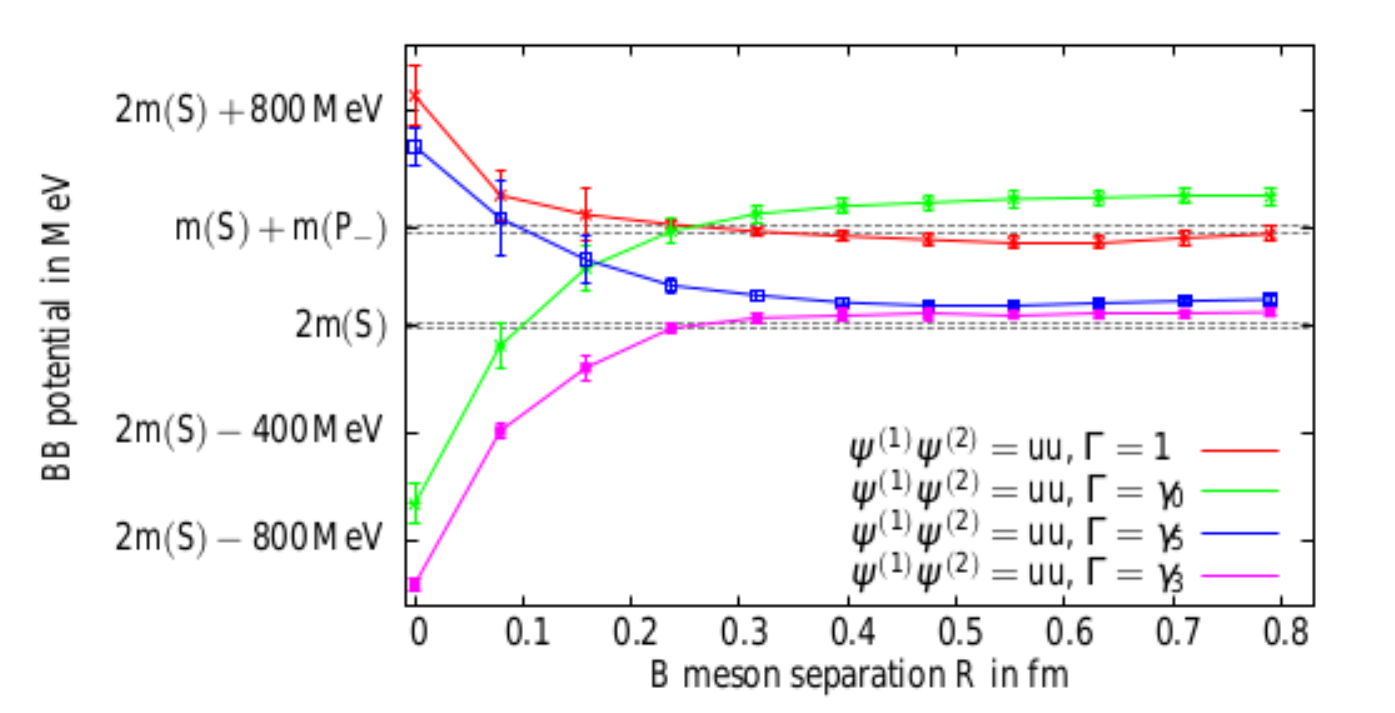}
\caption{Different potentials between two static-light mesons, in the two static antiquark and two dynamical quark case, from Refs.  \cite{Wagner:2010ad,Wagner:2011ev}
\label{fig:potentials}}
\end{centering}
\end{figure}

An even more economical approach, used for the computation of potentials, is the static quark approach, where the quarks are assumed to be infinitely heavy. As expected, an infinitely heavy quark sits in the same spatial point, just propagating in time.
A simple example is the Wilson Loop, a rectangular loop with a pair of spatial sides and a pair of temporal sides. This loop is used to compute static potentials where the spatial distance is the distance between static charges and the temporal distance is the time used in the evolution  of Eqs. \ref{eq:time} and \ref{eq:plateau}. Besides, it is possible to compute the spin dependent potentials as well, as detailed in Subsection \ref{sec:diquark}.

\subsection{
Summary}


The goal of this work is to review the progress of lattice QCD in the study of tetraquarks and pentaquarks with one or more heavy quarks,  and to compare the lattice QCD state of the art with the experimental results and with the theoretical paradigms of hadronic physics. The lattice QCD techniques are presented with a minimal detail, for the experimentalists and theorists to understand how the lattice QCD results are computed. We review all the different results so far obtained, for the experts in lattice QCD to comprehend the recent progress of this topic.

In this section I, we briefly review the onset of the multiquark studies, all the experimental results, the main theoretical ideas, and the essential tools of lattice QCD.

In Section II we review in detail the different lattice QCD approaches to address exotic tetraquarks with one or more heavy quarks. The tetraquarks have a section of their own because they are the simplest multiquarks and also the ones most studied in lattice QCD.

We leave the pentaquark review for Section III together other lattice QCD studies, with some affinity to multiquarks with heavy quarks. We also include hybrids, since these exotics may also have the same quantum numbers of tetraquarks. The only exotic systems we leave out of our review are the glueballs which have no quarks, and the hexaquarks which broadly include nuclei such as the deuteron and thus have a wide literature of their own.

In Section IV we conclude comparing lattice QCD results on tetraquarks and pentaquarks with the experimental results. We review the existing difficulties of lattice QCD and present our outlook on expected future lattice QCD results.

\section{Review of the different lattice QCD approaches to address exotic tetraquarks with heavy quarks \label{sec:difapp}}

\subsection{
Static potentials and colour field densities for tetraquarks and other multiquarks \label{sec:static}}

The first computations for tetraquarks in lattice QCD used static quarks only, since  static quarks are less expensive than dynamical quarks, and can be computed in pure gauge QCD. This effort was contemporary with the claimed observation of the $\theta^+$ in 2003  \cite{LEPS:2003wug}, which at the time motivated more multiquark studies.

Lattice QCD computed potentials for tetraquarks and pentaquarks using static quarks, by Okiharu et al. \cite{Okiharu:2004wy,Okiharu:2004ve} and Alexandrou et al. \cite{Alexandrou:2004ak}. 
The computations employ extended Wilson loops, including in each tetraquark operator two Levi-Civita symbols (one for each Steiner junction) 
to account for the diquark-antidiquark system, as in Fig. \ref{fig:Okiharu2004wy}. 
Pentaquark operators have three Levi-Civita symbols.
In particular Ref. \cite{Okiharu:2004ve} analyses in great detail the potential for different geometries of the four-quark system. The static potential is computed with generalized Wilson loops depicted in Fig. \ref{fig:Okiharu2004wy}.

%
\begin{figure}[t]
\begin{centering}
\includegraphics[width=0.45\columnwidth,trim=60pt 605pt 300pt 100pt, clip]{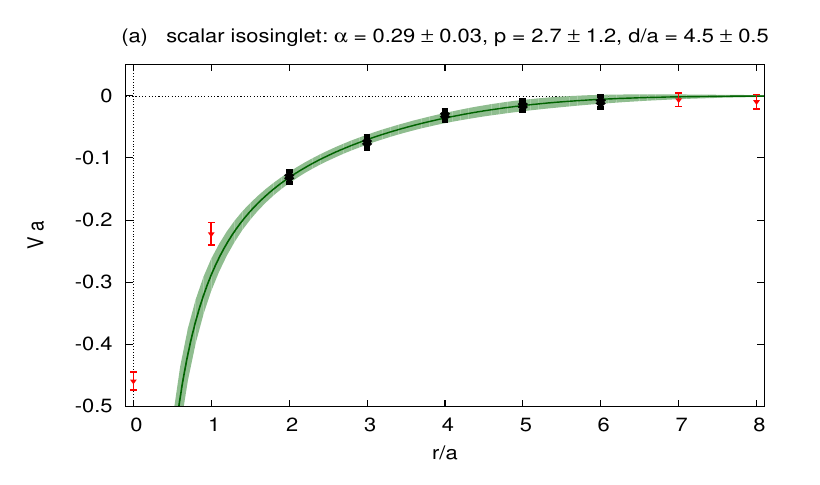} \ \
\includegraphics[width=0.45\columnwidth,trim=60pt 470pt 300pt 235pt, clip]{1.pdf}\caption{The lattice QCD potentials fitted in Ref. \cite{Bicudo:2012qt}, with quantum numbers for the light quark pair (left) scalar-isoscalar and (right) vector-isovector. Notice, with static quarks, the potentials don't depend on the heavy antiquark spins, since the heavy antiquarks are in the infinite mass limit. 
\label{fig:pot2010}}
\end{centering}
\end{figure}

The authors find that this potential extends the 
the static $Q\rm \bar{Q}$ potential 
which is well known from lattice QCD studies to be fitted by the expression 
\cite{Bali:1994de}
\begin{eqnarray}
V_{\rm Q\bar{Q}}
&=&-\frac{\alpha_{\rm Q\bar{Q}}}{r}
             +\sigma_{\rm Q\bar{Q}}r+C_{\rm Q\bar{Q}},
\label{QQbar}
\end{eqnarray}
and the static $QQQ$ potential which is presently understood to be 
the sum of the OGE Coulomb term 
and the Y or mercedes-benz star type linear confinement term as
\begin{eqnarray}
V_{\rm 3Q}=-\alpha_{\rm 3Q}\sum_{i<j}\frac{1}{|{\bf r}_i-{\bf r}_j|}
             +\sigma_{\rm 3Q}L_{\rm min}+C_{\rm 3Q}.
\label{eq:3body}
\end{eqnarray}
It was found by Fermat centuries ago that the minimal path $L_{\rm min}$ connecting the three charges corresponds to an angle of $2 \pi /3$ (unless the charge triangle is too flat for that) between the three linear string segments, meeting at a Steiner junction. Initially there was a discussion on whether the confining potential was a sum of two-body linear potentials (a Delta shaped potential) 
\cite{Sommer:1984xq,Sommer:1985da,Bali:2000gf,Alexandrou:2001ip}
or a single three body potential, since the difference between both potentials is subtle \cite{Dmitrasinovic:2013jta,Dmitrasinovic:2018irg}. The last lattice QCD
\cite{Takahashi:2000te,Takahashi:2002bw,Koma:2017hcm} results  show evidence for the string $Y$ shaped potential of Eq. (\ref{eq:3body}). Nevertheless it more lattice QCD results are welcome to definitely settle this discussion \cite{Dmitrasinovic:2013jta,Dmitrasinovic:2018irg}.

For the $QQ\bar Q \bar Q$ potential the authors find that the
potential $V_{\rm 4Q}$ is described by 
the OGE Coulomb plus a four-body potential $V_{\rm c4Q}$, with a double-Y or H extension of the three-body linear potential,
\begin{eqnarray}
V_{\rm 4Q}
&=&-\alpha_{\rm 4Q}\{(\frac{1}{r_{12}}+\frac{1}{r_{34}})
    +\frac{1}{2}(\frac{1}{r_{13}}+\frac{1}{r_{14}} \nonumber \\
& &             +\frac{1}{r_{23}}+\frac{1}{r_{24}}) \}
   +\sigma_{\rm 4Q}L_{\rm min}+C_{\rm 4Q} 
\label{Vc4Q}
\end{eqnarray}
with $r_{ij} \equiv |{\bf r}_i-{\bf r}_j|$ and 
$L_{\rm min}$ being the minimal value of the total flux-tube length.
Here, ${\bf r}_i$ denotes the location of $i$th (anti)quark.
For the $QQQQ \bar Q$ potentials the authors find,
\begin{eqnarray}
V_{\rm 5Q}
&=&-\alpha_{\rm 5Q}\{ ( \frac1{r_{12}}  + \frac1{r_{34}}) 
+\frac12(\frac1{r_{15}} +\frac1{r_{25}} +\frac1{r_{35}} +\frac1{r_{45}}) \nonumber \\
&+&\frac14(\frac1{r_{13}} +\frac1{r_{14}} +\frac1{r_{23}} +\frac1{r_{24}}) \}
+\sigma_{\rm 5Q} L_{\rm min}+C_{\rm 5Q},
\label{V5Qth}
\end{eqnarray}
Together with previous studies, the authors find these potentials are compatible, for the Coulomb potentials with the Casimir scaling present in the One Gluon Exchange potential
and the OGE result for the Coulomb coefficient as 
\begin{eqnarray}
\alpha_{\rm Q\bar{Q}} \simeq 2\alpha_{\rm 3Q} \simeq 
2\alpha_{\rm 4Q} \simeq 2\alpha_{\rm 5Q} \simeq 0.27 \ ,
\end{eqnarray}
and for the Linear potentials with the flux tube picture,
being fitted with universal parameters,
\begin{eqnarray}
\sigma_{\rm Q\bar{Q}} \simeq \sigma_{\rm 3Q} \simeq 
\sigma_{\rm 4Q} \simeq \sigma_{\rm 5Q} \simeq (420{\rm MeV})^2 \ .
\end{eqnarray}
As for the constant shifts of each potential, they
are non-scaling nonphysical quantities appearing in the lattice regularisation $a$, 
and we find an approximate relation as 
\begin{eqnarray}
\frac{C_{\rm Q\bar Q}}{2} \simeq \frac{C_{\rm 3Q}}{3} 
\simeq \frac{C_{\rm 4Q} }{4} \simeq \frac{C_{\rm 5Q} }{5} \simeq 0.32 a^{-1} \ .
\end{eqnarray}

Then, after the $Z_c$ and the $Z_b$ tetraquarks were discovered with a large experimental significance, there was a new and more lasting interest in tetraquarks from lattice QCD. 
The colour electric and colour magnetic square field densities were studied with four static quarks, by Cardoso et al. \cite{Cardoso:2011fq,Cardoso:2012rb}. 
In Fig. \ref{fig:fields} we show the QCD flux tube for a static tetraquark system. Clearly, It is a double-y  potential with two Steiner junctions.

{
\begin{table}[t!]
\begin{center}
\begin{tabular}{ccccc}
\toprule[1pt]
channel & $\alpha$ & $d/a$ & $p$ & $\chi^2 / \textrm{dof}$ \\
\hline
scalar isosinglet  &  $0.293(33)$  &  $4.51(54)$  &  $2.74(1.20)$  &  $0.35$ \\
vector isotriplet  &  $0.201(77)$  &  $2.48(69)$  &  $2.0$ (fixed) &  $0.06$ \\
\bottomrule[1pt]
\end{tabular}
\caption{ \label{ab:fit AEK} $\chi^2$ minimizing fit results of the ansatz (\ref{eq:ansatz}) to the lattice static antiquark-antiquark potential; fitting range $2 \leq r/a \leq 6$; lattice spacing $a \approx 0.079 \, \textrm{fm}$}
\end{center}
\end{table}
}

\subsection{
String flip-flop potentials with static quarks and the colour of mesons pairs and tetraquarks  \label{sec:flipflop}}

In Subsection \ref{sec:static}, we only considered the potential for tetraquarks corresponding to a $\bar 3  \, 3$ {\em diquark} tetraquarks.
However, in the string picture, the lowest energy string, corresponding to the lowest energy potential, may also depend on the geometry of the system. This is considered in the string flip-flop potentials, utilized to study the decay of resonances. In some particular geometries, the lowest energy colour singlet may be the colour $1 \, 1$ of a meson-meson system, whereas in other geometries the lowest energy colour singlet may be the colour $\bar 3  \, 3$ of a diquark-antidiquark system. For instance we expect the meson-meson potential is energetically favoured when a quark is close to an antiquark, and the {\em diquark} tetraquark potential is favoured when the quarks (antiquarks) are close to each other.

In lattice QCD, the dependence of the potential on the positions of four quark systems has been studied. The string flip-flop was already observed in the first study of tetraquark potentials \cite{Okiharu:2004ve}. Then Cardoso et al 
\cite{Cardoso:2012uka}
confirmed this flip-flop studying the flip-flop of the colour field densities, see Fig. \ref{fig:flipflop}. 

Moreover, to study a quantum system, not only the groundstate is important, the excitations are important as well. We also notice it is not sufficient to consider the spatial dependence of the potentials, it is necessary to consider colour as well.
With such a system of two quarks and two antiquarks, there are two independent colour singlets. The orthogonal colour singlet to the $\bar 3  \, 3$ is the $6 \, \bar 6$. And the orthogonal colour singlet to the $1 \, 1 $ is the $8 \, 8$. Since, depending on the geometry of our system, we can have three different groundstate - two different $1 \, 1$ and one $ \bar 3 \, 3$, the next excited state can also be in different colour states. 
Recently the computation, not only of the groundstate potential, but also of the first excited potential was performed \cite{Bicudo:2017usw}. Indeed, the flip-flop was observed both in the groundstate and in the excited potential.  The mixing of both potentials in the transition region was also observed.

Importantly the flip-flop was studied with dynamical QCD configurations  \cite{Bicudo:2017usw}. The results, obtained both quenched and with dynamical QCD configurations are similar.

However, it turns out that, when applying the string flip-flop potential in the Schrödinger equation for such a four body system
\cite{Bicudo:2015bra}, 
more bound state tetraquarks of the type $T_{bb}$, discussed in Subsection  \ref{sec:staticdy} are obtained 
than the ones computed with lattice QCD. 
Possibly this excess of states occurs because the potentials extracted from lattice QCD don't include yet the spin-dependent terms, which would disfavour the isovector tetraquark which light quarks have a combined spin1. The computation of such terms in lattice QCD  is discussed in Subsection \ref{sec:diquark}.

\subsection{
Potentials with static heavy quarks $\bar Q \bar Q$ and dynamical light quarks for the $T_{QQ}$ family of tetraquarks   \label{sec:staticdy}}

The first confirmation of observed tetraquarks, of the $z$ family shown in Table \ref{tab:Tetra}, led to new efforts to compute tetraquarks with lattice QCD. The first lattice QCD evidence of tetraquarks was obtained  for the  $T_{QQ}$ family of tetraquarks, which is less difficult technically because the channels are just B-B channels, we don't have to deal with the mixing of different channels. The new lattice QCD evidence motivated many studies of the $T_{bb}, \, T_{bbs}, \, T_{cb}, \, T_{cc} $ \dots reported in Table \ref{tab:boundTetra}.
The first lattice computations for $T_{bb}$ and related states utilized static potentials computed with static heavy quarks and dynamical light quarks. The dynamics provided to the heavy quarks in the Schrödinger equation with the Born-Oppenheimer approximation 
\cite{Born:1927} 
led to the prediction of boundstates.

%
\begin{figure}[t]
\begin{centering}
\includegraphics[width=0.7\columnwidth]{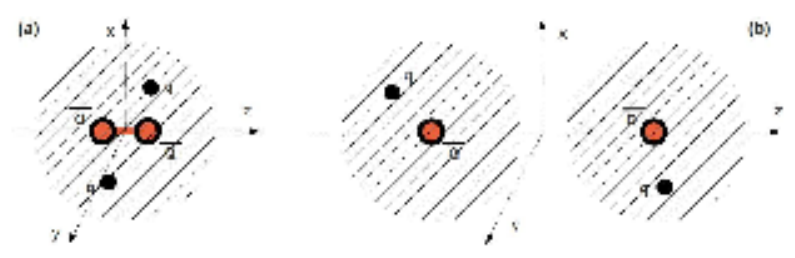}
\caption{The picture of perturbative one-gluon exchange at short distances and of meson wavefunction screening at large distances
proposed in 
Ref. \cite{Bicudo:2012qt}, 
consistent with the best ansatz in Eq. (\ref{eq:ansatz}).
\label{fig:screening}}
\end{centering}
\end{figure}

Static quarks have a fixed position, they are convenient for the study of potentials as a function of distance. They provide a good approximation in the case of  heavy quarks. A very interesting system to study is the family of $T_{QQ}$ tetraquarks with two heavy antiquarks and two light quarks, of the class proposed by Richard many years ago. These tetraquarks couple to systems of two heavy-light mesons. Using static heavy antiquarks, they couple to two static-light mesons. 

In lattice QCD, the study of this family of potentials was initiated with quenched quarks for the light quarks 
\cite{Mihaly:1996ue,Michael:1999nq,Doi:2006kx,Detmold:2007wk}. 
More recently, Wagner and others Refs. \cite{Wagner:2010ad,Wagner:2011ev,Bali:2010xa,Brown:2012tm},
extended this to dynamical light quarks, obtained with a comprehensive set of four quark operators of for instance of the form
\be
\label{EQN003} (\mathcal{C} \Gamma)_{A B} \Big(\bar{Q}_C(\mathbf{r}_1) \psi_A^{(1)}(\mathbf{r}_1)\Big) \Big(\bar{Q}_C(\mathbf{r}_2) \psi_B^{(2)}(\mathbf{r}_2)\Big) ,
\ee
where $\bar{Q}$ denotes a static quark operator, $\psi$ a light antiquark operator, $A$, $B$ and $C$ are spin indices and $\mathcal{C} = \gamma_0 \gamma_2$ is the charge conjugation matrix \cite{Wagner:2010ad,Wagner:2011ev}. With these operators it is possible to consider different light isospin $I$ and spin $S$ for the light quarks, with the dynamical potentials we can address full QCD.  More recently heavy quark spin effects were partly included
\cite{Bicudo:2016ooe},
the extrapolation to the chiral limit was performed
\cite{Bicudo:2015kna}
and the light quarks were extended from the $u, \, d$ flavours to the $s$ and $c$ flavours
\cite{Bicudo:2015vta}.

We will now analyse in detail the results with dynamical quarks of Refs. 
\cite{Wagner:2010ad,Wagner:2011ev}, 
illustrated in Fig. \ref{fig:potentials}. To be applied to the Schrödinger Equation, it is convenient to fit with an ansatz the points and error bars resulting from the lattice QCD computation 
\cite{Bicudo:2012qt}. 
The result of the fit is illustrated in Fig. \ref{fig:pot2010}.

To fit the potential points with a curve an ansatz was needed.  Educated by  what is known from  QCD, asymptotic freedom at small distances and infrared slavery at large distances, a Coulomb potential is expected at short distances. The the two heavy antiquarks are then in the groundstate if they are in a triplet colour state $3$, s-wave and thus must have a symmetric spin 1. At large distances, a screening typical of the static-light wavefunction is expected. It is sketched in Fig. \ref{fig:screening} from Ref. 
\cite{Bicudo:2012qt}, 
say Gaussian or exponential. The ansatz with best fit was found as,

\be
V(r) = -{\alpha \over r} \exp \left(-\left(r \over d\right)^p\right) .
\label{eq:ansatz} 
\ee
The error bar of the fit is reflected in the band in the plot of Fig. \ref{fig:pot2010}. Results are summarized in Table \ref{ab:fit AEK}.

To search for boundstates and compute the respective wavefunction  the potentials are included in the Schrödinger equation. The infinite mass limit is relaxed with a finite heavy mass, say $m_b$ or (less satisfactory) $m_c$, either the quark mass (bare current mass or the constituent mass used in quark models) or the B meson mass which are relatively similar.
Extrapolating to the physical pion mass, 
\cite{Bicudo:2012qt,Bicudo:2015vta,Bicudo:2015kna,Bicudo:2016ooe,Bicudo:2017szl,Bicudo:2021qxj},
a binding energy of $-90 \pm 43$ MeV was predicted. Partly including the effect of the spin of the heavy antiquarks reduced binding slightly to $-60 \pm 45 $ MeV, in the binding channel of a $B B^*$.
Other tetraquarks such as $ls \bar b \bar b$, $lc \bar b \bar b$, $sc \bar b \bar b$, $ll \bar c \bar b$, $ll \bar c \bar c$ did not bind in this approach, but a p-wave resonance with the same flavour was also predicted, if one ignores the heavy spin effects.
 
Importantly, the quantum numbers predicted for the $T_{bb}$ are $I(J^{P})=0(1^+)$. To understand the quantum numbers, It is sufficient to consider the small distance $r$, where the heavy quarks have spin 1. The light quarks are in a scalar-isoscalar state. Moreover the parity, with two quarks and an antiquark in s-waves is $+$.

\subsection{
Tetraquark and meson-meson bounstate structure with a generalized eigenvalue problem   \label{sec:GEVP}}

%
\begin{figure}[t!]
\begin{centering}
\includegraphics[width=0.48\columnwidth]{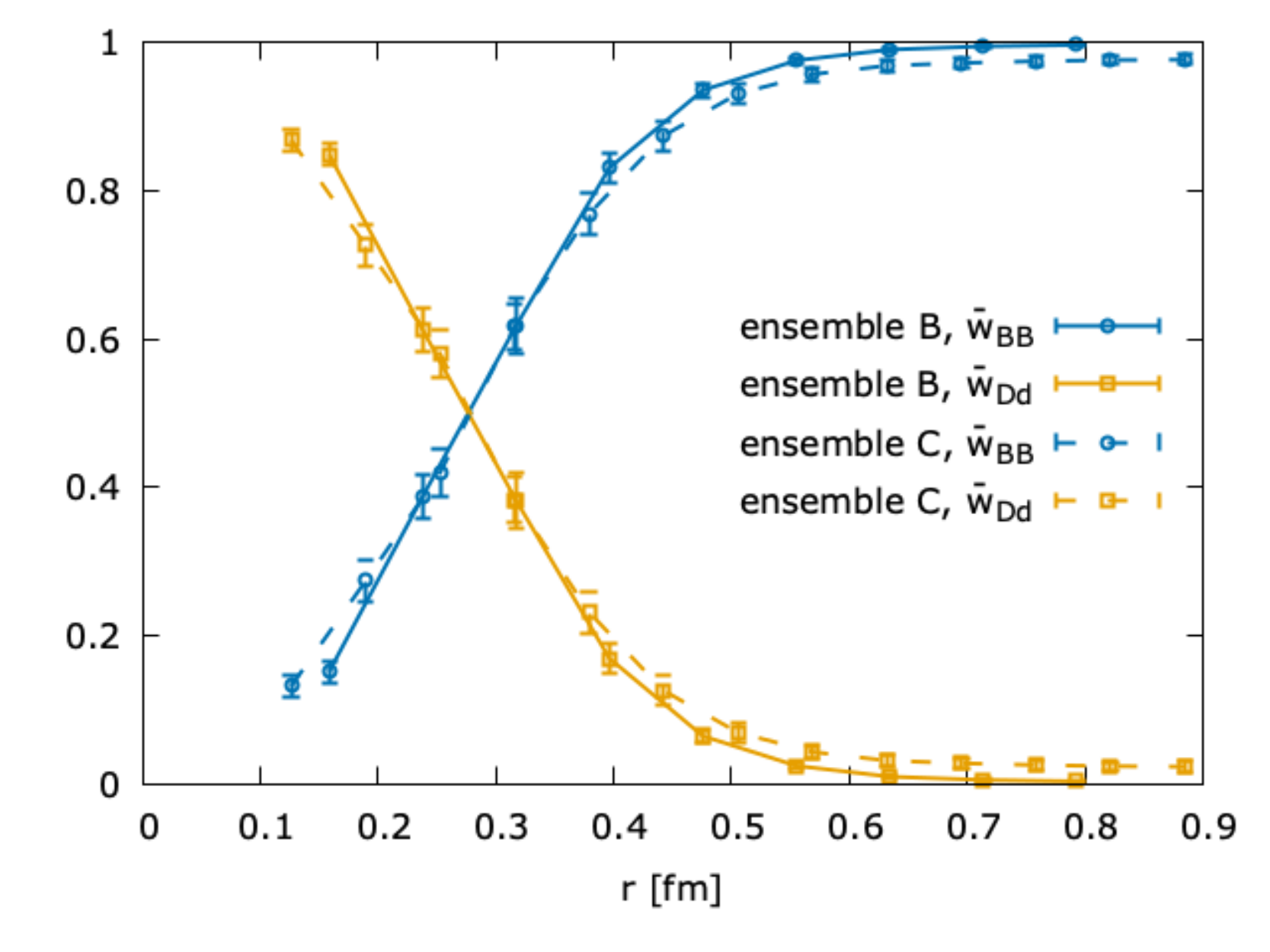}
\quad
\includegraphics[width=0.48\columnwidth]{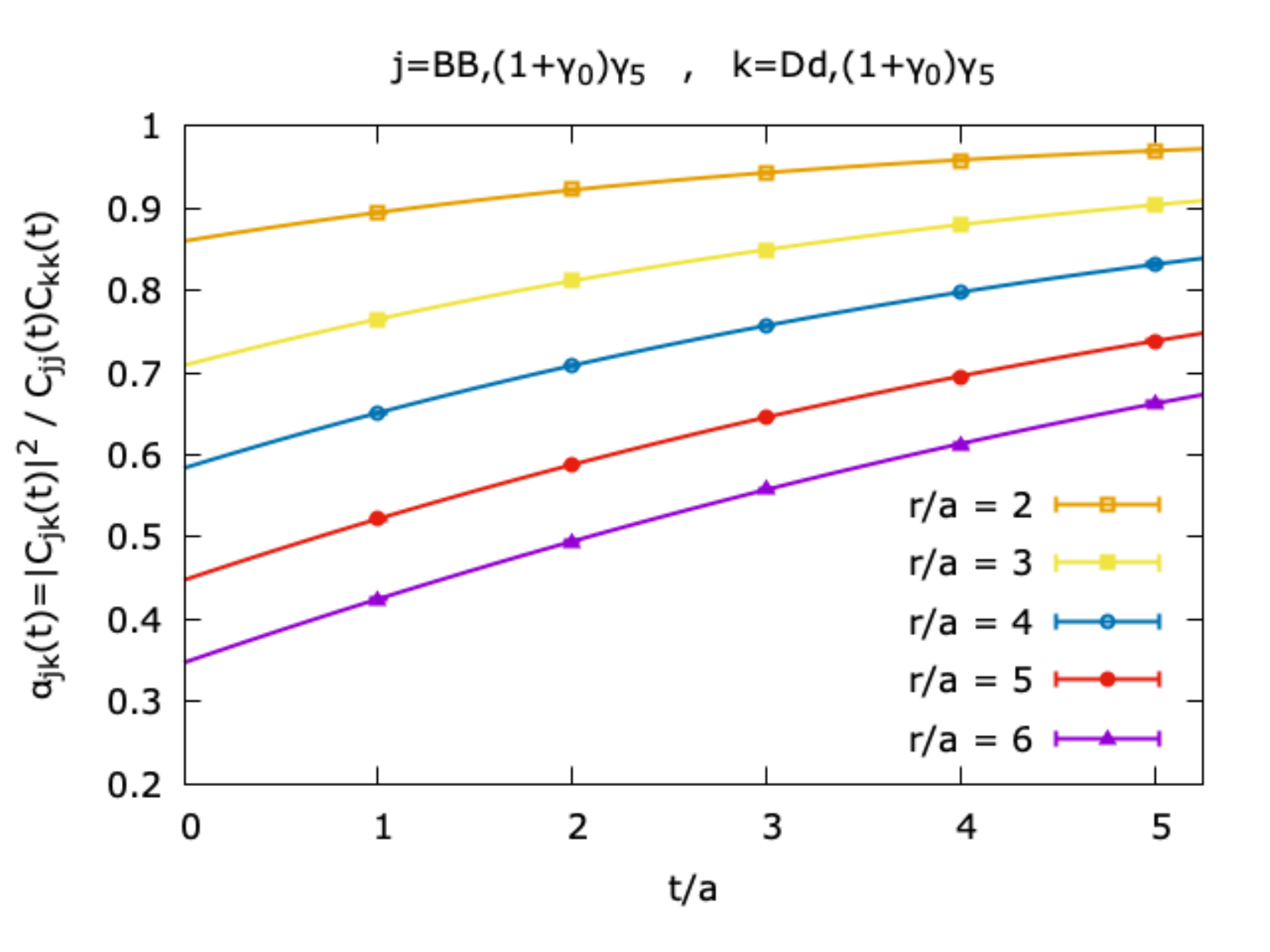}
\caption{Comparing the {\em diquark} tetraquark versus the {\em molecular} tetraquark according to Ref. \cite{Bicudo:2021qxj}. 
(left) Fitted normalized absolute squares of the coefficients, {\em diquark} $\bar w_{BB}$ and {\em molecular} $\bar w_{Dd}$ of the optimized trial state, as functions of the distance $r$ between the heavy antiquarks, for two different ensembles of configurations. 
(right) The squared overlap $\alpha_{j k}$  of the corresponding normalized trial states as a function of $t$ for several fixed $r$ for ensemble B40.24.}
\label{fig:realmol}
\end{centering}
\end{figure}

In Ref. \cite{Bicudo:2021qxj}, lattice QCD was utilized to compare two frequently discussed competing structures for a  tetraquark: the {\em diquark} tetraquark versus the {\em molecular} one. 

The considered tetraquark is the exotic $\bar b \bar b u d$ system with quantum numbers
$I(J^P)=0(1^+)$. This boundstate has been confirmed by different first principles lattice QCD computations, as refered in Subsections \ref{sec:staticdy}, \ref{sec:ressta}, \ref{sec:resdyn}.

In Ref.  \cite{Bicudo:2021qxj},  the authors considered  meson-meson as well as diquark-antidiquark creation operators. They also used the static-light approximation of Subsections \ref{sec:staticdy} and \ref{sec:ressta}, where the two $\bar b$ quarks are assumed to be infinitely heavy with frozen positions, while the light u and d quarks are fully relativistic, with dynamical configurations.

By minimizing effective energies and by solving a generalized eigenvalue problem (GEVP), they
determined the importance of the meson-meson and the diquark-antidiquark creation operators with respect
to the ground state. 
The first type of creation operators, also used in \cite{Wagner:2010ad,Wagner:2011ev,Bicudo:2015vta,Bicudo:2015kna}, excites two $B$ mesons at separation $r$,
\begin{eqnarray}
\label{EQN_BB} \mathcal{O}_{BB,\Gamma} = 2 N_{BB}
(\mathcal{C} \Gamma)_{AB} (\mathcal{C} \tilde{\Gamma})_{CD} \Big(\bar{Q}_C^a(\mathbf{r}_1) \psi^{(f) a}_A(\mathbf{r}_1)\Big) \Big(\bar{Q}_D^b(\mathbf{r}_2) \psi^{(f') b}_B(\mathbf{r}_2)\Big)
\end{eqnarray}
where $r = |\mathbf{r}_2 - \mathbf{r}_1|$ is the separation between the heavy antiquarks. The colour indices are $a,b$, spin indices are $A,B,C,D$ and $\psi^{(f)} \psi^{(f')} = u d - d u$. $N_{BB}$ is a normalisation. There are two independent choices for the light spin matrix $\Gamma$ consistent with $(j_z,\mathcal{P},\mathcal{P}_x) = (0,-,+)$. $\Gamma = (1+\gamma_0) \gamma_5$ predominantly excites two negative parity ground state mesons $B^{(\ast)} B^{(\ast)}$, while $\Gamma = (1-\gamma_0) \gamma_5$ mostly generates two positive parity excited mesons $B_{0,1}^\ast B_{0,1}^\ast$. Since static spins have no effect on energy levels, the heavy spin matrix is irrelevant and can be chosen arbitrarily, $\tilde{\Gamma} \in \{ (1-\gamma_0) \gamma_5 , (1-\gamma_0) \gamma_j \}$.
The second type of creation operators, resembles a diquark-antidiquark pair with heavy quarks separated by $r$ and connected by a gluonic string,
\begin{eqnarray}
\mathcal{O}_{Dd,\Gamma} = -N_{Dd}
\epsilon^{abc} \Big(\psi^{(f) b}_A(\mathbf{z}) (\mathcal{C} \Gamma)_{AB} \psi^{(f') c}_B(\mathbf{z})\Big)
\epsilon^{ade} \Big(\bar{Q}^f_C(\mathbf{r}_1) U^{fd}(\mathbf{r}_1;\mathbf{z}) (\mathcal{C} \tilde{\Gamma})_{CD} \bar{Q}^g_D(\mathbf{r}_2) U^{ge}(\mathbf{r}_2;\mathbf{z})\Big) .
\label{eq:Dd}
\end{eqnarray}
Again $N_{Dd}$ is a normalisation and the allowed light and heavy spin matrices are the same as for the operator $\mathcal{O}_{BB,\Gamma}$, i.e.\ $\Gamma \in \{ (1-\gamma_0) \gamma_5 , (1+\gamma_0) \gamma_5 \}$ and $\tilde{\Gamma} \in \{ (1-\gamma_0) \gamma_5 , (1-\gamma_0) \gamma_j \}$. For definiteness, $\tilde{\Gamma} = (1-\gamma_0) \gamma_3$ was chosen. 

Notice these operators are not orthogonal, they have the same quantum numbers. At $R=0$ the diquark-antidiquark operator is exactly identical to the meson-meson operator and when distance increases, the overlap decreases.  For instance, to compute the energy of the tetraquark, they are not both necessary, since in lattice QCD it is sufficient to have an operator with a good overlap with the wavefunction of the studied physical state. Nevertheless the relevance of each operator for the wavefunction can be determined. Wen we have non-orthogonal operators, we solve a GEVP. The application of a GEVP to lattice QCD problems is reviewed in Ref. \cite{Blossier:2009kd}.

Ref. \cite{Bicudo:2021qxj} concludes the diquark-antidiquark structure  of the tetraquark dominates for $\bar b \bar b$ separations r $\leq$ 0.25 fm, whereas it becomes increasingly more irrelevant for larger separations, where the $I(J^P)=0(1^+)$ tetraquark is mostly a meson-meson system. This is shown in Fig. \ref{fig:realmol}. The estimated meson-meson to diquark-antidiquark ratio of this tetraquark is around 60\% to 40\%. Moreover for the light quarks, the spinor matrix $(1 + \gamma_0)\gamma_5$ dominates.

%
\begin{figure}[t]
\begin{centering}
\includegraphics[width=0.48\columnwidth,trim=50pt 50pt 50pt 350pt, clip]{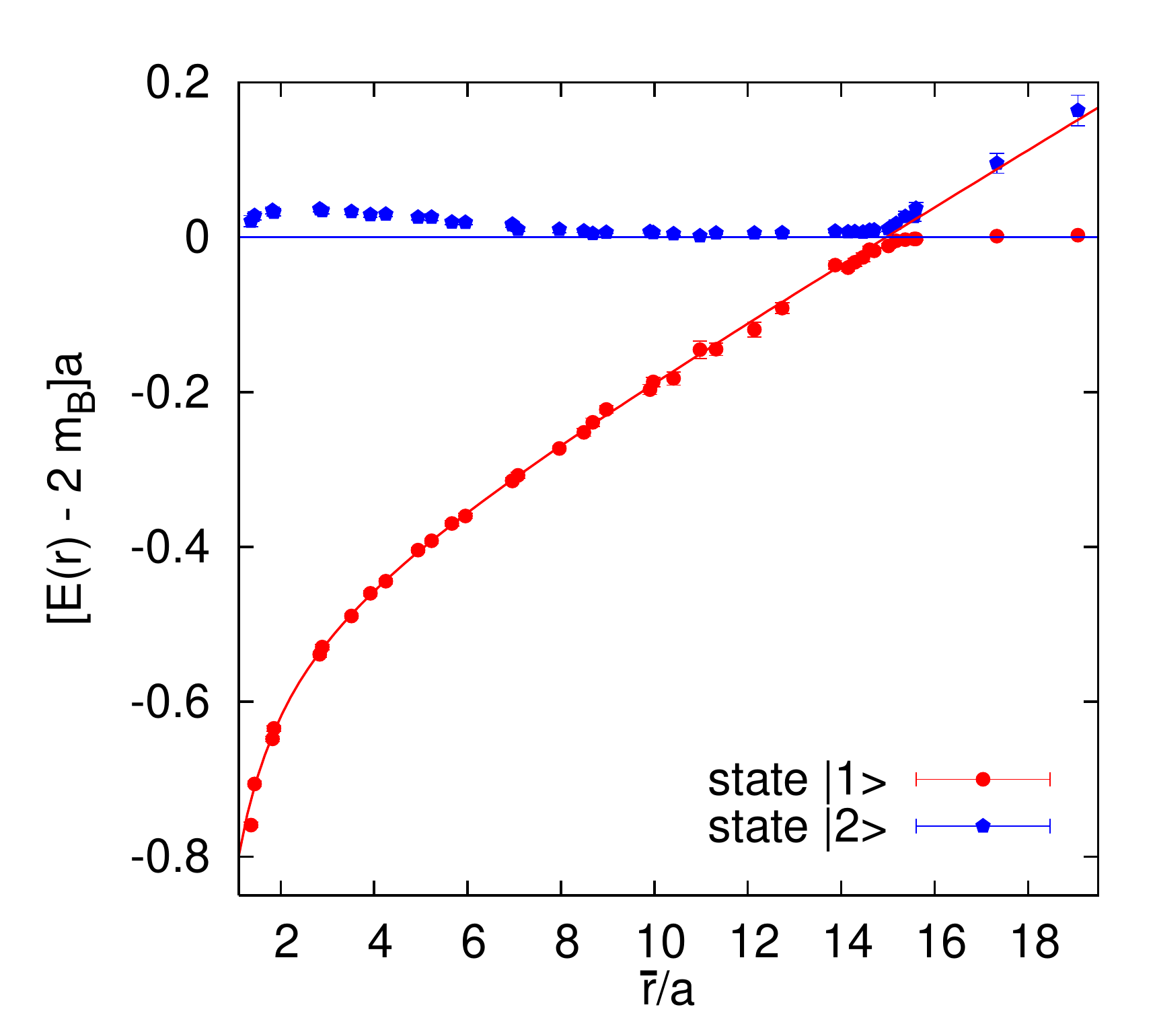} 
\quad
\includegraphics[width=0.48\columnwidth,trim=50pt 50pt 50pt 350pt, clip]{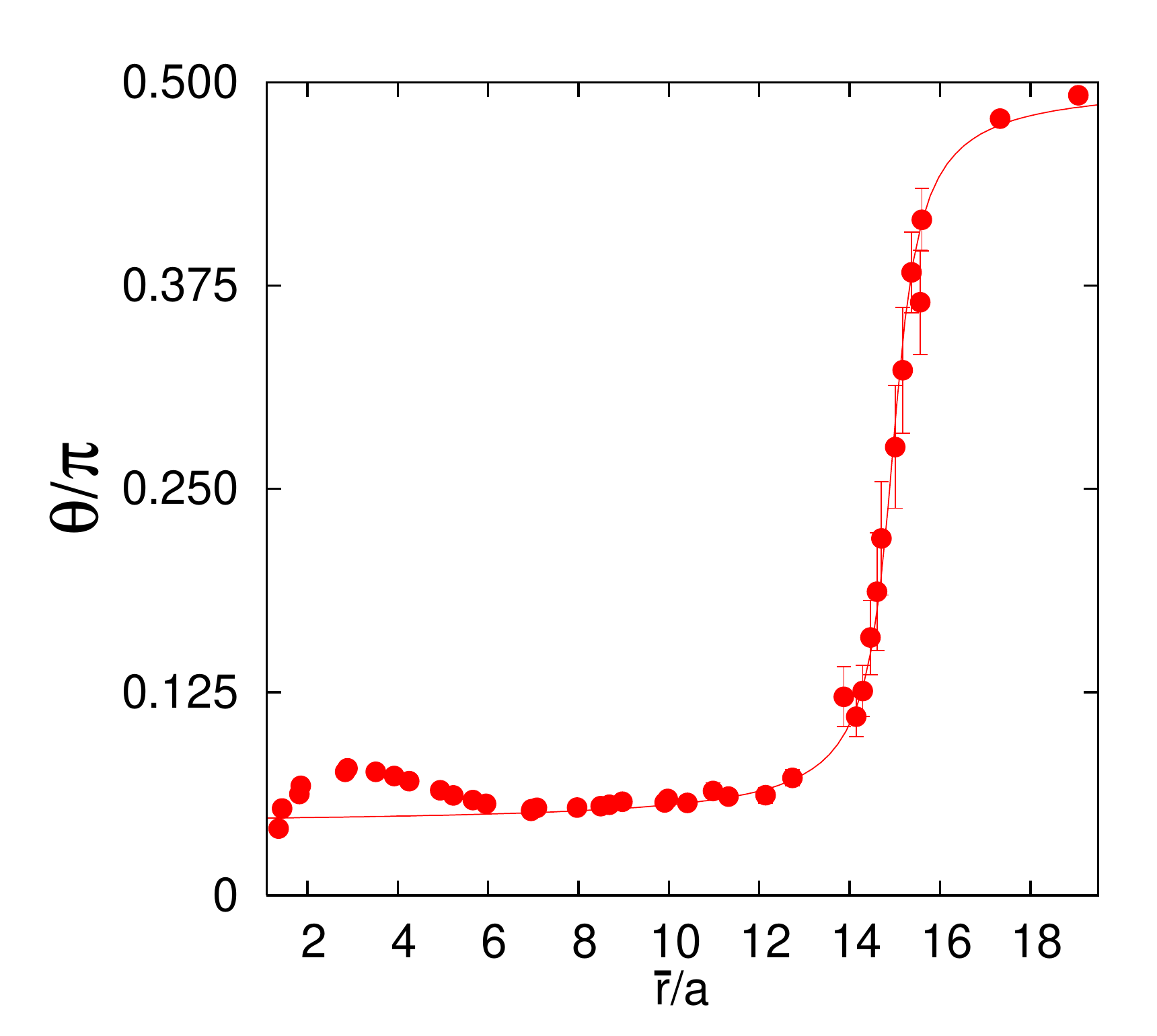}
\caption{
(left) Potential eigenvalues for a quarkonium and two static-light mesons computed in Ref. \cite{Bali:2005fu}.
(right) Mixing angle between the $Q \bar Q$ and $ M \bar M$.
\label{fig:stringbreak}}
\end{centering}\end{figure}

In what concerns the type of tetraquark, the attraction is provided when the system is at shortest distances in a {\em diquark} tetraquark, according to Eq. (\ref{eq:ansatz}) and Table \ref{ab:fit AEK} but the wavefunction also extends outside this short distance region and then the system is in a {\em molecular} tetraquark.

\subsection{
Potentials with static quarks $Q \bar Q$ and dynamical quarks for string breaking  and quarkonium decay  \label{sec:breaking}}

String breaking is an important aspect of the decay of hadrons and of hadronisation. When two quarks are pulled to large distances, the string picture fails. It is as though the string would break and at the breaking point a quark-antiquark pair would be created. 

String breaking in lattice gauge theory
\cite{Philipsen:1998de,Knechtli:1998gf,Drummond:1998ir,Detar:1998qa,Stewart:1998yj,Trottier:1998nk,Stephenson:1999kh,Philipsen:1999wf,Gliozzi:1999wq,deForcrand:1999kr,Kallio:2000jc,Knechtli:2000df,Duncan:2000kr,Bernard:2001tz,Chernodub:2002ym,Chernodub:2002en,Kratochvila:2003zj,Chernodub:2003sy,Wellegehausen:2010ai,Kuhn:2015zqa,Buyens:2015tea,Bonati:2020orj}
was first observed in simpler theories than QCD. However, it initially failed to be observed in the static quark-antiquark potential with SU(3) gauge fields and with dynamical quarks. Just using a simple Wilson loop may be insufficient for the string breaking to be visible in the quark-antiquark potential. 

To be clearly observed, the string breaking studies needed to explicitly include the coupling of the two static quark systems to a four quark system (two static and two dynamical). Although this system doesn't have exotic quantum numbers, it is useful to study cypto-exotic tetraquarks. The coupled potentials between in a static quark-antiquark channel and two static-light mesons have been computed by Bali et al. \cite{Bali:2005fu} with two light $u$ and $d$ flavours and by Bulava et al.  
\cite{Bulava:2019iut} 
with three light $u$, $d$ and $s$ flavours. The corresponding string breaking potential for Ref. 
\cite{Bali:2005fu}
and first excited state, showing the crossing of the quarkonium and meson-meson potentials,
 is  illustrated in Fig. \ref{fig:stringbreak} (left).

\begin{figure}[t]
\begin{centering}
\includegraphics[width=0.45\columnwidth]{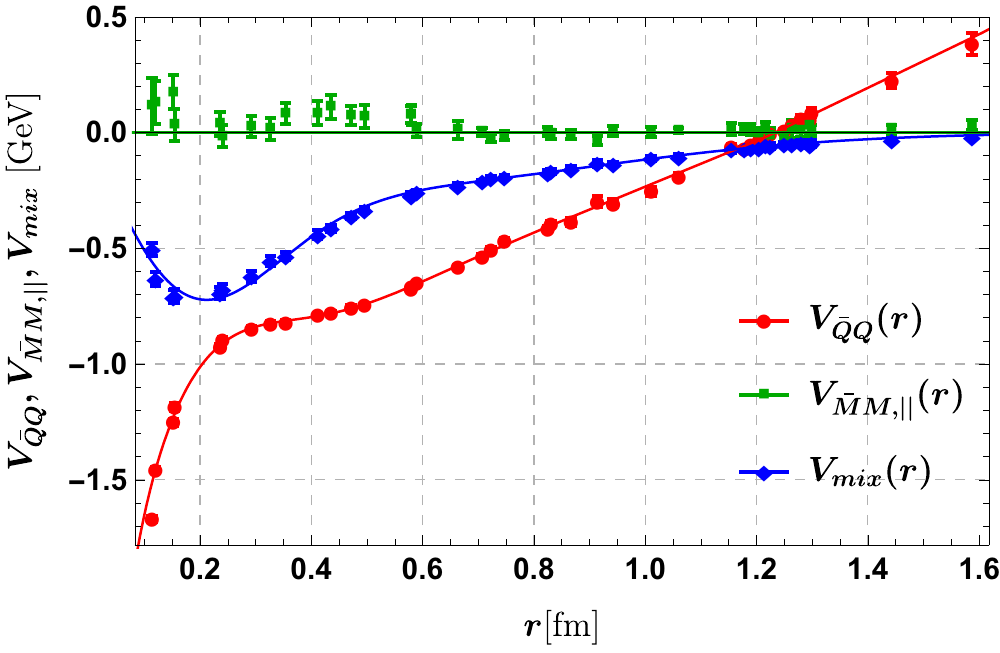}
\quad
\includegraphics[width=0.5\columnwidth]{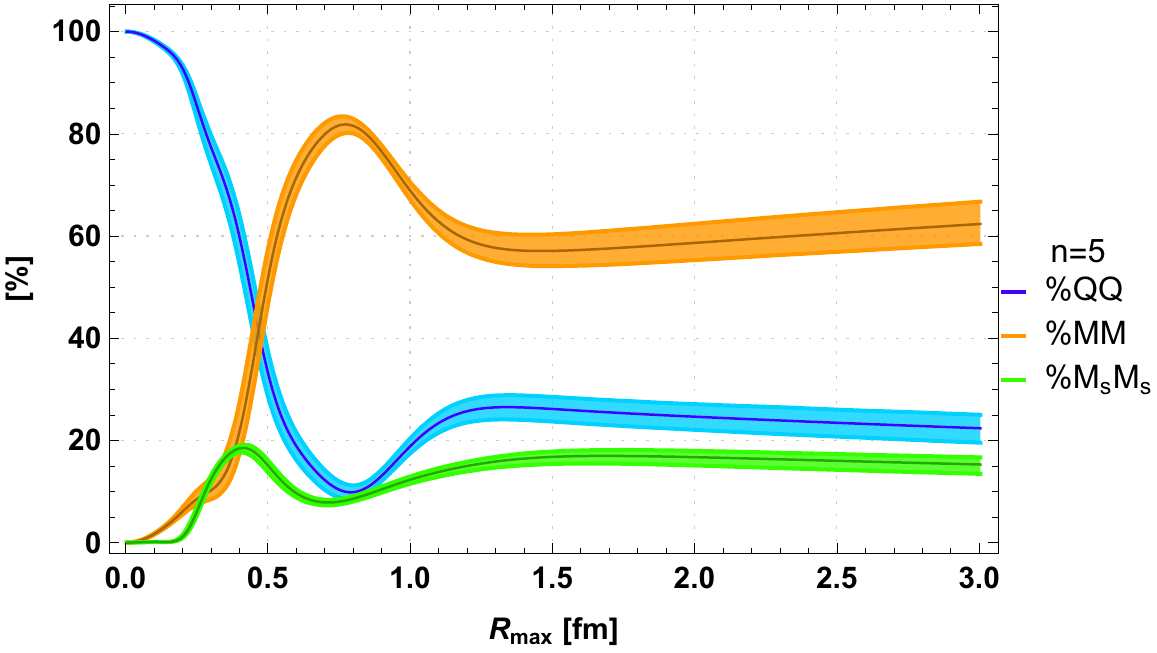}
\caption{(left) The matrix elements of the potential in Ref.  results \cite{Bicudo:2019ymo,Bicudo:2020qhp} and (right) the composition of the bottomonium state $\Upsilon(10753)$
\label{fig:polesstringbrea}.}
\end{centering}
\end{figure}

Moreover, Bali et al. 
\cite{Bali:2005fu}
 showed that the static quarkonium operator and the meson-meson operator have a very small overlap. This enables the computation of the mixing angle between the quarkonium and meson-meson basis and the basis of the two eigenvectors. With the mixing angle, it is possible to compute the potentials for the quarkonium, for the meson-meson system and the respective mixing potential between these two channels, illustrated in Fig. \ref{fig:polesstringbrea}. These potentials constitute the matrix elements of a non-diagonal matrix potentials, in the Fock space, including the quarkonium and the meson-meson subspaces. For instance for s-wave quarkonium,
\be
 V_0(r) = \left(\begin{array}{cc} V_{\bar{Q} Q}(r) & V_\textrm{mix}(r) \\ V_\textrm{mix}(r) & V_{\bar{M} M,\parallel}(r) \end{array}\right)
 \label{eq:mix}
\ee
Using the Born-Oppenheimer diabatic approximation, extending Subsection \ref{sec:staticdy}, this matrix of potentials has been used to study the decay of quarkonium to meson-meson channels, determining the poles of the bottomonium resonances 
\cite{Bicudo:2019ymo,Bicudo:2020qhp,Bicudo:2022ihz,Bicudo:2022jep}.

The advantage of this approach compared to having a full lattice QCD computation with non-relativistic quarks is that many levels of quarkonium can be addressed, and  that we are able to explore the $s$ matrix in the Riemann sheets of complex energies. 
Possibly this technique can be extended to study the $I=1$ channel, including the $Z_b$  family of tetraquarks, extending the technique used in Ref.  \cite{Cardoso:2008dd} for the $Z_c$ family, with potentials detailed in Subsection 
\ref{sec:staticdy2}. 

The drawback is that, in lattice QCD, we have not yet accounted for the heavy quark spin effects, as detailed in Subsection \ref{sec:diquark}. These effects may shift up to 100 MeV the masses computed with static heavy quarks. Besides, with static potentials, only the bottomonium spectrum can be addressed, whereas for the charmonium spectrum, including the very interesting $X(3872)$ resonance, more terms in the relativistic expansion are necessary.  Besides, new more precise lattice QCD computations of the potentials are necessary to determine in more detail the bottomonium spectrum with this technique.

Importantly, compared to the pure quarkonium spectrum, two extra states are found, in the $S$ and $D$ spectra
\cite{Bicudo:2022ihz,Bicudo:2022jep}. Moreover, both these states have a dominant meson-meson component shown in Fig. \ref{fig:polesstringbrea} and masses compatible with the recent $\Upsilon(10753)$ resonance observed at BELLE
\cite{Belle:2019cbt}, referred in Table \ref{tab:cryptoTetra}. 

These states are then {\em s pole} tetraquarks. Notice, with the potential matrix in Eq. (\ref{eq:mix}), it is possible to derive the Lippmann-Schwinger series for the $T$ matrix of Fig. \ref{fig:Spole}. This approach is expected to produce results comparable with the Lüscher technique for phase shifts, say for the study of the $X(3872$ in Subsection \ref{sec:phaseshifts}.

\subsection{
Potentials with static heavy quarks $Q \bar Q$ and dynamical light quarks for the $Z_b$ family of tetraquarks   \label{sec:staticdy2}}

%
\begin{figure}[t!]
\begin{center}
\includegraphics[width=0.47\textwidth]{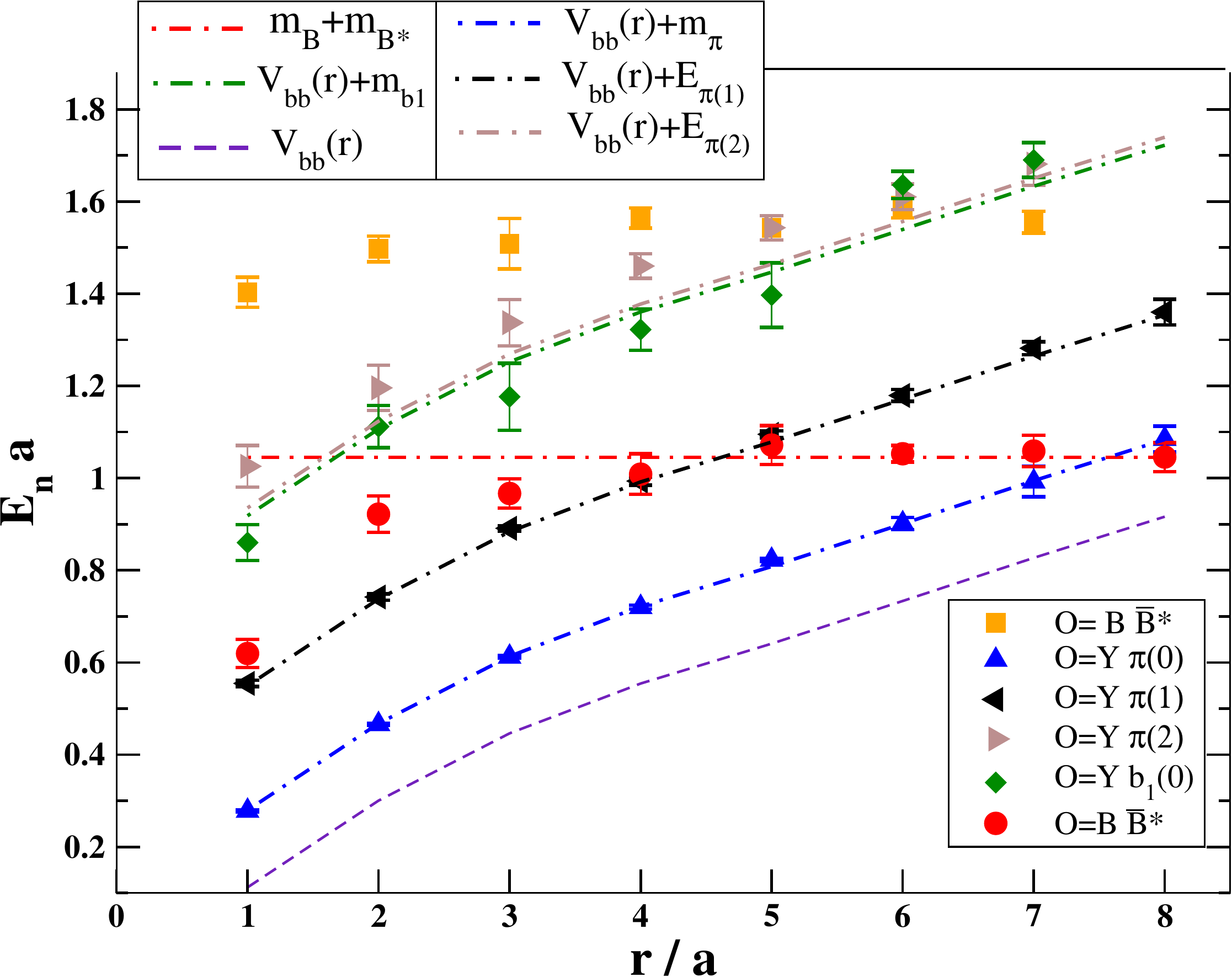}   
\caption{ Eigen-energies of  $\bar bb\bar du$  system, computed in Ref. 
\cite{Prelovsek:2019ywc},
for various separations $r$ between  static quarks $b$ and $\bar b$ are shown by points. The  label  indicates which two-hadron component dominates each eigenstate. The dot-dashed lines represent related two-hadron energies $E^{n.i.}$  when two hadrons   do not interact. The eigenstate dominated by $B\bar B^*$ (red circles) has energy  significantly below $m_B+m_{B^*}$ and  shows sizeable  attraction. Lattice spacing is $a\simeq 0.124~$fm.  }
\label{fig:En}
\end{center}
\end{figure}

The approach of computing, in a first step, potentials in lattice QCD using static quarks for the heavy quarks, as in Subsection \ref{sec:staticdy}, has also been applied to the case where the heavy pair is a $Q \bar Q$. This is the case of the $Z_b$ family of tetraquarks, see Table \ref{tab:Tetra}, with flavour $u \bar d b \bar b $. Then, if the potentials are clearly identified, the Born-Oppenheimer can also be applied and the respective boundstates or resonances can be studied by solving the quantum mechanics equations. However this $Z_b$ case is more difficult to address than the case of \ref{sec:staticdy} where the heavy pair is a $\bar Q \bar Q$, say for a tetraquark with flavour $u d \bar b \bar b $. Then, at large distances where the strong interactions between hadrons are screened, we only had one meson-meson pair, with flavour $u \bar b$ and $d \bar b$, a ${B^+}^{(*)} {B^0}^{(*)}$ pair.

In the present $Z_b$ case we have two possible flavours for the meson-meson pairs, either the pair $u \bar b={B^+}^{(*)}$ and $ \bar d b=\bar {{B}^0}^{(*)}$ or the pair $u \bar d$ and $b \bar b$, say corresponding to a meson $\pi^+$ and a quarkonium meson such as an $\Upsilon$. We must have at least two coupled channels, one with the pair ${B^+}^{(*)}\, \bar {{B}^0}^{(*)}$ and another with the pair $\pi^+ \, \Upsilon$.
The difficulty resides in identifying the different open coupled channels, and in the number of coupled channels, because the different channels are not orthogonal.

The first study was worked out by Peters et al considering the channel of the  $B \bar B$ meson pair only 
\cite{Peters:2016wjm,Peters:2017hon}. The authors found evidence for a boundstate due to an attractive potential between the two mesons. However to complete the study it would be necessary to include the second coupled channel, with a pion and a bottomonium state. This second channel is the open channel where the resonance decays to.

The second study already includes the two meson-meson coupled channels, by Prelovsek et al. 
\cite{Prelovsek:2019yae,Prelovsek:2019ywc,Sadl:2021bme}. 
The first step is the computation of the potentials, as a function of the distance between the heavy $Q \bar Q$, considered static. This is similar to the technique of Section \ref{sec:breaking}, where we aim to get the potential for each channel and the mixing potential between the considered channels. Then this can be addressed with quantum mechanics techniques to compute the scattering matrix. 

Another interesting lattice QCD approach  by Alberti et al
\cite{Alberti:2016dru} considered the composed system denominated hadroquarkonium
\cite{Dubynskiy:2008di}. 
This combines a quarkonium composed by a static quark-antiquark pair together with an extra light hadron composed of dynamical quarks. In this Ref.  \cite{Alberti:2016dru}, several cases of mesons - not just the pion - and baryons are considered as the extra hadron.

But having an extra light meson leads to an extra difficulty, which in a sense is overlooked in the hadroquarkonium: the light hadron may have any possible momentum. 
Refs.  
\cite{Prelovsek:2019yae,Prelovsek:2019ywc,Sadl:2021bme} 
go one step further by considering some of the first momenta of the light meson $\pi$, as shown in Fig. \ref{fig:En} . 
Moreover in 
\cite{Sadl:2021bme}, 
the four different quantum numbers $\Sigma_g^+, \, \Sigma_g^-, \, \Sigma_u^+, \, \Sigma_u^-$ in the notation of the symmetry group of homonuclear molecules, defined in Subsection \ref{sec:hybrid}, resulting from the coupling of the light meson and the static quarkonium are considered, resulting in four different types of potentials.

Nevertheless there is still one technical detail to be solved in this approach. To apply quantum mechanical techniques, the non-orthogonality of the different channels and the mixing angle between them and the eigenvalues of the potential matrix remain to be fully addressed.

\subsection{
HAL QCD non-static potentials for non-relativistic heavy quarks and  light dynamical quarks  \label{sec:HALqcd}}

\begin{figure}[t!]
\begin{center}
\includegraphics[width=0.65\textwidth,clip]{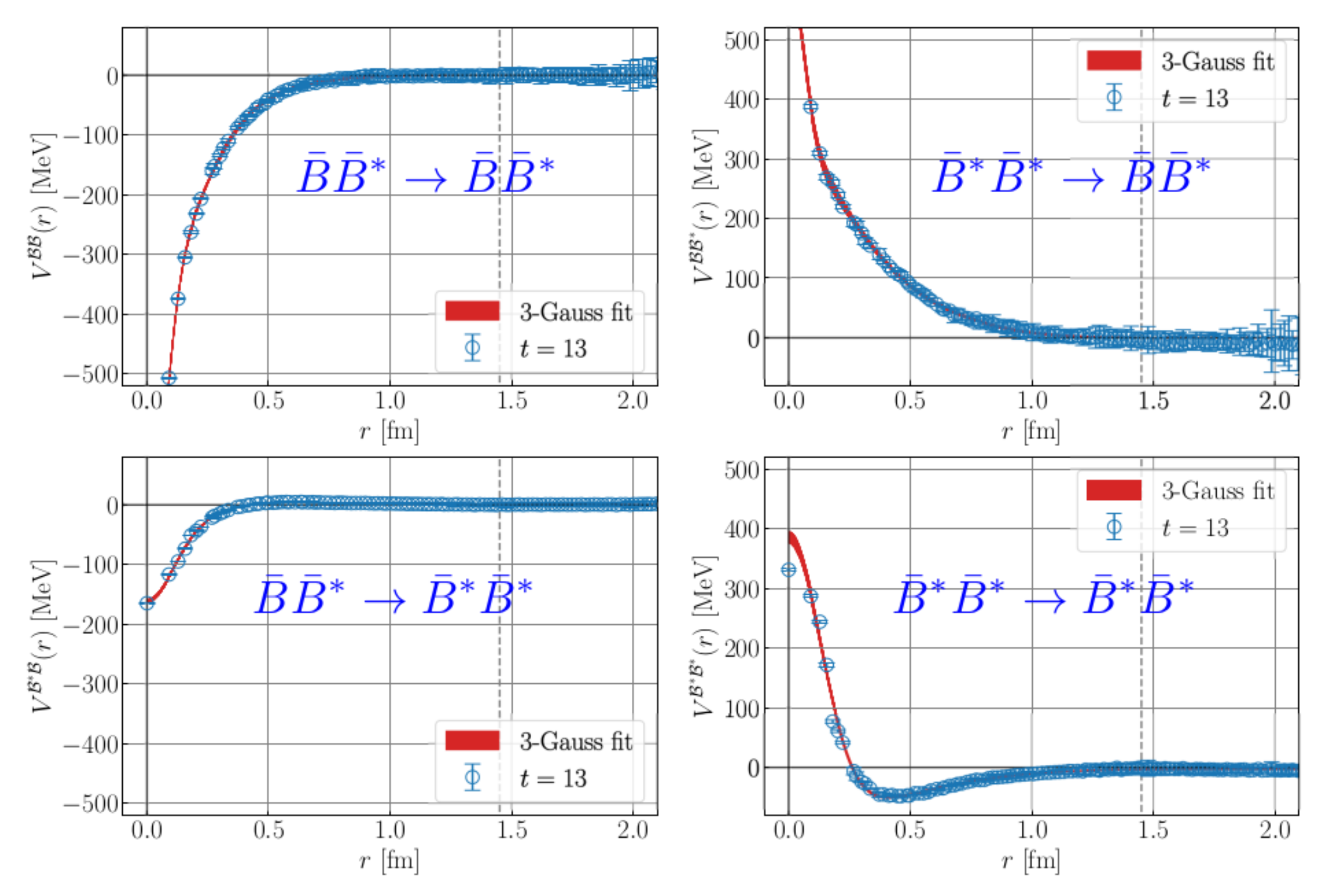} 
\end{center}
\caption{Coupled channel potentials obtained in Ref 
\cite{Aoki:2022lat}
with the HAL QCD method using dynamical light quarks and non-relativistic heavy quarks.  
\label{fig:HALpot}}
\end{figure}

With static sources one computed directly the position dependent potentials in lattice QCD, see Subsections \ref{sec:staticdy} and \ref{sec:staticdy2}.  The HAL QCD method extends this technique to quarks with dynamics   
\cite{Ishii:2006ec,Aoki:2011gt,Ishii:2012ssm}.
 This has been previously applied to study the $\pi-\pi$ scattering and the $N-N$ interaction. 
 In particular the short range repulsive core, the medium range attraction, and the long range OPEP discussed in Subsection \ref{sec:theory} have been computed, in agreement with the phenomenological $N-N$ interactions.
 Very recently it has been applied to study the  $T_{bb}$ tetraquark by Aoki
  \cite{Aoki:2022lat}.

The first step consists in computing the wavefunction / Nambu-Bethe-Salpeter amplitude for systems of two hadrons, each one composed of quarks. On the lattice, this is computed with a pair of quark operators. Then the respective potential is derived from the BS amplitude. To understand the concept, consider for instance the Schrödinger equation, which is a non-relativistic and instantaneous limit of the Nambu-Bethe-Salpeter equation,
\be
\left[ - {\Delta \phi(\bm r)  \over 2 m} + V(\bm r) \right] \, \phi(\bm r) = E\, \phi(\bm r) \ ,
\ee
it implies we can determine the potential from the wavefunction $\phi$ of the hadron pair and its energy $E$,
\be
V(\bm r) = {\Delta \phi(\bm r)  \over 2 m \phi(\bm r) }+ E \ .
\ee
From the basic principles of this example, it is possible to extend the technique to the relativistic and time dependent systems considered by HAL QCD. 

It is first necessary to get the best possible wavefunction for each one of the hadrons. This is obtained variationally, with an ansatze for the lattice QCD operator with the correct quantum numbers of the respective hadron. Then the parameters in the ansatze, for instance the smearing radius, are optimized to  produce the largest signal for the groundstate exponential in Eq. (\ref{eq:project}), corresponding to the largest overlap $c_0$. This also corresponds to the best signal over noise ratio in the effective mass plot for the groundstate energy in Eq. (\ref{eq:plateau}). Then these wavefunctions are used as a source or sink for the hadron-hadron wavefunction $\phi(\bm r)$ at separation $\bm r$.

In Ref. 
 \cite{Aoki:2022lat}
 for the $T_{bb}$ study, a $2 \times2$ time dependent coupled channel was studied. The coupled channels employed asymptotic channels with pairs of $B$ and $B^*$ mesons. A diquark-antidiquark operator was also added to the operator set.  For the light quarks, dynamical fermions were used and and for the bottom quarks, non-relativistic fermions were used, as in Subsection \ref{sec:effective}. In this work, chiral extrapolation to the physical pion mass was also performed.

\begin{figure}[t!]
\begin{center}
\includegraphics[width=0.65\textwidth,clip]{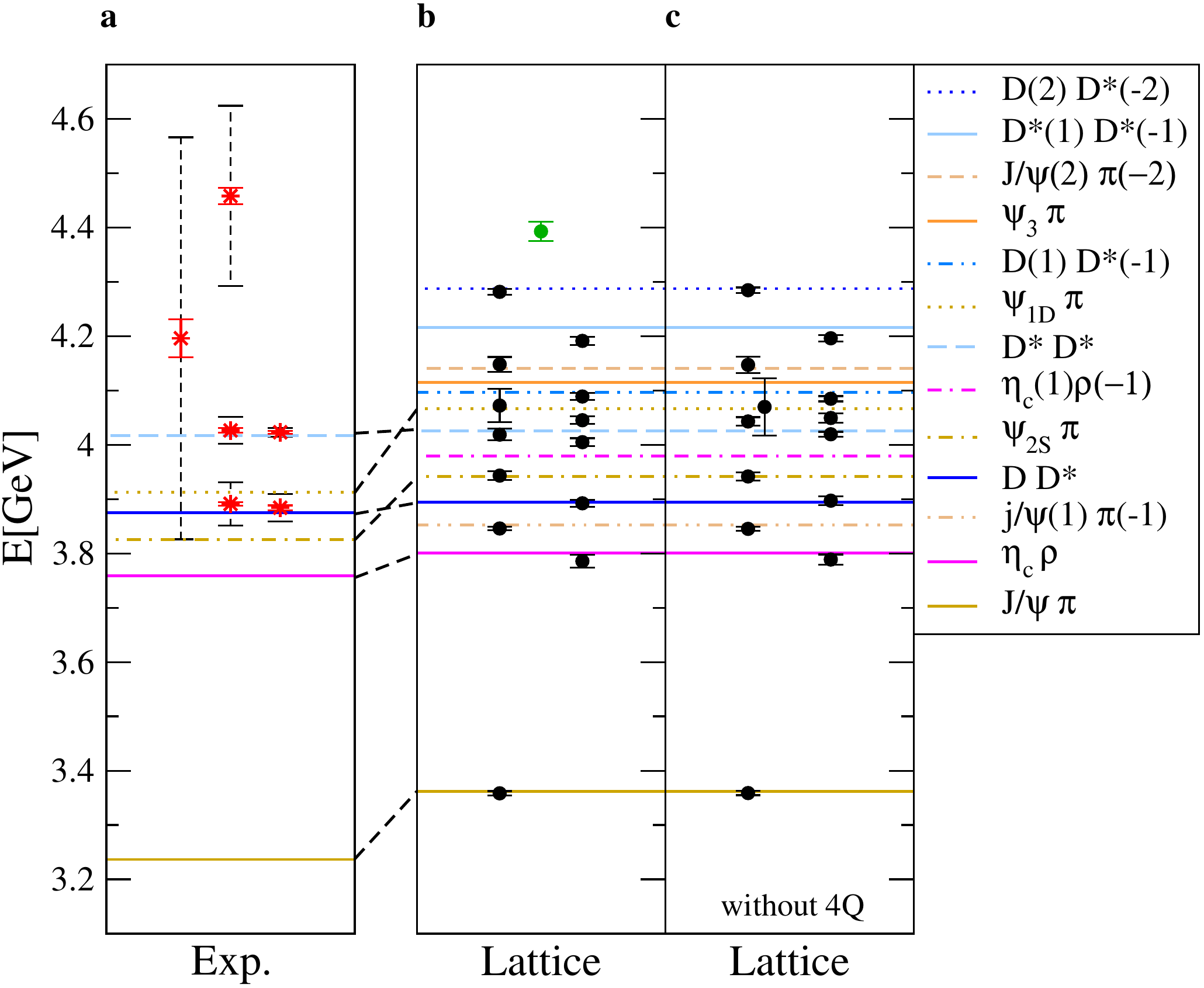} 
\end{center}
\caption{The spectrum for quantum numbers $ I^G(J^{PC})=1^+(1^{+-})$ from Ref. 
\cite{Prelovsek:2014swa}. (a) Position of the experimental $Z_c^+$ candidates \cite{Brambilla:2019esw}. (b,c)  The discrete energy spectrum from our lattice simulation: (b) shows energies based on complete $22\times 22$  matrix  of interpolators,  (c)  is based on the $18\times 18$ correlator matrix without diquark-antidiquark interpolating fields ${\cal O}^{4q}_{1-4}$. The thirteen   lowest lattice energy levels (black circles) are interpreted as two-particle states, which are inevitably  present in a dynamical lattice QCD simulation. No additional   candidate for the exotic $Z_c^+$ is found below $4.2~$GeV.   
The dashed vertical lines indicate twice the experimental widths to illustrate the energy range in which the additional energy level due to $Z_c$ might be expected. \label{fig:effenes}}
\end{figure}

 This work was able to obtain the potentials with dynamic light quarks and non-relativistic heavy quarks.  The potentials are shown in Fig. \ref{fig:HALpot}, for the $2 \times 2$ coupled channel system of a $B B^*$ pair and a $B^* B^*$ pair . They are in general comparable to the potentials obtained with static quarks in the same $2 \times 2$ coupled channel system of Ref.
 \cite{Bicudo:2016ooe}, 
 where they are denominated $V_j$ and $V_5$, as detailed in Subsection \ref{sec:diquark} and shown in Fig. \ref{fig:pots} (right).
Notice the potentials of Ref. 
  \cite{Aoki:2022lat}
  are quite recent, with more points, smaller errors and a wider range of distances $r$. Observing in detail the HAL QCD potentials, there is evidence for some OPEP attraction in the $B^* B^*$ pair, which has not yet been seen in the static potentials. This is an interesting result, possibly showing evidence of forces of both gluon-exchange OGEP and meson-exchange OPEP type.

\subsection{
Search for tetraquark resonances high in the spectrum with four dynamical quarks  \label{sec:scpectra}}

After $Z_c$ was discovered in three different experiments, full dynamical lattice QCD calculations were performed, searching for new energy levels in the excited spectrum by Prelovsek et al. \cite{Prelovsek:2013xba,Prelovsek:2014swa}, Guerrieri et al. \cite{Guerrieri:2014nxa} and Cheung et al. \cite{Cheung:2017tnt}. The authors used the technique of comparing the spectrum just with several meson-meson operators, and the spectrum after adding diquark-antidiquark operators. 
Representative examples of the many employed interpolators are
\be
\label{O_main}
 {\cal O}_1^{\psi(0)\pi(0)}=\bar c \gamma_i c(0)~\bar d\gamma_5 u(0)
\ee
for meson meson operators, and  for the diquark-antidiquark operators, similar to Eq. (\ref{eq:Dd}), 
\be
{\cal O}^{4q}_1\propto \epsilon_{abc} \epsilon_{ab'c'}(\bar c_{b}C \gamma_5\bar d_{c}~   c_{b'}\gamma_{i} C u_{c'}- \bar c_{b}C \gamma_i\bar d_{c}~  c_{b'}\gamma_{5} C u_{c'}) \ .
\ee
Notice, in other cases this same technique may succeed in identifying hadrons, for instance Ref. 
\cite{Padmanath:2015era}
were able to identify the $X(3872)$ state in the spectrum.

However no evidence for the $Z_c$ was found in the spectrum. For instance the authors of Ref.  \cite{Prelovsek:2013xba} searched for this state on the lattice by simulating the channel with $J^{PC} = 1^{+-}$  and $I = 1$, but they did not ﬁnd a candidate for ${Z_c}^+ (3900)$. Instead, they only  find discrete scattering states $D \bar D^*$ and  $J /\psi \pi^+$, shown in Fig. \ref{fig:effenes}, which inevitably have to be present in a dynamical QCD computation.

Notice the $Z_c$ tetraquarks are resonances, with a decay width corresponding to a complex pole of the $S$ matrix, and it is not a priori certain that they can be identified in the midst of a real energy spectrum. Moreover they are high in the spectrum and are coupled to a large number of decaying channels which makes them more difficult to identify.

Thus the numerous tetraquarks and pentaquarks of the $Z$ family so far escape evidence from lattice QCD computations.
A possible explanation is in Fig. \ref{fig:realmol} (right) we borrow from Ref. \cite{Bicudo:2021qxj}. In this example, there is a large overlap between the meson-meson operators and the diquark-antidiquark operator. This may explain why adding  diquark-antidiquark operators to the operator set does not change the spectrum in a noticeable way.

\subsection{
Tetraquark boundstate search with non-relativistic bottom quarks and light dynamical quarks \label{sec:effective}}

%
\begin{figure}[t!]
\begin{center}
   \includegraphics[width=1.0\columnwidth]{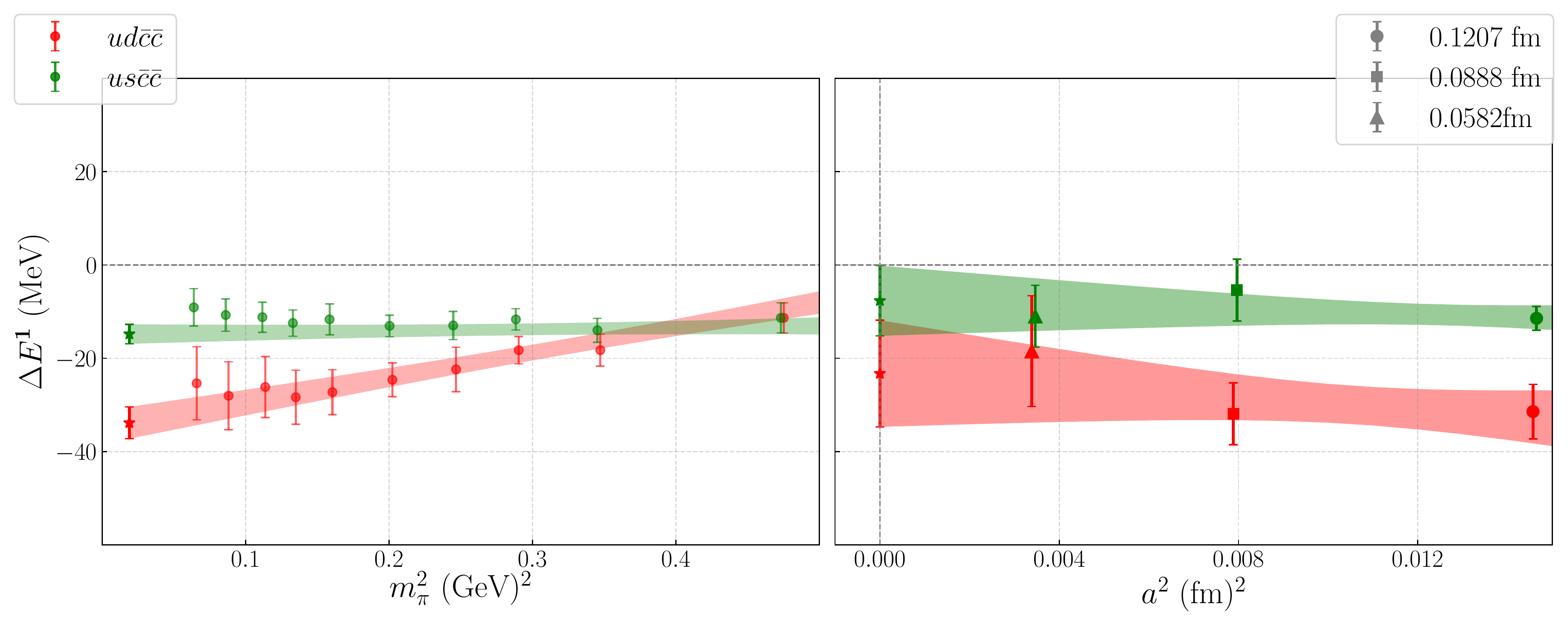}
\end{center}
   \caption{Results of Ref. 
   \cite{Junnarkar:2018twb} 
   for the $ud\bar{c}\bar{c}$ and $us\bar{c}\bar{c}$ doubly charm tetraquark states color coded in red and green in both panels.  Left panel: chiral extrapolation for several pion masses at $a = 0.1207$ fm for each of the states. Right panel: Continuum extrapolation at the chiral extrapolation to the physical pion mass.
   \label{fig:tccbind}}
  \end{figure}

While the computation of a lattice QCD spectrum, based in a set of operators as in Subsection \ref{sec:scpectra}, has not yet proved successful for the study resonances, it is effective for the study of boundstates. Indeed the simplest state to identify in a spectrum is the groundstate, such as boundstate lying below the continuum states.

After the $T_{bb}$ lattice QCD predictions by Bicudo and Wagner
\cite{Bicudo:2012qt} 
using static quarks and the Born-Oppenheimer adiabatic approximation, a more precise approach was applied to search for bound states in the $T_{bb}$ system with two light quarks and two heavy bottom antiquarks. Notice the standard lattice QCD dynamical quark technique cannot yet be applied to bottom quarks who are too heavy. The mesh of the present lattices is too wide for full dynamical bottom quarks. Bottom quarks are then implemented in lattice QCD with a non-relativistic approximation, which solves two problems of the static quarks. The potentials computed with static quarks are purely spin scalar, and important potentials such as the hyperfine potentials are not computed. Besides when providing a kinetic energy to the heavy quarks, it is not clear wether to use the quark mass from quark models or the meson mass.

The $T_{bb}$ fully exotic tetraquark has been studied by Francis et al. \cite{Francis:2018jyb}, Leskovec et al. \cite{Leskovec:2019ioa} and Junnarkar et al. \cite{Junnarkar:2018twb}. These authors uses the NRQCD lattice action \cite{Thacker:1990bm,Lepage:1992tx,Manohar:1997qy,Bali:2000gf,Brown:2014ena} to calculate bottom quark propagators. For instance, the Hamiltonian used in Ref.  \cite{Francis:2018jyb} is \cite{Davies:1994mp,Lewis:1998ka,Lewis:2008fu}
\begin{equation}
\begin{aligned}
H = &-\frac{\Delta^{(2)}}{2M_0}-c_1\frac{(\Delta^{(2)})^2}{8M_0^3}
+ \frac{c_2}{U_0^4}\frac{ig}{8M_0^2}(\bm{\tilde\Delta\cdot\tilde{E}}-\bm{\tilde{E}\cdot\tilde\Delta}) \\
& - \frac{c_3}{U_0^4}\frac{g}{8M_0^2}\bm{\sigma\cdot}(\bm{\tilde\Delta\times\tilde{E}}-\bm{\tilde{E}\times\tilde\Delta})\\
&-\frac{c_4}{U_0^4}\frac{g}{2M_0}\bm{\sigma\cdot\tilde{B}}
+ c_5\frac{a^2\Delta^{(4)}}{24M_0}
- c_6\frac{a(\Delta^{(2)})^2}{16nM_0^2}\;,
\end{aligned}
\label{eq:nrqcdhamilton}
\end{equation}
with tadpole improvement and correct up to order $o(a^3)$. The first term should provide results similar to the ones computed with static potentials.

\begin{figure}[t!]
\begin{center}
\includegraphics[width=0.7\columnwidth]{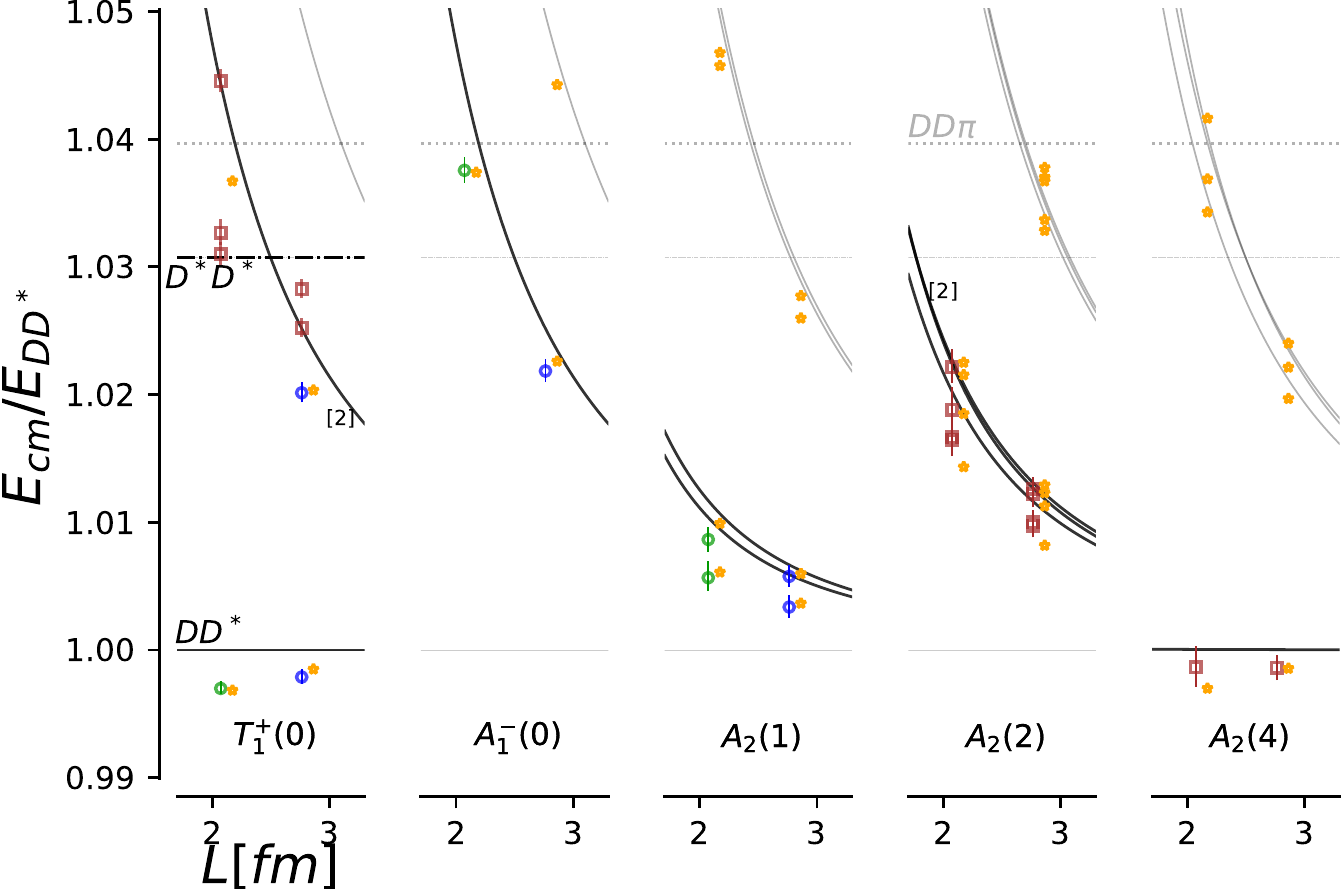}  
\caption{The center-of-momentum energy $E_{cm}=(E^2-\vec P^2)^{1/2}$ of the $cc\bar u\bar d$ system 
normalized by $E_{DD^*}\equiv m_D+m_{D^*}$, in various finite-volume 
irreps, utilized to compute the $T_{cc}$ pole. For a complete legend of the different data, see Ref.  \cite{Padmanath:2022cvl}.
The non-interacting $DD^*$ 
energies are shown by lines. }
\label{fig:Ecm}
\end{center}
\end{figure}

All the lattice QCD computations agree that there is a bound $T_{bb}$ tetraquark, and most likely a bound $T_{bbs}$ as well. While there there was some tension in the initial value of the binding energy, presently the results of the different lattice QCD groups tend to agree, see Table \ref{tab:boundTetra}. Where there is still some tension is in the other quantum numbers of this family of tetraquarks, where in general Ref. \cite{Junnarkar:2018twb} tends to find binding in $T_{cc}$, $T_{ccs}$ and $T_{bbc}$ where the other groups don't. The later group employed the chiral extrapolation to the physical $m_\pi$, and the continuum limit to vanishing $a$ , only the thermodynamic limit with large  volumes remain to be checked. This is illustrated in Fig. \ref{fig:tccbind} for the tetraquarks $T_{cc}$ and $T_{ccs}$. 

Also notice Hughes et al found a null evidence for a full bottom tetraquark boundstate in Ref. 
\cite{Hughes:2017xie}. In the case of $T_{bc}$ the situation is not clear yet on wether Lattice QCD predicts binding or not.
\cite{Pflaumer:2022lgp,Wagner:2022qwg}.

\subsection{
Scattering study of tetraquarks with the Lüscher method for phase shifts with four dynamical quarks \label{sec:phaseshifts}}

\begin{figure}[t!]
\begin{center}
\includegraphics[width=0.6\columnwidth]{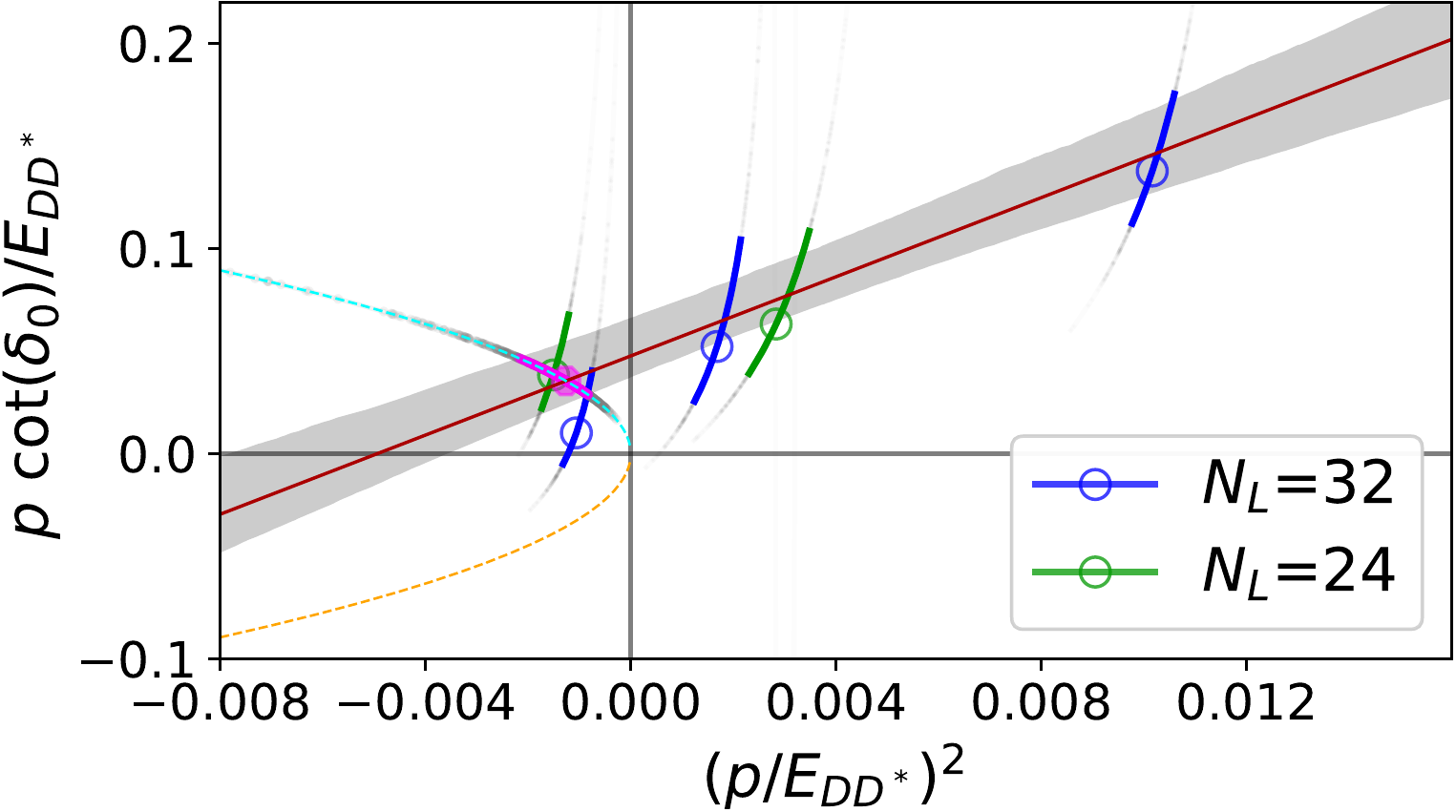}
\caption{$p \cot \delta_{l=0}^{(J=1)}$ for $DD^*$ scattering, computed in Ref.  \cite{Padmanath:2022cvl}, at the heavier charm quark mass (red line) 
and $ip=+|p|$ (cyan line) versus $p^2$, all normalized to $E_{DD^*}\equiv m_D+m_{D^*}$. The $T_{cc}$ virtual bound 
state occurs at the momenta indicated by the magenta octagon, where two curves intersect. }
\label{fig:pcotdel}
\end{center}
\end{figure}   

In the previous Subsections, the main computation to study tetraquarks consisted in computing a energy spectrum, with the effective mass plot technique of Eq. (\ref{eq:plateau}). Either we would directly compute the energy of the tetraquark or first compute a potential (energy of static quarks) to use it with the Born-Oppenheimer approximation in a Schrödinger equation. 

However the previous techniques do not allow to study directly scattering and resonances in lattice QCD. 
Here we review another technique to compute the phase shifts of the scattering matrix with lattice QCD. As a basic example, let us consider for instance a non-relativistic particle with a potential $V$ in a D=1 dimensional space. In the continuum the particle wave-function suffers a phase shift, as a function of its momentum $k=\sqrt{2 m E}$ due to the potential. At large enough distance for the potential (assumed compact) to be negligible, the wavefunction is proportional to $\phi_k(x) = \exp[ i k x + i \delta(k) ]$, where  $\delta(k)$ is the phase shift. 
Clearly when $V$ is attractive the phase shift is negative (we have more oscillations within the potential zone) and vice-versa. The scattering length is defined as $a= {d \over dk}  \delta(k)$, in the case of a repulsive potential it provides the effective size of a hard target equivalent to the potential. 

Let us now sketch how this can be of interest in the case of a lattice (assuming a small lattice spacing), say like a finite box of length $L$ with periodic boundary conditions. In the free and non-relativistic case, with no potential, the periodic wavefunctions are
\be
\phi_n(x)=\exp[ i k_n x ] \ , \ \ k_n= n {2 \pi \over L} \ , 
\ee
corresponding to an energy
\be
E_n= {k^2 \over 2 m} = n^2 { 2 \pi^2 \over m L^2}
\ee
however, with the potential, the wavefunctions at large distance from the potential have a phase shift and then the periodic boundary condition forces the standing waves to comply with the condition
\bea
k L + 2 \delta(k) &=& n 2 \pi
\non \\
E_n &=&  n^2 { 2 \pi^2 \over m (L+ 2 \delta(k))^2} \ .
\eea
Thus, in this trivial example, once we measure the energy levels, we are able to compute the phase shifts for a discrete set of energies, determined by the box size. An order of magnitude of the splitting between the levels, say for a mass $m$ of the order of 1 GeV, and a lattice size of 1 to 2 fm, is 0.19 $n^2$ GeV to 0.77 $n^2$ GeV. To determine the phase shift at several energies one has to consider different box sizes.

In the real case of QCD one has at least two particles interacting. Moreover this basic idea has successfully be extended to 3 spatial dimensions, with angular momentum and spin, and to relativistic particles, by Lüscher et al and then by others,
\cite{Luscher:1985dn,Luscher:1986pf,Luscher:1985dn,Luscher:1990ck,Luscher:1990ux,Briceno:2014oea,Padmanath:2018tuc}. 
Moreover the extension to several channels and to three particle resonances is under development by Hansen et al,
\cite{Hansen:2012tf,Hansen:2014eka,Hansen:2015zga}.
For instance in the case of a single channel with two particles, using the effective range approach \cite{Chew:1955zz}
 it is possible to measure in lattice QCD
\be
k^{2l+1} \cot \delta_l (k) = {-1 \over a_l} + { r_l \over 2} \, k^2 +  o(k^4)
\label{eq:scalen}
\ee
and parametrize it with the scattering length $a_l$ and the effective range $r_l$ \cite{Morningstar:2017spu}.
where the condition for having a pole in the $S= e ^{i \, 2 \delta_l}$ matrix is equivalent to $ \cot \delta_l +i=0$.
Once this has been implemented, and the statistical and systematic errors are under control, the first scattering cases have been studied in lattice QCD.

%
\begin{figure}[t!]
\begin{centering}
\includegraphics[width=0.6\columnwidth]{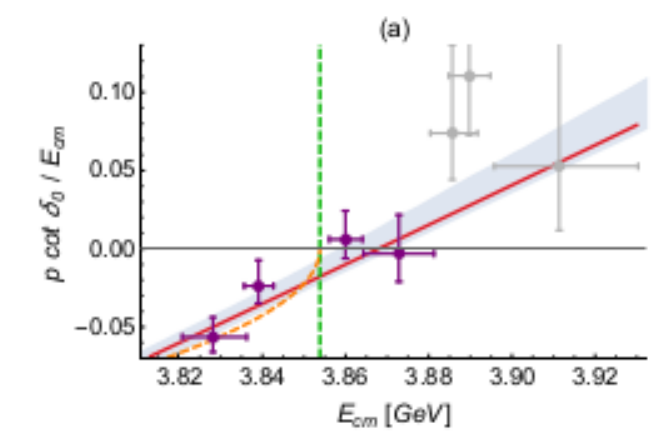}
\caption{$ D \bar D$ scattering in partial wave $l =$ 0 near threshold for the study of the charmonium spectrum and decay widths
\cite{Prelovsek:2020eiw}.
The green dashed line denotes the
$ D \bar D$ threshold in the simulation. The violet crosses show the quantity
$p \cot \delta /E_\text{cm}$ , computed with the  Lüscher method.
The red line indicates the linear fit corresponding to the effective range method. The orange line represents
$i p / E_\text{cm}$. A bound state is located at the energy where the red and orange curves intersect.
\label{fig:phashi}}
\end{centering}
\end{figure}

Very recently the first fully exotic tetraquark to be observed, the $T_{cc}$ was studied in the channel $ D D^*$, see Ref. \cite{Padmanath:2022cvl} 
by Padmanath and Prelovsek. 
Well before its observation, it was already studied with the phase shift technique by 
Ykeda et al \cite{Ikeda:2013vwa}
who did not find any boundstate or resonance.
The determination of the channels of Ref. \cite{Padmanath:2022cvl}
 is illustrated in Fig. \ref{fig:Ecm}.
However, using the effective range approach, shown in Fig. \ref{fig:pcotdel},  so far only a virtual boundstate pole was found, not yet the $T_{cc}$ experimental resonance pole, because both parameters $a_0$ and $r_0$ in Eq. (\ref{eq:scalen}) are positive. In this case there is a solution for a pole with purely imaginary momentum, but with Im$(k)<0$, whereas for a boundstate we should have Im$(k)>0$.

Moreover, 
In Refs. 
 \cite{Aoki:2022lat,Wagner:2022qwg}
 the scattering of a pair of  pairs of $B$ and $B^*$ mesons was also studied, and in these cases a boundstate for the $T_{bb}$ was clearly obtained with the effective range approach. The binding energy is compatible with the results shown in Subsections \ref{sec:static} and \ref{sec:effective}.

\subsection{
Meson-meson scattering phase shift with the Lüscher method for crypto-exotics and non-exotics  \label{sec:Luschercrypto}}

Besides the studies we reviewed of the exotic tetraquarks $T_{cc}$ and $T_{bb}$ with the Lüscher phase shift technique in Subsection \ref{sec:phaseshifts},
there are as well some very interesting studies of systems with crypto-exotic tetraquarks using this technique.

There are two different types of crypto-exotic systems with four quark components, as detailed in Subsection \ref{sec:nonpert}, the {\em molecular} tetraquarks and the {\em s pole} tetraquarks. We define the crypto-exotics as systems with a dominant four quark component, but with quantum numbers that are not exotic. While the {\em molecular} tetraquarks are essentially a two-meson system, the {\em s pole} tetraquarks also have a simple meson component, but this is component is small.

Moreover there as as well non-exotics. We define the non-exotics as systems where the meson system is dominant, but it couples to meson-meson channels. In particular when it is above the respective threshold it may decay to a meson-meson channel.
For instance, in Subsection \ref{sec:breaking}, using string breaking potentials, it was concluded the  $\Upsilon(10753)$ observed at Belle is a crypto-exotic because it is mostly composed of $B^{(*)}\bar B^{(*)}$ and $B^{(*)}_s \bar B^{(*)}_s$, while the other $\Upsilon({nS})$ are non-exotic bottomonium because they are mostly composed of $b\bar b$ components.

In what concerns non-exotic charmonium-like resonances, they were studied with $D \bar D$ coupled channels first for  $J^{PC}=1^{--}$ and $3^{--}$ \cite{Piemonte:2019cbi} in agreement with experiment. Then, in the coupled channel of $D \bar D$ and $D_s \bar D_s$, the charmonium-like resonances with $J^{PC}=0^{++}$ and $2^{++}$ were studied in Ref. 
\cite{Prelovsek:2020eiw}, finding interesting new resonances. 
 The authors found a previously unobserved $D \bar D$ bound state just below threshold and a $D \bar D$ resonance likely related to 
 $\chi_{c0}(3860)$, which is believed to be $\chi_{c0}(2P)$. In addition, they found an indication for a narrow $0^{++}$ resonance just below the $D_s \bar D_s$ threshold with a large coupling to $D_s \bar D_s$ and a very small coupling to $D \bar D$. This resonance is possibly related to the narrow $X(3915)/ \chi_{c0}(3930)$ observed in experiment also just below $D_s \bar D_s$. The partial wave $l=2$ features a resonance likely related to $ \chi_{c2}(3930$. The search for a pole using the phase shift in the effective range approach is illustrated in Fig. \ref{fig:phashi}.

The most famous candidate for crypto-exotic tetraquark $X(3872)$ was studied in the coupled channels $ D D^{-*}$ and $J/\Psi \omega$ by Prelovsek et al  \cite{Prelovsek:2013cra}. A candidate for the charmonium-like state $X(3872)$ was found $11 \pm 7$ MeV below the $ D D^{-*}$  threshold using dynamical $N_f$=2 lattice simulation with $J^{PC}=1^{++}$ and $I=0$, similar in mass to another lattice QCD search \cite{Lee:2014uta}. This state was again studied together with the resonance $Y(4140)$ 
 in Ref. \cite{Padmanath:2015era}. This computation confirmed the $X(3872)$ with a mass closer to the threshold, but found no evidence for the $Y(4140)$.

%
\begin{figure}[t!]
\begin{center}
    \includegraphics[width=0.6\columnwidth]{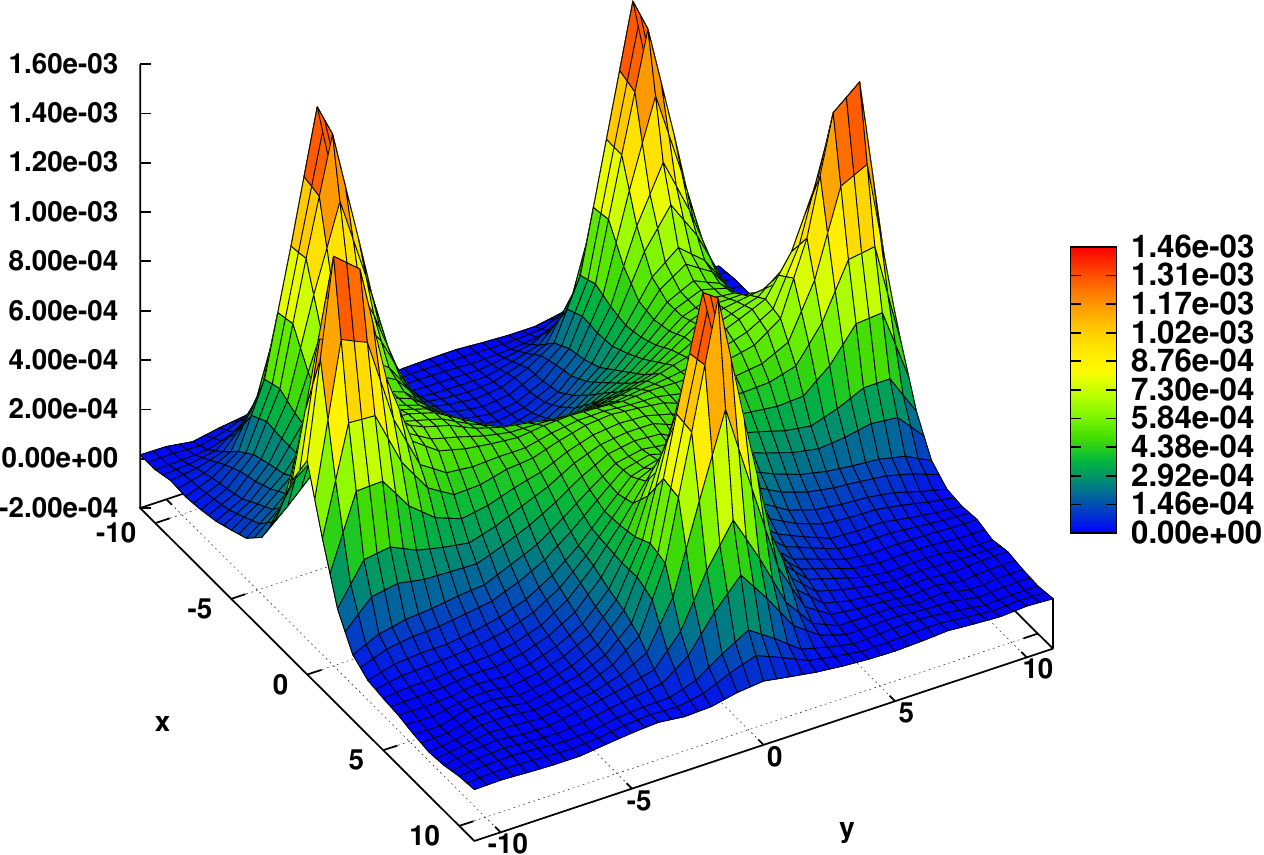}
\end{center}
    \caption{Lagrangian density 3D plot for a pentaquark in a favorable geometry, from Ref. \cite{Cardoso:2012rb},
showing a clear a pentaquark triple-Y flux tube. The results are presented in lattice spacing units (colour online).
    \label{fig:pentafields}}
\end{figure}

The candidates for charmed $\pi$ and $K$ molecule candidates with $D$ and $D_s$ quantum numbers have also been studied  by Mohler et al \cite{Mohler:2012na}, Refs. \cite{Mohler:2013rwa,Lang:2014yfa}. 
Studying scattering in the $D \pi$ channels, the authors computed the masses and widths of the broad scalar D0*(2400) and the axial D1(2430), and reproduced the experiment. 
In the $DK$ scattering, the $D_{s0}^*(2317)$ is found 37(17) MeV below the $DK$ threshold, close to the experiment value of 45 MeV, in a lattice simulation of the $J^P=0^+$ channel using both $DK$ as well as $s\bar c$ interpolating fields. The $D_{s1}(2460)$ is also found  as a strong interaction bound state 44(10) MeV below the $D^*K$ threshold, which is in agreement with the experiment. The narrow resonances $D_{s1}(2536)$ and $D_{s2}^*(2573)$ are also found close to the experimental masses. 
These studies have been further continued by Refs. 
\cite{Moir:2016srx,Bali:2017pdv}.

However, more recently and using as well the scattering technique, Alexandrou et al 
\cite{Alexandrou:2019tmk}
found that the dominant component of the $D_{s0}^*(2317)$ is a $q \bar q$ component, not a tetraquark one. Thus they find it is a regular meson and not a {\em s pole} tetraquark.

Searching for bottomed $K$ molecules with $B_s$ quantum numbers, Lang et al \cite{Lang:2015hza}
first verified they reproduced the experimentally observed states $B_{s1}(5830)$ and ${B_{s2}^*}(5840)$. They also predicted a new molecule:  a $J^P=0^+$ bound state $B_{s0}$ with mass of $5.711 \pm 32$ GeV. This state is a bottomed partner of the charmed one.

\section{Brief review of other related lattice QCD studies of exotic systems with heavy quarks}\label{sec:other}

\subsection{
Pentaquarks with heavy quarks \label{sec:penta}}

The all-static potentials studies of pentaquarks are already referred in Subsection \ref{sec:static}. They have been computed at the same time as the all static potentials for tetraquarks. These static potentials 
\cite{Okiharu:2004ve,Alexandrou:2004ak}
are also consistent with the potentials computed for mesons and baryons: they have a sum of two-body Coulomb terms with a Casimir operator scaling, a constant term proportional to the number of quarks, and a confining potential with three junctions, detailed in Eq. (\ref{V5Qth}). In Fig. \ref{fig:pentafields} we represent the Lagrangian field density produced by pentaquark static sources.

Many lattice QCD studies of light pentaquarks
\cite{Csikor:2003ng,Mathur:2004jr,Alexandrou:2004ws,Sasaki:2004vz,Lasscock:2005tt,Csikor:2005xb,Alexandrou:2005gc,Takahashi:2005uk,Holland:2005yt,Lasscock:2005kx,Alexandrou:2005ek,Jahn:2005zb,Takahashi:2005mr,Doi:2005mn,Chiu:2005ge,Tariq:2007ck}
were performed just after the $\theta^+$ event in 2003,
\cite{LEPS:2003wug}
which, however, was finally not confirmed by other experiments.
These works, after searching for resonances in different $K-N$ channels, reached different conclusions, most of them preliminary, on the $\theta^+$ resonance existence:
yes 
\cite{Csikor:2003ng,Alexandrou:2004ws,Takahashi:2005uk,Holland:2005yt,Lasscock:2005kx,Takahashi:2005mr},
maybe 
\cite{Sasaki:2004vz,Alexandrou:2005gc,Alexandrou:2005ek,Jahn:2005zb,Chiu:2005ge},
no 
\cite{Mathur:2004jr,Lasscock:2005tt,Csikor:2005xb,Doi:2005mn,Tariq:2007ck}.
Most of theses studies utilized the technique, referred in subsection \ref{sec:scpectra}, of searching for states directly in the midst of the meson-baryon spectrum, which often is inconclusive.

As discussed for the tetraquarks in Section \ref{sec:difapp}, there are two more able techniques to identify resonances: either the computation of potentials for static heavy quarks and dynamical light quarks, or the study of meson-baryon scattering with the Lüscher phase shift technique. 

We now review the techniques and results applied to the to the pentaquarks of the $P_c$ family, which include a $c \bar c$ heavy quark pair. These pentaquarks started to be observed recently, as summarized in Table \ref{tab:Penta}.

Using the
hadroquarkonium picture
\cite{Alberti:2016dru}, 
already referred in Subsection
\ref{sec:staticdy2}, Alberti et al considered the 
  static quark-antiquark $Q \bar Q$  ($m_Q\to \infty$ closer to $b$ than to $c$) as function of distance between $\bar Q$ and $Q$  in presence of the nucleon. 
The energy was shifted down by  a few MeV, compatible with a molecular boundstate.

Beane et al  \cite{Beane:2014sda}, considering a very heavy $m_\pi\simeq 800$ MeV in  $J=\frac{1}{2}^-$ ($G_1^-$) channel of $N\eta_c$, found a boundstate with binding  energy  $\Delta E\approx -20~ \text{MeV}$, which was almost independent of the volume $L\simeq 3.4 - 6.7$ fm  
\cite{Beane:2014sda}.  Such a boundstate would be very interesting but it remains to be confirmed for a physical pion mass.

\begin{figure}
\begin{center}
\includegraphics[width =0.5 \textwidth]{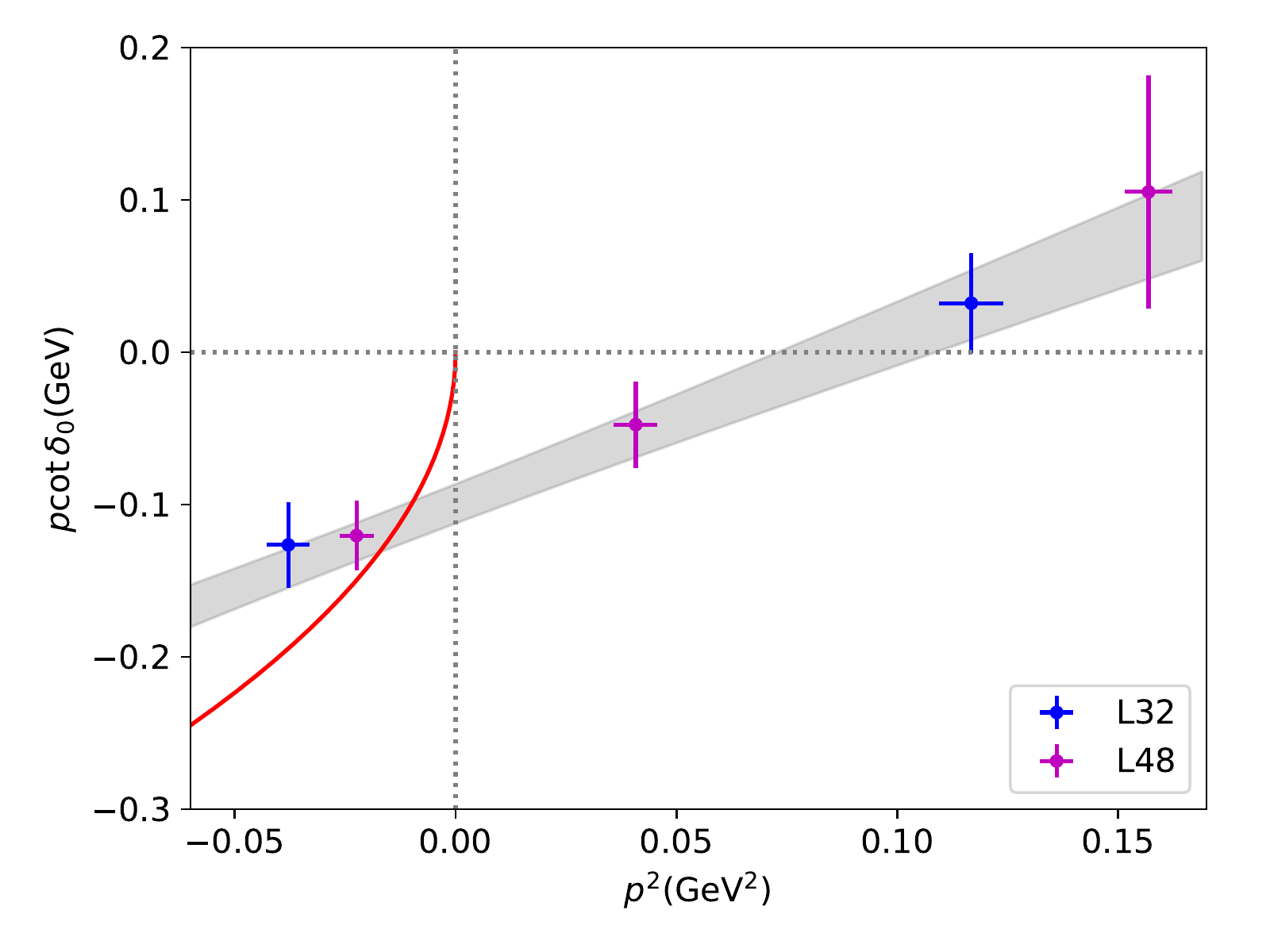}
\end{center}
\caption{Fit of the $p\cot \delta_0$ to the effective range expansion for the $\Sigma_c \bar{D}$ scattering of Ref. \cite{Xing:2022ijm} . The red curve is $ip=-|p|$ versus $p^2$. The bound state pole occurs at the intersection of the fitted band (the grey band) and the red curve.}
\label{fig:pcot_SigmacD}
\end{figure}

Skerbis and Prelovsek \cite{Skerbis:2018unn,Skerbis:2018lew} utilized the technique of studying the spectrum with a large set of correlators. 
They simulated the $NJ/\psi$ and $N\eta_c$ levels in the lattice at $m_\pi\simeq 266~$MeV in channels with all possible $J^P$. This  includes $J^P=3/2^\pm$ and $5/2^\pm$ where LHCb discovered $P_c(4380)$ and $P_c(4450)$ pentaquark states in proton$-J/\psi$ decay. 
However they found no evidence for any $P_c$ pentaquark in the spectrum amongst these two-body channels. However, they have not yet applied the Lüscher phase shift technique to their scattering states.

Two studies computed the potential between $N$ and $J/\psi$ or $\eta_c$ in a s-wave as a function of distance using the HAL QCD method,
 in  a quenched  
 \cite{Kawanai:2010ev}  
 and in a dynamical   
 \cite{Sugiura:2017vks} 
 simulation. The light-quark mass was  larger than physical with $m_\pi=640-870~$MeV in 
 \cite{Kawanai:2010ev} and the nucleon mass $m_N\simeq 1.8~$GeV in 
 \cite{Sugiura:2017vks}.  
They find weakly-attractive interaction near threshold in three channels explored: $J^P=\frac{1}{2}^-,\frac{3}{2}^-$ for $NJ/\psi$ and $\frac{1}{2}^-$ for $N\eta_c$.   The resulting interaction was not  strong enough to form bound states or resonances, but the most interesting experimental region $4.3-4.5~$GeV was not explored. 

The s-wave scattering between a nucleon N and $J/\psi(\eta_c)$ was already studied some time ago, using the   L{\"u}scher formalism in quenched  
\cite{Yokokawa:2006td} 
and dynamical 
\cite{Liu:2008rza} 
QCD.  All calculated scattering lengths $a_0$, were found consistent with zero but only within $1$ or $2$ sigma, while implying a small attractive interaction.

Very recently, Xing et al 
\cite{Xing:2022ijm} 
finally claimed tetraquark resonances with the L\"uscher scattering technique.
The s-wave scattering of $\Sigma_c \bar{D}$ and $\Sigma_c \bar{D}^*$ in the $I(J^P) = \frac{1}{2}(\frac{1}{2}^-)$ channel was calculated with different volumes and already a lower pion mass $M_\pi \sim 294\mathrm{MeV}$ than previous studies. The effective range method, illistrated in Fig. \ref{fig:pcot_SigmacD} led to bound state poles in both $\Sigma_c \bar{D}$ and $\Sigma_c \bar{D}^*$ channels with respective binding energies of $6 \pm 4$MeV   and $7 \pm 4$MeV. However these channels have the same quantum numbers as the channels $p J/\psi$ or $p \eta_c$ or $\Lambda_c \bar{D}^*$ which have a lighter energy threshold; this study may possibly be observing an evidence of the neglected channels.  Thus this study is not absolutely conclusive, a study with all coupled channels should be the next step for a definitive result.

Nevertheless the pentaquark studies in lattice QCD are perhaps more promising than the tetraquark ones, since for instance the $P_c$ decay channels are much closer in energy and reduced mass that the decay channels of the $Z_c$ which may include a $\pi$.

\subsection{
Spin dependent potentials and spin effects in diquarks  \label{sec:diquark}}

There is a missing ingredient in the lattice QCD potentials studies for tetraquarks: the spin dependent interactions. While they have already been computed for mesons, they have not yet been computed for baryons or tetraquarks. Nevertheless we details the studies of spin dependent potentials for mesons, since quark-antiquark pairs are relevant for tetraquarks.

There are also studies of diquarks, who indirectly probe the spin effects in quark-quark pairs. In this subsection we also address the diquark studies in lattice QCD.

%
\begin{figure}[t]
\centering
\includegraphics[width=0.45\columnwidth]{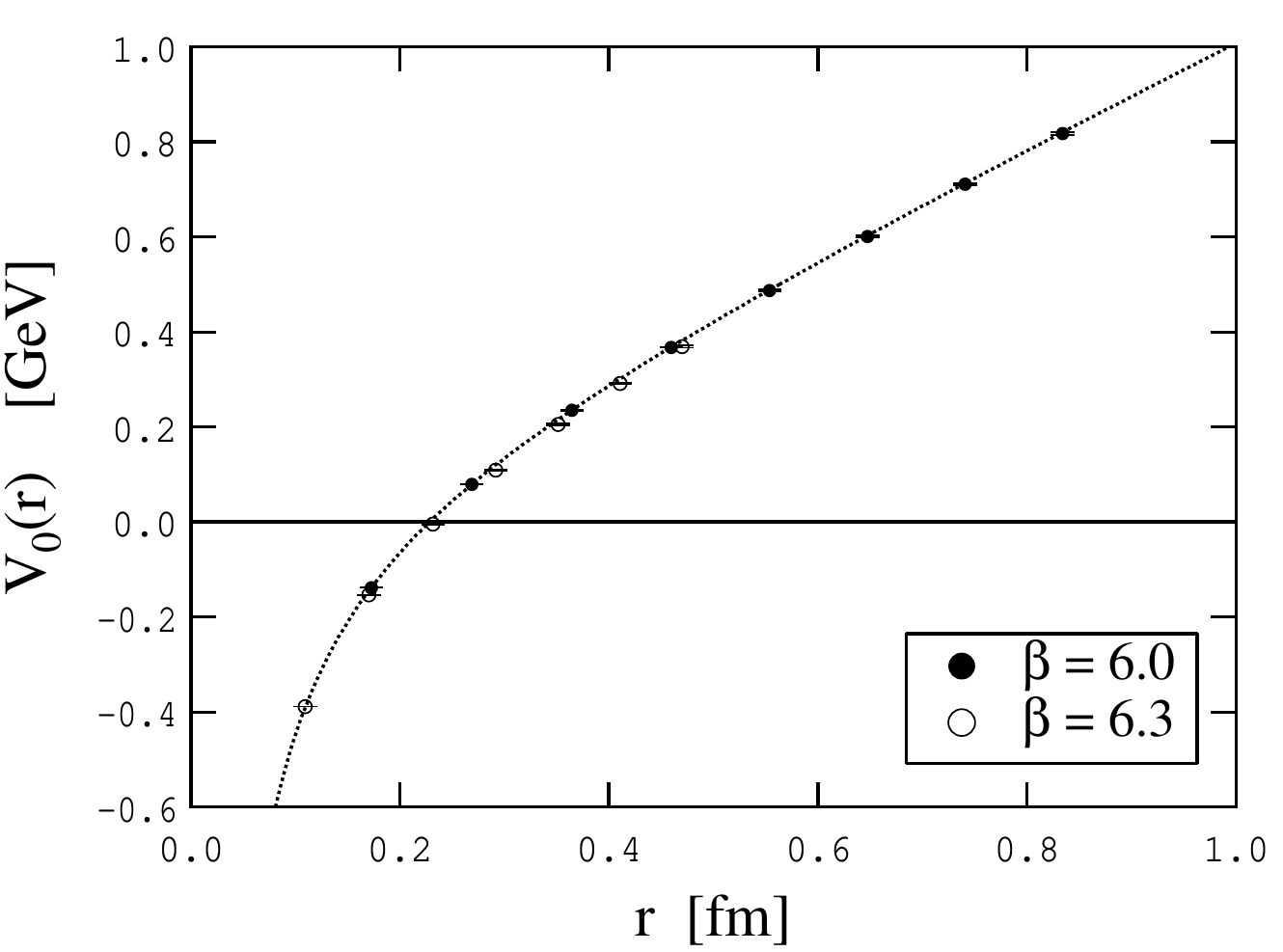}
\includegraphics[width=0.46\columnwidth]{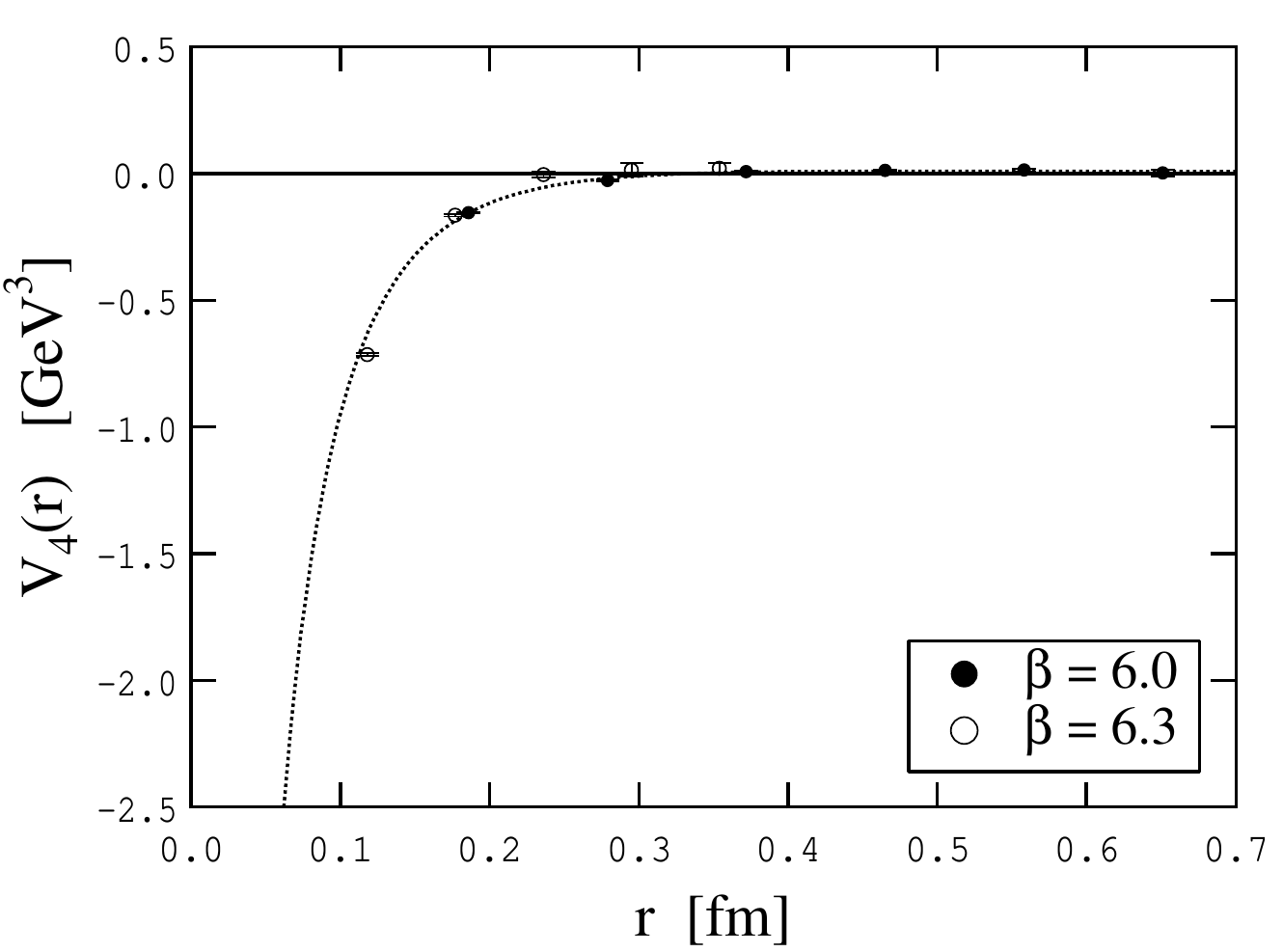}
\caption{
Comparing (left) the spin independent potential $V_{0}(r)$ with 
(right) the spin-spin potential $V_{4}(r)$ at $\beta=6.0$ 
and $\beta=6.3$ computed in Ref. 
\cite{Koma:2006fw}.
The dotted line is the fit curve Eq.~\eqref{eqn:v4-fit},
applied to the data of $\beta=6.0$.}
\label{fig:v4}
\end{figure}

There are some lattice QCD studies of spin-dependent interactions in mesonic quark-antiquark systems
\cite{Campbell:1987nv,Campostrini:1986ki,Perantonis:1988uz,Huntley:1986de,Michael:1985rh,Michael:1985wf,Campostrini:1984uj,deForcrand:1985zc,Lucha:1991vn,Lucha:1992rq,Drummond:1999db,Koma:2006fw,Murano:2013xxa,Bicudo:2016ooe}.
Most of the observed tetraquark candidates are expected to be groundstate tetraquarks, in s-waves. In this case, of all the spin-dependent potentials, only the hyperfine term contributes. This hyperfine term of the quark-antiquark potential in mesons is,
\be
V_{\rm hyp}(r) =
 +
 \frac{\vec{s}_{1}\cdot \vec{s}_{2}}{3m_{1}m_{2}}
 \left (
 c_{F}^{(1)}c_{F}^{(2)}V_{4}(r) -48 \pi C_{F} \alpha_{s} d_{v} 
 \delta^{(3)} (r) 
\right ) \  ,
\label{eqn:potential}
\ee 
where $\alpha_{s}=g^2/(4\pi)$ is the strong coupling, $C_{F}=4/3$ is the Casimir charge of the fundamental representation and $d_{v}$ is the mixing coefficient of the four-quark operator in the (p)NRQCD Lagrangian.
In Eq. (\ref{eqn:potential}),
the Dirac delta term is due to the perturbative OGEP already mentioned in Subsection \ref{sec:theory}. The delta term cannot be computed in lattice QCD which is adequate only at distances larger than the lattice spacing $a$. Moreover it is puzzling since, in the Schrödinger equation, say added to a funnel potential, the delta term produces a solution collapsing to origin, with minus infinity energy. Thus the effective quark models either only use the delta term perturbatively (just shifting the energy of the groundstate obtained with the funnel potential) or use a smeared version of the delta potential \cite{Bhaduri:1981pn} or of the Coulomb potential \cite{Godfrey:1985xj}.

Besides, in Eq. (\ref{eqn:potential}),
$c_{F}^{(i)}(\mu, m_{i})$ ($i=1,2$) is 
the matching coefficient in the (p)NRQCD
Lagrangian which multiplies the term 
$\vec{\sigma}\cdot\vec{B}/(2m_{i})$ and this
coefficient plays an important role when connecting
QCD at a scale $\mu$ with (p)NRQCD at scales~$m_{i}$.
When the matching is performed at tree-level of perturbation theory, the coefficient is $c_{F}^{(i)}=1$
\cite{Manohar:1997qy}
and Eq. (\ref{eqn:potential}) is
reduced to the expression given 
in Refs.
\cite{Eichten:1979pu,Eichten:1980mw,Gromes:1983pm}.

The $V_4(r)$ contribution to the potential has been computed in lattice QCD by Koma and Koma \cite{Koma:2006fw}, the result is shown in Fig. \ref{fig:v4}. To fit the lattice data, the authors used the ansatz,
\bea
V_{\rm 4; fit}(r)=- g' m_{g}^2 \frac{e^{-m_{g}r}}{r} 
+4 \frac{\sigma_{\rm v4}}{r}  \; ,
\label{eqn:v4-fit}
\eea
with a repulsive Coulomb term and an attractive screened Coulomb term motivated by the exchange of a pseudoscalar glueball with assumed $m_{g}= 2.47$~GeV, which is taken from a lattice study 
of the glueball masses 
\cite{Chen:2005mg}.
Treating $g'$ and $\sigma_{\rm v4}$ as free parameters.
The result was $g'=0.292(12)$ 
and $\sigma_{\rm v4} a^2 =0.0015(3)$
with $\chi_{\rm min}^{2}/N_{\mathrm{df}}=5.1$,
and the corresponding curve is put in Fig.~\ref{fig:v4}
(if $m_{g}$ is relaxed to be a free parameter,
$\chi_{\rm min}^2/N_{\rm df}$ is significantly reduced).
The authors also computed the spectrum of heavy quarkonium 
\cite{Koma:2012bc} 
with the spin-dependent potentials up to order $1/m_i$ (which does not yet include the hyperfine potential) and found good agreement with the bottonomium spectrum. The extension of the spin-dependent potentials to tetraquarks would be very interesting.

Another approach to address the spin dependence, is to study diquarks.
Diquarks are important parts, not only of baryons, also of tetraquarks. Besides, they are also relevant for condensed quark matter. For the presently studied tetraquarks, we are mostly interested in groundstate diquarks, where the two quarks are in a relative s-wave. There are two types of such diquarks, with spin 1 and spin 0. In quark models, these diquarks are different due to the spin-spin, or hyperfine, interaction.
While the hyperfine potential in quark-quark pairs has not been computed in lattice QCD, diquarks have been studied in different frameworks 
\cite{Karsch:1998xd,Hess:1998sd,Gockeler:2002sb,Babich:2005ay,Orginos:2005vr,Fodor:2005qx,Alexandrou:2005zn,Alexandrou:2006cq,Liu:2006zi,Babich:2006eu,Babich:2007ah,Gottlieb:2007ay,Gottlieb:2007jm,Bissey:2009gw,Green:2010vc,Watanabe:2021oyv,Francis:2021vrr,Watanabe:2021nwe}.

\begin{figure}[t!]
\centering
\includegraphics[width=0.48\columnwidth]{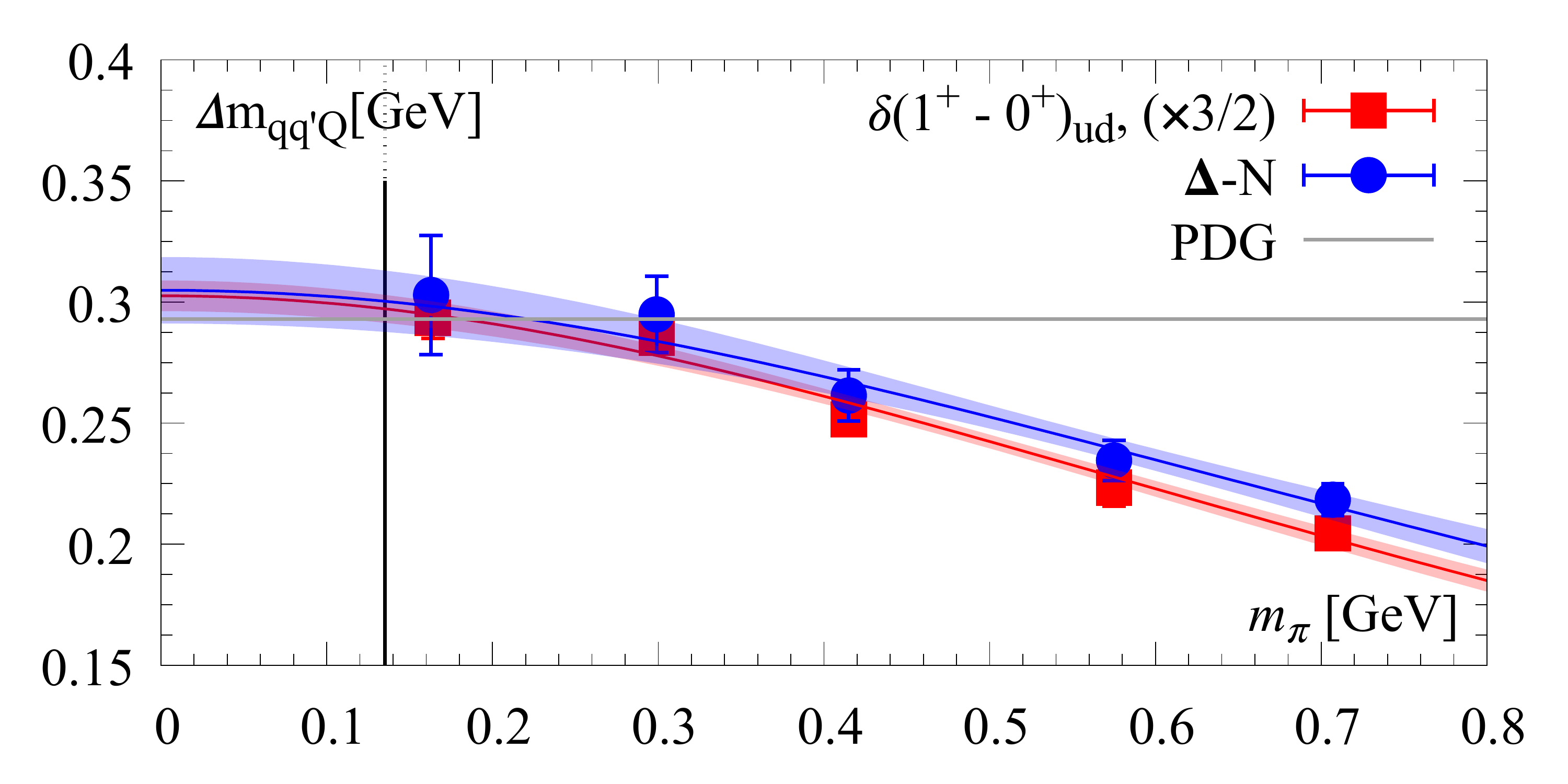}
\quad
\includegraphics[width=0.48\columnwidth]{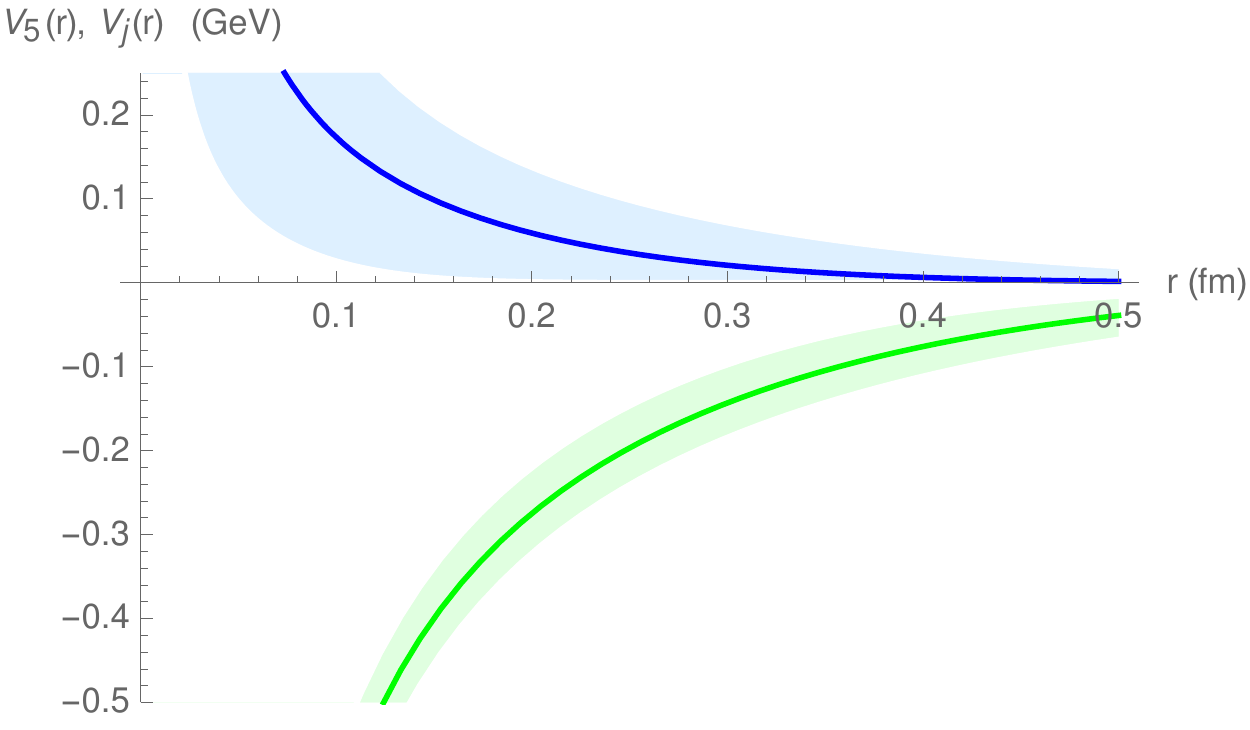}
\caption{
Indirect ways of estimating the hyperfine splitting in tetraquarks.
(left) Agreement 
 \cite{Francis:2021vrr}
of the bad-good diquark mass splitting with the prediction~\cite{Jaffe:2004ph}, 
$\delta(\Delta-N)=3/2 \times \delta(1^+-0^+)_{ud}$.
(right) The attractive $I = 0$ potential $V_5(r)$ (green) and the repulsive $I = 0$ potential $V_j(r)$ (blue) computed in Ref. 
\cite{Bicudo:2015kna},
the error bands reflect the uncertainties of the fits.
}
\label{fig:pots}
\end{figure}

Of the different studies, we highlight a recent study of diquarks on the lattice 
\cite{Francis:2021vrr}
in the background of a 
static quark, in a gauge-invariant formalism with quark masses 
down to almost physical $m_\pi$. 
With more details than the previous studies, it determines mass differences 
between diquark channels as well as 
diquark-quark mass differences. The lightest 
and next-to-lightest diquarks have
{\em good} scalar, $\bar{3}_F$, $\bar{3}_c$, $J^P=0^+$,
{\em bad} axial vector, $6_F$, $\bar{3}_c$, $J^P=1^+$, 
quantum numbers.
The bad-good mass difference for $ud$ flavours, $198(4)~\rm{MeV}$, is
in agreement with phenomenological 
determinations, see Fig. \ref{fig:pots}.
As a very interesting observation, quark-quark attraction is 
found only in the {\em good} diquark channel,
suggesting the hyperfine potential is stronger than the Coulomb potential.
A first exploration of the {\em good} diquark shape,
shows it to be spherical, with a size of $\sim 0.6~\rm{fm}$ .

In a third different approach, the quark-quark hyperfine splitting has also been included in tetraquarks, completing the potential study of Subsection \ref{sec:staticdy}, with an indirect method, in Ref. \cite{Bicudo:2016ooe}. The potential study was performed using bilinear light quark-quark $qq$ fields for the creation operators. To understand the details of the meson-meson structure generated by the creation operators, one has to express them in terms of static-light meson bilinears $\bar Q \Gamma q$.

The $T_{bb}$ is then studied in an s-wave, with a $2 \time 2$ coupled channel equation of the $BB^*$ and $B^* B^*$ channels,
\be
\nonumber  \hspace{-0.7cm}
\left[
\begin{pmatrix}
m_{B^\ast} + m_B & 0 \\
0 & 2 m_{B^\ast}
\end{pmatrix}
- \frac{\hbar}{2 \mu} \frac{\text{d}^2}{\text{d}r^2} \, {1}_{2 \times 2} 
 +
\begin{pmatrix}
V_j(r) + V_5(r) & V_j(r) - V_5(r) \\
V_j(r) - V_5(r) & V_j(r) + V_5(r)
\end{pmatrix}
\right] \chi(r) = E \chi(r) ,
\label{eq:radSG}
\ee
where $\mu = m_b / 2$ and the potentials, attractive $V_5$ and repulsive $V_j$, computed in 
\cite{Bicudo:2015kna}
are depicted in Fig. \ref{fig:pots}. The hyperfine splitting including a heavy quark is now present in the experimental mass difference of the $B$ and $B^*$ mesons. If these masses were degenerate, we could diagonalize the potential, and one of the eigenvalues would be $V_5$; in this case we would recover the result of Ref. 
\cite{Bicudo:2015kna}
where the heavy quark spin effects were neglected.

The existence of the $u d \bar b \bar b$ tetraquark, is confirmed with a binding energy with regards to the lowest channel $BB^*$  reduced to $59 \pm 38$ MeV,
but compared to the median between this channel and $B^*B^*$ the binding energy is essentially unchanged.

\subsection{
String and hybrid excitations with static quarks   \label{sec:hybrid}}

The systems where a quark-antiquark meson is supplemented by the gluon, or flux tube, degrees of freedom are denominated hybrid hadrons. They have a very rich spectrum. Besides, a gluon has the same quantum numbers of a quark-antiquark pair in the $8$ representation of SU(3), and in constituent gluon models, the gluon has an effective gluon mass of the order of two constituent quark masses. Thus exotic hybrids can be confused with tetraquarks who may have the same quantum numbers or masses. In this Subsection we address the lattice QCD flux tubes when the two heavy quarks are static.

%
\begin{figure}[t!]
\begin{centering}
\includegraphics[width=0.24\columnwidth]{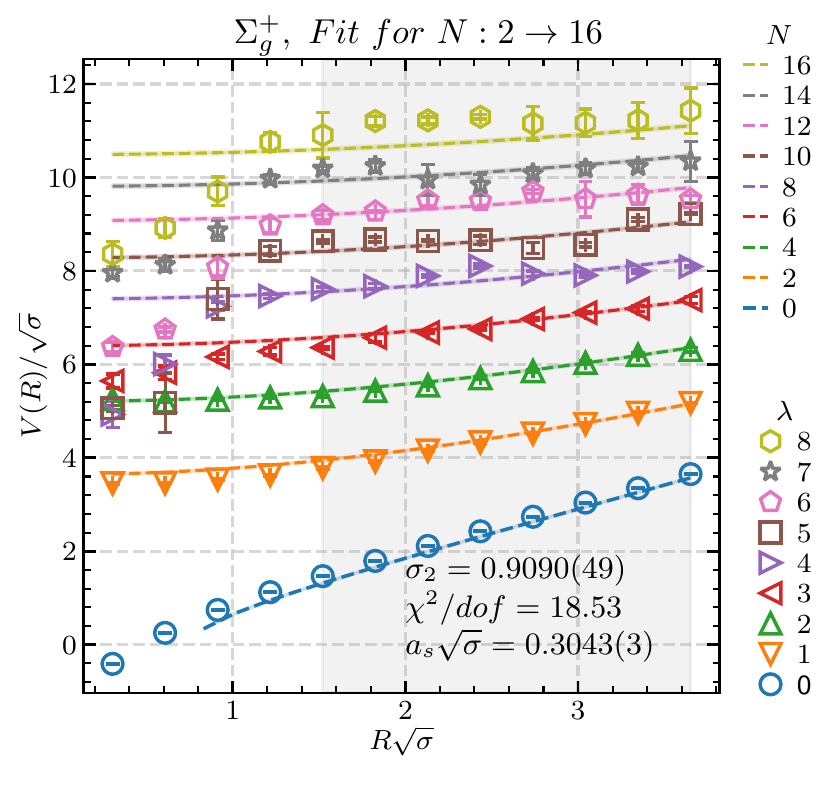}
\
\includegraphics[width=0.24\columnwidth]{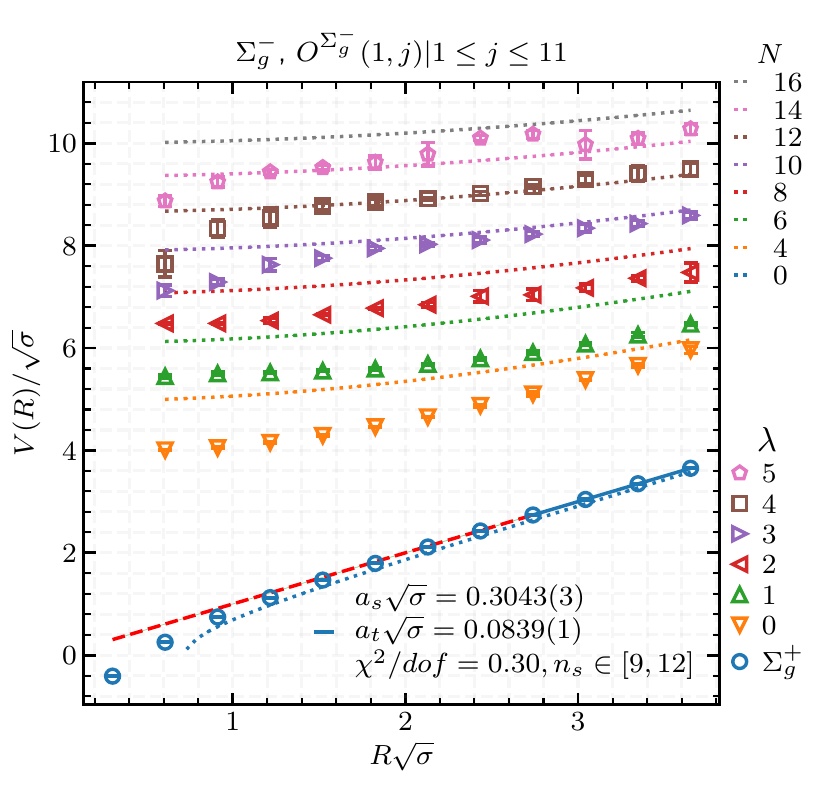}
\
\includegraphics[width=0.24\columnwidth]{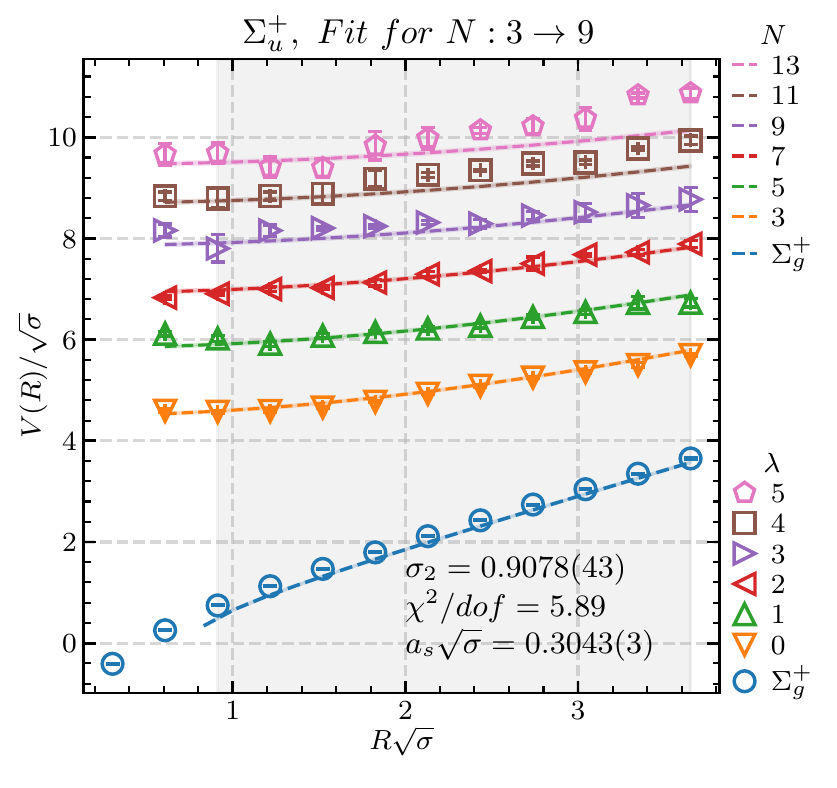}
\
\includegraphics[width=0.24\columnwidth]{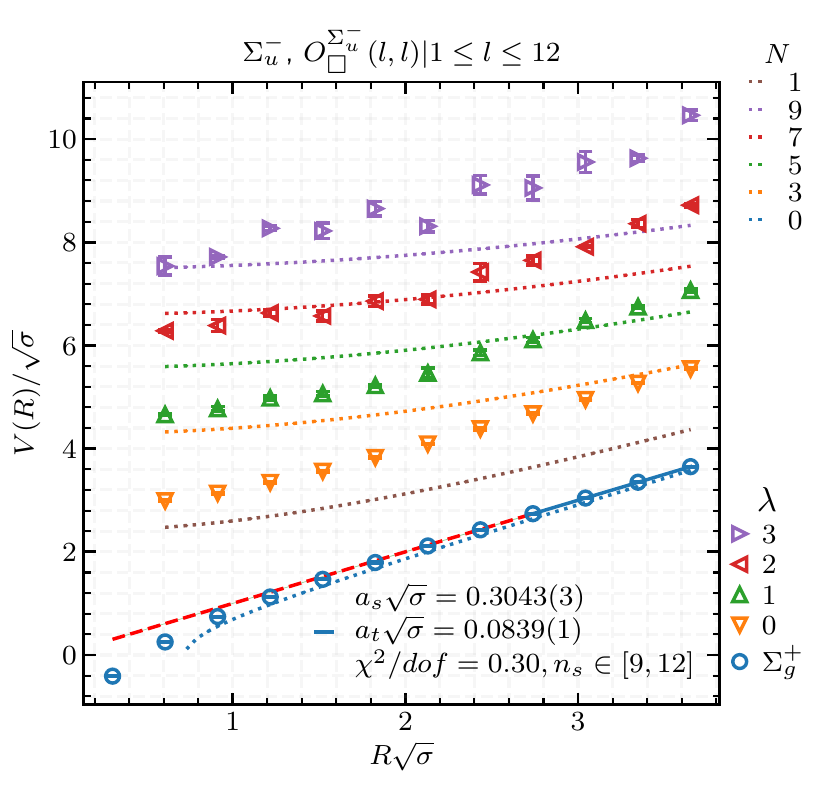}
\\
\includegraphics[width=0.24\columnwidth]{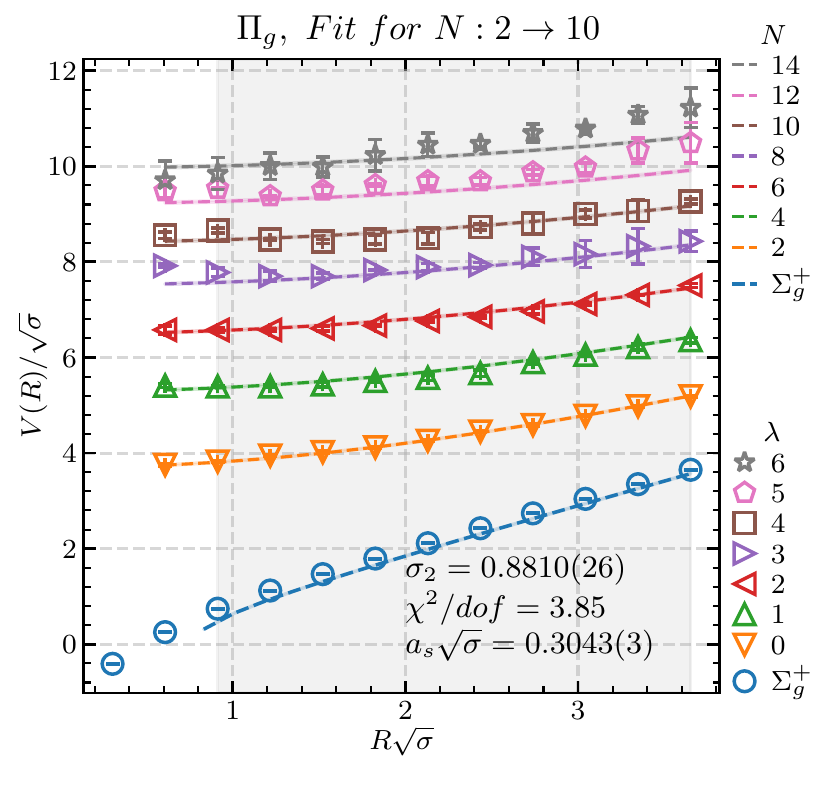}
\
\includegraphics[width=0.24\columnwidth]{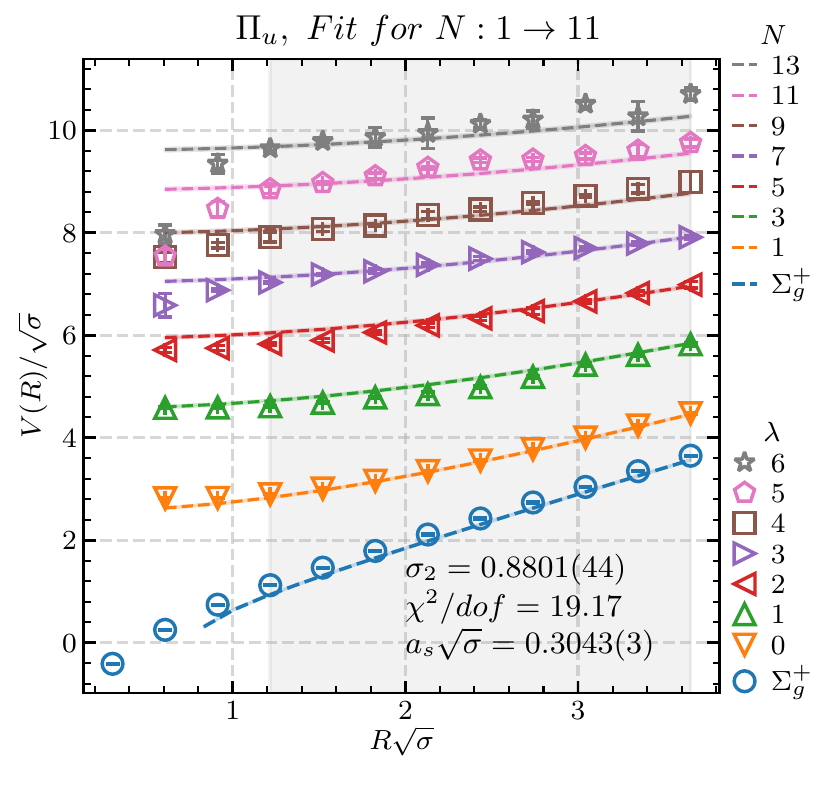}
\
\includegraphics[width=0.24\columnwidth]{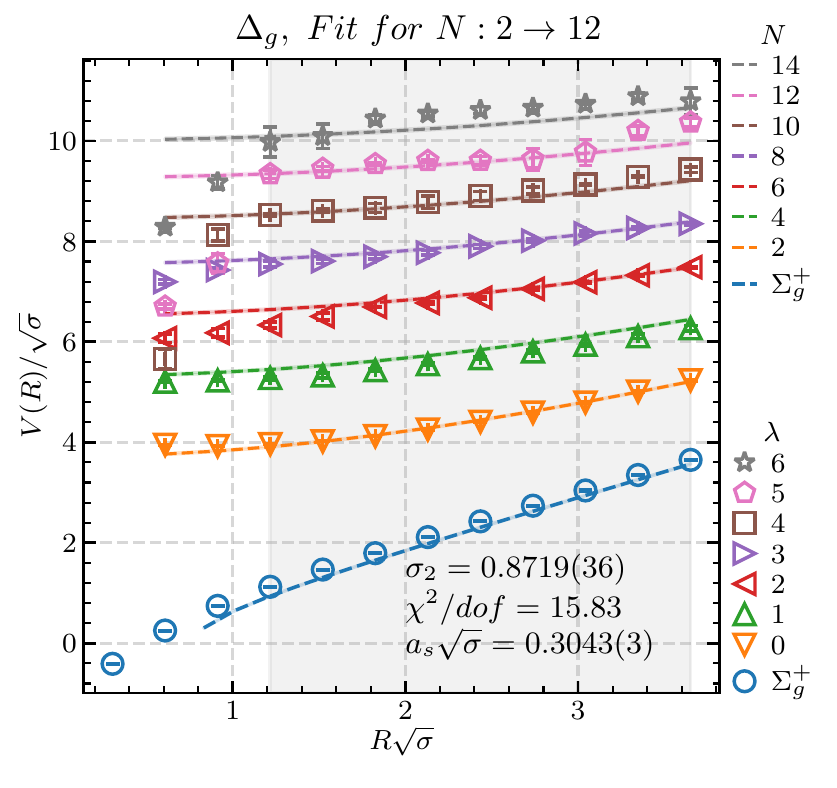}
\
\includegraphics[width=0.24\columnwidth]{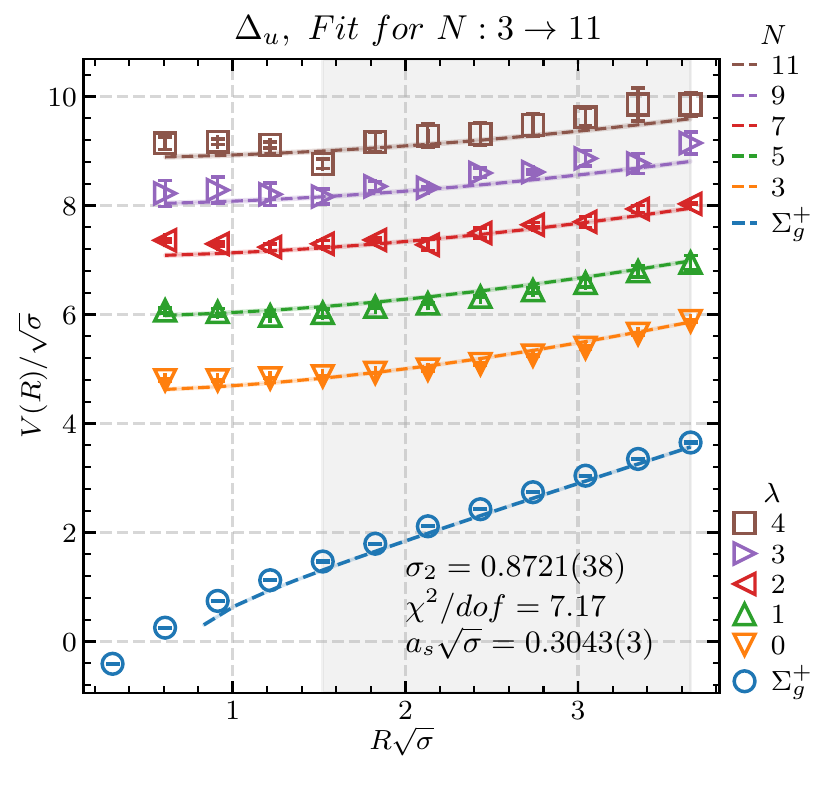}
\end{centering}
\caption{The many excited states of the first eight spectra recently computed in lattice QCD,
dashed lines show the spectrum of the modified Nambu-Goto ansatz
\cite{Bicudo:2022lat}.
\label{fig:eightspectra}}
\end{figure}

The symmetry group of flux tubes is equivalent to the group of the molecular orbitals of homonuclear diatomic molecules. It is the point group denominated $D_{\infty h}$, which nomenclature was as well adopted in the lattice QCD studies of QCD flux tube excitations \cite{Campbell:1987nv,Perantonis:1990dy,Lacock:1996ny,Lacock:1996vy,Juge:1999ie,Juge:2002br,Reisinger:2017btr,Bicudo:2018yhk,Bicudo:2018jbb,Mueller:2019mkh,Bicudo:2021tsc}. 
$D_{\infty h}$ has three symmetry sub-groups, and they determine three quantum numbers.

The two-dimensional rotation about the charge axis corresponds to the quantum angular number, projected in the unit vector of the charge axis  $\Lambda = \left| {\bf J}_g\!\cdot \hat e_z \right| $. The capital Greek
letters $\Sigma, \Pi, \Delta, \Phi, \Gamma \dots$ indicate as usually states
with $\Lambda=0,1,2,3,4 \dots$, respectively. The notation is reminiscent of the $s, \, p, \, d \cdots $ waves in atomic physics.  In the case of two-dimensional rotations there are only two projections ${\bf J}_g\!\cdot \hat e_z = \pm \Lambda$. They are degenerate in energy, and thus we skip this $\pm$ projection in the spectrum notation.

The permutation of the quark and the antiquark static charges is equivalent to the combined operations of
charge conjugation and spatial inversion about the origin. Its eigenvalue is denoted 
$\eta_{CP}$.  States with $\eta_{CP}=1 (-1)$ are denoted
by the subscripts $g$ ($u$), short notation for {\em gerade} ({\em ungerade}).  

Moreover there is a third quantum number, due to the planar, and not three-dimensional, angular momentum. There is an additional label for the s-wave
$\Sigma$ states only. $\Sigma$ states which
are even (odd) under the reflection about a plane containing the molecular
axis are denoted by a superscript $+$ $(-)$. 

With these quantum numbers, the energy levels of the flux tubes are labeled  as $\Sigma_g^+$, $\Sigma_g^-$, $\Sigma_u^+$, $\Sigma_u^-$,
$\Pi_u$, $\Pi_g$, $\Delta_g$, $\Delta_u \cdots$

The first lattice QCD simulations in the literature had only reported,  for the  $\Sigma_g^+$ spectrum up to two levels by Juge et al \cite{Juge:1999ie,Juge:2002br},
and for the other spectra only the groundstate level had been reported in the literature \cite{Campbell:1987nv,Perantonis:1990dy,Lacock:1996ny,Lacock:1996vy,Juge:1999ie,Juge:2002br,Reisinger:2017btr}. Recently, several excitations, up to eight for the $\Sigma_g^+$, have been computed 
\cite{Bicudo:2021tsc,Bicudo:2022lat}.

The simplest Effective String Theory (EST), is the Nambu-Goto model, which action is the area of the transverse bosonic  string surface in time and space, and is classically equivalent to the Polyakov action. Its  spectrum for an open string with ends fixed at distance $R$ with Dirichlet boundary conditions is given by the Arvis potential  
\cite{Arvis:1983fp},
\begin{equation}
V_{n}(R) = 
\sigma\sqrt{
R^{2}
+\frac{2\pi}{\sigma}
\left(
\sum_{a_i=1}^\infty n_{a_i}
-\frac{D-2}{24}
\right)
} \ .
\label{eq:Arvis}
\end{equation}
which has a tachyon in the groundstate. The lattice QCD groundstate is better reproduced by the large distance expansion of Eq. (\ref{eq:Arvis}) , including a 
Coulomb term: the  Lüscher term 
\cite{Luscher:1980ac} 
$\frac{\pi}{R}\left(\sum_{a_i=1}^\infty n_{a_i}-\frac{D-2}{24} \right)$.
$a_i$ are the principal transverse modes, and the continuum field theory computation of the zero mode energy is obtained using the Riemann Zeta regularisation 
\cite{Lovelace:1971fa,Brower:1972wj},
which in lattice QCD is provided by the lattice regularisation.

Fig. \ref{fig:eightspectra} shows the many excited states of the first eight spectra recently computed in lattice QCD,
where the dashed lines show the spectrum of the modified version of the Nambu-Goto ansatz.
\cite{Bicudo:2022lat}.

Also notice there is lattice QCD evidence for an intrinsic width of the QCD flux tube \cite{Cardoso:2013lla} superposed with the quantum vibration width predicted by Lüscher \cite{Luscher:1980iy}.
The QCD flux tube has a rich structure in chromoelectric and chromomagnetic field densities \cite{Bicudo:2018jbb,Mueller:2019mkh}.

\subsection{
Tetraquarks with light quarks only  \label{sec:light}}

%
\begin{figure}[t!]
\begin{center}
\includegraphics[width=0.55\columnwidth]{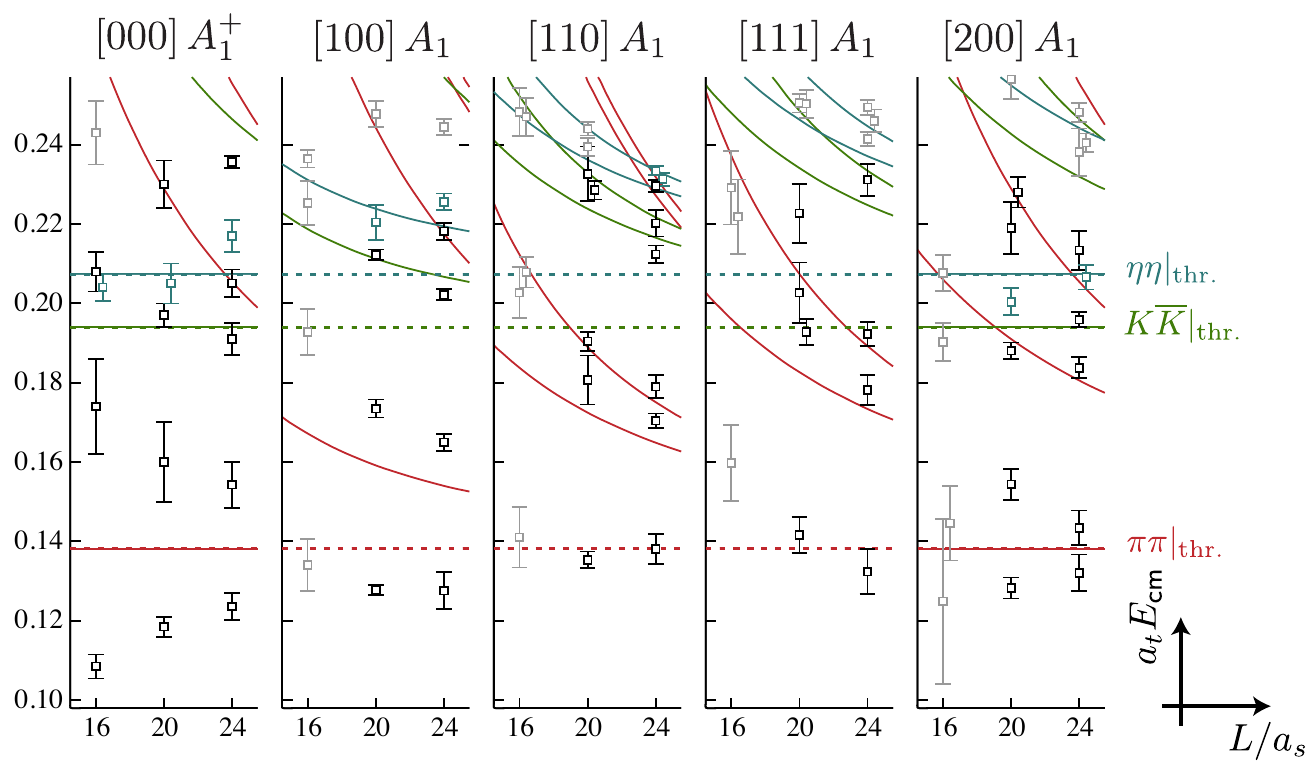}
\includegraphics[width = 0.4 \columnwidth,trim=0pt 20pt 0pt 0pt, clip]{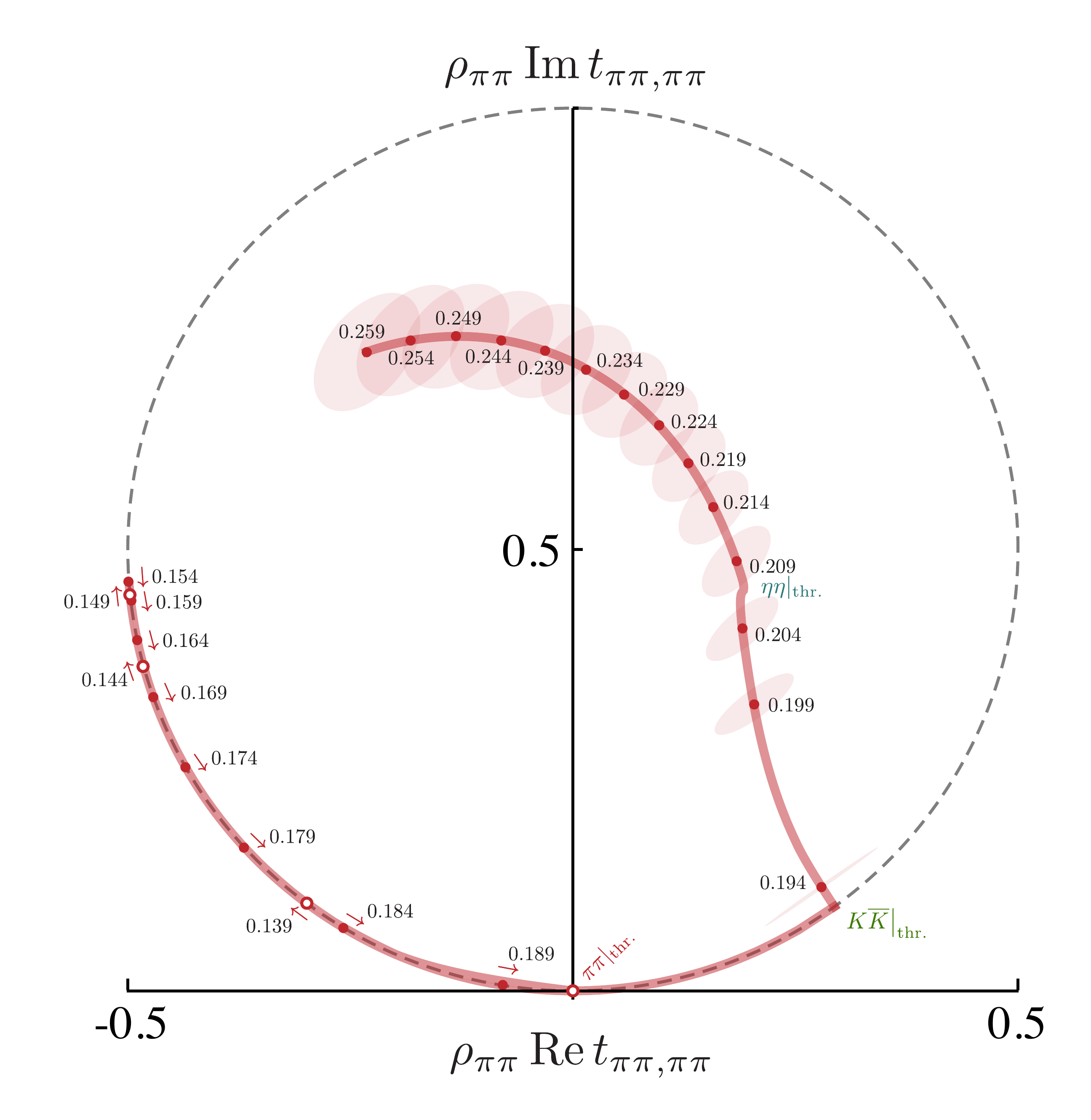}
\caption{ Plots of Ref. 
\cite{Briceno:2017qmb} for the scattering with three coupled channels. 
(left) Finite volume spectra obtained in $A_1$ irreps.
Dashed lines represent the thresholds and the solid curves the meson-meson energies in the absence of interactions: red indicates $\pi \pi$, green is $K \bar K$ and blue is $\eta \eta$. 
(right) Argand diagram representation of the $t_{\pi\pi, \pi\pi}$ element of the amplitude. The amplitude initially moves clockwise from $\pi\pi$ threshold (open circles) before doubling back upon itself (closed circles). The energy is in $a_t\simeq 0.04fm$ units.}\label{fig:argandBriceno}
\end{center}
\end{figure} 

There are many studies of tetraquarks with light quarks only, with a very high state of the art. The first realistic study of a hadron resonance with the Lüscher phase shift technique was for $\pi\pi \to \rho $ scattering in the $J^ {PC}=1^{--}$ channel, leading to the computation of the decay with of the rho meson 
\cite{McNeile:2002fh,CP-PACS:2007wro,Gockeler:2008kc,Budapest-Marseille-Wuppertal:2010gis,Feng:2010es,PACS-CS:2010dxu,Lang:2011mn} 
from first principles QCD.   

The technique was since then applied to the study of the $\sigma$ meson and its nonet including the $\kappa$
\cite{Prelovsek:2010kg,Dudek:2013yja,Fu:2013ffa,Wakayama:2014gpa,Howarth:2015caa,Briceno:2016mjc,Wilson:2014cna}
contributing to clarify its subtle nature of a {\em s pole} tetraquark nonet 
\cite{Alford:2000mm}, and the $a_0$
\cite{Berlin:2014qyu,Berlin:2016zci,Dudek:2016cru, Alexandrou:2017itd}
where the four quark component is important as well to complement the $q \bar q$ component. 

So far, the more advanced study is the one of Briceno et al
\cite{Briceno:2017qmb}, who succeeded in studying the three channel case of the  $\pi\pi$, $K \bar K$ and $\eta\eta$. Fig. \ref {fig:argandBriceno} illustrates some of their achievements. Nevertheless the efforts to perfect the technique must be continued 
\cite{Briceno:2017max}
to be hopefully applied to cases such as the $Z_c$ and $Z_b$ that have many coupled channels below their mass.

\section{ Conclusion and outlook on the predictions, difficulties and expected progress of lattice QCD \label{sec:conclu}}

\subsection{
{\em diquark} tetraquark boundstates  $T_{QQ}$ with two heavy quarks(antiquarks): predictions and challenges \label{sec:resTqq}}

The tetraquark boundstates with two heavy quarks (antiquarks) ind evidence in lattice QCD computations, see Table \ref{tab:boundTetra}, but so far are not observed in the experiments.

Heavy Ion collisions have shown to be able to produce loosely bound systems such as small nuclei or the $T_{cc}$. This is possible because, during hadronisation very large numbers of hadrons are produced, including nearly co-moving hadrons, who may from boundstates.

 However, for a strong boundstate we need at least an heavier $b$ quark, but there are fewer $b$ quarks produced at LHC than $c$ quarks, not enough to have sufficient b close enough in phase space to form boundstates. Nevertheless, these predicted heavier tetraquarks, at least the $T_{bc}$ are expected to be observable in the future with the High Luminosity Large Hadron collider (HL-LHC) at CERN
\cite{Apollinari:2017lan}, 
including detections expected in the future by the experimental collaborations LHCb and CMS. 

The core of these tetraquarks, in the region where there is attraction, is of type {\em diquark} tetraquark, whereas the tail of their wavefunction is of the meson-meson type.
We now review the lattice QCD results on bound tetraquarks, with techniques referred in subsections \ref{sec:static}, \ref{sec:flipflop}, \ref{sec:staticdy}, \ref{sec:GEVP}, \ref{sec:HALqcd}, \ref{sec:effective} and \ref{sec:phaseshifts}.

\subsubsection{
Using static potentials for the heavy quarks together with dynamical light quarks  \label{sec:ressta}}

Using static heavy  quarks, the only predictions so far are just for $T_{bb}$ boundstates, as reviewed in subsections \ref{sec:staticdy} and \ref{sec:GEVP}. The $T_{bb}$ predictions are coincident with the results of Section \ref{sec:resdyn}. This is a very firm prediction and as soon as the experimental collaborations will be able to measure it in the future, we expect this state to be observed.

No other tetraquarks of this class have been obtained with static heavy quarks. Notice only one lattice computation of potentials was performed in 2010 \cite{Wagner:2010ad,Wagner:2011ev}, but with the progress of lattice QCD more precise potentials could presently be computed.

\subsubsection{
Using heavy quark effective theory together with dynamical light quarks   \label{sec:resdyn}}

Using heavy quark effective theory and dynamical light quarks, the predictions of a $T_{bb}1^+$ and a $T_{bbs}1^+$ boundstate are reliable, see Subsection \ref{sec:effective}.

For the  $T_{bb}$, in the earliest computations, there was some tension with binding energies from 83 MeV up to 189 MeV. However the last computations are now agreeing on a binding energy of the order of 110 to 150 MeV. 

However there is still some tension on other tetraquarks, since binding for the tetraquarks $T_{cc}$, $T_{ccs}$, $T_{bb}0^+$, $T_{bbc}$ and $T_{bbcs}$ is predicted in the computations of Ref. \cite{Junnarkar:2018twb} only. Possibly, this group has not yet performed the large volume limit (thermodynamic limit), this might be revised in the future.

It is necessary to extrapolate to three different limits to have reliable results: the continuum limit of a fine mesh in the limit of lattice spacing $a \to 0$, the thermodynamic limit of a volume $V \to \infty$, and the near chiral limit of $m_\pi \to 140$ MeV. 

In lattice QCD, these limits are impossible to reach. Nevertheless, if one gets close enough, an extrapolation of these parameters in the correct direction is possible, and this is the approach used in lattice QCD, see for instance Figs. \ref{fig:tccbind}, \ref{fig:pots}.

\subsubsection{
Future precision lattice QCD studies of {\em diquark} tetraquark boundsates  }

The  $T_{bb}$ and $T_{bbs}$ tetraquarks are already well established and we expect them to provide benchmarks for future lattice QCD computations, with a large engagement of the community to reduce the small tensions between the different approaches used. 

Detecting the $T_{bc}$ will be a major goal of future experiments.  In principle it should form a boundstate because it is more likely to bind than $T_{cc}$ which was already found at LHCb. It is more likely to bind for two different reasons: the $B D^* - D B^*$ system has a larger reduced mass, intermediate between the  $B B^*$ and the $D D^*$, and moreover the heavy quarks can form a {\em good} singlet diquark.

Presently, the  $T_{bc}$ has not yet been confirmed by lattice QCD since the error bar is larger than  the binding energy. We expect the lattice community to invest in more precise simulations, to conclude on the binding/unbinding of  $T_{bc}$. 

All the different approaches can be improved, already with the presently available lattice QCD techniques,  for instance spin dependence can be included in the static approach. Very recently, new approaches started to be used.
In Ref.  \cite{Hudspith:2022lat},
machine learning was used to improve the spin terms in the non-relativistic effective heavy quark formalism for the $b$ quark.
We also expect new efforts from the community to compute more potentials for static quarks.
Ref.
\cite{Hollwieser:2022pov}
is using Laplacian eigenmodes to compute static potentials in the presence of dynamical quarks.
In the case of shallow boundstates, large lattices may be necessary, this is detailed in Subsection \ref{sec:resolight}.

\subsection{
Full-heavy {\em diquark} tetraquarks: no boundstates, resonances will be difficult to study }

The full-heavy tetraquarks, with flavours $c c \bar c \bar c$ or $b b \bar b \bar b$ and any intermediate combination with $c$ and $b$ quarks or tetraquarks, can only be of the type {\em diquark} tetraquark because they cannot exchange light mesons in the $t$ Mandelstam channel and the annihilation in the $s$ Mandelstam channel is negligible.

Experimentally, there is no evidence so far for boundstates of full heavy-tetraquarks. 
While states with a pair $bb$ of heavy quarks (antiquarks) has not been accessible experimentally, systems with $b \bar b$, $c \bar c$, $cc$ or $ccc$ have been observed. In particular a pair of full-heavy tetraquark has been observed, as shown in Table \ref{tab:cryptoTetra}. The lightest one is the $X(6600)$ with flavour $c c \bar c \bar c$ which is of the order of 700 MeV above the lowest possible decay channel of a pair of $\eta_c$ mesons, it is a resonance, clearly not a boundstate. This suggests that indeed there is no full-heavy tetraquark boundstate, at least of the type  $c c \bar c \bar c$.

The full heavy $b b \bar b \bar b$ tetraquarks have been studied by the  lattice QCD group of Huges et al. 
\cite{Hughes:2017xie}.
Indeed no evidence was found for a boundstate. 
In Table \ref{tab:Fierz}, table II in Ref. \cite{Hughes:2017xie}, the authors present the quantum numbers studied:  $0^{++}$, $1^{+-}$ and $2^{++}$  and show that the meson-meson systems overlap with the diquark-antidiquark system.

\begin{table}[t]
\begin{center}
  \begin{tabular}{l|c|c}
    \hline \hline
    $J^{PC}$ & Diquark-AntiDiquark & Two-Meson \\\hline
    $0^{++}$ & $\bar{3}_c\times 3_c$ & $-\frac{1}{2}|0;\Upsilon\Upsilon\rangle + \frac{\sqrt{3}}{2}|0;\eta_b\eta_b\rangle$ \\
    $0^{++}$ & $6_c\times \bar{6}_c$ & $\frac{\sqrt{3}}{2}|0;\Upsilon\Upsilon\rangle + \frac{1}{2}|0;\eta_b\eta_b\rangle$ \\
    $1^{+-}$ & $\bar{3}_c\times 3_c$ & $\frac{1}{\sqrt{2}}\left(|1;\Upsilon\eta_b\rangle + |1;\eta_b\Upsilon\rangle\right)$ \\
    $2^{++}$ & $\bar{3}_c\times 3_c$ & $|2;\Upsilon\Upsilon\rangle$\\\hline \hline
  \end{tabular}
\end{center}
  \caption{Fierz relations in the $\bar{b}\bar{b}bb$ system relating the two-meson and the diquark-antidiquark bilinears, used in full-heavy tetraquarks by \cite{Hughes:2017xie}. These relations are consistent with Fig. \ref{fig:realmol} (right), showing the {\em diquark} and {\em molecular} tetraquarks are not orthogonal.}
\label{tab:Fierz}
\end{table}

Moreover Junnarkar et al have studied several different types of tetraquarks, 
\cite{Junnarkar:2018twb}
at least double-heavy, and the closest to a full heavy boundstate they find is a $u c \bar b \bar b$. In principle, since they addressed the $c$ quark as both one of the heaviest or lightest quarks, they could have studied different sorts of full-heavy tetraquarks. This suggests that they did not find boundstates with flavours  $c c \bar c \bar c$ or $c c \bar b \bar b$.

Having no boundstate of this type is also suggested by the quark model. Let us focus on a diquark. Notice a bounstate should be in principle in a s-wave, with a symmetric spatial wavefunction. The colour of a $qq$ pair should be in an anti-triplet which is antisymmetric.  If we have two identical flavours, say $cc$ or $bb$, the flavour is symmetric. The antisymmetry of fermion wavefunctions forces the spin to be symmetric, corresponding to a spin 1 system, and this increases the mass of our system. Moreover, after separating the centre of mass of each hadron, each meson has one Jacobi coordinate, and we count two Jacobi coordinates for a free meson-meson system. However a bound tetraquark would have three Jacobi coordinates. Due to the zero mode energy of each coordinate in a quantum system, this also increases the energy of the full-heavy tetraquark, and thus it is not expectable that the full heavy tetraquark binds, at least when all the for quarks have the same flavour. This is what Richard et al 
\cite{Ader:1981db,Ballot:1983iv,Zouzou:1986qh,Gignoux:1987cn}
have been advocating since the eighties, pointing that the heavy-light tetraquarks are the preferred candidates for boundstates, although others 
\cite{Heupel:2012ua}
are able to bind full heavy tetraquarks.

Thus lattice QCD, in these full-heavy boundstate studies, agrees with the existing experimental observations and with some of the models: there is no evidence for full-heavy boundstate tetraquarks.

The difficult challenge is then to study the full heavy resonances in lattice QCD. The experimentally observed resonances are detailed in Table \ref{tab:cryptoTetra}, there is the $X(6900)$ observed at LHCb and CMS, and the   $X(6600)$ observed at CMS. Notice these are very excited resonances, 600 MeV to 1000 MeV above threshold. 
To reach their energy, Since a $D$ meson is heavy, with a small gap between energy levels, many levels of energy in the meson-meson channels must be computed.
They are observed in the $J/\psi J/\psi$ channel but are expected to couple as well to the $\eta_c \eta_c$ and $\psi^* \psi^*$ channels.
Thus there are too many meson-meson channels, with too many energy levels, to be simulable  with the present lattice QCD techniques, which state of the art is referred in Subsection \ref{sec:light}. 

Nevertheless these tetraquark resonances are closer to the state of the art than tetraquarks even higher in the spectrum such as the ones of Subsection  \ref{sec:veryhigh}. We expect, in the future, this challenge will be addressed by the lattice QCD community.

\subsection{
{\em s pole} tetraquarks: charmonium and bottonomium extra states are amenable to lattice QCD }

The {\em s pole} tetraquark type. may exist when a meson-meson system has a non-perturbative coupling to a single meson/quarkonium, with a string breaking/annihilation. This is presently amenable to lattice QCD simulations. The coupling needs to be strong enough to produce extra resonances, with dominant meson-meson content.

In the case of extra resonances with at least one charm quark, using the Lüscher scattering technique with one $D$ meson and another meson, lattice QCD was able to reproduce several resonances as detailed in Subsection  \ref{sec:Luschercrypto}, including the $X(3872)$.

In the case of extra resonances with bottom quarks, using lattice QCD string breaking static potentials as in Subsection \ref{sec:breaking}, and the Born-Oppenheimer approximation, it is possible to study resonances high in the bottonomium spectrum. In this case a pair of new {\em s pole} tetraquarks were identified, compatible with the $\Upsilon(10753)$.

It is interesting lattice QCD is able to address these resonances, but not yet the ones of Subsection \ref{sec:veryhigh}.
Nevertheless some developments are still necessary. 

The effective range approach used in  Subsection  \ref{sec:Luschercrypto} is only approximate, and needs to be improved for very excited resonances. It has not yet been applied to bottonomium, say with the technique of Ref. \cite{Pflaumer:2022lgp}.

The string breaking potentials used in Subsection \ref{sec:breaking} need improvement, with more precise data for potentials and mixing angles, with a clarification of the quantum  numbers and with spin dependent terms as well.

\subsection{
Lattice QCD expected progress in reproducing the tetraquark resonances close to lightest thresholds   \label{sec:resolight}}

\begin{figure}[t!]
\begin{center}
\includegraphics[width=0.5\columnwidth]{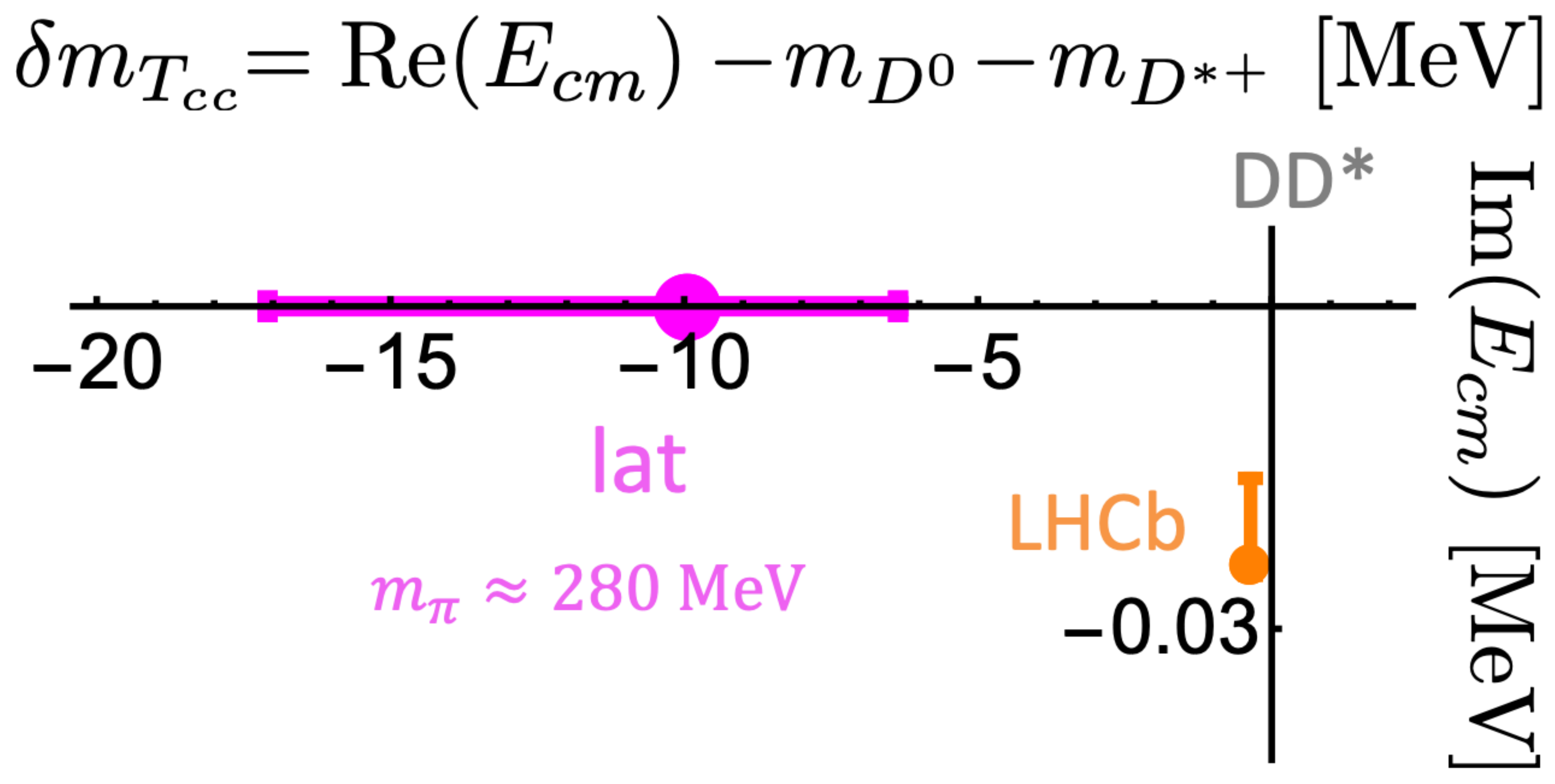}
\caption{The pole in the scattering amplitude related to $T_{cc}$ in the complex energy plane: the lattice 
result of Ref.  \cite{Padmanath:2022cvl} at the heavier charm quark mass (magenta) and the LHCb result (orange). We expect future lattice QCD computations to improve the pion mass and check whether the virtual pole moves closer to the physical one observed at LHCb }
\label{fig:pole}
\end{center}
\end{figure} 

Lattice QCD shows some difficulty to find tetraquark boundstates close to the threshold, as we discussed in Subsection  \ref{sec:phaseshifts}.
Lattice QCD is able to study the $s pole$ type of tetraquark, but  not the {\em molecular} or {\em diquark}. Such is the case of the $T_{cc}$, which is both similar to the $T_{bb}$ and to the deuteron.  The $T_{cc}$ is a very challenging tetraquark. In the future new lattice QCD studies are expected to comprehend it.

For the first time, a very recent computation \cite{Padmanath:2022cvl} studied the $T_{cc}$  tetraquark with the Lüscher phase shift, method as 
detailed in Table \ref{tab:boundTetra}, computing the scattering of two $D$ and $\bar D$ mesons. However no resonance was found close to the thresholds, only a virtual pole was observed in the $T$ matrix, as shown in Fig. \ref{fig:pole}.  Notice the experimental pole is just of the order of 360 KeV below the $D^*{}^+ \, D^0$ threshold but is a resonance above the  $D^0 \, D^0 \, \pi^+$ channel.  If the $\pi^+$ mass used in the lattice computation is heavier than the physical one, what we would expect is a boundstate. 

This indicates that there is still room for improvement in lattice QCD computations. 
Possibly the OPEP, described in Subsection \ref{sec:nuclearTh} is important for the $T_{cc}$.
Notice the OPEP is attractive in isoscalar systems, for instance in the famous case of the deuteron. The $T_{cc}$ is an isoscalar as well, at least according to the lattice QCD computation of potentials
for tetraquarks with two heavy antiquarks \cite{Bicudo:2012qt}.
 The OPEP needs not only a light pion, but also a large enough volume. 
Moreover, the $T_{cc}$ 
\cite{LHCb:2021vvq,LHCb:2021auc}
is reported experimentally as a loosely bound state with a large size of the order of 7 fm.  
The range of the Yukawa potential is of the order of $1 / m_\pi$, considering an average mass of 138 MeV we get a range of the order of 1.43 fm (actually the double of this since we should count the diameter and not the radius). Thus, in the case of loosely bound molecules, we would need a lattice with a size of a some fm to get all the binding provided by OPEP. Although lattice QCD with dynamical fermions is very expensive, (with small ensembles) is is now possible to use very large lattices. With the master-field approach, it is presently possible to use lattices of size $192^ 4$ \cite{Fritzsch:2021klm,Francis:2019muy,Luscher:2017cjh,Ce::2022lat}.

In Ref. \cite{Padmanath:2022cvl}, the lattice spacing is $a=0.08636(98)(40)$~fm, $m_u$ and 
$m_d$ are degenerate and heavier than in nature, corresponding to   $m_\pi\!=\!280(3)~$MeV, utilizing  
255 configurations on spatial volume $ N_L^3= 24^3$ and 492 configurations on $ 32^3$ 
\cite{Bruno:2016plf} with periodic boundary conditions in space. This pion mass is twice as heavy as the physical one, possibly this decreases the binding produced by the OPEP. Moreover a size $a \, N_L$ of 2.1 to 2.8 fm would be insufficient to harvest the full binding power of the OPEP, had the pion the physical mass. On the other hand, the experimental LHCb Ref. \cite{LHCb:2021auc} mentions that an excessive binding may lead to a virtual pole. 
Decreasing the mass of $m_\pi$ closer to the physical one and further increasing the volume should move the pole closer to the experimental value. With state of the art techniques it is already possible to determine the precise path of the pole as a function of these parameters. 

Moreover the extension to the three-body scattering in the decay channel $D D \pi$ should be possible with the techniques of Hansen et al,
\cite{Hansen:2012tf,Hansen:2014eka,Hansen:2015zga,Hansen:2022lat}.

\subsection{
New techniques needed for the {\em diquark} or {\em molecular} multiquark resonances high in the spectrum   \label{sec:veryhigh}}

So far lattice QCD  has tried with no success to reproduce tetraquark resonances of the types {\em diquark} or {\em molecular} high in the spectrum. The large efforts to reproduce the first observed tetraquarks, the $Z_b, \ Z_c$ \dots family, high in the spectrum, and with many decay coupled channels, so far have failed. There is a recent positive result on the first observed pentaquarks of the $P_c$ family, but it may need to be completed with a coupled channel study.

The first technique utilized was the one of Subsection \ref{sec:scpectra}, of studying the spectrum with different operators. It failed to find any evidence of the new tetraquarks. This possibly happens because the operator basis is over-complete, as discussed in Subsection \ref{sec:GEVP}.

Then, using potentials approximating heavy quarks as static ones is a promising approximation, but in fully exotic systems we have not yet succeeded in identifying the mixing angle between the different channels.

The most rigorous approach to study resonances in lattice QCD is the Lüscher phase shift method. However this technique, presented in Subsection \ref{sec:phaseshifts}, so far only can be applied to resonances with few open coupled channels of two mesons.

In a sense, lattice QCD is suffering from what led some of its successes. The groundstate energy is naturally obtained, as discussed in Subsection
\ref{sec:latticeQCD}, in Eq. (\ref{eq:plateau}), even if we have a not so good operator. To use the Lüscher technique high in the spectrum,  we need to determine the spectrum of the different two meson or three meson states in a torus, and the corresponding eigenstates with a good precision. Eq. (\ref{eq:plateau}) can indeed be extended to a matrix equation where we can get several excited states.

However, we still face two problems: the noise increases with the number of excited states since lattice QCD is a statistical approach, and moreover it is difficult to determine the wavefunctions of each state. The difficulty with a large number of coupled channels, lies in the disentanglement of these channels. We would need operators extremely close to the eigenfunctions, in order to clearly identify the energy levels.

This is different from the experimental observations. For instance, a $Z_c$ tetraquark resonance decays to many channels, but we can study the mass and width just from the lightest and more convenient energy channel, the $J/\psi \pi$, although the respective partial decay width is much smaller than the total decay width. This success has been possible because the decay products, say $J/\psi$ and  $\pi^+$ can be excellently tagged for instance in muon detectors, $J/\psi \to \mu^+ + \mu^-$ and $\pi^+ \to \mu^++\nu_\mu$.
In lattice QCD we are still missing a successful tagging technique.

\subsection{
Brief conclusion   \label{sec:bcon}}

As a brief conclusion, lattice QCD succeeded in predicting {\em diquark} tetraquark  boundstates, and {\em s pole} resonances. It is expected that soon the shallow boundstates or very narrow states close to light thresholds will be successfully computed as well.

The robust results of lattice QCD can be used by models and effective theories of QCD, to calibrate their parameters. Moreover lattice QCD can address theoretical questions such as whether the {\em diquark} versus {\em molecular} operators are important for tetraquarks.

So far there are few lattice QCD results on pentaquarks, but the available techniques already applied to tetraquarks are expected to produce more pentaquark and hexaquark results in the future. 

Moreover, new avenues for tetraquark and pentaquark studies still need to be developed, to study the multiquark systems excited high in the spectrum. We expect the subtle tetraquarks and pentaquarks will become a priority for the lattice QCD community, with more computations, increased precision and new techniques to map this new world of hadronic physics.

%
\begin{figure}[t!]
\begin{centering}
\includegraphics[width=0.5\columnwidth]{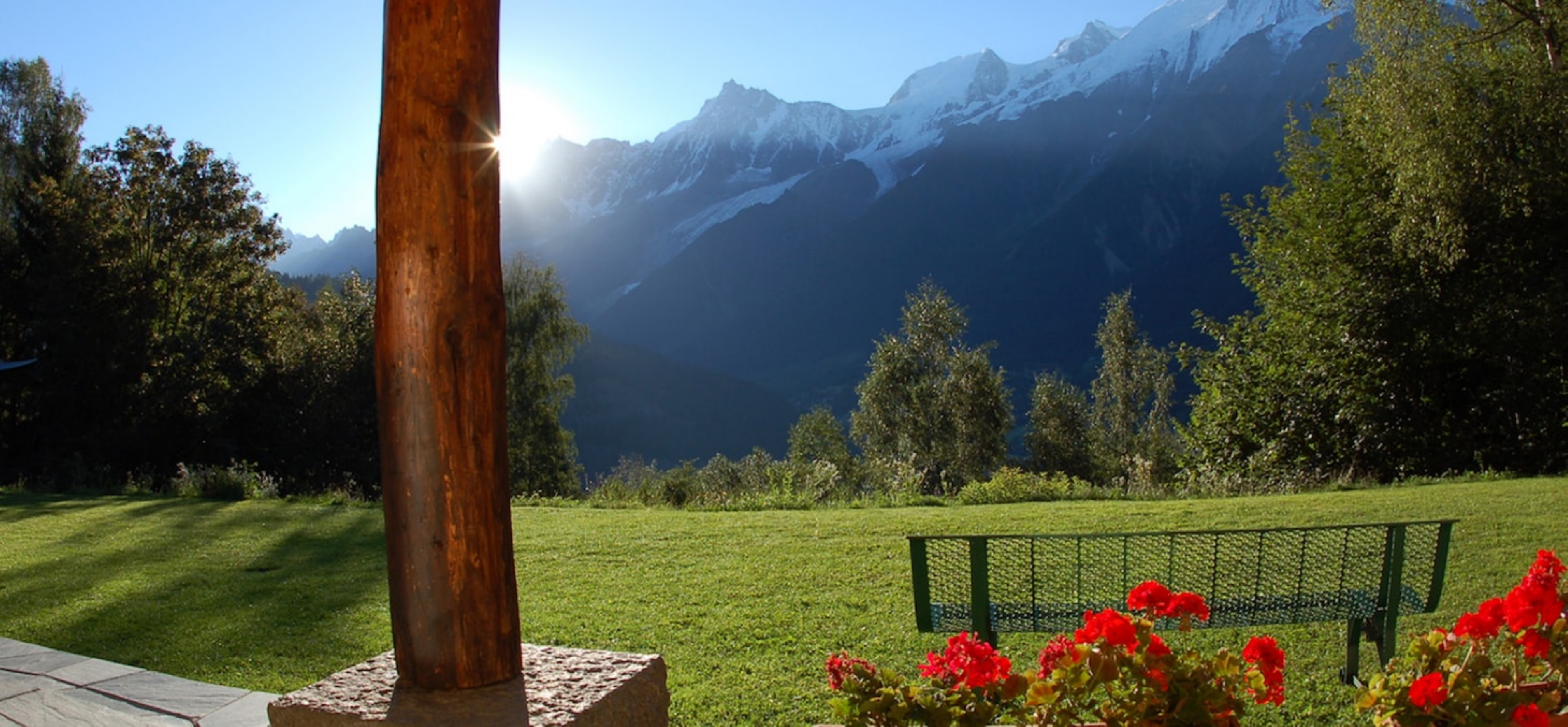}
\caption{View of the Alps peaks from Les Houches School of Physics
\label{fig:Houches}}
\end{centering}
\end{figure}

\section*{Acknowledgements \label{sec:ackno}}
\addcontentsline{toc}{section}{Acknowledgements}

This review on multiquarks with heavy quarks in lattice QCD was first sketched for the workshop {\em Double charm tetraquark and other exotics} 
\cite{richard_uras:2021}
organized by Jean-Marc Richard and co-organizers at IP2I, Claude Bernard University Lyon, in November 2021 after the $T_{cc}$ discovery at LHCb.  P. B. first met Jean-Marc Richard at one of the meetings where he proposed his tetraquark mechanism, held at Les Houches School of Physics, with the inspiring landscape of Fig. \ref{fig:Houches}. The excellent presentations and lively discussions at these meeting inspired this review. 
P. B. also thanks discussions with Gunnar Bali, Marco Cardoso, Nuno Cardoso, Christof Gattringer, Martin Lüscher, Marina Marinkovic, Colin Morningstar, Lasse Müller, Herbert Neuberger, Fumiko Okiharu, Orlando Oliveira, Antje Peters, Sasa Prelovsek, José Emílio Ribeiro, Hideo Suganuma and Marc Wagner. 
P.B.  acknowledges the support of CeFEMA, a research centre at IST of Lisbon University, under the FCT contract for R\&D Units UIDB/04540/2020, the
Pauli Institute for Theoretical Studies Visiting Researcher program
and the hospitality of Institute for Theoretical Physics of ETH Zurich.

\bibliographystyle{elsarticle-num}
\bibliography{xxxxlat,xxBB_e}

\begin{thebibliography}{100}
\expandafter\ifx\csname url\endcsname\relax
  \def\url#1{\texttt{#1}}\fi
\expandafter\ifx\csname urlprefix\endcsname\relax\def\urlprefix{URL }\fi
\expandafter\ifx\csname href\endcsname\relax
  \def\href#1#2{#2} \def\path#1{#1}\fi

\bibitem{Gell-Mann:1964ewy}
M.~Gell-Mann, {A Schematic Model of Baryons and Mesons}, Phys. Lett. 8 (1964)
  214--215.
\newblock \href {https://doi.org/10.1016/S0031-9163(64)92001-3}
  {\path{doi:10.1016/S0031-9163(64)92001-3}}.

\bibitem{Zweig:1964ruk}
G.~Zweig, {An SU(3) model for strong interaction symmetry and its breaking.
  Version 1} (1 1964).

\bibitem{Zweig:1964jf}
G.~Zweig, {An SU(3) model for strong interaction symmetry and its breaking.
  Version 2}, 1964, pp. 22--101.

\bibitem{Glashow:1970gm}
S.~L. Glashow, J.~Iliopoulos, L.~Maiani, {Weak Interactions with Lepton-Hadron
  Symmetry}, Phys. Rev. D 2 (1970) 1285--1292.
\newblock \href {https://doi.org/10.1103/PhysRevD.2.1285}
  {\path{doi:10.1103/PhysRevD.2.1285}}.

\bibitem{SLAC-SP-017:1974ind}
J.~E. Augustin, et~al., {Discovery of a Narrow Resonance in $e^+ e^-$
  Annihilation}, Phys. Rev. Lett. 33 (1974) 1406--1408.
\newblock \href {https://doi.org/10.1103/PhysRevLett.33.1406}
  {\path{doi:10.1103/PhysRevLett.33.1406}}.

\bibitem{E598:1974sol}
J.~J. Aubert, et~al., {Experimental Observation of a Heavy Particle $J$}, Phys.
  Rev. Lett. 33 (1974) 1404--1406.
\newblock \href {https://doi.org/10.1103/PhysRevLett.33.1404}
  {\path{doi:10.1103/PhysRevLett.33.1404}}.

\bibitem{tHooft:1972tcz}
G.~'t~Hooft, M.~J.~G. Veltman, {Regularization and Renormalization of Gauge
  Fields}, Nucl. Phys. B 44 (1972) 189--213.
\newblock \href {https://doi.org/10.1016/0550-3213(72)90279-9}
  {\path{doi:10.1016/0550-3213(72)90279-9}}.

\bibitem{Gross:1973id}
D.~J. Gross, F.~Wilczek, {Ultraviolet Behavior of Nonabelian Gauge Theories},
  Phys. Rev. Lett. 30 (1973) 1343--1346.
\newblock \href {https://doi.org/10.1103/PhysRevLett.30.1343}
  {\path{doi:10.1103/PhysRevLett.30.1343}}.

\bibitem{Politzer:1973fx}
H.~D. Politzer, {Reliable Perturbative Results for Strong Interactions?}, Phys.
  Rev. Lett. 30 (1973) 1346--1349.
\newblock \href {https://doi.org/10.1103/PhysRevLett.30.1346}
  {\path{doi:10.1103/PhysRevLett.30.1346}}.

\bibitem{Wilson:1974sk}
K.~G. Wilson, {Confinement of Quarks}, Phys. Rev. D 10 (1974) 2445--2459.
\newblock \href {https://doi.org/10.1103/PhysRevD.10.2445}
  {\path{doi:10.1103/PhysRevD.10.2445}}.

\bibitem{Jaffe:1976ig}
R.~L. Jaffe, {Multi-Quark Hadrons. 1. The Phenomenology of (2 Quark 2
  anti-Quark) Mesons}, Phys.Rev. D15 (1977) 267.
\newblock \href {https://doi.org/10.1103/PhysRevD.15.267}
  {\path{doi:10.1103/PhysRevD.15.267}}.

\bibitem{Jaffe:1976ih}
R.~L. Jaffe, {Multi-Quark Hadrons. 2. Methods}, Phys. Rev. D 15 (1977) 281.
\newblock \href {https://doi.org/10.1103/PhysRevD.15.281}
  {\path{doi:10.1103/PhysRevD.15.281}}.

\bibitem{Jaffe:1976yi}
R.~L. Jaffe, {Perhaps a Stable Dihyperon}, Phys. Rev. Lett. 38 (1977) 195--198,
  [Erratum: Phys.Rev.Lett. 38, 617 (1977)].
\newblock \href {https://doi.org/10.1103/PhysRevLett.38.195}
  {\path{doi:10.1103/PhysRevLett.38.195}}.

\bibitem{Diakonov:1997mm}
D.~Diakonov, V.~Petrov, M.~V. Polyakov, {Exotic anti-decuplet of baryons:
  Prediction from chiral solitons}, Z. Phys. A 359 (1997) 305--314.
\newblock \href {http://arxiv.org/abs/hep-ph/9703373}
  {\path{arXiv:hep-ph/9703373}}, \href {https://doi.org/10.1007/s002180050406}
  {\path{doi:10.1007/s002180050406}}.

\bibitem{LEPS:2003wug}
T.~Nakano, et~al., {Evidence for a narrow S = +1 baryon resonance in
  photoproduction from the neutron}, Phys. Rev. Lett. 91 (2003) 012002.
\newblock \href {http://arxiv.org/abs/hep-ex/0301020}
  {\path{arXiv:hep-ex/0301020}}, \href
  {https://doi.org/10.1103/PhysRevLett.91.012002}
  {\path{doi:10.1103/PhysRevLett.91.012002}}.

\bibitem{Ader:1981db}
J.~P. Ader, J.~M. Richard, P.~Taxil, {DO NARROW HEAVY MULTI - QUARK STATES
  EXIST?}, Phys. Rev. D25 (1982) 2370.
\newblock \href {https://doi.org/10.1103/PhysRevD.25.2370}
  {\path{doi:10.1103/PhysRevD.25.2370}}.

\bibitem{Ballot:1983iv}
J.~l. Ballot, J.~M. Richard, {FOUR QUARK STATES IN ADDITIVE POTENTIALS}, Phys.
  Lett. B123 (1983) 449--451.
\newblock \href {https://doi.org/10.1016/0370-2693(83)90991-7}
  {\path{doi:10.1016/0370-2693(83)90991-7}}.

\bibitem{Zouzou:1986qh}
S.~Zouzou, B.~Silvestre-Brac, C.~Gignoux, J.~Richard, {FOUR QUARK BOUND
  STATES}, Z.Phys. C30 (1986) 457.
\newblock \href {https://doi.org/10.1007/BF01557611}
  {\path{doi:10.1007/BF01557611}}.

\bibitem{Gignoux:1987cn}
C.~Gignoux, B.~Silvestre-Brac, J.~M. Richard, {Possibility of Stable Multi -
  Quark Baryons}, Phys. Lett. B 193 (1987) 323.
\newblock \href {https://doi.org/10.1016/0370-2693(87)91244-5}
  {\path{doi:10.1016/0370-2693(87)91244-5}}.

\bibitem{LHCb:2021vvq}
R.~Aaij, et~al., {Observation of an exotic narrow doubly charmed tetraquark} (9
  2021).
\newblock \href {http://arxiv.org/abs/2109.01038} {\path{arXiv:2109.01038}}.

\bibitem{LHCb:2021auc}
R.~Aaij, et~al., {Study of the doubly charmed tetraquark $T_{cc}^+$} (9 2021).
\newblock \href {http://arxiv.org/abs/2109.01056} {\path{arXiv:2109.01056}}.

\bibitem{Belle:2003nnu}
S.~K. Choi, et~al., {Observation of a narrow charmonium-like state in exclusive
  $B^\pm \to K^\pm \pi^+ \pi^- J/\psi$ decays}, Phys. Rev. Lett. 91 (2003)
  262001.
\newblock \href {http://arxiv.org/abs/hep-ex/0309032}
  {\path{arXiv:hep-ex/0309032}}, \href
  {https://doi.org/10.1103/PhysRevLett.91.262001}
  {\path{doi:10.1103/PhysRevLett.91.262001}}.

\bibitem{CDF:2003cab}
D.~Acosta, et~al., {Observation of the narrow state $X(3872) \to J/\psi \pi^+
  \pi^-$ in $\bar{p}p$ collisions at $\sqrt{s} = 1.96$ TeV}, Phys. Rev. Lett.
  93 (2004) 072001.
\newblock \href {http://arxiv.org/abs/hep-ex/0312021}
  {\path{arXiv:hep-ex/0312021}}, \href
  {https://doi.org/10.1103/PhysRevLett.93.072001}
  {\path{doi:10.1103/PhysRevLett.93.072001}}.

\bibitem{CMS:2021znk}
A.~M. Sirunyan, et~al., {Evidence for X(3872) in Pb-Pb Collisions and Studies
  of its Prompt Production at $\sqrt {s_{NN}}$=5.02\,\,TeV}, Phys. Rev. Lett.
  128~(3) (2022) 032001.
\newblock \href {http://arxiv.org/abs/2102.13048} {\path{arXiv:2102.13048}},
  \href {https://doi.org/10.1103/PhysRevLett.128.032001}
  {\path{doi:10.1103/PhysRevLett.128.032001}}.

\bibitem{LHCb:2022bly}
{Observation of sizeable $\omega$ contribution to
  $\chi_{c1}(3872)\to\pi^+\pi^-J/\psi$ decays} (4 2022).
\newblock \href {http://arxiv.org/abs/2204.12597} {\path{arXiv:2204.12597}}.

\bibitem{CDF:2009jgo}
T.~Aaltonen, et~al., {Evidence for a Narrow Near-Threshold Structure in the
  $J/\psi\phi$ Mass Spectrum in $B^+\to J/\psi\phi K^+$ Decays}, Phys. Rev.
  Lett. 102 (2009) 242002.
\newblock \href {http://arxiv.org/abs/0903.2229} {\path{arXiv:0903.2229}},
  \href {https://doi.org/10.1103/PhysRevLett.102.242002}
  {\path{doi:10.1103/PhysRevLett.102.242002}}.

\bibitem{D0:2013jvp}
V.~M. Abazov, et~al., {Search for the $X$(4140) state in $B^+ \to
  $J$_{\psi,\phi}K^+$ decays with the D0 Detector}, Phys. Rev. D 89~(1) (2014)
  012004.
\newblock \href {http://arxiv.org/abs/1309.6580} {\path{arXiv:1309.6580}},
  \href {https://doi.org/10.1103/PhysRevD.89.012004}
  {\path{doi:10.1103/PhysRevD.89.012004}}.

\bibitem{CMS:2013jru}
S.~Chatrchyan, et~al., {Observation of a Peaking Structure in the $J/\psi \phi$
  Mass Spectrum from $B^{\pm} \to J/\psi \phi K^{\pm}$ Decays}, Phys. Lett. B
  734 (2014) 261--281.
\newblock \href {http://arxiv.org/abs/1309.6920} {\path{arXiv:1309.6920}},
  \href {https://doi.org/10.1016/j.physletb.2014.05.055}
  {\path{doi:10.1016/j.physletb.2014.05.055}}.

\bibitem{LHCb:2016axx}
R.~Aaij, et~al., {Observation of $J/\psi\phi$ structures consistent with exotic
  states from amplitude analysis of $B^+\to J/\psi \phi K^+$ decays}, Phys.
  Rev. Lett. 118~(2) (2017) 022003.
\newblock \href {http://arxiv.org/abs/1606.07895} {\path{arXiv:1606.07895}},
  \href {https://doi.org/10.1103/PhysRevLett.118.022003}
  {\path{doi:10.1103/PhysRevLett.118.022003}}.

\bibitem{LHCb:2021uow}
R.~Aaij, et~al., {Observation of New Resonances Decaying to $J/\psi K^+$+ and
  $J/\psi \phi$}, Phys. Rev. Lett. 127~(8) (2021) 082001.
\newblock \href {http://arxiv.org/abs/2103.01803} {\path{arXiv:2103.01803}},
  \href {https://doi.org/10.1103/PhysRevLett.127.082001}
  {\path{doi:10.1103/PhysRevLett.127.082001}}.

\bibitem{LHCb:2020coc}
R.~Aaij, et~al., {Study of $B^0_s \to J\psi \pi^+\pi^-K^+K^-$ decays}, JHEP 02
  (2021) 024, [Erratum: JHEP 04, 170 (2021)].
\newblock \href {http://arxiv.org/abs/2011.01867} {\path{arXiv:2011.01867}},
  \href {https://doi.org/10.1007/JHEP02(2021)024}
  {\path{doi:10.1007/JHEP02(2021)024}}.

\bibitem{LHCb:2020bwg}
R.~Aaij, et~al., {Observation of structure in the $J /\psi$ -pair mass
  spectrum}, Sci. Bull. 65~(23) (2020) 1983--1993.
\newblock \href {http://arxiv.org/abs/2006.16957} {\path{arXiv:2006.16957}},
  \href {https://doi.org/10.1016/j.scib.2020.08.032}
  {\path{doi:10.1016/j.scib.2020.08.032}}.

\bibitem{CMS:2022yhl}
{Observation of new structures in the $\mathrm{J}/\psi \mathrm{J}/\psi$ mass
  spectrum in $\mathrm{p}\mathrm{p}$ collisions at $\sqrt{s} =
  13$\textbackslash{},TeV} (2022).

\bibitem{Belle:2019cbt}
R.~Mizuk, et~al., {Observation of a new structure near 10.75 GeV in the energy
  dependence of the ${e^+} {e^-} \to \Upsilon (nS) {\pi^+} {\pi^-} \ (n = 1, 2,
  3)$ cross sections}, JHEP 10 (2019) 220.
\newblock \href {http://arxiv.org/abs/1905.05521} {\path{arXiv:1905.05521}},
  \href {https://doi.org/10.1007/JHEP10(2019)220}
  {\path{doi:10.1007/JHEP10(2019)220}}.

\bibitem{Ketzer:2012vn}
B.~Ketzer, {Hybrid Mesons}, PoS QNP2012 (2012) 025.
\newblock \href {http://arxiv.org/abs/1208.5125} {\path{arXiv:1208.5125}},
  \href {https://doi.org/10.22323/1.157.0025} {\path{doi:10.22323/1.157.0025}}.

\bibitem{JPAC:2018zyd}
A.~Rodas, et~al., {Determination of the pole position of the lightest hybrid
  meson candidate}, Phys. Rev. Lett. 122~(4) (2019) 042002.
\newblock \href {http://arxiv.org/abs/1810.04171} {\path{arXiv:1810.04171}},
  \href {https://doi.org/10.1103/PhysRevLett.122.042002}
  {\path{doi:10.1103/PhysRevLett.122.042002}}.

\bibitem{LHCb:2020bls}
R.~Aaij, et~al., {A model-independent study of resonant structure in $B^+\to
  D^+D^-K^+$ decays}, Phys. Rev. Lett. 125 (2020) 242001.
\newblock \href {http://arxiv.org/abs/2009.00025} {\path{arXiv:2009.00025}},
  \href {https://doi.org/10.1103/PhysRevLett.125.242001}
  {\path{doi:10.1103/PhysRevLett.125.242001}}.

\bibitem{LHCb:2020pxc}
R.~Aaij, et~al., {Amplitude analysis of the $B^+\to D^+D^-K^+$ decay}, Phys.
  Rev. D 102 (2020) 112003.
\newblock \href {http://arxiv.org/abs/2009.00026} {\path{arXiv:2009.00026}},
  \href {https://doi.org/10.1103/PhysRevD.102.112003}
  {\path{doi:10.1103/PhysRevD.102.112003}}.

\bibitem{BESIII:2020qkh}
M.~Ablikim, et~al., {Observation of a Near-Threshold Structure in the $K^+$
  Recoil-Mass Spectra in $e^+e^- \rightarrow K^+(D_s^-D^{*0}+D_s^{*-}D^0$)},
  Phys. Rev. Lett. 126~(10) (2021) 102001.
\newblock \href {http://arxiv.org/abs/2011.07855} {\path{arXiv:2011.07855}},
  \href {https://doi.org/10.1103/PhysRevLett.126.102001}
  {\path{doi:10.1103/PhysRevLett.126.102001}}.

\bibitem{BESIII:2022qzr}
M.~Ablikim, et~al., {Evidence for a Neutral Near-Threshold Structure in the
  $K^{0}_{S}$ recoil-mass spectra in $e^+e^-\rightarrow K^{0}_{S}D_s^+D^{*-}$
  and $e^+e^-\rightarrow K^{0}_{S}D_s^{*+}D^{-}$}, Phys. Rev. Lett. 129~(11)
  (2022) 112003.
\newblock \href {http://arxiv.org/abs/2204.13703} {\path{arXiv:2204.13703}},
  \href {https://doi.org/10.1103/PhysRevLett.129.112003}
  {\path{doi:10.1103/PhysRevLett.129.112003}}.

\bibitem{BESIII:2013ris}
M.~Ablikim, et~al., {Observation of a Charged Charmoniumlike Structure in
  $e^+e^- \to \pi^+\pi^- J/\psi$ at $\sqrt{s}$ =4.26 GeV}, Phys. Rev. Lett. 110
  (2013) 252001.
\newblock \href {http://arxiv.org/abs/1303.5949} {\path{arXiv:1303.5949}},
  \href {https://doi.org/10.1103/PhysRevLett.110.252001}
  {\path{doi:10.1103/PhysRevLett.110.252001}}.

\bibitem{Belle:2013yex}
Z.~Q. Liu, et~al., {Study of $e^+e^- \to n^+ n^- J/\Psi$ and Observation of a
  Charged Charmoniumlike State at Belle}, Phys. Rev. Lett. 110 (2013) 252002,
  [Erratum: Phys.Rev.Lett. 111, 019901 (2013)].
\newblock \href {http://arxiv.org/abs/1304.0121} {\path{arXiv:1304.0121}},
  \href {https://doi.org/10.1103/PhysRevLett.110.252002}
  {\path{doi:10.1103/PhysRevLett.110.252002}}.

\bibitem{Xiao:2013iha}
T.~Xiao, S.~Dobbs, A.~Tomaradze, K.~K. Seth, {Observation of the Charged Hadron
  $Z\_c^{\pm}(3900)$ and Evidence for the Neutral $Z\_c^0(3900)$ in $e^+e^-\to
  \pi\pi J/\psi $ at $\sqrt{s}=4170$ MeV}, Phys. Lett. B727 (2013) 366--370.
\newblock \href {http://arxiv.org/abs/1304.3036} {\path{arXiv:1304.3036}},
  \href {https://doi.org/10.1016/j.physletb.2013.10.041}
  {\path{doi:10.1016/j.physletb.2013.10.041}}.

\bibitem{BESIII:2013qmu}
M.~Ablikim, et~al., {Observation of a charged $(D\bar{D}^{*})^\pm$ mass peak in
  $e^{+}e^{-} \to \pi D\bar{D}^{*}$ at $\sqrt{s} =$ 4.26 GeV}, Phys. Rev. Lett.
  112~(2) (2014) 022001.
\newblock \href {http://arxiv.org/abs/1310.1163} {\path{arXiv:1310.1163}},
  \href {https://doi.org/10.1103/PhysRevLett.112.022001}
  {\path{doi:10.1103/PhysRevLett.112.022001}}.

\bibitem{BESIII:2015pqw}
M.~Ablikim, et~al., {Confirmation of a charged charmoniumlike state
  $Z_c(3885)^{\mp}$ in $e^+e^-\to\pi^{\pm}(D\bar{D}^*)^\mp$ with double $D$
  tag}, Phys. Rev. D 92~(9) (2015) 092006.
\newblock \href {http://arxiv.org/abs/1509.01398} {\path{arXiv:1509.01398}},
  \href {https://doi.org/10.1103/PhysRevD.92.092006}
  {\path{doi:10.1103/PhysRevD.92.092006}}.

\bibitem{BESIII:2013ouc}
M.~Ablikim, et~al., {Observation of a Charged Charmoniumlike Structure
  $Z_c$(4020) and Search for the $Z_c$(3900) in $e^+e^- \to n^+ n^- h_c$},
  Phys. Rev. Lett. 111~(24) (2013) 242001.
\newblock \href {http://arxiv.org/abs/1309.1896} {\path{arXiv:1309.1896}},
  \href {https://doi.org/10.1103/PhysRevLett.111.242001}
  {\path{doi:10.1103/PhysRevLett.111.242001}}.

\bibitem{BESIII:2013mhi}
M.~Ablikim, et~al., {Observation of a charged charmoniumlike structure in
  $e^+e^- \to (D^{*} \bar{D}^{*})^{\pm} \pi^\mp$ at $\sqrt{s}=4.26$GeV}, Phys.
  Rev. Lett. 112~(13) (2014) 132001.
\newblock \href {http://arxiv.org/abs/1308.2760} {\path{arXiv:1308.2760}},
  \href {https://doi.org/10.1103/PhysRevLett.112.132001}
  {\path{doi:10.1103/PhysRevLett.112.132001}}.

\bibitem{Belle:2008qeq}
R.~Mizuk, et~al., {Observation of two resonance-like structures in the pi+
  chi(c1) mass distribution in exclusive anti-B0 ---\ensuremath{>} K- pi+
  chi(c1) decays}, Phys. Rev. D 78 (2008) 072004.
\newblock \href {http://arxiv.org/abs/0806.4098} {\path{arXiv:0806.4098}},
  \href {https://doi.org/10.1103/PhysRevD.78.072004}
  {\path{doi:10.1103/PhysRevD.78.072004}}.

\bibitem{LHCb:2018oeg}
R.~Aaij, et~al., {Evidence for an $\eta _c(1S) \pi ^-$ resonance in $B^0
  \rightarrow \eta _c(1S) K^+\pi ^-$ decays}, Eur. Phys. J. C 78~(12) (2018)
  1019.
\newblock \href {http://arxiv.org/abs/1809.07416} {\path{arXiv:1809.07416}},
  \href {https://doi.org/10.1140/epjc/s10052-018-6447-z}
  {\path{doi:10.1140/epjc/s10052-018-6447-z}}.

\bibitem{Belle:2014nuw}
K.~Chilikin, et~al., {Observation of a new charged charmoniumlike state in
  $\bar{B}^0 \to J/ \Psi \Psi^-n^+$ decays}, Phys. Rev. D 90~(11) (2014)
  112009.
\newblock \href {http://arxiv.org/abs/1408.6457} {\path{arXiv:1408.6457}},
  \href {https://doi.org/10.1103/PhysRevD.90.112009}
  {\path{doi:10.1103/PhysRevD.90.112009}}.

\bibitem{Belle:2007hrb}
S.~K. Choi, et~al., {Observation of a resonance-like structure in the $pi^\pm
  \psi^\prime$ mass distribution in exclusive $B \to K \pi^\pm \psi^\prime$
  decays}, Phys. Rev. Lett. 100 (2008) 142001.
\newblock \href {http://arxiv.org/abs/0708.1790} {\path{arXiv:0708.1790}},
  \href {https://doi.org/10.1103/PhysRevLett.100.142001}
  {\path{doi:10.1103/PhysRevLett.100.142001}}.

\bibitem{Belle:2009lvn}
R.~Mizuk, et~al., {Dalitz analysis of B ---\ensuremath{>} K pi+ psi-prime
  decays and the Z(4430)+}, Phys. Rev. D 80 (2009) 031104.
\newblock \href {http://arxiv.org/abs/0905.2869} {\path{arXiv:0905.2869}},
  \href {https://doi.org/10.1103/PhysRevD.80.031104}
  {\path{doi:10.1103/PhysRevD.80.031104}}.

\bibitem{Belle:2013shl}
K.~Chilikin, et~al., {Experimental constraints on the spin and parity of the
  $Z$(4430)$^+$}, Phys. Rev. D 88~(7) (2013) 074026.
\newblock \href {http://arxiv.org/abs/1306.4894} {\path{arXiv:1306.4894}},
  \href {https://doi.org/10.1103/PhysRevD.88.074026}
  {\path{doi:10.1103/PhysRevD.88.074026}}.

\bibitem{LHCb:2014zfx}
R.~Aaij, et~al., {Observation of the resonant character of the $Z(4430)^-$
  state}, Phys. Rev. Lett. 112~(22) (2014) 222002.
\newblock \href {http://arxiv.org/abs/1404.1903} {\path{arXiv:1404.1903}},
  \href {https://doi.org/10.1103/PhysRevLett.112.222002}
  {\path{doi:10.1103/PhysRevLett.112.222002}}.

\bibitem{LHCb:2015sqg}
R.~Aaij, et~al., {Model-independent confirmation of the $Z(4430)^-$ state},
  Phys. Rev. D 92~(11) (2015) 112009.
\newblock \href {http://arxiv.org/abs/1510.01951} {\path{arXiv:1510.01951}},
  \href {https://doi.org/10.1103/PhysRevD.92.112009}
  {\path{doi:10.1103/PhysRevD.92.112009}}.

\bibitem{Belle:2011aa}
A.~Bondar, et~al., {Observation of two charged bottomonium-like resonances in
  Y(5S) decays}, Phys. Rev. Lett. 108 (2012) 122001.
\newblock \href {http://arxiv.org/abs/1110.2251} {\path{arXiv:1110.2251}},
  \href {https://doi.org/10.1103/PhysRevLett.108.122001}
  {\path{doi:10.1103/PhysRevLett.108.122001}}.

\bibitem{Junnarkar:2018twb}
P.~Junnarkar, N.~Mathur, M.~Padmanath, {Study of doubly heavy tetraquarks in
  Lattice QCD}, Phys. Rev. D 99~(3) (2019) 034507.
\newblock \href {http://arxiv.org/abs/1810.12285} {\path{arXiv:1810.12285}},
  \href {https://doi.org/10.1103/PhysRevD.99.034507}
  {\path{doi:10.1103/PhysRevD.99.034507}}.

\bibitem{Padmanath:2022cvl}
M.~Padmanath, S.~Prelovsek, {Evidence for a doubly charm tetraquark pole in
  $DD^*$ scattering on the lattice} (2 2022).
\newblock \href {http://arxiv.org/abs/2202.10110} {\path{arXiv:2202.10110}}.

\bibitem{Wagner:2022qwg}
M.~Wagner, S.~Meinel, M.~Pflauman, {Search for $\bar b \bar bus$ and $\bar b
  \bar cud$ tetraquark bound states using lattice QCD}.

\bibitem{Pflaumer:2022lgp}
M.~Pflaumer, C.~Alexandrou, J.~Finkenrath, T.~Leontiou, S.~Meinel, M.~Wagner,
  {Antiheavy-Antiheavy-Light-Light Four-Quark Bound States}, in: {39th
  International Symposium on Lattice Field Theory}, 2022.
\newblock \href {http://arxiv.org/abs/2211.00951} {\path{arXiv:2211.00951}}.

\bibitem{Bicudo:2012qt}
P.~Bicudo, M.~Wagner, {Lattice QCD signal for a bottom-bottom tetraquark},
  Phys.Rev. D87~(11) (2013) 114511.
\newblock \href {http://arxiv.org/abs/1209.6274} {\path{arXiv:1209.6274}},
  \href {https://doi.org/10.1103/PhysRevD.87.114511}
  {\path{doi:10.1103/PhysRevD.87.114511}}.

\bibitem{Bicudo:2015vta}
P.~Bicudo, K.~Cichy, A.~Peters, B.~Wagenbach, M.~Wagner, {Evidence for the
  existence of $u d \bar{b} \bar{b}$ and the non-existence of $s s \bar{b}
  \bar{b}$ and $c c \bar{b} \bar{b}$ tetraquarks from lattice QCD} (2015).
\newblock \href {http://arxiv.org/abs/1505.00613} {\path{arXiv:1505.00613}}.

\bibitem{Bicudo:2015kna}
P.~Bicudo, K.~Cichy, A.~Peters, M.~Wagner, {BB interactions with static bottom
  quarks from Lattice QCD}, Phys. Rev. D 93~(3) (2016) 034501.
\newblock \href {http://arxiv.org/abs/1510.03441} {\path{arXiv:1510.03441}},
  \href {https://doi.org/10.1103/PhysRevD.93.034501}
  {\path{doi:10.1103/PhysRevD.93.034501}}.

\bibitem{Bicudo:2017szl}
P.~Bicudo, M.~Cardoso, A.~Peters, M.~Pflaumer, M.~Wagner, {$u d \bar{b}
  \bar{b}$ tetraquark resonances with lattice QCD potentials and the
  Born-Oppenheimer approximation}, Phys. Rev. D 96~(5) (2017) 054510.
\newblock \href {http://arxiv.org/abs/1704.02383} {\path{arXiv:1704.02383}},
  \href {https://doi.org/10.1103/PhysRevD.96.054510}
  {\path{doi:10.1103/PhysRevD.96.054510}}.

\bibitem{Bicudo:2021qxj}
P.~Bicudo, A.~Peters, S.~Velten, M.~Wagner, {Importance of meson-meson and of
  diquark-antidiquark creation operators for a $\bar{b} \bar{b} u d$
  tetraquark}, Phys. Rev. D 103~(11) (2021) 114506.
\newblock \href {http://arxiv.org/abs/2101.00723} {\path{arXiv:2101.00723}},
  \href {https://doi.org/10.1103/PhysRevD.103.114506}
  {\path{doi:10.1103/PhysRevD.103.114506}}.

\bibitem{Bicudo:2016ooe}
P.~Bicudo, J.~Scheunert, M.~Wagner, {Including heavy spin effects in the
  prediction of a $\bar{b} \bar{b} u d$ tetraquark with lattice QCD
  potentials}, Phys. Rev. D 95~(3) (2017) 034502.
\newblock \href {http://arxiv.org/abs/1612.02758} {\path{arXiv:1612.02758}},
  \href {https://doi.org/10.1103/PhysRevD.95.034502}
  {\path{doi:10.1103/PhysRevD.95.034502}}.

\bibitem{Francis:2016hui}
A.~Francis, R.~J. Hudspith, R.~Lewis, K.~Maltman, {Lattice Prediction for
  Deeply Bound Doubly Heavy Tetraquarks}, Phys. Rev. Lett. 118~(14) (2017)
  142001.
\newblock \href {http://arxiv.org/abs/1607.05214} {\path{arXiv:1607.05214}},
  \href {https://doi.org/10.1103/PhysRevLett.118.142001}
  {\path{doi:10.1103/PhysRevLett.118.142001}}.

\bibitem{Francis:2018jyb}
A.~Francis, R.~J. Hudspith, R.~Lewis, K.~Maltman, {Evidence for charm-bottom
  tetraquarks and the mass dependence of heavy-light tetraquark states from
  lattice QCD}, Phys. Rev. D 99~(5) (2019) 054505.
\newblock \href {http://arxiv.org/abs/1810.10550} {\path{arXiv:1810.10550}},
  \href {https://doi.org/10.1103/PhysRevD.99.054505}
  {\path{doi:10.1103/PhysRevD.99.054505}}.

\bibitem{Hudspith:2020tdf}
R.~J. Hudspith, B.~Colquhoun, A.~Francis, R.~Lewis, K.~Maltman, {A lattice
  investigation of exotic tetraquark channels}, Phys. Rev. D 102 (2020) 114506.
\newblock \href {http://arxiv.org/abs/2006.14294} {\path{arXiv:2006.14294}},
  \href {https://doi.org/10.1103/PhysRevD.102.114506}
  {\path{doi:10.1103/PhysRevD.102.114506}}.

\bibitem{Leskovec:2019ioa}
L.~Leskovec, S.~Meinel, M.~Pflaumer, M.~Wagner, {Lattice QCD investigation of a
  doubly-bottom $\bar{b} \bar{b} u d$ tetraquark with quantum numbers $I(J^P) =
  0(1^+)$}, Phys. Rev. D 100~(1) (2019) 014503.
\newblock \href {http://arxiv.org/abs/1904.04197} {\path{arXiv:1904.04197}},
  \href {https://doi.org/10.1103/PhysRevD.100.014503}
  {\path{doi:10.1103/PhysRevD.100.014503}}.

\bibitem{Colquhoun:2022dte}
B.~Colquhoun, A.~Francis, R.~J. Hudspith, R.~Lewis, K.~Maltman, {Investigating
  exotic heavy-light tetraquarks with 2+1 flavour lattice QCD}, PoS LATTICE2021
  (2022) 144.
\newblock \href {https://doi.org/10.22323/1.396.0144}
  {\path{doi:10.22323/1.396.0144}}.

\bibitem{Aoki:2022lat}
S.~Aoki, T.~Aoki, {Lattice study on a tetra-quark state Tbb in the HALQCD
  method}, PoS LATTICE2022 (2023).

\bibitem{Wagner:2022bff}
M.~Wagner, C.~Alexandrou, J.~Finkenrath, T.~Leontiou, S.~Meinel, M.~Pflaumer,
  {Lattice QCD study of antiheavy-antiheavy-light-light tetraquarks based on
  correlation functions with scattering interpolating operators both at the
  source and at the sink}, in: {39th International Symposium on Lattice Field
  Theory}, 2022.
\newblock \href {http://arxiv.org/abs/2210.09281} {\path{arXiv:2210.09281}}.

\bibitem{Brown:2012tm}
Z.~S. Brown, K.~Orginos, {Tetraquark bound states in the heavy-light
  heavy-light system}, Phys.Rev. D86 (2012) 114506.
\newblock \href {http://arxiv.org/abs/1210.1953} {\path{arXiv:1210.1953}},
  \href {https://doi.org/10.1103/PhysRevD.86.114506}
  {\path{doi:10.1103/PhysRevD.86.114506}}.

\bibitem{Pflaumer:2021ong}
M.~Pflaumer, L.~Leskovec, S.~Meinel, M.~Wagner, {Existence and Non-Existence of
  Doubly Heavy Tetraquark Bound States}, in: {38th International Symposium on
  Lattice Field Theory}, 2021.
\newblock \href {http://arxiv.org/abs/2108.10704} {\path{arXiv:2108.10704}}.

\bibitem{Gershon:2022xnn}
T.~Gershon, {Exotic hadron naming convention} (6 2022).
\newblock \href {http://arxiv.org/abs/2206.15233} {\path{arXiv:2206.15233}}.

\bibitem{LHCparticlesurl}
P.~Koppenburg, \href{https://www.nikhef.nl/~pkoppenb/particles.html}{New
  particles discovered at the lhc}.
\newline\urlprefix\url{https://www.nikhef.nl/~pkoppenb/particles.html}

\bibitem{ParticleDataGroup:2022pth}
R.~L. Workman, et~al., {Review of Particle Physics}, PTEP 2022 (2022) 083C01.
\newblock \href {https://doi.org/10.1093/ptep/ptac097}
  {\path{doi:10.1093/ptep/ptac097}}.

\bibitem{Chen:2016qju}
H.-X. Chen, W.~Chen, X.~Liu, S.-L. Zhu, {The hidden-charm pentaquark and
  tetraquark states}, Phys. Rept. 639 (2016) 1--121.
\newblock \href {http://arxiv.org/abs/1601.02092} {\path{arXiv:1601.02092}},
  \href {https://doi.org/10.1016/j.physrep.2016.05.004}
  {\path{doi:10.1016/j.physrep.2016.05.004}}.

\bibitem{LHCb:2015yax}
R.~Aaij, et~al., {Observation of $J/\psi p$ Resonances Consistent with
  Pentaquark States in $\Lambda_b^0 \to J/\psi K^- p$ Decays}, Phys. Rev. Lett.
  115 (2015) 072001.
\newblock \href {http://arxiv.org/abs/1507.03414} {\path{arXiv:1507.03414}},
  \href {https://doi.org/10.1103/PhysRevLett.115.072001}
  {\path{doi:10.1103/PhysRevLett.115.072001}}.

\bibitem{Yukawa:1935xg}
H.~Yukawa, {On the Interaction of Elementary Particles I}, Proc. Phys. Math.
  Soc. Jap. 17 (1935) 48--57.
\newblock \href {https://doi.org/10.1143/PTPS.1.1}
  {\path{doi:10.1143/PTPS.1.1}}.

\bibitem{Reid:1968sq}
R.~V. Reid, Jr., {Local phenomenological nucleon-nucleon potentials}, Annals
  Phys. 50 (1968) 411--448.
\newblock \href {https://doi.org/10.1016/0003-4916(68)90126-7}
  {\path{doi:10.1016/0003-4916(68)90126-7}}.

\bibitem{Nagels:1975fb}
M.~M. Nagels, T.~A. Rijken, J.~J. de~Swart, {Baryon Baryon Scattering in an
  OBEP Approach. 1. Nucleon-Nucleon Scattering}, Phys. Rev. D 12 (1975) 744.
\newblock \href {https://doi.org/10.1103/PhysRevD.12.744}
  {\path{doi:10.1103/PhysRevD.12.744}}.

\bibitem{Lacombe:1980dr}
M.~Lacombe, B.~Loiseau, J.~M. Richard, R.~Vinh~Mau, J.~Cote, P.~Pires,
  R.~De~Tourreil, {Parametrization of the Paris n n Potential}, Phys. Rev. C 21
  (1980) 861--873.
\newblock \href {https://doi.org/10.1103/PhysRevC.21.861}
  {\path{doi:10.1103/PhysRevC.21.861}}.

\bibitem{Machleidt:1987hj}
R.~Machleidt, K.~Holinde, C.~Elster, {The Bonn Meson Exchange Model for the
  Nucleon Nucleon Interaction}, Phys. Rept. 149 (1987) 1--89.
\newblock \href {https://doi.org/10.1016/S0370-1573(87)80002-9}
  {\path{doi:10.1016/S0370-1573(87)80002-9}}.

\bibitem{Wiringa:1994wb}
R.~B. Wiringa, V.~G.~J. Stoks, R.~Schiavilla, {An Accurate nucleon-nucleon
  potential with charge independence breaking}, Phys. Rev. C 51 (1995) 38--51.
\newblock \href {http://arxiv.org/abs/nucl-th/9408016}
  {\path{arXiv:nucl-th/9408016}}, \href
  {https://doi.org/10.1103/PhysRevC.51.38} {\path{doi:10.1103/PhysRevC.51.38}}.

\bibitem{Guo:2017jvc}
F.-K. Guo, C.~Hanhart, U.-G. Mei\ss{}ner, Q.~Wang, Q.~Zhao, B.-S. Zou,
  {Hadronic molecules}, Rev. Mod. Phys. 90~(1) (2018) 015004, [Erratum:
  Rev.Mod.Phys. 94, 029901 (2022)].
\newblock \href {http://arxiv.org/abs/1705.00141} {\path{arXiv:1705.00141}},
  \href {https://doi.org/10.1103/RevModPhys.90.015004}
  {\path{doi:10.1103/RevModPhys.90.015004}}.

\bibitem{LHCb:2019kea}
R.~Aaij, et~al., {Observation of a narrow pentaquark state, $P_c(4312)^+$, and
  of two-peak structure of the $P_c(4450)^+$}, Phys. Rev. Lett. 122~(22) (2019)
  222001.
\newblock \href {http://arxiv.org/abs/1904.03947} {\path{arXiv:1904.03947}},
  \href {https://doi.org/10.1103/PhysRevLett.122.222001}
  {\path{doi:10.1103/PhysRevLett.122.222001}}.

\bibitem{LHCb:2021chn}
R.~Aaij, et~al., {Evidence for a new structure in the $J/\psi p$ and $J/\psi
  \bar{p}$ systems in $B_s^0 \to J/\psi p \bar{p}$ decays} (8 2021).
\newblock \href {http://arxiv.org/abs/2108.04720} {\path{arXiv:2108.04720}}.

\bibitem{LHCb:2020jpq}
R.~Aaij, et~al., {Evidence of a $J/\psi\Lambda$ structure and observation of
  excited $\Xi^-$ states in the $\Xi^-_b \to J/\psi\Lambda K^-$ decay}, Sci.
  Bull. 66 (2021) 1391.
\newblock \href {http://arxiv.org/abs/2012.10380} {\path{arXiv:2012.10380}},
  \href {https://doi.org/10.1016/j.scib.2021.02.030}
  {\path{doi:10.1016/j.scib.2021.02.030}}.

\bibitem{Nambu:1961tp}
Y.~Nambu, G.~Jona-Lasinio, {Dynamical Model of Elementary Particles Based on an
  Analogy with Superconductivity. 1.}, Phys. Rev. 122 (1961) 345--358.
\newblock \href {https://doi.org/10.1103/PhysRev.122.345}
  {\path{doi:10.1103/PhysRev.122.345}}.

\bibitem{Goldstone:1961eq}
J.~Goldstone, {Field Theories with Superconductor Solutions}, Nuovo Cim. 19
  (1961) 154--164.
\newblock \href {https://doi.org/10.1007/BF02812722}
  {\path{doi:10.1007/BF02812722}}.

\bibitem{Gell-Mann:1960mvl}
M.~Gell-Mann, M.~Levy, {The axial vector current in beta decay}, Nuovo Cim. 16
  (1960) 705.
\newblock \href {https://doi.org/10.1007/BF02859738}
  {\path{doi:10.1007/BF02859738}}.

\bibitem{Skyrme:1961vq}
T.~H.~R. Skyrme, {A Nonlinear field theory}, Proc. Roy. Soc. Lond. A 260 (1961)
  127--138.
\newblock \href {https://doi.org/10.1098/rspa.1961.0018}
  {\path{doi:10.1098/rspa.1961.0018}}.

\bibitem{Weinberg:1991um}
S.~Weinberg, {Effective chiral Lagrangians for nucleon - pion interactions and
  nuclear forces}, Nucl. Phys. B 363 (1991) 3--18.
\newblock \href {https://doi.org/10.1016/0550-3213(91)90231-L}
  {\path{doi:10.1016/0550-3213(91)90231-L}}.

\bibitem{Kamerlingh-Onnes:1911}
H.~Kamerlingh-Onnes, {The resistance of pure mercury at helium temperatures},
  Commun. Phys. Lab. Univ. Leiden 122b (1911).

\bibitem{Meissner:1933}
W.~Meissner, R.~Ochsenfeld, {Ein neuer Effekt bei Eintritt der
  Supraleitfähigkeit}, Naturwissenschaften 211 (1933) 787--788.
\newblock \href {https://doi.org/https://doi.org/10.1007/BF01504252}
  {\path{doi:https://doi.org/10.1007/BF01504252}}.

\bibitem{Rjabinin:1935}
S.~L. Rjabinin~J.N., {Magnetic properties and critical currents of
  superconducting alloys}, Phys. Z. Sowjet V.7, No.1 (1935) 122--125.

\bibitem{Ginzburg:1950sr}
V.~L. Ginzburg, L.~D. Landau, {On the Theory of superconductivity}, Zh. Eksp.
  Teor. Fiz. 20 (1950) 1064--1082.

\bibitem{Abrikosov:1956sx}
A.~A. Abrikosov, {On the Magnetic properties of superconductors of the second
  group}, Sov. Phys. JETP 5 (1957) 1174--1182.

\bibitem{Maldacena:1997re}
J.~M. Maldacena, {The Large N limit of superconformal field theories and
  supergravity}, Adv. Theor. Math. Phys. 2 (1998) 231--252.
\newblock \href {http://arxiv.org/abs/hep-th/9711200}
  {\path{arXiv:hep-th/9711200}}, \href
  {https://doi.org/10.1023/A:1026654312961}
  {\path{doi:10.1023/A:1026654312961}}.

\bibitem{Chodos:1974je}
A.~Chodos, R.~L. Jaffe, K.~Johnson, C.~B. Thorn, V.~F. Weisskopf, {A New
  Extended Model of Hadrons}, Phys. Rev. D 9 (1974) 3471--3495.
\newblock \href {https://doi.org/10.1103/PhysRevD.9.3471}
  {\path{doi:10.1103/PhysRevD.9.3471}}.

\bibitem{Bardeen:1974wr}
W.~A. Bardeen, M.~S. Chanowitz, S.~D. Drell, M.~Weinstein, T.-M. Yan, {Heavy
  Quarks and Strong Binding: A Field Theory of Hadron Structure}, Phys. Rev. D
  11 (1975) 1094.
\newblock \href {https://doi.org/10.1103/PhysRevD.11.1094}
  {\path{doi:10.1103/PhysRevD.11.1094}}.

\bibitem{Friedberg:1978sc}
R.~Friedberg, T.~D. Lee, {QCD and the Soliton Model of Hadrons}, Phys. Rev. D
  18 (1978) 2623.
\newblock \href {https://doi.org/10.1103/PhysRevD.18.2623}
  {\path{doi:10.1103/PhysRevD.18.2623}}.

\bibitem{Nielsen:1978tr}
H.~B. Nielsen, M.~Ninomiya, {A Bound on Bag Constant and Nielsen-Olesen
  Unstable Mode in QCD}, Nucl. Phys. B 156 (1979) 1--28.
\newblock \href {https://doi.org/10.1016/0550-3213(79)90490-5}
  {\path{doi:10.1016/0550-3213(79)90490-5}}.

\bibitem{DeRujula:1975qlm}
A.~De~Rujula, H.~Georgi, S.~L. Glashow, {Hadron Masses in a Gauge Theory},
  Phys. Rev. D 12 (1975) 147--162.
\newblock \href {https://doi.org/10.1103/PhysRevD.12.147}
  {\path{doi:10.1103/PhysRevD.12.147}}.

\bibitem{Godfrey:1985xj}
S.~Godfrey, N.~Isgur, {Mesons in a Relativized Quark Model with
  Chromodynamics}, Phys.Rev. D32 (1985) 189--231.
\newblock \href {https://doi.org/10.1103/PhysRevD.32.189}
  {\path{doi:10.1103/PhysRevD.32.189}}.

\bibitem{Henriques:1976jd}
A.~B. Henriques, B.~H. Kellett, R.~G. Moorhouse, {Radiative Transitions and the
  p Wave Levels in Charmonium}, Phys. Lett. B 64 (1976) 85--92.
\newblock \href {https://doi.org/10.1016/0370-2693(76)90364-6}
  {\path{doi:10.1016/0370-2693(76)90364-6}}.

\bibitem{Jaffe:2003sg}
R.~L. Jaffe, F.~Wilczek, {Diquarks and exotic spectroscopy}, Phys. Rev. Lett.
  91 (2003) 232003.
\newblock \href {http://arxiv.org/abs/hep-ph/0307341}
  {\path{arXiv:hep-ph/0307341}}, \href
  {https://doi.org/10.1103/PhysRevLett.91.232003}
  {\path{doi:10.1103/PhysRevLett.91.232003}}.

\bibitem{Maiani:2004vq}
L.~Maiani, F.~Piccinini, A.~D. Polosa, V.~Riquer, {Diquark-antidiquarks with
  hidden or open charm and the nature of X(3872)}, Phys. Rev. D 71 (2005)
  014028.
\newblock \href {http://arxiv.org/abs/hep-ph/0412098}
  {\path{arXiv:hep-ph/0412098}}, \href
  {https://doi.org/10.1103/PhysRevD.71.014028}
  {\path{doi:10.1103/PhysRevD.71.014028}}.

\bibitem{Lenz:1985jk}
F.~Lenz, J.~Londergan, E.~Moniz, R.~Rosenfelder, M.~Stingl, et~al., {Quark
  Confinement and Hadronic Interactions}, Annals Phys. 170 (1986) 65.
\newblock \href {https://doi.org/10.1016/0003-4916(86)90088-6}
  {\path{doi:10.1016/0003-4916(86)90088-6}}.

\bibitem{Bicudo:2015bra}
P.~Bicudo, M.~Cardoso, {Tetraquark bound states and resonances in the unitary
  and microscopic triple string flip-flop quark model, the
  light-light-antiheavy-antiheavy $q q \bar Q\bar Q$ case study}, Phys. Rev. D
  94~(9) (2016) 094032.
\newblock \href {http://arxiv.org/abs/1509.04943} {\path{arXiv:1509.04943}},
  \href {https://doi.org/10.1103/PhysRevD.94.094032}
  {\path{doi:10.1103/PhysRevD.94.094032}}.

\bibitem{Ribeiro:1978gx}
J.~E.~T. Ribeiro, {Microscopic Calculation of the Repulsive Core in the Elastic
  Nucleon Nucleon Scattering}, Z. Phys. C 5 (1980) 27.
\newblock \href {https://doi.org/10.1007/BF01546954}
  {\path{doi:10.1007/BF01546954}}.

\bibitem{Wheeler:1937zz}
J.~A. Wheeler, {On the Mathematical Description of Light Nuclei by the Method
  of Resonating Group Structure}, Phys. Rev. 52 (1937) 1107--1122.
\newblock \href {https://doi.org/10.1103/PhysRev.52.1107}
  {\path{doi:10.1103/PhysRev.52.1107}}.

\bibitem{Bicudo:1987tz}
P.~Bicudo, J.~E. Ribeiro, {$K - N$ Exotic S Channel I = 1, 0 Phase Shifts: A
  Test of the Nonannihilating Quark Quark Potential}, Z. Phys. C 38 (1988) 453.
\newblock \href {https://doi.org/10.1007/BF01584396}
  {\path{doi:10.1007/BF01584396}}.

\bibitem{Adler:1984ri}
S.~L. Adler, A.~C. Davis, {Chiral Symmetry Breaking in Coulomb Gauge QCD},
  Nucl. Phys. B 244 (1984) 469.
\newblock \href {https://doi.org/10.1016/0550-3213(84)90324-9}
  {\path{doi:10.1016/0550-3213(84)90324-9}}.

\bibitem{Bicudo:1989sh}
P.~J.~A. Bicudo, J.~E. Ribeiro, {Current Quark Model in a $p$ Wave Triplet
  Condensed Vacuum. 1. The Dynamical Breaking of Chiral Symmetry}, Phys.Rev.
  D42 (1990) 1611--1624.
\newblock \href {https://doi.org/10.1103/PhysRevD.42.1611}
  {\path{doi:10.1103/PhysRevD.42.1611}}.

\bibitem{LeYaouanc:1984ntu}
A.~Le~Yaouanc, L.~Oliver, S.~Ono, O.~Pene, J.~C. Raynal, {A Quark Model of
  Light Mesons with Dynamically Broken Chiral Symmetry}, Phys. Rev. D 31 (1985)
  137--159.
\newblock \href {https://doi.org/10.1103/PhysRevD.31.137}
  {\path{doi:10.1103/PhysRevD.31.137}}.

\bibitem{Bicudo:1989si}
P.~J.~A. Bicudo, J.~E. Ribeiro, {Current Quark Model in a $p$ Wave Triplet
  Condensed Vacuum. 2. Salpeter Equations: $\pi$, $K$, $\rho$, $\phi$ as $q
  \bar{q}$ Bound States}, Phys.Rev. D42 (1990) 1625--1634.
\newblock \href {https://doi.org/10.1103/PhysRevD.42.1625}
  {\path{doi:10.1103/PhysRevD.42.1625}}.

\bibitem{Roberts:1987xc}
C.~D. Roberts, R.~T. Cahill, J.~Praschifka, {The Effective Action for the
  Goldstone Modes in a Global Color Symmetry Model of {QCD}}, Annals Phys. 188
  (1988) 20.
\newblock \href {https://doi.org/10.1016/0003-4916(88)90090-5}
  {\path{doi:10.1016/0003-4916(88)90090-5}}.

\bibitem{Williams:1989tv}
A.~G. Williams, G.~Krein, C.~D. Roberts, {Quark Propagator in an Ansatz
  Approach to QCD}, Annals Phys. 210 (1991) 464--485.
\newblock \href {https://doi.org/10.1016/0003-4916(91)90051-9}
  {\path{doi:10.1016/0003-4916(91)90051-9}}.

\bibitem{vonSmekal:1997ohs}
L.~von Smekal, R.~Alkofer, A.~Hauck, {The Infrared behavior of gluon and ghost
  propagators in Landau gauge QCD}, Phys. Rev. Lett. 79 (1997) 3591--3594.
\newblock \href {http://arxiv.org/abs/hep-ph/9705242}
  {\path{arXiv:hep-ph/9705242}}, \href
  {https://doi.org/10.1103/PhysRevLett.79.3591}
  {\path{doi:10.1103/PhysRevLett.79.3591}}.

\bibitem{Feuchter:2004mk}
C.~Feuchter, H.~Reinhardt, {Variational solution of the Yang-Mills Schrodinger
  equation in Coulomb gauge}, Phys. Rev. D 70 (2004) 105021.
\newblock \href {http://arxiv.org/abs/hep-th/0408236}
  {\path{arXiv:hep-th/0408236}}, \href
  {https://doi.org/10.1103/PhysRevD.70.105021}
  {\path{doi:10.1103/PhysRevD.70.105021}}.

\bibitem{Szczepaniak:1995cw}
A.~Szczepaniak, E.~S. Swanson, C.-R. Ji, S.~R. Cotanch, {Glueball spectroscopy
  in a relativistic many body approach to hadron structure}, Phys. Rev. Lett.
  76 (1996) 2011--2014.
\newblock \href {http://arxiv.org/abs/hep-ph/9511422}
  {\path{arXiv:hep-ph/9511422}}, \href
  {https://doi.org/10.1103/PhysRevLett.76.2011}
  {\path{doi:10.1103/PhysRevLett.76.2011}}.

\bibitem{Bicudo:2001jq}
P.~Bicudo, S.~Cotanch, F.~J. Llanes-Estrada, P.~Maris, E.~Ribeiro,
  A.~Szczepaniak, {Chirally symmetric quark description of low-energy pi pi
  scattering}, Phys. Rev. D 65 (2002) 076008.
\newblock \href {http://arxiv.org/abs/hep-ph/0112015}
  {\path{arXiv:hep-ph/0112015}}, \href
  {https://doi.org/10.1103/PhysRevD.65.076008}
  {\path{doi:10.1103/PhysRevD.65.076008}}.

\bibitem{Bicudo:2003fp}
P.~Bicudo, {Analytic proof that the quark model complies with partially
  conserved axial current theorems}, Phys. Rev. C 67 (2003) 035201.
\newblock \href {http://arxiv.org/abs/hep-ph/0311277}
  {\path{arXiv:hep-ph/0311277}}, \href
  {https://doi.org/10.1103/PhysRevC.67.035201}
  {\path{doi:10.1103/PhysRevC.67.035201}}.

\bibitem{Weinberg:1966kf}
S.~Weinberg, {Pion scattering lengths}, Phys. Rev. Lett. 17 (1966) 616--621.
\newblock \href {https://doi.org/10.1103/PhysRevLett.17.616}
  {\path{doi:10.1103/PhysRevLett.17.616}}.

\bibitem{Bicudo:1989sj}
P.~J.~A. Bicudo, J.~E. Ribeiro, {Current Quark Model in $p$ Wave Triplet
  Condensed Vacuum. 3. Generalized r.$G$.m. Equations: The $\phi$ and $\rho$
  Resonances}, Phys.Rev. D42 (1990) 1635--1650.
\newblock \href {https://doi.org/10.1103/PhysRevD.42.1635}
  {\path{doi:10.1103/PhysRevD.42.1635}}.

\bibitem{Garcia-Martin:2011iqs}
R.~Garcia-Martin, R.~Kaminski, J.~R. Pelaez, J.~Ruiz~de Elvira, F.~J. Yndurain,
  {The Pion-pion scattering amplitude. IV: Improved analysis with once
  subtracted Roy-like equations up to 1100 MeV}, Phys. Rev. D 83 (2011) 074004.
\newblock \href {http://arxiv.org/abs/1102.2183} {\path{arXiv:1102.2183}},
  \href {https://doi.org/10.1103/PhysRevD.83.074004}
  {\path{doi:10.1103/PhysRevD.83.074004}}.

\bibitem{Tornqvist:1982yv}
N.~A. Tornqvist, {Scalar Mesons in the Unitarized Quark Model}, Phys. Rev.
  Lett. 49 (1982) 624--627.
\newblock \href {https://doi.org/10.1103/PhysRevLett.49.624}
  {\path{doi:10.1103/PhysRevLett.49.624}}.

\bibitem{vanBeveren:1986ea}
E.~van Beveren, T.~A. Rijken, K.~Metzger, C.~Dullemond, G.~Rupp, J.~E. Ribeiro,
  {A Low Lying Scalar Meson Nonet in a Unitarized Meson Model}, Z. Phys. C 30
  (1986) 615--620.
\newblock \href {http://arxiv.org/abs/0710.4067} {\path{arXiv:0710.4067}},
  \href {https://doi.org/10.1007/BF01571811} {\path{doi:10.1007/BF01571811}}.

\bibitem{Ribeiro:1981fk}
J.~E. Ribeiro, {Matrix Elements of the Exchange Operator for Arbitrary Angular
  Momentum Two Meson States}, Phys. Rev. D 25 (1982) 2406.
\newblock \href {https://doi.org/10.1103/PhysRevD.25.2406}
  {\path{doi:10.1103/PhysRevD.25.2406}}.

\bibitem{vanBeveren:1982sj}
E.~van Beveren, {Recoupling Matrix Elements and Decay}, Z. Phys. C 17 (1983)
  135.
\newblock \href {http://arxiv.org/abs/hep-ph/0602248}
  {\path{arXiv:hep-ph/0602248}}, \href {https://doi.org/10.1007/BF01574179}
  {\path{doi:10.1007/BF01574179}}.

\bibitem{Bernard:1990ye}
V.~Bernard, U.~G. Meissner, A.~Blin, B.~Hiller, {Four point functions in quark
  flavor dynamics: Meson meson scattering}, Phys. Lett. B 253 (1991) 443--450.
\newblock \href {https://doi.org/10.1016/0370-2693(91)91749-L}
  {\path{doi:10.1016/0370-2693(91)91749-L}}.

\bibitem{Creutz:1984mg}
M.~Creutz, {Quarks, gluons and lattices}, Cambridge Monographs on Mathematical
  Physics, Cambridge Univ. Press, Cambridge, UK, 1985.

\bibitem{Gattringer:2010zz}
C.~Gattringer, C.~B. Lang, {Quantum chromodynamics on the lattice}, Vol. 788,
  Springer, Berlin, 2010.
\newblock \href {https://doi.org/10.1007/978-3-642-01850-3}
  {\path{doi:10.1007/978-3-642-01850-3}}.

\bibitem{Creutz:1979zg}
M.~Creutz, L.~Jacobs, C.~Rebbi, {Monte Carlo Study of Abelian Lattice Gauge
  Theories}, Phys. Rev. D 20 (1979) 1915.
\newblock \href {https://doi.org/10.1103/PhysRevD.20.1915}
  {\path{doi:10.1103/PhysRevD.20.1915}}.

\bibitem{Creutz:1979dw}
M.~Creutz, {Confinement and the Critical Dimensionality of Space-Time}, Phys.
  Rev. Lett. 43 (1979) 553--556, [Erratum: Phys.Rev.Lett. 43, 890 (1979)].
\newblock \href {https://doi.org/10.1103/PhysRevLett.43.553}
  {\path{doi:10.1103/PhysRevLett.43.553}}.

\bibitem{Creutz:1980zw}
M.~Creutz, {Monte Carlo Study of Quantized SU(2) Gauge Theory}, Phys. Rev. D 21
  (1980) 2308--2315.
\newblock \href {https://doi.org/10.1103/PhysRevD.21.2308}
  {\path{doi:10.1103/PhysRevD.21.2308}}.

\bibitem{Creutz:1980wj}
M.~Creutz, {Asymptotic Freedom Scales}, Phys. Rev. Lett. 45 (1980) 313.
\newblock \href {https://doi.org/10.1103/PhysRevLett.45.313}
  {\path{doi:10.1103/PhysRevLett.45.313}}.

\bibitem{McLerran:1980pk}
L.~D. McLerran, B.~Svetitsky, {A Monte Carlo Study of SU(2) Yang-Mills Theory
  at Finite Temperature}, Phys. Lett. B 98 (1981) 195.
\newblock \href {https://doi.org/10.1016/0370-2693(81)90986-2}
  {\path{doi:10.1016/0370-2693(81)90986-2}}.

\bibitem{Takahashi:2000te}
T.~T. Takahashi, H.~Matsufuru, Y.~Nemoto, H.~Suganuma, {The Three quark
  potential in the SU(3) lattice QCD}, Phys. Rev. Lett. 86 (2001) 18--21.
\newblock \href {http://arxiv.org/abs/hep-lat/0006005}
  {\path{arXiv:hep-lat/0006005}}, \href
  {https://doi.org/10.1103/PhysRevLett.86.18}
  {\path{doi:10.1103/PhysRevLett.86.18}}.

\bibitem{Takahashi:2002bw}
T.~T. Takahashi, H.~Suganuma, Y.~Nemoto, H.~Matsufuru, {Detailed analysis of
  the three quark potential in SU(3) lattice QCD}, Phys. Rev. D 65 (2002)
  114509.
\newblock \href {http://arxiv.org/abs/hep-lat/0204011}
  {\path{arXiv:hep-lat/0204011}}, \href
  {https://doi.org/10.1103/PhysRevD.65.114509}
  {\path{doi:10.1103/PhysRevD.65.114509}}.

\bibitem{Okiharu:2004ve}
F.~Okiharu, H.~Suganuma, T.~T. Takahashi, {Detailed analysis of the tetraquark
  potential and flip-flop in SU(3) lattice QCD}, Phys.Rev. D72 (2005) 014505.
\newblock \href {http://arxiv.org/abs/hep-lat/0412012}
  {\path{arXiv:hep-lat/0412012}}, \href
  {https://doi.org/10.1103/PhysRevD.72.014505}
  {\path{doi:10.1103/PhysRevD.72.014505}}.

\bibitem{Cardoso:2011fq}
N.~Cardoso, M.~Cardoso, P.~Bicudo, {Colour Fields Computed in SU(3) Lattice QCD
  for the Static Tetraquark System}, Phys.Rev. D84 (2011) 054508.
\newblock \href {http://arxiv.org/abs/1107.1355} {\path{arXiv:1107.1355}},
  \href {https://doi.org/10.1103/PhysRevD.84.054508}
  {\path{doi:10.1103/PhysRevD.84.054508}}.

\bibitem{Brower:1981vt}
R.~Brower, P.~Rossi, C.-I. Tan, {The External Field Problem for {QCD}}, Nucl.
  Phys. B 190 (1981) 699--718.
\newblock \href {https://doi.org/10.1016/0550-3213(81)90046-8}
  {\path{doi:10.1016/0550-3213(81)90046-8}}.

\bibitem{Parisi:1983hm}
G.~Parisi, R.~Petronzio, F.~Rapuano, {A Measurement of the String Tension Near
  the Continuum Limit}, Phys. Lett. B 128 (1983) 418--420.
\newblock \href {https://doi.org/10.1016/0370-2693(83)90930-9}
  {\path{doi:10.1016/0370-2693(83)90930-9}}.

\bibitem{APE:1987ehd}
M.~Albanese, et~al., {Glueball Masses and String Tension in Lattice QCD}, Phys.
  Lett. B 192 (1987) 163--169.
\newblock \href {https://doi.org/10.1016/0370-2693(87)91160-9}
  {\path{doi:10.1016/0370-2693(87)91160-9}}.

\bibitem{Luscher:2001up}
M.~Luscher, P.~Weisz, {Locality and exponential error reduction in numerical
  lattice gauge theory}, JHEP 09 (2001) 010.
\newblock \href {http://arxiv.org/abs/hep-lat/0108014}
  {\path{arXiv:hep-lat/0108014}}, \href
  {https://doi.org/10.1088/1126-6708/2001/09/010}
  {\path{doi:10.1088/1126-6708/2001/09/010}}.

\bibitem{Cardoso:2013lla}
N.~Cardoso, M.~Cardoso, P.~Bicudo, {Inside the SU(3) quark-antiquark QCD flux
  tube: screening versus quantum widening}, Phys. Rev. D 88 (2013) 054504.
\newblock \href {http://arxiv.org/abs/1302.3633} {\path{arXiv:1302.3633}},
  \href {https://doi.org/10.1103/PhysRevD.88.054504}
  {\path{doi:10.1103/PhysRevD.88.054504}}.

\bibitem{Cardoso:2012uka}
M.~Cardoso, N.~Cardoso, P.~Bicudo, {Variational study of the flux tube
  recombination in the two quarks and two quarks system in Lattice QCD},
  Phys.Rev. D86 (2012) 014503.
\newblock \href {http://arxiv.org/abs/1204.5131} {\path{arXiv:1204.5131}},
  \href {https://doi.org/10.1103/PhysRevD.86.014503}
  {\path{doi:10.1103/PhysRevD.86.014503}}.

\bibitem{Ginsparg:1981bj}
P.~H. Ginsparg, K.~G. Wilson, {A Remnant of Chiral Symmetry on the Lattice},
  Phys. Rev. D 25 (1982) 2649.
\newblock \href {https://doi.org/10.1103/PhysRevD.25.2649}
  {\path{doi:10.1103/PhysRevD.25.2649}}.

\bibitem{Davies:1994mp}
C.~T.~H. Davies, K.~Hornbostel, A.~Langnau, G.~P. Lepage, A.~Lidsey,
  J.~Shigemitsu, J.~H. Sloan, {Precision Upsilon spectroscopy from
  nonrelativistic lattice QCD}, Phys. Rev. D 50 (1994) 6963--6977.
\newblock \href {http://arxiv.org/abs/hep-lat/9406017}
  {\path{arXiv:hep-lat/9406017}}, \href
  {https://doi.org/10.1103/PhysRevD.50.6963}
  {\path{doi:10.1103/PhysRevD.50.6963}}.

\bibitem{Gray:2005ur}
A.~Gray, I.~Allison, C.~T.~H. Davies, E.~Dalgic, G.~P. Lepage, J.~Shigemitsu,
  M.~Wingate, {The Upsilon spectrum and m(b) from full lattice QCD}, Phys. Rev.
  D 72 (2005) 094507.
\newblock \href {http://arxiv.org/abs/hep-lat/0507013}
  {\path{arXiv:hep-lat/0507013}}, \href
  {https://doi.org/10.1103/PhysRevD.72.094507}
  {\path{doi:10.1103/PhysRevD.72.094507}}.

\bibitem{Wagner:2010ad}
M.~Wagner, {Forces between static-light mesons}, PoS LATTICE2010 (2010) 162.
\newblock \href {http://arxiv.org/abs/1008.1538} {\path{arXiv:1008.1538}}.

\bibitem{Wagner:2011ev}
M.~Wagner, {Static-static-light-light tetraquarks in lattice QCD}, Acta Phys.
  Polon. Supp. 4 (2011) 747--752.
\newblock \href {http://arxiv.org/abs/1103.5147} {\path{arXiv:1103.5147}},
  \href {https://doi.org/10.5506/APhysPolBSupp.4.747}
  {\path{doi:10.5506/APhysPolBSupp.4.747}}.

\bibitem{Okiharu:2004wy}
F.~Okiharu, H.~Suganuma, T.~T. Takahashi, {First study for the pentaquark
  potential in SU(3) lattice QCD}, Phys. Rev. Lett. 94 (2005) 192001.
\newblock \href {http://arxiv.org/abs/hep-lat/0407001}
  {\path{arXiv:hep-lat/0407001}}, \href
  {https://doi.org/10.1103/PhysRevLett.94.192001}
  {\path{doi:10.1103/PhysRevLett.94.192001}}.

\bibitem{Alexandrou:2004ak}
C.~Alexandrou, G.~Koutsou, {The Static tetraquark and pentaquark potentials},
  Phys.Rev. D71 (2005) 014504.
\newblock \href {http://arxiv.org/abs/hep-lat/0407005}
  {\path{arXiv:hep-lat/0407005}}, \href
  {https://doi.org/10.1103/PhysRevD.71.014504}
  {\path{doi:10.1103/PhysRevD.71.014504}}.

\bibitem{Bali:1994de}
G.~S. Bali, K.~Schilling, C.~Schlichter, {Observing long color flux tubes in
  SU(2) lattice gauge theory}, Phys. Rev. D 51 (1995) 5165--5198.
\newblock \href {http://arxiv.org/abs/hep-lat/9409005}
  {\path{arXiv:hep-lat/9409005}}, \href
  {https://doi.org/10.1103/PhysRevD.51.5165}
  {\path{doi:10.1103/PhysRevD.51.5165}}.

\bibitem{Sommer:1984xq}
R.~Sommer, J.~Wosiek, {Baryonic Loops and the Confinement in the Three Quark
  Channel}, Phys. Lett. B 149 (1984) 497--500.
\newblock \href {https://doi.org/10.1016/0370-2693(84)90374-5}
  {\path{doi:10.1016/0370-2693(84)90374-5}}.

\bibitem{Sommer:1985da}
R.~Sommer, J.~Wosiek, {BARYONIC STRINGS ON A LATTICE}, Nucl. Phys. B 267 (1986)
  531--538.
\newblock \href {https://doi.org/10.1016/0550-3213(86)90129-X}
  {\path{doi:10.1016/0550-3213(86)90129-X}}.

\bibitem{Bali:2000gf}
G.~S. Bali, {QCD forces and heavy quark bound states}, Phys. Rept. 343 (2001)
  1--136.
\newblock \href {http://arxiv.org/abs/hep-ph/0001312}
  {\path{arXiv:hep-ph/0001312}}, \href
  {https://doi.org/10.1016/S0370-1573(00)00079-X}
  {\path{doi:10.1016/S0370-1573(00)00079-X}}.

\bibitem{Alexandrou:2001ip}
C.~Alexandrou, P.~De~Forcrand, A.~Tsapalis, {The Static three quark SU(3) and
  four quark SU(4) potentials}, Phys. Rev. D 65 (2002) 054503.
\newblock \href {http://arxiv.org/abs/hep-lat/0107006}
  {\path{arXiv:hep-lat/0107006}}, \href
  {https://doi.org/10.1103/PhysRevD.65.054503}
  {\path{doi:10.1103/PhysRevD.65.054503}}.

\bibitem{Dmitrasinovic:2013jta}
V.~Dmitra\v{s}inovi\'c, I.~Salom, {Differentiating Between ${\Delta }$- and
  Y-string Confinement: Can One See the Difference in Baryon Spectra?}, Acta
  Phys. Polon. Supp. 6~(3) (2013) 905--910.
\newblock \href {https://doi.org/10.5506/APhysPolBSupp.6.905}
  {\path{doi:10.5506/APhysPolBSupp.6.905}}.

\bibitem{Dmitrasinovic:2018irg}
V.~Dmitra\v{s}inovi\'c, I.~Salom, {$O(6)$ algebraic theory of three
  nonrelativistic quarks bound by spin-independent interactions}, Phys. Rev. D
  97~(9) (2018) 094011.
\newblock \href {http://arxiv.org/abs/1805.00386} {\path{arXiv:1805.00386}},
  \href {https://doi.org/10.1103/PhysRevD.97.094011}
  {\path{doi:10.1103/PhysRevD.97.094011}}.

\bibitem{Koma:2017hcm}
Y.~Koma, M.~Koma, {Precise determination of the three-quark potential in SU(3)
  lattice gauge theory}, Phys. Rev. D 95~(9) (2017) 094513.
\newblock \href {http://arxiv.org/abs/1703.06247} {\path{arXiv:1703.06247}},
  \href {https://doi.org/10.1103/PhysRevD.95.094513}
  {\path{doi:10.1103/PhysRevD.95.094513}}.

\bibitem{Cardoso:2012rb}
N.~Cardoso, P.~Bicudo, {Color fields of the static pentaquark system computed
  in SU(3) lattice QCD}, Phys. Rev. D 87~(3) (2013) 034504.
\newblock \href {http://arxiv.org/abs/1209.1532} {\path{arXiv:1209.1532}},
  \href {https://doi.org/10.1103/PhysRevD.87.034504}
  {\path{doi:10.1103/PhysRevD.87.034504}}.

\bibitem{Bicudo:2017usw}
P.~Bicudo, M.~Cardoso, O.~Oliveira, P.~J. Silva, {Lattice QCD static potentials
  of the meson-meson and tetraquark systems computed with both quenched and
  full QCD}, Phys. Rev. D 96~(7) (2017) 074508.
\newblock \href {http://arxiv.org/abs/1702.07789} {\path{arXiv:1702.07789}},
  \href {https://doi.org/10.1103/PhysRevD.96.074508}
  {\path{doi:10.1103/PhysRevD.96.074508}}.

\bibitem{Born:1927}
M.~Born, R.~Oppenheimer, {Zur Quantentheorie der Molekeln}, Annalen der Physik
  389 (1927) 457.

\bibitem{Mihaly:1996ue}
A.~Mihaly, H.~R. Fiebig, H.~Markum, K.~Rabitsch, {Interactions between heavy -
  light mesons in lattice QCD}, Phys. Rev. D 55 (1997) 3077--3081.
\newblock \href {https://doi.org/10.1103/PhysRevD.55.3077}
  {\path{doi:10.1103/PhysRevD.55.3077}}.

\bibitem{Michael:1999nq}
C.~Michael, P.~Pennanen, {Two heavy - light mesons on a lattice}, Phys. Rev. D
  60 (1999) 054012.
\newblock \href {http://arxiv.org/abs/hep-lat/9901007}
  {\path{arXiv:hep-lat/9901007}}, \href
  {https://doi.org/10.1103/PhysRevD.60.054012}
  {\path{doi:10.1103/PhysRevD.60.054012}}.

\bibitem{Doi:2006kx}
T.~Doi, T.~T. Takahashi, H.~Suganuma, {Meson-meson and meson-baryon
  interactions in lattice QCD}, AIP Conf. Proc. 842~(1) (2006) 246--248.
\newblock \href {http://arxiv.org/abs/hep-lat/0601008}
  {\path{arXiv:hep-lat/0601008}}, \href {https://doi.org/10.1063/1.2220239}
  {\path{doi:10.1063/1.2220239}}.

\bibitem{Detmold:2007wk}
W.~Detmold, K.~Orginos, M.~J. Savage, {BB Potentials in Quenched Lattice QCD},
  Phys. Rev. D 76 (2007) 114503.
\newblock \href {http://arxiv.org/abs/hep-lat/0703009}
  {\path{arXiv:hep-lat/0703009}}, \href
  {https://doi.org/10.1103/PhysRevD.76.114503}
  {\path{doi:10.1103/PhysRevD.76.114503}}.

\bibitem{Bali:2010xa}
G.~Bali, M.~Hetzenegger, {Static-light meson-meson potentials}, PoS LATTICE2010
  (2010) 142.
\newblock \href {http://arxiv.org/abs/1011.0571} {\path{arXiv:1011.0571}},
  \href {https://doi.org/10.22323/1.105.0142} {\path{doi:10.22323/1.105.0142}}.

\bibitem{Blossier:2009kd}
B.~Blossier, M.~Della~Morte, G.~von Hippel, T.~Mendes, R.~Sommer, {On the
  generalized eigenvalue method for energies and matrix elements in lattice
  field theory}, JHEP 04 (2009) 094.
\newblock \href {http://arxiv.org/abs/0902.1265} {\path{arXiv:0902.1265}},
  \href {https://doi.org/10.1088/1126-6708/2009/04/094}
  {\path{doi:10.1088/1126-6708/2009/04/094}}.

\bibitem{Bali:2005fu}
G.~S. Bali, H.~Neff, T.~Duessel, T.~Lippert, K.~Schilling, {Observation of
  string breaking in QCD}, Phys. Rev. D71 (2005) 114513.
\newblock \href {http://arxiv.org/abs/hep-lat/0505012}
  {\path{arXiv:hep-lat/0505012}}, \href
  {https://doi.org/10.1103/PhysRevD.71.114513}
  {\path{doi:10.1103/PhysRevD.71.114513}}.

\bibitem{Philipsen:1998de}
O.~Philipsen, H.~Wittig, {String breaking in nonAbelian gauge theories with
  fundamental matter fields}, Phys. Rev. Lett. 81 (1998) 4056--4059, [Erratum:
  Phys.Rev.Lett. 83, 2684 (1999)].
\newblock \href {http://arxiv.org/abs/hep-lat/9807020}
  {\path{arXiv:hep-lat/9807020}}, \href
  {https://doi.org/10.1103/PhysRevLett.81.4056}
  {\path{doi:10.1103/PhysRevLett.81.4056}}.

\bibitem{Knechtli:1998gf}
F.~Knechtli, R.~Sommer, {String breaking in SU(2) gauge theory with scalar
  matter fields}, Phys. Lett. B 440 (1998) 345--352, [Erratum: Phys.Lett.B 454,
  399--399 (1999)].
\newblock \href {http://arxiv.org/abs/hep-lat/9807022}
  {\path{arXiv:hep-lat/9807022}}, \href
  {https://doi.org/10.1016/S0370-2693(98)01098-3}
  {\path{doi:10.1016/S0370-2693(98)01098-3}}.

\bibitem{Drummond:1998ir}
I.~T. Drummond, {Mixing scenarios for lattice string breaking}, Phys. Lett. B
  442 (1998) 279--290.
\newblock \href {http://arxiv.org/abs/hep-lat/9808014}
  {\path{arXiv:hep-lat/9808014}}, \href
  {https://doi.org/10.1016/S0370-2693(98)01222-2}
  {\path{doi:10.1016/S0370-2693(98)01222-2}}.

\bibitem{Detar:1998qa}
C.~E. Detar, O.~Kaczmarek, F.~Karsch, E.~Laermann, {String breaking in lattice
  quantum chromodynamics}, Phys. Rev. D 59 (1999) 031501.
\newblock \href {http://arxiv.org/abs/hep-lat/9808028}
  {\path{arXiv:hep-lat/9808028}}, \href
  {https://doi.org/10.1103/PhysRevD.59.031501}
  {\path{doi:10.1103/PhysRevD.59.031501}}.

\bibitem{Stewart:1998yj}
C.~Stewart, R.~Koniuk, {String breaking in quenched QCD}, Phys. Rev. D 59
  (1999) 114503.
\newblock \href {http://arxiv.org/abs/hep-lat/9811012}
  {\path{arXiv:hep-lat/9811012}}, \href
  {https://doi.org/10.1103/PhysRevD.59.114503}
  {\path{doi:10.1103/PhysRevD.59.114503}}.

\bibitem{Trottier:1998nk}
H.~D. Trottier, {String breaking by dynamical fermions in three-dimensional
  lattice QCD}, Phys. Rev. D 60 (1999) 034506.
\newblock \href {http://arxiv.org/abs/hep-lat/9812021}
  {\path{arXiv:hep-lat/9812021}}, \href
  {https://doi.org/10.1103/PhysRevD.60.034506}
  {\path{doi:10.1103/PhysRevD.60.034506}}.

\bibitem{Stephenson:1999kh}
P.~W. Stephenson, {Breaking of the adjoint string in (2+1)-dimensions}, Nucl.
  Phys. B 550 (1999) 427--448.
\newblock \href {http://arxiv.org/abs/hep-lat/9902002}
  {\path{arXiv:hep-lat/9902002}}, \href
  {https://doi.org/10.1016/S0550-3213(99)00210-2}
  {\path{doi:10.1016/S0550-3213(99)00210-2}}.

\bibitem{Philipsen:1999wf}
O.~Philipsen, H.~Wittig, {String breaking in SU(2) Yang-Mills theory with
  adjoint sources}, Phys. Lett. B 451 (1999) 146--154.
\newblock \href {http://arxiv.org/abs/hep-lat/9902003}
  {\path{arXiv:hep-lat/9902003}}, \href
  {https://doi.org/10.1016/S0370-2693(99)00183-5}
  {\path{doi:10.1016/S0370-2693(99)00183-5}}.

\bibitem{Gliozzi:1999wq}
F.~Gliozzi, P.~Provero, {The Confining string and its breaking in QCD}, Nucl.
  Phys. B 556 (1999) 76--88.
\newblock \href {http://arxiv.org/abs/hep-lat/9903013}
  {\path{arXiv:hep-lat/9903013}}, \href
  {https://doi.org/10.1016/S0550-3213(99)00330-2}
  {\path{doi:10.1016/S0550-3213(99)00330-2}}.

\bibitem{deForcrand:1999kr}
P.~de~Forcrand, O.~Philipsen, {Adjoint string breaking in 4-D SU(2) Yang-Mills
  theory}, Phys. Lett. B 475 (2000) 280--288.
\newblock \href {http://arxiv.org/abs/hep-lat/9912050}
  {\path{arXiv:hep-lat/9912050}}, \href
  {https://doi.org/10.1016/S0370-2693(00)00117-9}
  {\path{doi:10.1016/S0370-2693(00)00117-9}}.

\bibitem{Kallio:2000jc}
K.~Kallio, H.~D. Trottier, {Adjoint 'quarks' on coarse anisotropic lattices:
  Implications for string breaking in full QCD}, Phys. Rev. D 66 (2002) 034503.
\newblock \href {http://arxiv.org/abs/hep-lat/0001020}
  {\path{arXiv:hep-lat/0001020}}, \href
  {https://doi.org/10.1103/PhysRevD.66.034503}
  {\path{doi:10.1103/PhysRevD.66.034503}}.

\bibitem{Knechtli:2000df}
F.~Knechtli, R.~Sommer, {String breaking as a mixing phenomenon in the SU(2)
  Higgs model}, Nucl. Phys. B 590 (2000) 309--328.
\newblock \href {http://arxiv.org/abs/hep-lat/0005021}
  {\path{arXiv:hep-lat/0005021}}, \href
  {https://doi.org/10.1016/S0550-3213(00)00470-3}
  {\path{doi:10.1016/S0550-3213(00)00470-3}}.

\bibitem{Duncan:2000kr}
A.~Duncan, E.~Eichten, H.~Thacker, {String breaking in four-dimensional lattice
  QCD}, Phys. Rev. D 63 (2001) 111501.
\newblock \href {http://arxiv.org/abs/hep-lat/0011076}
  {\path{arXiv:hep-lat/0011076}}, \href
  {https://doi.org/10.1103/PhysRevD.63.111501}
  {\path{doi:10.1103/PhysRevD.63.111501}}.

\bibitem{Bernard:2001tz}
C.~W. Bernard, T.~A. DeGrand, C.~E. Detar, P.~Lacock, S.~A. Gottlieb, U.~M.
  Heller, J.~Hetrick, K.~Orginos, D.~Toussaint, R.~L. Sugar, {Zero temperature
  string breaking in lattice quantum chromodynamics}, Phys. Rev. D 64 (2001)
  074509.
\newblock \href {http://arxiv.org/abs/hep-lat/0103012}
  {\path{arXiv:hep-lat/0103012}}, \href
  {https://doi.org/10.1103/PhysRevD.64.074509}
  {\path{doi:10.1103/PhysRevD.64.074509}}.

\bibitem{Chernodub:2002ym}
M.~N. Chernodub, E.-M. Ilgenfritz, A.~Schiller, {String breaking and monopoles:
  A Case study in the 3-D Abelian Higgs model}, Phys. Lett. B 547 (2002)
  269--277.
\newblock \href {http://arxiv.org/abs/hep-lat/0207020}
  {\path{arXiv:hep-lat/0207020}}, \href
  {https://doi.org/10.1016/S0370-2693(02)02761-2}
  {\path{doi:10.1016/S0370-2693(02)02761-2}}.

\bibitem{Chernodub:2002en}
M.~N. Chernodub, E.-M. Ilgenfritz, A.~Schiller, {More on string breaking in the
  3-D Abelian Higgs model: The Photon propagator}, Phys. Lett. B 555 (2003)
  206--214.
\newblock \href {http://arxiv.org/abs/hep-lat/0212005}
  {\path{arXiv:hep-lat/0212005}}, \href
  {https://doi.org/10.1016/S0370-2693(03)00059-5}
  {\path{doi:10.1016/S0370-2693(03)00059-5}}.

\bibitem{Kratochvila:2003zj}
S.~Kratochvila, P.~de~Forcrand, {Observing string breaking with Wilson loops},
  Nucl. Phys. B 671 (2003) 103--132.
\newblock \href {http://arxiv.org/abs/hep-lat/0306011}
  {\path{arXiv:hep-lat/0306011}}, \href
  {https://doi.org/10.1016/j.nuclphysb.2003.08.014}
  {\path{doi:10.1016/j.nuclphysb.2003.08.014}}.

\bibitem{Chernodub:2003sy}
M.~N. Chernodub, K.~Hashimoto, T.~Suzuki, {Matter degrees of freedom and string
  breaking in Abelian projected quenched SU(2) QCD}, Phys. Rev. D 70 (2004)
  014506.
\newblock \href {http://arxiv.org/abs/hep-lat/0311026}
  {\path{arXiv:hep-lat/0311026}}, \href
  {https://doi.org/10.1103/PhysRevD.70.014506}
  {\path{doi:10.1103/PhysRevD.70.014506}}.

\bibitem{Wellegehausen:2010ai}
B.~H. Wellegehausen, A.~Wipf, C.~Wozar, {Casimir Scaling and String Breaking in
  G(2) Gluodynamics}, Phys. Rev. D 83 (2011) 016001.
\newblock \href {http://arxiv.org/abs/1006.2305} {\path{arXiv:1006.2305}},
  \href {https://doi.org/10.1103/PhysRevD.83.016001}
  {\path{doi:10.1103/PhysRevD.83.016001}}.

\bibitem{Kuhn:2015zqa}
S.~K\"uhn, J.~I. Cirac, M.~C. Ba\~nuls, {Non-Abelian string breaking phenomena
  with Matrix Product States}, JHEP 07 (2015) 130.
\newblock \href {http://arxiv.org/abs/1505.04441} {\path{arXiv:1505.04441}},
  \href {https://doi.org/10.1007/JHEP07(2015)130}
  {\path{doi:10.1007/JHEP07(2015)130}}.

\bibitem{Buyens:2015tea}
B.~Buyens, J.~Haegeman, H.~Verschelde, F.~Verstraete, K.~Van~Acoleyen,
  {Confinement and string breaking for QED$_2$ in the Hamiltonian picture},
  Phys. Rev. X 6~(4) (2016) 041040.
\newblock \href {http://arxiv.org/abs/1509.00246} {\path{arXiv:1509.00246}},
  \href {https://doi.org/10.1103/PhysRevX.6.041040}
  {\path{doi:10.1103/PhysRevX.6.041040}}.

\bibitem{Bonati:2020orj}
C.~Bonati, S.~Morlacchi, {Flux tubes and string breaking in three dimensional
  SU(2) Yang-Mills theory}, Phys. Rev. D 101~(9) (2020) 094506.
\newblock \href {http://arxiv.org/abs/2003.07244} {\path{arXiv:2003.07244}},
  \href {https://doi.org/10.1103/PhysRevD.101.094506}
  {\path{doi:10.1103/PhysRevD.101.094506}}.

\bibitem{Bulava:2019iut}
J.~Bulava, B.~H\"orz, F.~Knechtli, V.~Koch, G.~Moir, C.~Morningstar,
  M.~Peardon, {String breaking by light and strange quarks in QCD}, Phys. Lett.
  B 793 (2019) 493--498.
\newblock \href {http://arxiv.org/abs/1902.04006} {\path{arXiv:1902.04006}},
  \href {https://doi.org/10.1016/j.physletb.2019.05.018}
  {\path{doi:10.1016/j.physletb.2019.05.018}}.

\bibitem{Bicudo:2019ymo}
P.~Bicudo, M.~Cardoso, N.~Cardoso, M.~Wagner, {Bottomonium resonances with $I =
  0$ from lattice QCD correlation functions with static and light quarks},
  Phys. Rev. D 101~(3) (2020) 034503.
\newblock \href {http://arxiv.org/abs/1910.04827} {\path{arXiv:1910.04827}},
  \href {https://doi.org/10.1103/PhysRevD.101.034503}
  {\path{doi:10.1103/PhysRevD.101.034503}}.

\bibitem{Bicudo:2020qhp}
P.~Bicudo, N.~Cardoso, L.~M\"uller, M.~Wagner, {Computation of the quarkonium
  and meson-meson composition of the $\Upsilon(nS)$ states and of the new
  $\Upsilon(10753)$ Belle resonance from lattice QCD static potentials}, Phys.
  Rev. D 103~(7) (2021) 074507.
\newblock \href {http://arxiv.org/abs/2008.05605} {\path{arXiv:2008.05605}},
  \href {https://doi.org/10.1103/PhysRevD.103.074507}
  {\path{doi:10.1103/PhysRevD.103.074507}}.

\bibitem{Bicudo:2022ihz}
P.~Bicudo, N.~Cardoso, L.~Mueller, M.~Wagner, {Study of $I=0$ bottomonium bound
  states and resonances in $S$, $P$, $D$ and $F$ waves with lattice QCD
  static-static-light-light potentials} (5 2022).
\newblock \href {http://arxiv.org/abs/2205.11475} {\path{arXiv:2205.11475}}.

\bibitem{Bicudo:2022jep}
P.~Bicudo, N.~Cardoso, L.~Mueller, M.~Wagner, {Study of $I=0$ bottomonium bound
  states and resonances based on lattice QCD static potentials}, in: {39th
  International Symposium on Lattice Field Theory}, 2022.
\newblock \href {http://arxiv.org/abs/2210.13284} {\path{arXiv:2210.13284}}.

\bibitem{Cardoso:2008dd}
M.~Cardoso, P.~Bicudo, {Microscopic calculation of the decay of Jaffe-Wilczek
  tetraquarks, and the Z(4433)}, AIP Conf.Proc. 1030 (2008) 352.
\newblock \href {http://arxiv.org/abs/0805.2260} {\path{arXiv:0805.2260}},
  \href {https://doi.org/10.1063/1.2973526} {\path{doi:10.1063/1.2973526}}.

\bibitem{Prelovsek:2019ywc}
S.~Prelovsek, H.~Bahtiyar, J.~Petkovic, {Zb tetraquark channel from lattice QCD
  and Born-Oppenheimer approximation}, Phys. Lett. B 805 (2020) 135467.
\newblock \href {http://arxiv.org/abs/1912.02656} {\path{arXiv:1912.02656}},
  \href {https://doi.org/10.1016/j.physletb.2020.135467}
  {\path{doi:10.1016/j.physletb.2020.135467}}.

\bibitem{Peters:2016wjm}
A.~Peters, P.~Bicudo, K.~Cichy, M.~Wagner, {Investigation of $B\bar B$
  four-quark systems using lattice QCD}, J. Phys. Conf. Ser. 742~(1) (2016)
  012006.
\newblock \href {http://arxiv.org/abs/1602.07621} {\path{arXiv:1602.07621}},
  \href {https://doi.org/10.1088/1742-6596/742/1/012006}
  {\path{doi:10.1088/1742-6596/742/1/012006}}.

\bibitem{Peters:2017hon}
A.~Peters, P.~Bicudo, M.~Wagner, {$b\bar b u\bar d$ four-quark systems in the
  Born-Oppenheimer approximation: prospects and challenges}, EPJ Web Conf. 175
  (2018) 14018.
\newblock \href {http://arxiv.org/abs/1709.03306} {\path{arXiv:1709.03306}},
  \href {https://doi.org/10.1051/epjconf/201817514018}
  {\path{doi:10.1051/epjconf/201817514018}}.

\bibitem{Prelovsek:2019yae}
S.~Prelovsek, H.~Bahtiyar, J.~Petkovic, {$Z_b$ tetraquark channel and $B\bar
  B^*$ interaction}, PoS LATTICE2019 (2019) 012.
\newblock \href {http://arxiv.org/abs/1909.02356} {\path{arXiv:1909.02356}},
  \href {https://doi.org/10.22323/1.363.0012} {\path{doi:10.22323/1.363.0012}}.

\bibitem{Sadl:2021bme}
M.~Sadl, S.~Prelovsek, {Tetraquark systems $\bar bb \bar du$ in the static
  limit and lattice QCD}, Phys. Rev. D 104~(11) (2021) 114503.
\newblock \href {http://arxiv.org/abs/2109.08560} {\path{arXiv:2109.08560}},
  \href {https://doi.org/10.1103/PhysRevD.104.114503}
  {\path{doi:10.1103/PhysRevD.104.114503}}.

\bibitem{Alberti:2016dru}
M.~Alberti, G.~S. Bali, S.~Collins, F.~Knechtli, G.~Moir, W.~S\"oldner,
  {Hadroquarkonium from lattice QCD}, Phys. Rev. D 95~(7) (2017) 074501.
\newblock \href {http://arxiv.org/abs/1608.06537} {\path{arXiv:1608.06537}},
  \href {https://doi.org/10.1103/PhysRevD.95.074501}
  {\path{doi:10.1103/PhysRevD.95.074501}}.

\bibitem{Dubynskiy:2008di}
S.~Dubynskiy, A.~Gorsky, M.~B. Voloshin, {Holographic Hadro-Quarkonium}, Phys.
  Lett. B 671 (2009) 82--86.
\newblock \href {http://arxiv.org/abs/0804.2244} {\path{arXiv:0804.2244}},
  \href {https://doi.org/10.1016/j.physletb.2008.11.040}
  {\path{doi:10.1016/j.physletb.2008.11.040}}.

\bibitem{Ishii:2006ec}
N.~Ishii, S.~Aoki, T.~Hatsuda, {The Nuclear Force from Lattice QCD}, Phys. Rev.
  Lett. 99 (2007) 022001.
\newblock \href {http://arxiv.org/abs/nucl-th/0611096}
  {\path{arXiv:nucl-th/0611096}}, \href
  {https://doi.org/10.1103/PhysRevLett.99.022001}
  {\path{doi:10.1103/PhysRevLett.99.022001}}.

\bibitem{Aoki:2011gt}
S.~Aoki, N.~Ishii, T.~Doi, T.~Hatsuda, Y.~Ikeda, T.~Inoue, K.~Murano,
  H.~Nemura, K.~Sasaki, {Extraction of Hadron Interactions above Inelastic
  Threshold in Lattice QCD}, Proc. Japan Acad. B 87 (2011) 509--517.
\newblock \href {http://arxiv.org/abs/1106.2281} {\path{arXiv:1106.2281}},
  \href {https://doi.org/10.2183/pjab.87.509} {\path{doi:10.2183/pjab.87.509}}.

\bibitem{Ishii:2012ssm}
N.~Ishii, S.~Aoki, T.~Doi, T.~Hatsuda, Y.~Ikeda, T.~Inoue, K.~Murano,
  H.~Nemura, K.~Sasaki, {Hadron\textendash{}hadron interactions from
  imaginary-time Nambu\textendash{}Bethe\textendash{}Salpeter wave function on
  the lattice}, Phys. Lett. B 712 (2012) 437--441.
\newblock \href {http://arxiv.org/abs/1203.3642} {\path{arXiv:1203.3642}},
  \href {https://doi.org/10.1016/j.physletb.2012.04.076}
  {\path{doi:10.1016/j.physletb.2012.04.076}}.

\bibitem{Prelovsek:2014swa}
S.~Prelovsek, C.~B. Lang, L.~Leskovec, D.~Mohler, {Study of the $Z_c^+$ channel
  using lattice QCD}, Phys. Rev. D 91~(1) (2015) 014504.
\newblock \href {http://arxiv.org/abs/1405.7623} {\path{arXiv:1405.7623}},
  \href {https://doi.org/10.1103/PhysRevD.91.014504}
  {\path{doi:10.1103/PhysRevD.91.014504}}.

\bibitem{Brambilla:2019esw}
N.~Brambilla, S.~Eidelman, C.~Hanhart, A.~Nefediev, C.-P. Shen, C.~E. Thomas,
  A.~Vairo, C.-Z. Yuan, {The $XYZ$ states: experimental and theoretical status
  and perspectives}, Phys. Rept. 873 (2020) 1--154.
\newblock \href {http://arxiv.org/abs/1907.07583} {\path{arXiv:1907.07583}},
  \href {https://doi.org/10.1016/j.physrep.2020.05.001}
  {\path{doi:10.1016/j.physrep.2020.05.001}}.

\bibitem{Prelovsek:2013xba}
S.~Prelovsek, L.~Leskovec, {Search for $Z^{+}_{c}$(3900) in the $1^{+-}$
  Channel on the Lattice}, Phys. Lett. B 727 (2013) 172--176.
\newblock \href {http://arxiv.org/abs/1308.2097} {\path{arXiv:1308.2097}},
  \href {https://doi.org/10.1016/j.physletb.2013.10.009}
  {\path{doi:10.1016/j.physletb.2013.10.009}}.

\bibitem{Guerrieri:2014nxa}
A.~L. Guerrieri, M.~Papinutto, A.~Pilloni, A.~D. Polosa, N.~Tantalo, {Flavored
  tetraquark spectroscopy}, PoS LATTICE2014 (2015) 106.
\newblock \href {http://arxiv.org/abs/1411.2247} {\path{arXiv:1411.2247}}.

\bibitem{Cheung:2017tnt}
G.~K.~C. Cheung, C.~E. Thomas, J.~J. Dudek, R.~G. Edwards, {Tetraquark
  operators in lattice QCD and exotic flavour states in the charm sector}, JHEP
  11 (2017) 033.
\newblock \href {http://arxiv.org/abs/1709.01417} {\path{arXiv:1709.01417}},
  \href {https://doi.org/10.1007/JHEP11(2017)033}
  {\path{doi:10.1007/JHEP11(2017)033}}.

\bibitem{Padmanath:2015era}
M.~Padmanath, C.~B. Lang, S.~Prelovsek, {X(3872) and Y(4140) using
  diquark-antidiquark operators with lattice QCD}, Phys. Rev. D 92~(3) (2015)
  034501.
\newblock \href {http://arxiv.org/abs/1503.03257} {\path{arXiv:1503.03257}},
  \href {https://doi.org/10.1103/PhysRevD.92.034501}
  {\path{doi:10.1103/PhysRevD.92.034501}}.

\bibitem{Thacker:1990bm}
B.~A. Thacker, G.~P. Lepage, {Heavy quark bound states in lattice QCD}, Phys.
  Rev. D 43 (1991) 196--208.
\newblock \href {https://doi.org/10.1103/PhysRevD.43.196}
  {\path{doi:10.1103/PhysRevD.43.196}}.

\bibitem{Lepage:1992tx}
G.~P. Lepage, L.~Magnea, C.~Nakhleh, U.~Magnea, K.~Hornbostel, {Improved
  nonrelativistic QCD for heavy quark physics}, Phys. Rev. D 46 (1992)
  4052--4067.
\newblock \href {http://arxiv.org/abs/hep-lat/9205007}
  {\path{arXiv:hep-lat/9205007}}, \href
  {https://doi.org/10.1103/PhysRevD.46.4052}
  {\path{doi:10.1103/PhysRevD.46.4052}}.

\bibitem{Manohar:1997qy}
A.~V. Manohar, {The HQET / NRQCD Lagrangian to order alpha / m-3}, Phys. Rev. D
  56 (1997) 230--237.
\newblock \href {http://arxiv.org/abs/hep-ph/9701294}
  {\path{arXiv:hep-ph/9701294}}, \href
  {https://doi.org/10.1103/PhysRevD.56.230}
  {\path{doi:10.1103/PhysRevD.56.230}}.

\bibitem{Brown:2014ena}
Z.~S. Brown, W.~Detmold, S.~Meinel, K.~Orginos, {Charmed bottom baryon
  spectroscopy from lattice QCD}, Phys. Rev. D 90~(9) (2014) 094507.
\newblock \href {http://arxiv.org/abs/1409.0497} {\path{arXiv:1409.0497}},
  \href {https://doi.org/10.1103/PhysRevD.90.094507}
  {\path{doi:10.1103/PhysRevD.90.094507}}.

\bibitem{Lewis:1998ka}
R.~Lewis, R.~M. Woloshyn, {O(1 / M**3) effects for heavy - light mesons in
  lattice NRQCD}, Phys. Rev. D 58 (1998) 074506.
\newblock \href {http://arxiv.org/abs/hep-lat/9803004}
  {\path{arXiv:hep-lat/9803004}}, \href
  {https://doi.org/10.1103/PhysRevD.58.074506}
  {\path{doi:10.1103/PhysRevD.58.074506}}.

\bibitem{Lewis:2008fu}
R.~Lewis, R.~M. Woloshyn, {Bottom baryons from a dynamical lattice QCD
  simulation}, Phys. Rev. D 79 (2009) 014502.
\newblock \href {http://arxiv.org/abs/0806.4783} {\path{arXiv:0806.4783}},
  \href {https://doi.org/10.1103/PhysRevD.79.014502}
  {\path{doi:10.1103/PhysRevD.79.014502}}.

\bibitem{Hughes:2017xie}
C.~Hughes, E.~Eichten, C.~T.~H. Davies, {Searching for beauty-fully bound
  tetraquarks using lattice nonrelativistic QCD}, Phys. Rev. D 97~(5) (2018)
  054505.
\newblock \href {http://arxiv.org/abs/1710.03236} {\path{arXiv:1710.03236}},
  \href {https://doi.org/10.1103/PhysRevD.97.054505}
  {\path{doi:10.1103/PhysRevD.97.054505}}.

\bibitem{Luscher:1985dn}
M.~Lüscher, {Volume Dependence of the Energy Spectrum in Massive Quantum Field
  Theories. 1. Stable Particle States}, Commun.Math.Phys. 104 (1986) 177.
\newblock \href {https://doi.org/10.1007/BF01211589}
  {\path{doi:10.1007/BF01211589}}.

\bibitem{Luscher:1986pf}
M.~Lüscher, {Volume Dependence of the Energy Spectrum in Massive Quantum Field
  Theories. 2. Scattering States}, Commun.Math.Phys. 105 (1986) 153--188.
\newblock \href {https://doi.org/10.1007/BF01211097}
  {\path{doi:10.1007/BF01211097}}.

\bibitem{Luscher:1990ck}
M.~Luscher, U.~Wolff, {How to Calculate the Elastic Scattering Matrix in
  Two-dimensional Quantum Field Theories by Numerical Simulation}, Nucl. Phys.
  B 339 (1990) 222--252.
\newblock \href {https://doi.org/10.1016/0550-3213(90)90540-T}
  {\path{doi:10.1016/0550-3213(90)90540-T}}.

\bibitem{Luscher:1990ux}
M.~Luscher, {Two particle states on a torus and their relation to the
  scattering matrix}, Nucl. Phys. B 354 (1991) 531--578.
\newblock \href {https://doi.org/10.1016/0550-3213(91)90366-6}
  {\path{doi:10.1016/0550-3213(91)90366-6}}.

\bibitem{Briceno:2014oea}
R.~A. Briceno, {Two-particle multichannel systems in a finite volume with
  arbitrary spin}, Phys. Rev. D 89~(7) (2014) 074507.
\newblock \href {http://arxiv.org/abs/1401.3312} {\path{arXiv:1401.3312}},
  \href {https://doi.org/10.1103/PhysRevD.89.074507}
  {\path{doi:10.1103/PhysRevD.89.074507}}.

\bibitem{Padmanath:2018tuc}
M.~Padmanath, S.~Collins, D.~Mohler, S.~Piemonte, S.~Prelovsek, A.~Sch\"afer,
  S.~Weishaeupl, {Identifying spin and parity of charmonia in flight with
  lattice QCD}, Phys. Rev. D 99~(1) (2019) 014513.
\newblock \href {http://arxiv.org/abs/1811.04116} {\path{arXiv:1811.04116}},
  \href {https://doi.org/10.1103/PhysRevD.99.014513}
  {\path{doi:10.1103/PhysRevD.99.014513}}.

\bibitem{Hansen:2012tf}
M.~T. Hansen, S.~R. Sharpe, {Multiple-channel generalization of
  Lellouch-Luscher formula}, Phys. Rev. D 86 (2012) 016007.
\newblock \href {http://arxiv.org/abs/1204.0826} {\path{arXiv:1204.0826}},
  \href {https://doi.org/10.1103/PhysRevD.86.016007}
  {\path{doi:10.1103/PhysRevD.86.016007}}.

\bibitem{Hansen:2014eka}
M.~T. Hansen, S.~R. Sharpe, {Relativistic, model-independent, three-particle
  quantization condition}, Phys. Rev. D 90~(11) (2014) 116003.
\newblock \href {http://arxiv.org/abs/1408.5933} {\path{arXiv:1408.5933}},
  \href {https://doi.org/10.1103/PhysRevD.90.116003}
  {\path{doi:10.1103/PhysRevD.90.116003}}.

\bibitem{Hansen:2015zga}
M.~T. Hansen, S.~R. Sharpe, {Expressing the three-particle finite-volume
  spectrum in terms of the three-to-three scattering amplitude}, Phys. Rev. D
  92~(11) (2015) 114509.
\newblock \href {http://arxiv.org/abs/1504.04248} {\path{arXiv:1504.04248}},
  \href {https://doi.org/10.1103/PhysRevD.92.114509}
  {\path{doi:10.1103/PhysRevD.92.114509}}.

\bibitem{Chew:1955zz}
G.~F. Chew, F.~E. Low, {Effective range approach to the low-energy p wave pion
  - nucleon interaction}, Phys. Rev. 101 (1956) 1570--1579.
\newblock \href {https://doi.org/10.1103/PhysRev.101.1570}
  {\path{doi:10.1103/PhysRev.101.1570}}.

\bibitem{Morningstar:2017spu}
C.~Morningstar, J.~Bulava, B.~Singha, R.~Brett, J.~Fallica, A.~Hanlon,
  B.~H\"orz, {Estimating the two-particle $K$-matrix for multiple partial waves
  and decay channels from finite-volume energies}, Nucl. Phys. B 924 (2017)
  477--507.
\newblock \href {http://arxiv.org/abs/1707.05817} {\path{arXiv:1707.05817}},
  \href {https://doi.org/10.1016/j.nuclphysb.2017.09.014}
  {\path{doi:10.1016/j.nuclphysb.2017.09.014}}.

\bibitem{Prelovsek:2020eiw}
S.~Prelovsek, S.~Collins, D.~Mohler, M.~Padmanath, S.~Piemonte,
  {Charmonium-like resonances with J$^{PC}$ = 0$^{++}$, 2$^{++}$ in coupled $
  \mathrm{D}\overline{\mathrm{D}} $, $
  {\mathrm{D}}_{\mathrm{s}}{\overline{\mathrm{D}}}_{\mathrm{s}} $ scattering on
  the lattice}, JHEP 06 (2021) 035.
\newblock \href {http://arxiv.org/abs/2011.02542} {\path{arXiv:2011.02542}},
  \href {https://doi.org/10.1007/JHEP06(2021)035}
  {\path{doi:10.1007/JHEP06(2021)035}}.

\bibitem{Ikeda:2013vwa}
Y.~Ikeda, B.~Charron, S.~Aoki, T.~Doi, T.~Hatsuda, T.~Inoue, N.~Ishii,
  K.~Murano, H.~Nemura, K.~Sasaki, {Charmed tetraquarks $T_{cc}$ and $T_{cs}$
  from dynamical lattice QCD simulations}, Phys. Lett. B 729 (2014) 85--90.
\newblock \href {http://arxiv.org/abs/1311.6214} {\path{arXiv:1311.6214}},
  \href {https://doi.org/10.1016/j.physletb.2014.01.002}
  {\path{doi:10.1016/j.physletb.2014.01.002}}.

\bibitem{Piemonte:2019cbi}
S.~Piemonte, S.~Collins, D.~Mohler, M.~Padmanath, S.~Prelovsek, {Charmonium
  resonances with $J^{PC}=1^{--}$ and $3^{--}$ from $\bar DD$ scattering on the
  lattice}, Phys. Rev. D 100~(7) (2019) 074505.
\newblock \href {http://arxiv.org/abs/1905.03506} {\path{arXiv:1905.03506}},
  \href {https://doi.org/10.1103/PhysRevD.100.074505}
  {\path{doi:10.1103/PhysRevD.100.074505}}.

\bibitem{Prelovsek:2013cra}
S.~Prelovsek, L.~Leskovec, {Evidence for X(3872) from DD* scattering on the
  lattice}, Phys. Rev. Lett. 111 (2013) 192001.
\newblock \href {http://arxiv.org/abs/1307.5172} {\path{arXiv:1307.5172}},
  \href {https://doi.org/10.1103/PhysRevLett.111.192001}
  {\path{doi:10.1103/PhysRevLett.111.192001}}.

\bibitem{Lee:2014uta}
S.-h. Lee, C.~DeTar, D.~Mohler, H.~Na, {Searching for the $X(3872)$ and
  $Z_c^+(3900)$ on HISQ Lattices} (11 2014).
\newblock \href {http://arxiv.org/abs/1411.1389} {\path{arXiv:1411.1389}}.

\bibitem{Mohler:2012na}
D.~Mohler, S.~Prelovsek, R.~M. Woloshyn, {$D \pi$ scattering and $D$ meson
  resonances from lattice QCD}, Phys. Rev. D 87~(3) (2013) 034501.
\newblock \href {http://arxiv.org/abs/1208.4059} {\path{arXiv:1208.4059}},
  \href {https://doi.org/10.1103/PhysRevD.87.034501}
  {\path{doi:10.1103/PhysRevD.87.034501}}.

\bibitem{Mohler:2013rwa}
D.~Mohler, C.~B. Lang, L.~Leskovec, S.~Prelovsek, R.~M. Woloshyn,
  {$D_{s0}^*(2317)$ Meson and $D$-Meson-Kaon Scattering from Lattice QCD},
  Phys. Rev. Lett. 111~(22) (2013) 222001.
\newblock \href {http://arxiv.org/abs/1308.3175} {\path{arXiv:1308.3175}},
  \href {https://doi.org/10.1103/PhysRevLett.111.222001}
  {\path{doi:10.1103/PhysRevLett.111.222001}}.

\bibitem{Lang:2014yfa}
C.~B. Lang, L.~Leskovec, D.~Mohler, S.~Prelovsek, R.~M. Woloshyn, {Ds mesons
  with DK and D*K scattering near threshold}, Phys. Rev. D 90~(3) (2014)
  034510.
\newblock \href {http://arxiv.org/abs/1403.8103} {\path{arXiv:1403.8103}},
  \href {https://doi.org/10.1103/PhysRevD.90.034510}
  {\path{doi:10.1103/PhysRevD.90.034510}}.

\bibitem{Moir:2016srx}
G.~Moir, M.~Peardon, S.~M. Ryan, C.~E. Thomas, D.~J. Wilson, {Coupled-Channel
  $D\pi$, $D\eta$ and $D_{s}\bar{K}$ Scattering from Lattice QCD}, JHEP 10
  (2016) 011.
\newblock \href {http://arxiv.org/abs/1607.07093} {\path{arXiv:1607.07093}},
  \href {https://doi.org/10.1007/JHEP10(2016)011}
  {\path{doi:10.1007/JHEP10(2016)011}}.

\bibitem{Bali:2017pdv}
G.~S. Bali, S.~Collins, A.~Cox, A.~Sch\"afer, {Masses and decay constants of
  the $D_{s0}^*(2317)$ and $D_{s1}(2460)$ from $N_f=2$ lattice QCD close to the
  physical point}, Phys. Rev. D 96~(7) (2017) 074501.
\newblock \href {http://arxiv.org/abs/1706.01247} {\path{arXiv:1706.01247}},
  \href {https://doi.org/10.1103/PhysRevD.96.074501}
  {\path{doi:10.1103/PhysRevD.96.074501}}.

\bibitem{Alexandrou:2019tmk}
C.~Alexandrou, J.~Berlin, J.~Finkenrath, T.~Leontiou, M.~Wagner, {Tetraquark
  interpolating fields in a lattice QCD investigation of the
  $D_{s0}^\ast(2317)$ meson}, Phys. Rev. D 101~(3) (2020) 034502.
\newblock \href {http://arxiv.org/abs/1911.08435} {\path{arXiv:1911.08435}},
  \href {https://doi.org/10.1103/PhysRevD.101.034502}
  {\path{doi:10.1103/PhysRevD.101.034502}}.

\bibitem{Lang:2015hza}
C.~B. Lang, D.~Mohler, S.~Prelovsek, R.~M. Woloshyn, {Predicting positive
  parity B$_s$ mesons from lattice QCD}, Phys. Lett. B 750 (2015) 17--21.
\newblock \href {http://arxiv.org/abs/1501.01646} {\path{arXiv:1501.01646}},
  \href {https://doi.org/10.1016/j.physletb.2015.08.038}
  {\path{doi:10.1016/j.physletb.2015.08.038}}.

\bibitem{Csikor:2003ng}
F.~Csikor, Z.~Fodor, S.~D. Katz, T.~G. Kovacs, {Pentaquark hadrons from lattice
  QCD}, JHEP 11 (2003) 070.
\newblock \href {http://arxiv.org/abs/hep-lat/0309090}
  {\path{arXiv:hep-lat/0309090}}, \href
  {https://doi.org/10.1088/1126-6708/2003/11/070}
  {\path{doi:10.1088/1126-6708/2003/11/070}}.

\bibitem{Mathur:2004jr}
N.~Mathur, et~al., {A Study of pentaquarks on the lattice with overlap
  fermions}, Phys. Rev. D 70 (2004) 074508.
\newblock \href {http://arxiv.org/abs/hep-ph/0406196}
  {\path{arXiv:hep-ph/0406196}}, \href
  {https://doi.org/10.1103/PhysRevD.70.074508}
  {\path{doi:10.1103/PhysRevD.70.074508}}.

\bibitem{Alexandrou:2004ws}
C.~Alexandrou, G.~Koutsou, A.~Tsapalis, {The Pentaquark potential, mass and
  density-density correlator}, Nucl. Phys. B Proc. Suppl. 140 (2005) 275--277.
\newblock \href {http://arxiv.org/abs/hep-lat/0409065}
  {\path{arXiv:hep-lat/0409065}}, \href
  {https://doi.org/10.1016/j.nuclphysbps.2004.11.270}
  {\path{doi:10.1016/j.nuclphysbps.2004.11.270}}.

\bibitem{Sasaki:2004vz}
S.~Sasaki, {Pentaquarks: Status and perspectives for lattice calculations},
  Nucl. Phys. B Proc. Suppl. 140 (2005) 127--133.
\newblock \href {http://arxiv.org/abs/hep-lat/0410016}
  {\path{arXiv:hep-lat/0410016}}, \href
  {https://doi.org/10.1016/j.nuclphysbps.2004.11.354}
  {\path{doi:10.1016/j.nuclphysbps.2004.11.354}}.

\bibitem{Lasscock:2005tt}
B.~G. Lasscock, J.~N. Hedditch, W.~Kamleh, D.~B. Leinweber, W.~Melnitchouk,
  A.~W. Thomas, A.~G. Williams, R.~D. Young, J.~M. Zanotti, {Search for the
  pentaquark resonance signature in lattice QCD}, Phys. Rev. D 72 (2005)
  014502.
\newblock \href {http://arxiv.org/abs/hep-lat/0503008}
  {\path{arXiv:hep-lat/0503008}}, \href
  {https://doi.org/10.1103/PhysRevD.72.014502}
  {\path{doi:10.1103/PhysRevD.72.014502}}.

\bibitem{Csikor:2005xb}
F.~Csikor, Z.~Fodor, S.~D. Katz, T.~G. Kovacs, B.~C. Toth, {A Comprehensive
  search for the Theta+ pentaquark on the lattice}, Phys. Rev. D 73 (2006)
  034506.
\newblock \href {http://arxiv.org/abs/hep-lat/0503012}
  {\path{arXiv:hep-lat/0503012}}, \href
  {https://doi.org/10.1103/PhysRevD.73.034506}
  {\path{doi:10.1103/PhysRevD.73.034506}}.

\bibitem{Alexandrou:2005gc}
C.~Alexandrou, A.~Tsapalis, {A Lattice study of the pentaquark state}, Phys.
  Rev. D 73 (2006) 014507.
\newblock \href {http://arxiv.org/abs/hep-lat/0503013}
  {\path{arXiv:hep-lat/0503013}}, \href
  {https://doi.org/10.1103/PhysRevD.73.014507}
  {\path{doi:10.1103/PhysRevD.73.014507}}.

\bibitem{Takahashi:2005uk}
T.~T. Takahashi, T.~Umeda, T.~Onogi, T.~Kunihiro, {Search for the possible S=+1
  pentaquark states in quenched lattice QCD}, Phys. Rev. D 71 (2005) 114509.
\newblock \href {http://arxiv.org/abs/hep-lat/0503019}
  {\path{arXiv:hep-lat/0503019}}, \href
  {https://doi.org/10.1103/PhysRevD.71.114509}
  {\path{doi:10.1103/PhysRevD.71.114509}}.

\bibitem{Holland:2005yt}
K.~Holland, K.~J. Juge, {Absence of evidence for pentaquarks on the lattice},
  Phys. Rev. D 73 (2006) 074505.
\newblock \href {http://arxiv.org/abs/hep-lat/0504007}
  {\path{arXiv:hep-lat/0504007}}, \href
  {https://doi.org/10.1103/PhysRevD.73.074505}
  {\path{doi:10.1103/PhysRevD.73.074505}}.

\bibitem{Lasscock:2005kx}
B.~G. Lasscock, D.~B. Leinweber, W.~Melnitchouk, A.~W. Thomas, A.~G. Williams,
  R.~D. Young, J.~M. Zanotti, {Spin 3/2 pentaquark resonance signature in
  lattice QCD}, Phys. Rev. D 72 (2005) 074507.
\newblock \href {http://arxiv.org/abs/hep-lat/0504015}
  {\path{arXiv:hep-lat/0504015}}, \href
  {https://doi.org/10.1103/PhysRevD.72.074507}
  {\path{doi:10.1103/PhysRevD.72.074507}}.

\bibitem{Alexandrou:2005ek}
C.~Alexandrou, A.~Tsapalis, {The Volume dependence of spectral weights and the
  pentaquark state}, PoS LAT2005 (2006) 023.
\newblock \href {http://arxiv.org/abs/hep-lat/0509139}
  {\path{arXiv:hep-lat/0509139}}, \href {https://doi.org/10.22323/1.020.0023}
  {\path{doi:10.22323/1.020.0023}}.

\bibitem{Jahn:2005zb}
O.~Jahn, J.~W. Negele, D.~Sigaev, {The Quark structure of pentaquarks}, PoS
  LAT2005 (2006) 069.
\newblock \href {http://arxiv.org/abs/hep-lat/0509102}
  {\path{arXiv:hep-lat/0509102}}, \href {https://doi.org/10.22323/1.020.0069}
  {\path{doi:10.22323/1.020.0069}}.

\bibitem{Takahashi:2005mr}
T.~T. Takahashi, T.~Umeda, T.~Onogi, T.~Kunihiro, {Search for the S=+1
  pentaquarks in quenched lattice QCD}, PoS LAT2005 (2006) 071.
\newblock \href {http://arxiv.org/abs/hep-lat/0509148}
  {\path{arXiv:hep-lat/0509148}}, \href {https://doi.org/10.22323/1.020.0071}
  {\path{doi:10.22323/1.020.0071}}.

\bibitem{Doi:2005mn}
T.~Doi, N.~Ishii, Y.~Nemoto, M.~Oka, H.~Suganuma, {Anisotropic lattice QCD
  study of pentaquark baryons in spin 3/2 channel}, PoS LAT2005 (2006) 064.
\newblock \href {http://arxiv.org/abs/hep-lat/0509145}
  {\path{arXiv:hep-lat/0509145}}, \href {https://doi.org/10.22323/1.020.0064}
  {\path{doi:10.22323/1.020.0064}}.

\bibitem{Chiu:2005ge}
T.-W. Chiu, T.-H. Hsieh, {Mass spectra of pentaquarks: Overlap versus Wilson
  fermions}, PoS LAT2005 (2006) 065.
\newblock \href {http://arxiv.org/abs/hep-lat/0509175}
  {\path{arXiv:hep-lat/0509175}}, \href {https://doi.org/10.22323/1.020.0065}
  {\path{doi:10.22323/1.020.0065}}.

\bibitem{Tariq:2007ck}
A.~S.~B. Tariq, {Revisiting the pentaquark episode for lattice QCD}, PoS
  LATTICE2007 (2007) 136.
\newblock \href {http://arxiv.org/abs/0711.0566} {\path{arXiv:0711.0566}},
  \href {https://doi.org/10.22323/1.042.0136} {\path{doi:10.22323/1.042.0136}}.

\bibitem{Beane:2014sda}
S.~R. Beane, E.~Chang, S.~D. Cohen, W.~Detmold, H.~W. Lin, K.~Orginos,
  A.~Parre\~no, M.~J. Savage, {Quarkonium-Nucleus Bound States from Lattice
  QCD}, Phys. Rev. D 91~(11) (2015) 114503.
\newblock \href {http://arxiv.org/abs/1410.7069} {\path{arXiv:1410.7069}},
  \href {https://doi.org/10.1103/PhysRevD.91.114503}
  {\path{doi:10.1103/PhysRevD.91.114503}}.

\bibitem{Xing:2022ijm}
H.~Xing, J.~Liang, L.~Liu, P.~Sun, Y.-B. Yang, {First observation of the
  hidden-charm pentaquarks on lattice} (10 2022).
\newblock \href {http://arxiv.org/abs/2210.08555} {\path{arXiv:2210.08555}}.

\bibitem{Skerbis:2018unn}
U.~Skerbis, S.~Prelovsek, {$J/\psi$-nucleon scattering in $P_{c}^{+}$
  pentaquark channels}, PoS LATTICE2018 (2018) 087.
\newblock \href {http://arxiv.org/abs/1811.03580} {\path{arXiv:1811.03580}},
  \href {https://doi.org/10.22323/1.334.0087} {\path{doi:10.22323/1.334.0087}}.

\bibitem{Skerbis:2018lew}
U.~Skerbis, S.~Prelovsek, {Nucleon-$J/\psi$ and nucleon-$\eta_{c}$ scattering
  in $P_{c}$ pentaquark channels from LQCD}, Phys. Rev. D 99~(9) (2019) 094505.
\newblock \href {http://arxiv.org/abs/1811.02285} {\path{arXiv:1811.02285}},
  \href {https://doi.org/10.1103/PhysRevD.99.094505}
  {\path{doi:10.1103/PhysRevD.99.094505}}.

\bibitem{Kawanai:2010ev}
T.~Kawanai, S.~Sasaki, {Charmonium-nucleon potential from lattice QCD}, Phys.
  Rev. D 82 (2010) 091501.
\newblock \href {http://arxiv.org/abs/1009.3332} {\path{arXiv:1009.3332}},
  \href {https://doi.org/10.1103/PhysRevD.82.091501}
  {\path{doi:10.1103/PhysRevD.82.091501}}.

\bibitem{Sugiura:2017vks}
T.~Sugiura, Y.~Ikeda, N.~Ishii, {Charmonium-nucleon interactions from the
  time-dependent HAL QCD method}, EPJ Web Conf. 175 (2018) 05011.
\newblock \href {http://arxiv.org/abs/1711.11219} {\path{arXiv:1711.11219}},
  \href {https://doi.org/10.1051/epjconf/201817505011}
  {\path{doi:10.1051/epjconf/201817505011}}.

\bibitem{Yokokawa:2006td}
K.~Yokokawa, S.~Sasaki, T.~Hatsuda, A.~Hayashigaki, {First lattice study of
  low-energy charmonium-hadron interaction}, Phys. Rev. D 74 (2006) 034504.
\newblock \href {http://arxiv.org/abs/hep-lat/0605009}
  {\path{arXiv:hep-lat/0605009}}, \href
  {https://doi.org/10.1103/PhysRevD.74.034504}
  {\path{doi:10.1103/PhysRevD.74.034504}}.

\bibitem{Liu:2008rza}
L.~Liu, H.-W. Lin, K.~Orginos, {Charmed Hadron Interactions}, PoS LATTICE2008
  (2008) 112.
\newblock \href {http://arxiv.org/abs/0810.5412} {\path{arXiv:0810.5412}},
  \href {https://doi.org/10.22323/1.066.0112} {\path{doi:10.22323/1.066.0112}}.

\bibitem{Koma:2006fw}
Y.~Koma, M.~Koma, {Spin-dependent potentials from lattice QCD}, Nucl. Phys.
  B769 (2007) 79--107.
\newblock \href {http://arxiv.org/abs/hep-lat/0609078}
  {\path{arXiv:hep-lat/0609078}}, \href
  {https://doi.org/10.1016/j.nuclphysb.2007.01.033}
  {\path{doi:10.1016/j.nuclphysb.2007.01.033}}.

\bibitem{Campbell:1987nv}
N.~A. Campbell, A.~Huntley, C.~Michael, {Heavy Quark Potentials and Hybrid
  Mesons From SU(3) Lattice Gauge Theory}, Nucl. Phys. B306 (1988) 51--62.
\newblock \href {https://doi.org/10.1016/0550-3213(88)90170-8}
  {\path{doi:10.1016/0550-3213(88)90170-8}}.

\bibitem{Campostrini:1986ki}
M.~Campostrini, K.~Moriarty, C.~Rebbi, {Monte Carlo Calculation of the Spin
  Dependent Potentials for Heavy Quark Spectroscopy}, Phys. Rev. Lett. 57
  (1986) 44.
\newblock \href {https://doi.org/10.1103/PhysRevLett.57.44}
  {\path{doi:10.1103/PhysRevLett.57.44}}.

\bibitem{Perantonis:1988uz}
S.~Perantonis, A.~Huntley, C.~Michael, {Static Potentials From Pure SU(2)
  Lattice Gauge Theory}, Nucl. Phys. B 326 (1989) 544--556.
\newblock \href {https://doi.org/10.1016/0550-3213(89)90141-7}
  {\path{doi:10.1016/0550-3213(89)90141-7}}.

\bibitem{Huntley:1986de}
A.~Huntley, C.~Michael, {Spin Spin and Spin - Orbit Potentials From Lattice
  Gauge Theory}, Nucl. Phys. B 286 (1987) 211--230.
\newblock \href {https://doi.org/10.1016/0550-3213(87)90438-X}
  {\path{doi:10.1016/0550-3213(87)90438-X}}.

\bibitem{Michael:1985rh}
C.~Michael, {The Long Range Spin Orbit Potential}, Phys. Rev. Lett. 56 (1986)
  1219.
\newblock \href {https://doi.org/10.1103/PhysRevLett.56.1219}
  {\path{doi:10.1103/PhysRevLett.56.1219}}.

\bibitem{Michael:1985wf}
C.~Michael, P.~E.~L. Rakow, {Spin Dependence of Interquark Forces From Lattice
  Gauge Theory}, Nucl. Phys. B 256 (1985) 640--652.
\newblock \href {https://doi.org/10.1016/0550-3213(85)90412-2}
  {\path{doi:10.1016/0550-3213(85)90412-2}}.

\bibitem{Campostrini:1984uj}
M.~Campostrini, {The Spin Dependent Forces Between Heavy Quarks in Lattice
  Gauge Theories}, Nucl. Phys. B 256 (1985) 717--726.
\newblock \href {https://doi.org/10.1016/0550-3213(85)90417-1}
  {\path{doi:10.1016/0550-3213(85)90417-1}}.

\bibitem{deForcrand:1985zc}
P.~de~Forcrand, J.~D. Stack, {Spin Dependent Potentials in SU(3) Lattice Gauge
  Theory}, Phys. Rev. Lett. 55 (1985) 1254.
\newblock \href {https://doi.org/10.1103/PhysRevLett.55.1254}
  {\path{doi:10.1103/PhysRevLett.55.1254}}.

\bibitem{Lucha:1991vn}
W.~Lucha, F.~F. Schoberl, D.~Gromes, {Bound states of quarks}, Phys. Rept. 200
  (1991) 127--240.
\newblock \href {https://doi.org/10.1016/0370-1573(91)90001-3}
  {\path{doi:10.1016/0370-1573(91)90001-3}}.

\bibitem{Lucha:1992rq}
W.~Lucha, F.~F. Schoberl, {Quark anti-quark bound states: Relativistic versus
  nonrelativistic point of view}, Int. J. Mod. Phys. A 7 (1992) 6431--6456.
\newblock \href {https://doi.org/10.1142/S0217751X92002945}
  {\path{doi:10.1142/S0217751X92002945}}.

\bibitem{Drummond:1999db}
I.~T. Drummond, N.~A. Goodman, R.~R. Horgan, H.~P. Shanahan, L.~C. Storoni,
  {Spin effects in heavy hybrid mesons on an anisotropic lattice}, Phys. Lett.
  B 478 (2000) 151--160.
\newblock \href {http://arxiv.org/abs/hep-lat/9912041}
  {\path{arXiv:hep-lat/9912041}}, \href
  {https://doi.org/10.1016/S0370-2693(00)00225-2}
  {\path{doi:10.1016/S0370-2693(00)00225-2}}.

\bibitem{Murano:2013xxa}
K.~Murano, N.~Ishii, S.~Aoki, T.~Doi, T.~Hatsuda, Y.~Ikeda, T.~Inoue,
  H.~Nemura, K.~Sasaki, {Spin\textendash{}orbit force from lattice QCD}, Phys.
  Lett. B 735 (2014) 19--24.
\newblock \href {http://arxiv.org/abs/1305.2293} {\path{arXiv:1305.2293}},
  \href {https://doi.org/10.1016/j.physletb.2014.05.061}
  {\path{doi:10.1016/j.physletb.2014.05.061}}.

\bibitem{Bhaduri:1981pn}
R.~K. Bhaduri, L.~E. Cohler, Y.~Nogami, {A Unified Potential for Mesons and
  Baryons}, Nuovo Cim. A 65 (1981) 376--390.
\newblock \href {https://doi.org/10.1007/BF02827441}
  {\path{doi:10.1007/BF02827441}}.

\bibitem{Eichten:1979pu}
E.~Eichten, F.~L. Feinberg, {Spin Dependent Forces in Heavy Quark Systems},
  Phys. Rev. Lett. 43 (1979) 1205.
\newblock \href {https://doi.org/10.1103/PhysRevLett.43.1205}
  {\path{doi:10.1103/PhysRevLett.43.1205}}.

\bibitem{Eichten:1980mw}
E.~Eichten, F.~Feinberg, {Spin Dependent Forces in QCD}, Phys. Rev. D 23 (1981)
  2724.
\newblock \href {https://doi.org/10.1103/PhysRevD.23.2724}
  {\path{doi:10.1103/PhysRevD.23.2724}}.

\bibitem{Gromes:1983pm}
D.~Gromes, {Relativistic Corrections to the Long Range Quark Anti-quark
  Potential, Electric Flux Tubes, and Area Law}, Z. Phys. C 22 (1984) 265.
\newblock \href {https://doi.org/10.1007/BF01575791}
  {\path{doi:10.1007/BF01575791}}.

\bibitem{Chen:2005mg}
Y.~Chen, et~al., {Glueball spectrum and matrix elements on anisotropic
  lattices}, Phys. Rev. D 73 (2006) 014516.
\newblock \href {http://arxiv.org/abs/hep-lat/0510074}
  {\path{arXiv:hep-lat/0510074}}, \href
  {https://doi.org/10.1103/PhysRevD.73.014516}
  {\path{doi:10.1103/PhysRevD.73.014516}}.

\bibitem{Koma:2012bc}
Y.~Koma, M.~Koma, {Heavy quarkonium spectroscopy in pNRQCD with lattice QCD
  input}, PoS LATTICE2012 (2012) 140.
\newblock \href {http://arxiv.org/abs/1211.6795} {\path{arXiv:1211.6795}},
  \href {https://doi.org/10.22323/1.164.0140} {\path{doi:10.22323/1.164.0140}}.

\bibitem{Karsch:1998xd}
F.~Karsch, M.~Hess, E.~Laermann, I.~Wetzorke, {Constituent quarks, diquarks and
  the N - Delta mass splitting}, Nucl. Phys. B Proc. Suppl. 73 (1999) 213--215.
\newblock \href {http://arxiv.org/abs/hep-lat/9809011}
  {\path{arXiv:hep-lat/9809011}}, \href
  {https://doi.org/10.1016/S0920-5632(99)85026-9}
  {\path{doi:10.1016/S0920-5632(99)85026-9}}.

\bibitem{Hess:1998sd}
M.~Hess, F.~Karsch, E.~Laermann, I.~Wetzorke, {Diquark masses from lattice
  QCD}, Phys. Rev. D 58 (1998) 111502.
\newblock \href {http://arxiv.org/abs/hep-lat/9804023}
  {\path{arXiv:hep-lat/9804023}}, \href
  {https://doi.org/10.1103/PhysRevD.58.111502}
  {\path{doi:10.1103/PhysRevD.58.111502}}.

\bibitem{Gockeler:2002sb}
M.~Gockeler, R.~Horsley, B.~Klaus, D.~Pleiter, P.~E.~L. Rakow, S.~Schaefer,
  A.~Schafer, G.~Schierholz, {Applied lattice gauge calculations: Diquark
  content of the nucleon}, Nucl. Phys. A 711 (2002) 297--302.
\newblock \href {http://arxiv.org/abs/hep-lat/0208008}
  {\path{arXiv:hep-lat/0208008}}, \href
  {https://doi.org/10.1016/S0375-9474(02)01232-0}
  {\path{doi:10.1016/S0375-9474(02)01232-0}}.

\bibitem{Babich:2005ay}
R.~Babich, F.~Berruto, N.~Garron, C.~Hoelbling, J.~Howard, L.~Lellouch,
  C.~Rebbi, N.~Shoresh, {Light hadron and diquark spectroscopy in quenched QCD
  with overlap quarks on a large lattice}, JHEP 01 (2006) 086.
\newblock \href {http://arxiv.org/abs/hep-lat/0509027}
  {\path{arXiv:hep-lat/0509027}}, \href
  {https://doi.org/10.1088/1126-6708/2006/01/086}
  {\path{doi:10.1088/1126-6708/2006/01/086}}.

\bibitem{Orginos:2005vr}
K.~Orginos, {Diquark properties from lattice QCD}, PoS LAT2005 (2006) 054.
\newblock \href {http://arxiv.org/abs/hep-lat/0510082}
  {\path{arXiv:hep-lat/0510082}}, \href {https://doi.org/10.22323/1.020.0054}
  {\path{doi:10.22323/1.020.0054}}.

\bibitem{Fodor:2005qx}
Z.~Fodor, C.~Hoelbling, M.~Mechtel, K.~Szabo, {Nonperturbative investigation of
  the diquark potential}, PoS LAT2005 (2006) 310.
\newblock \href {http://arxiv.org/abs/hep-lat/0511032}
  {\path{arXiv:hep-lat/0511032}}, \href {https://doi.org/10.22323/1.020.0310}
  {\path{doi:10.22323/1.020.0310}}.

\bibitem{Alexandrou:2005zn}
C.~Alexandrou, P.~de~Forcrand, B.~Lucini, {Searching for diquarks in hadrons},
  PoS LAT2005 (2006) 053.
\newblock \href {http://arxiv.org/abs/hep-lat/0509113}
  {\path{arXiv:hep-lat/0509113}}, \href {https://doi.org/10.22323/1.020.0053}
  {\path{doi:10.22323/1.020.0053}}.

\bibitem{Alexandrou:2006cq}
C.~Alexandrou, P.~de~Forcrand, B.~Lucini, {Evidence for diquarks in lattice
  QCD}, Phys. Rev. Lett. 97 (2006) 222002.
\newblock \href {http://arxiv.org/abs/hep-lat/0609004}
  {\path{arXiv:hep-lat/0609004}}, \href
  {https://doi.org/10.1103/PhysRevLett.97.222002}
  {\path{doi:10.1103/PhysRevLett.97.222002}}.

\bibitem{Liu:2006zi}
Z.~Liu, T.~DeGrand, {Baryon correlators containing different diquarks from
  lattice simulations}, PoS LAT2006 (2006) 116.
\newblock \href {http://arxiv.org/abs/hep-lat/0609038}
  {\path{arXiv:hep-lat/0609038}}, \href {https://doi.org/10.22323/1.032.0116}
  {\path{doi:10.22323/1.032.0116}}.

\bibitem{Babich:2006eu}
R.~Babich, N.~Garron, C.~Hoelbling, J.~Howard, L.~Lellouch, C.~Rebbi, {Matrix
  elements and diquark correlations in quenched QCD with overlap fermions}, PoS
  LAT2006 (2006) 091.
\newblock \href {http://arxiv.org/abs/hep-lat/0610079}
  {\path{arXiv:hep-lat/0610079}}, \href {https://doi.org/10.22323/1.032.0091}
  {\path{doi:10.22323/1.032.0091}}.

\bibitem{Babich:2007ah}
R.~Babich, N.~Garron, C.~Hoelbling, J.~Howard, L.~Lellouch, C.~Rebbi, {Diquark
  correlations in baryons on the lattice with overlap quarks}, Phys. Rev. D 76
  (2007) 074021.
\newblock \href {http://arxiv.org/abs/hep-lat/0701023}
  {\path{arXiv:hep-lat/0701023}}, \href
  {https://doi.org/10.1103/PhysRevD.76.074021}
  {\path{doi:10.1103/PhysRevD.76.074021}}.

\bibitem{Gottlieb:2007ay}
S.~Gottlieb, H.~Na, K.~Nagata, {Diquark representations for singly heavy
  baryons with light staggered quarks}, Phys. Rev. D 77 (2008) 017505.
\newblock \href {http://arxiv.org/abs/0707.3537} {\path{arXiv:0707.3537}},
  \href {https://doi.org/10.1103/PhysRevD.77.017505}
  {\path{doi:10.1103/PhysRevD.77.017505}}.

\bibitem{Gottlieb:2007jm}
S.~Gottlieb, H.~Na, K.~Nagata, {Staggered diquarks for singly heavy baryons},
  PoS LATTICE2007 (2007) 125.
\newblock \href {http://arxiv.org/abs/0710.0347} {\path{arXiv:0710.0347}},
  \href {https://doi.org/10.22323/1.042.0125} {\path{doi:10.22323/1.042.0125}}.

\bibitem{Bissey:2009gw}
F.~Bissey, A.~I. Signal, D.~B. Leinweber, {Comparison of gluon flux-tube
  distributions for quark-diquark and quark-antiquark hadrons}, Phys. Rev. D 80
  (2009) 114506.
\newblock \href {http://arxiv.org/abs/0910.0958} {\path{arXiv:0910.0958}},
  \href {https://doi.org/10.1103/PhysRevD.80.114506}
  {\path{doi:10.1103/PhysRevD.80.114506}}.

\bibitem{Green:2010vc}
J.~Green, J.~Negele, M.~Engelhardt, P.~Varilly, {Spatial diquark correlations
  in a hadron}, PoS LATTICE2010 (2010) 140.
\newblock \href {http://arxiv.org/abs/1012.2353} {\path{arXiv:1012.2353}},
  \href {https://doi.org/10.22323/1.105.0140} {\path{doi:10.22323/1.105.0140}}.

\bibitem{Watanabe:2021oyv}
K.~Watanabe, N.~Ishii, {Building Diquark Model from Lattice QCD}, Few Body
  Syst. 62~(3) (2021) 45.
\newblock \href {http://arxiv.org/abs/2105.07969} {\path{arXiv:2105.07969}},
  \href {https://doi.org/10.1007/s00601-021-01627-y}
  {\path{doi:10.1007/s00601-021-01627-y}}.

\bibitem{Francis:2021vrr}
A.~Francis, P.~de~Forcrand, R.~Lewis, K.~Maltman, {Diquark properties from full
  QCD lattice simulations}, JHEP 05 (2022) 062.
\newblock \href {http://arxiv.org/abs/2106.09080} {\path{arXiv:2106.09080}},
  \href {https://doi.org/10.1007/JHEP05(2022)062}
  {\path{doi:10.1007/JHEP05(2022)062}}.

\bibitem{Watanabe:2021nwe}
K.~Watanabe, {Quark-diquark potential and diquark mass from lattice QCD}, Phys.
  Rev. D 105~(7) (2022) 074510.
\newblock \href {http://arxiv.org/abs/2111.15167} {\path{arXiv:2111.15167}},
  \href {https://doi.org/10.1103/PhysRevD.105.074510}
  {\path{doi:10.1103/PhysRevD.105.074510}}.

\bibitem{Jaffe:2004ph}
R.~L. Jaffe, {Exotica}, Phys. Rept. 409 (2005) 1--45.
\newblock \href {http://arxiv.org/abs/hep-ph/0409065}
  {\path{arXiv:hep-ph/0409065}}, \href
  {https://doi.org/10.1016/j.physrep.2004.11.005}
  {\path{doi:10.1016/j.physrep.2004.11.005}}.

\bibitem{Bicudo:2022lat}
P.~Bicudo, N.~Cardoso, S.~Alireza, {Eight spectra of very excited flux tubes in
  SU(3) gauge theory }.

\bibitem{Perantonis:1990dy}
S.~Perantonis, C.~Michael, {Static potentials and hybrid mesons from pure SU(3)
  lattice gauge theory}, Nucl. Phys. B347 (1990) 854--868.
\newblock \href {https://doi.org/10.1016/0550-3213(90)90386-R}
  {\path{doi:10.1016/0550-3213(90)90386-R}}.

\bibitem{Lacock:1996ny}
P.~Lacock, C.~Michael, P.~Boyle, P.~Rowland, {Hybrid mesons from quenched QCD},
  Phys. Lett. B401 (1997) 308--312.
\newblock \href {http://arxiv.org/abs/hep-lat/9611011}
  {\path{arXiv:hep-lat/9611011}}, \href
  {https://doi.org/10.1016/S0370-2693(97)00384-5}
  {\path{doi:10.1016/S0370-2693(97)00384-5}}.

\bibitem{Lacock:1996vy}
P.~Lacock, C.~Michael, P.~Boyle, P.~Rowland, {Orbitally excited and hybrid
  mesons from the lattice}, Phys. Rev. D54 (1996) 6997--7009.
\newblock \href {http://arxiv.org/abs/hep-lat/9605025}
  {\path{arXiv:hep-lat/9605025}}, \href
  {https://doi.org/10.1103/PhysRevD.54.6997}
  {\path{doi:10.1103/PhysRevD.54.6997}}.

\bibitem{Juge:1999ie}
K.~J. Juge, J.~Kuti, C.~J. Morningstar, {Ab initio study of hybrid anti-b g b
  mesons}, Phys. Rev. Lett. 82 (1999) 4400--4403.
\newblock \href {http://arxiv.org/abs/hep-ph/9902336}
  {\path{arXiv:hep-ph/9902336}}, \href
  {https://doi.org/10.1103/PhysRevLett.82.4400}
  {\path{doi:10.1103/PhysRevLett.82.4400}}.

\bibitem{Juge:2002br}
K.~J. Juge, J.~Kuti, C.~Morningstar, {Fine structure of the QCD string
  spectrum}, Phys. Rev. Lett. 90 (2003) 161601.
\newblock \href {http://arxiv.org/abs/hep-lat/0207004}
  {\path{arXiv:hep-lat/0207004}}, \href
  {https://doi.org/10.1103/PhysRevLett.90.161601}
  {\path{doi:10.1103/PhysRevLett.90.161601}}.

\bibitem{Reisinger:2017btr}
C.~Reisinger, S.~Capitani, O.~Philipsen, M.~Wagner,
  \href{https://inspirehep.net/record/1616731/files/arXiv:1708.05562.pdf}{{Computation
  of hybrid static potentials in SU(3) lattice gauge theory}}, in: {35th
  International Symposium on Lattice Field Theory (Lattice 2017) Granada,
  Spain, June 18-24, 2017}, 2017.
\newblock \href {http://arxiv.org/abs/1708.05562} {\path{arXiv:1708.05562}}.
\newline\urlprefix\url{https://inspirehep.net/record/1616731/files/arXiv:1708.05562.pdf}

\bibitem{Bicudo:2018yhk}
P.~Bicudo, M.~Cardoso, N.~Cardoso, {Colour fields of the quark-antiquark
  excited flux tube}, EPJ Web Conf. 175 (2018) 14009.
\newblock \href {http://arxiv.org/abs/1803.04569} {\path{arXiv:1803.04569}},
  \href {https://doi.org/10.1051/epjconf/201817514009}
  {\path{doi:10.1051/epjconf/201817514009}}.

\bibitem{Bicudo:2018jbb}
P.~Bicudo, N.~Cardoso, M.~Cardoso, {Color field densities of the
  quark-antiquark excited flux tubes in SU(3) lattice QCD}, Phys. Rev. D
  98~(11) (2018) 114507.
\newblock \href {http://arxiv.org/abs/1808.08815} {\path{arXiv:1808.08815}},
  \href {https://doi.org/10.1103/PhysRevD.98.114507}
  {\path{doi:10.1103/PhysRevD.98.114507}}.

\bibitem{Mueller:2019mkh}
L.~M\"uller, O.~Philipsen, C.~Reisinger, M.~Wagner, {Hybrid static potential
  flux tubes from SU(2) and SU(3) lattice gauge theory}, Phys. Rev. D 100~(5)
  (2019) 054503.
\newblock \href {http://arxiv.org/abs/1907.01482} {\path{arXiv:1907.01482}},
  \href {https://doi.org/10.1103/PhysRevD.100.054503}
  {\path{doi:10.1103/PhysRevD.100.054503}}.

\bibitem{Bicudo:2021tsc}
P.~Bicudo, N.~Cardoso, A.~Sharifian, {Spectrum of very excited
  \ensuremath{\Sigma}g+ flux tubes in SU(3) gauge theory}, Phys. Rev. D 104~(5)
  (2021) 054512.
\newblock \href {http://arxiv.org/abs/2105.12159} {\path{arXiv:2105.12159}},
  \href {https://doi.org/10.1103/PhysRevD.104.054512}
  {\path{doi:10.1103/PhysRevD.104.054512}}.

\bibitem{Arvis:1983fp}
J.~F. Arvis, {The Exact $q \bar{q}$ Potential in Nambu String Theory}, Phys.
  Lett. B 127 (1983) 106--108.
\newblock \href {https://doi.org/10.1016/0370-2693(83)91640-4}
  {\path{doi:10.1016/0370-2693(83)91640-4}}.

\bibitem{Luscher:1980ac}
M.~Luscher, {Symmetry Breaking Aspects of the Roughening Transition in Gauge
  Theories}, Nucl. Phys. B 180 (1981) 317--329.
\newblock \href {https://doi.org/10.1016/0550-3213(81)90423-5}
  {\path{doi:10.1016/0550-3213(81)90423-5}}.

\bibitem{Lovelace:1971fa}
C.~Lovelace, {Pomeron form-factors and dual Regge cuts}, Phys. Lett. B 34
  (1971) 500--506.
\newblock \href {https://doi.org/10.1016/0370-2693(71)90665-4}
  {\path{doi:10.1016/0370-2693(71)90665-4}}.

\bibitem{Brower:1972wj}
R.~C. Brower, {Spectrum generating algebra and no ghost theorem for the dual
  model}, Phys. Rev. D 6 (1972) 1655--1662.
\newblock \href {https://doi.org/10.1103/PhysRevD.6.1655}
  {\path{doi:10.1103/PhysRevD.6.1655}}.

\bibitem{Luscher:1980iy}
M.~Luscher, G.~Munster, P.~Weisz, {How Thick Are Chromoelectric Flux Tubes?},
  Nucl. Phys. B 180 (1981) 1--12.
\newblock \href {https://doi.org/10.1016/0550-3213(81)90151-6}
  {\path{doi:10.1016/0550-3213(81)90151-6}}.

\bibitem{Briceno:2017qmb}
R.~A. Briceno, J.~J. Dudek, R.~G. Edwards, D.~J. Wilson, {Isoscalar $\pi\pi,
  K\overline{K}, \eta\eta$ scattering and the $\sigma, f_0, f_2$ mesons from
  QCD}, Phys. Rev. D 97~(5) (2018) 054513.
\newblock \href {http://arxiv.org/abs/1708.06667} {\path{arXiv:1708.06667}},
  \href {https://doi.org/10.1103/PhysRevD.97.054513}
  {\path{doi:10.1103/PhysRevD.97.054513}}.

\bibitem{McNeile:2002fh}
C.~McNeile, C.~Michael, {Hadronic decay of a vector meson from the lattice},
  Phys. Lett. B 556 (2003) 177--184.
\newblock \href {http://arxiv.org/abs/hep-lat/0212020}
  {\path{arXiv:hep-lat/0212020}}, \href
  {https://doi.org/10.1016/S0370-2693(03)00130-8}
  {\path{doi:10.1016/S0370-2693(03)00130-8}}.

\bibitem{CP-PACS:2007wro}
S.~Aoki, et~al., {Lattice QCD Calculation of the rho Meson Decay Width}, Phys.
  Rev. D 76 (2007) 094506.
\newblock \href {http://arxiv.org/abs/0708.3705} {\path{arXiv:0708.3705}},
  \href {https://doi.org/10.1103/PhysRevD.76.094506}
  {\path{doi:10.1103/PhysRevD.76.094506}}.

\bibitem{Gockeler:2008kc}
M.~Gockeler, R.~Horsley, Y.~Nakamura, D.~Pleiter, P.~E.~L. Rakow,
  G.~Schierholz, J.~Zanotti, {Extracting the rho resonance from lattice QCD
  simulations at small quark masses}, PoS LATTICE2008 (2008) 136.
\newblock \href {http://arxiv.org/abs/0810.5337} {\path{arXiv:0810.5337}},
  \href {https://doi.org/10.22323/1.066.0136} {\path{doi:10.22323/1.066.0136}}.

\bibitem{Budapest-Marseille-Wuppertal:2010gis}
J.~Frison, et~al., {Rho decay width from the lattice}, PoS LATTICE2010 (2010)
  139.
\newblock \href {http://arxiv.org/abs/1011.3413} {\path{arXiv:1011.3413}},
  \href {https://doi.org/10.22323/1.105.0139} {\path{doi:10.22323/1.105.0139}}.

\bibitem{Feng:2010es}
X.~Feng, K.~Jansen, D.~B. Renner, {Resonance Parameters of the rho-Meson from
  Lattice QCD}, Phys. Rev. D 83 (2011) 094505.
\newblock \href {http://arxiv.org/abs/1011.5288} {\path{arXiv:1011.5288}},
  \href {https://doi.org/10.1103/PhysRevD.83.094505}
  {\path{doi:10.1103/PhysRevD.83.094505}}.

\bibitem{PACS-CS:2010dxu}
S.~Aoki, et~al., {Calculation of $\rho$ meson decay width from the PACS-CS
  configurations}, PoS LATTICE2010 (2010) 108.
\newblock \href {http://arxiv.org/abs/1011.1063} {\path{arXiv:1011.1063}},
  \href {https://doi.org/10.22323/1.105.0108} {\path{doi:10.22323/1.105.0108}}.

\bibitem{Lang:2011mn}
C.~Lang, D.~Mohler, S.~Prelovsek, M.~Vidmar, {Coupled channel analysis of the
  rho meson decay in lattice QCD}, Phys.Rev. D84~(5) (2011) 054503.
\newblock \href {http://arxiv.org/abs/1105.5636} {\path{arXiv:1105.5636}},
  \href {https://doi.org/10.1103/PhysRevD.89.059903,
  10.1103/PhysRevD.84.054503} {\path{doi:10.1103/PhysRevD.89.059903,
  10.1103/PhysRevD.84.054503}}.

\bibitem{Prelovsek:2010kg}
S.~Prelovsek, T.~Draper, C.~B. Lang, M.~Limmer, K.-F. Liu, N.~Mathur,
  D.~Mohler, {Lattice study of light scalar tetraquarks with I=0,2,1/2,3/2: Are
  \textbackslash{}sigma and \textbackslash{}kappa tetraquarks?}, Phys. Rev. D
  82 (2010) 094507.
\newblock \href {http://arxiv.org/abs/1005.0948} {\path{arXiv:1005.0948}},
  \href {https://doi.org/10.1103/PhysRevD.82.094507}
  {\path{doi:10.1103/PhysRevD.82.094507}}.

\bibitem{Dudek:2013yja}
J.~J. Dudek, R.~G. Edwards, P.~Guo, C.~E. Thomas, {Toward the excited isoscalar
  meson spectrum from lattice QCD}, Phys. Rev. D 88~(9) (2013) 094505.
\newblock \href {http://arxiv.org/abs/1309.2608} {\path{arXiv:1309.2608}},
  \href {https://doi.org/10.1103/PhysRevD.88.094505}
  {\path{doi:10.1103/PhysRevD.88.094505}}.

\bibitem{Fu:2013ffa}
Z.~Fu, {Lattice QCD study of the s-wave $\pi\pi $ scattering lengths in the I=0
  and 2 channels}, Phys. Rev. D 87~(7) (2013) 074501.
\newblock \href {http://arxiv.org/abs/1303.0517} {\path{arXiv:1303.0517}},
  \href {https://doi.org/10.1103/PhysRevD.87.074501}
  {\path{doi:10.1103/PhysRevD.87.074501}}.

\bibitem{Wakayama:2014gpa}
M.~Wakayama, T.~Kunihiro, S.~Muroya, A.~Nakamura, C.~Nonaka, M.~Sekiguchi,
  H.~Wada, {Lattice QCD study of four-quark components of the isosinglet scalar
  mesons: Significance of disconnected diagrams}, Phys. Rev. D 91~(9) (2015)
  094508.
\newblock \href {http://arxiv.org/abs/1412.3909} {\path{arXiv:1412.3909}},
  \href {https://doi.org/10.1103/PhysRevD.91.094508}
  {\path{doi:10.1103/PhysRevD.91.094508}}.

\bibitem{Howarth:2015caa}
D.~Howarth, J.~Giedt, {The sigma meson from lattice QCD with two-pion
  interpolating operators}, Int. J. Mod. Phys. C 28~(10) (2017) 1750124.
\newblock \href {http://arxiv.org/abs/1508.05658} {\path{arXiv:1508.05658}},
  \href {https://doi.org/10.1142/S0129183117501248}
  {\path{doi:10.1142/S0129183117501248}}.

\bibitem{Briceno:2016mjc}
R.~A. Briceno, J.~J. Dudek, R.~G. Edwards, D.~J. Wilson, {Isoscalar $\pi\pi$
  scattering and the $\sigma$ meson resonance from QCD}, Phys. Rev. Lett.
  118~(2) (2017) 022002.
\newblock \href {http://arxiv.org/abs/1607.05900} {\path{arXiv:1607.05900}},
  \href {https://doi.org/10.1103/PhysRevLett.118.022002}
  {\path{doi:10.1103/PhysRevLett.118.022002}}.

\bibitem{Wilson:2014cna}
D.~J. Wilson, J.~J. Dudek, R.~G. Edwards, C.~E. Thomas, {Resonances in coupled
  $\pi K, \eta K$ scattering from lattice QCD}, Phys. Rev. D 91~(5) (2015)
  054008.
\newblock \href {http://arxiv.org/abs/1411.2004} {\path{arXiv:1411.2004}},
  \href {https://doi.org/10.1103/PhysRevD.91.054008}
  {\path{doi:10.1103/PhysRevD.91.054008}}.

\bibitem{Alford:2000mm}
M.~G. Alford, R.~L. Jaffe, {Insight into the scalar mesons from a lattice
  calculation}, Nucl. Phys. B 578 (2000) 367--382.
\newblock \href {http://arxiv.org/abs/hep-lat/0001023}
  {\path{arXiv:hep-lat/0001023}}, \href
  {https://doi.org/10.1016/S0550-3213(00)00155-3}
  {\path{doi:10.1016/S0550-3213(00)00155-3}}.

\bibitem{Berlin:2014qyu}
J.~Berlin, A.~Abdel-Rehim, C.~Alexandrou, M.~Dalla~Brida, M.~Gravina,
  M.~Wagner, {Investigation of the tetraquark candidate $a_0(980)$: technical
  aspects and preliminary results}, PoS LATTICE2014 (2014) 104.
\newblock \href {http://arxiv.org/abs/1410.8757} {\path{arXiv:1410.8757}},
  \href {https://doi.org/10.22323/1.214.0104} {\path{doi:10.22323/1.214.0104}}.

\bibitem{Berlin:2016zci}
J.~Berlin, A.~Abdel-Rehim, C.~Alexandrou, M.~Dalla~Brida, J.~Finkenrath,
  M.~Gravina, T.~Leontiou, M.~Wagner, {Importance of closed quark loops for
  lattice QCD studies of tetraquarks}, PoS LATTICE2016 (2016) 128.
\newblock \href {http://arxiv.org/abs/1611.07762} {\path{arXiv:1611.07762}},
  \href {https://doi.org/10.22323/1.256.0128} {\path{doi:10.22323/1.256.0128}}.

\bibitem{Dudek:2016cru}
J.~J. Dudek, R.~G. Edwards, D.~J. Wilson, {An $a_0$ resonance in strongly
  coupled $\pi \eta$, $K\overline{K}$ scattering from lattice QCD}, Phys. Rev.
  D 93~(9) (2016) 094506.
\newblock \href {http://arxiv.org/abs/1602.05122} {\path{arXiv:1602.05122}},
  \href {https://doi.org/10.1103/PhysRevD.93.094506}
  {\path{doi:10.1103/PhysRevD.93.094506}}.

\bibitem{Alexandrou:2017itd}
C.~Alexandrou, J.~Berlin, M.~Dalla~Brida, J.~Finkenrath, T.~Leontiou,
  M.~Wagner, {Lattice QCD investigation of the structure of the $a_0(980)$
  meson}, Phys. Rev. D 97~(3) (2018) 034506.
\newblock \href {http://arxiv.org/abs/1711.09815} {\path{arXiv:1711.09815}},
  \href {https://doi.org/10.1103/PhysRevD.97.034506}
  {\path{doi:10.1103/PhysRevD.97.034506}}.

\bibitem{Briceno:2017max}
R.~A. Briceno, J.~J. Dudek, R.~D. Young, {Scattering processes and resonances
  from lattice QCD}, Rev. Mod. Phys. 90~(2) (2018) 025001.
\newblock \href {http://arxiv.org/abs/1706.06223} {\path{arXiv:1706.06223}},
  \href {https://doi.org/10.1103/RevModPhys.90.025001}
  {\path{doi:10.1103/RevModPhys.90.025001}}.

\bibitem{Apollinari:2017lan}
{High-Luminosity Large Hadron Collider (HL-LHC)}: {Technical Design Report V.
  0.1} 4/2017 (2017).
\newblock \href {https://doi.org/10.23731/CYRM-2017-004}
  {\path{doi:10.23731/CYRM-2017-004}}.

\bibitem{Hudspith:2022lat}
R.~Hudspith, D.~Mohler, {Non-perturbative heavy quark action tuning using
  machine learning}, PoS LATTICE2022 (2023).

\bibitem{Hollwieser:2022pov}
R.~H\"ollwieser, F.~Knechtli, M.~Peardon, {The static energy of a
  quark-antiquark pair from Laplacian eigenmodes}, in: {39th International
  Symposium on Lattice Field Theory}, 2022.
\newblock \href {http://arxiv.org/abs/2209.01239} {\path{arXiv:2209.01239}}.

\bibitem{Heupel:2012ua}
W.~Heupel, G.~Eichmann, C.~S. Fischer, {Tetraquark Bound States in a
  Bethe-Salpeter Approach}, Phys. Lett. B 718 (2012) 545--549.
\newblock \href {http://arxiv.org/abs/1206.5129} {\path{arXiv:1206.5129}},
  \href {https://doi.org/10.1016/j.physletb.2012.11.009}
  {\path{doi:10.1016/j.physletb.2012.11.009}}.

\bibitem{Fritzsch:2021klm}
P.~Fritzsch, J.~Bulava, M.~C\`e, A.~Francis, M.~L\"uscher, A.~Rago,
  {Master-field simulations of QCD}, PoS LATTICE2021 (2022) 465.
\newblock \href {http://arxiv.org/abs/2111.11544} {\path{arXiv:2111.11544}},
  \href {https://doi.org/10.22323/1.396.0465} {\path{doi:10.22323/1.396.0465}}.

\bibitem{Francis:2019muy}
A.~Francis, P.~Fritzsch, M.~L\"uscher, A.~Rago, {Master-field simulations of
  O($a$)-improved lattice QCD: Algorithms, stability and exactness}, Comput.
  Phys. Commun. 255 (2020) 107355.
\newblock \href {http://arxiv.org/abs/1911.04533} {\path{arXiv:1911.04533}},
  \href {https://doi.org/10.1016/j.cpc.2020.107355}
  {\path{doi:10.1016/j.cpc.2020.107355}}.

\bibitem{Luscher:2017cjh}
M.~L\"uscher, {Stochastic locality and master-field simulations of very large
  lattices}, EPJ Web Conf. 175 (2018) 01002.
\newblock \href {http://arxiv.org/abs/1707.09758} {\path{arXiv:1707.09758}},
  \href {https://doi.org/10.1051/epjconf/201817501002}
  {\path{doi:10.1051/epjconf/201817501002}}.

\bibitem{Ce::2022lat}
M.~Ce, {Hadronic observables from master-field simulations}, PoS LATTICE2022
  (2023).

\bibitem{Bruno:2016plf}
M.~Bruno, T.~Korzec, S.~Schaefer, {Setting the scale for the CLS $2 + 1$ flavor
  ensembles}, Phys. Rev. D 95~(7) (2017) 074504.
\newblock \href {http://arxiv.org/abs/1608.08900} {\path{arXiv:1608.08900}},
  \href {https://doi.org/10.1103/PhysRevD.95.074504}
  {\path{doi:10.1103/PhysRevD.95.074504}}.

\bibitem{Hansen:2022lat}
M.~Hansen, {Implementing the finite-volume scattering and decay formalism
  across all three-pion isospin channels}, PoS LATTICE2022 (2023).

\bibitem{richard_uras:2021}
A.~Ealet, J.-M. Richard, Y.~Gao, A.~Uras, A.~Valcarce, J.~Vijande,
  \href{https://indico.in2p3.fr/event/24937/}{Double charm tetraquark and other
  exotics} (Nov 2021).
\newline\urlprefix\url{https://indico.in2p3.fr/event/24937/}

\end{thebibliography}
\addcontentsline{toc}{section}{References}

\end{document}